Université du Québec
Institut National de la Recherche Scientifique
Centre Énergie, Matériaux et Télécommunications

# Magneto-photonic phenomena at terahertz frequencies

by

Mostafa Shalaby

A thesis submitted in partial fulfillment of the requirements for
the degree of Doctor of Philosophy

Present affiliation: SwissFEL, Paul Scherrer Institut, Villigen 5232, Switzerland
mostafa.shalaby@psi.ch; most.shalaby@gmail.com

The complete thesis was submitted for revision in December 2012. All the results presented here were completed, analyzed, and written before then.

A list of publications is included to credit the work of other contributors to each part of the thesis.

Sections 3.1 and 4.2 present a hereby copyrighted, but yet unpublished results. These contributions were presented in international conferences, given at the beginning of the thesis.

## ACKNOWLEDGEMENTS

I am delighted to *thank* all those who have helped me grow up and develop intellectually so far, whether on the professional or the personal side of my life. Some of them may not even know much about science-like my parents, random people I met on the course of my life, and even the amazing Varennes' inhabitants who used to pick up a poor student hitchhiking in the middle of nowhere after finishing work so late.

… and above all, big thanks to my funky awesome friends, who I consider the real treasure in my life.

I gratefully *acknowledge* the support of those who directly contributed to my PhD thesis research; Roberto Morandotti-research director who did his best to put me on the correct career path; Marco Peccianti-who was always there to ask; Yavuz Ozturk-with whom I spent countless nights working in the laboratory and shared so many disgusting T...'s and M...'s food at 3 am; Thomas Feurer (Bern Univ.) –with whom I had many brainstorming discussions; Quebec funding program (FQRNT) -which funded my PhD studies. Specific parts of my projects were performed with fruitful collaborations with: Hannes Merbold and Florian Enderli (Bern Univ.); Anja Weber, Hans Sigg, and Laura Heyderman (PSI, Zurich); Rob Helsten, Francois Vidal, Tsuneyuki Ozaki,Ibraheem Al-Naib, Luca Razzari, Matteo Clerici, Gargi Sharma, Bruno Schmidt, Mathieu Giguere, and François Legare(INRS); Antonio Lotti and Arnaud Couairon (CNRS); Daniele Faccio (HW Univ., Edinburgh); Anna Mazhorova, Maksim Skorobogatiy (Ecole Poly., Montreal); Mathias Klaui (EPFL, Lausanne).

Finally, I am deeply thankful to two persons; Larry Carr (BNL, NY) -who proposed a small experimentthat turned into a wonderful research to base my PhD experiments on; Bruce Patterson (PSI, Zurich) -who, in addition of being a trustful friend, simply taught me that science is beautiful!

# ABSTRACT

Magneto-terahertz phenomena are the main focus of the thesis. This work started as supporting research for the science of an X-ray laser (SwissFEL). X-ray lasers have recently drawn great attention as an unprecedented tool for scientific research on the ultrafast scale. A potential terahertz-pump / X-ray-probe experiment is foreseen to reveal the fundamentals of magnetic systems on the ultrafast time scale and benefits the ultrafast magnetic storage industry.

The main objective of this work was to find the conditions and prove that a suitable terahertz pulse can induce ultrafast magnetization dynamics on the picoseconds scale. To answer this fundamental question, we performed original numerical simulations using a coupled Landau-Lifshitz-Gilbert Maxwell model. Calculations showed us that terahertz pulses can trigger ultrafast dynamics, but highlighted the requirements of properly shaped pulses and beyond-current-technology peak field intensities.

Those requirements werethe motivations for the experiments performed in the second part of the thesis. To shape the terahertz pulses, we used time-resolved optical-pump / terahertz-probe of free carriers in semiconductors. We managed to temporally shape the terahertz pulses and even extend the technique to spectral shaping as well. Regarding the field intensities, we followed two approaches. The first deals with field enhancement in nanoslits arrays. We designed a sub-wavelength structure characterized by simultaneous high field enhancementand high transmission at terahertz frequencies to suit nonlinear sources. The second approach depended on up-scaling the generation from laser-induced plasma by increasing the pump wavelengths.

Numerical calculations have also brought to our attention the importance of linear magneto-terahertz effects. In particular, the simulations showed that the ultrafast dynamics could lead to significant rotation of the polarization plane of the triggering terahertz pulse. Motivated by this finding, we focused in the last part of the thesis on the linear effects. We performed three original studies coming out with first demonstrations of broadband non-reciprocal terahertz phase retarders, terahertzmagnetic modulators, and the non-reciprocal terahertz isolators. In the first two experiments, we extended the unique properties of the magnetic liquids (Ferrofluids) to the terahertz regime. In the latter experiment, we used a permanent magnet (Ferrite) to experimentally show complete isolation (unidirectional transmission) of the terahertz waves.

# TABLE OF CONTENTS

# TABLE OF CONTENTS

# LIST OF FIGURES

# LIST OF ACRYNOMS

| | |
|---|---|
| $h$ | Planck's constant |
| $\boldsymbol{J}$ | Free current density |
| $\varrho$ | Charge density |
| $\boldsymbol{D}$ | Electric displacement |
| $\boldsymbol{B}$ | Magnetic induction |
| $\boldsymbol{E}$ | Electric field |
| $\boldsymbol{H}$ | Magnetic field |
| $\boldsymbol{P}$ | Polarization |
| $\boldsymbol{M}$ | Magnetization |
| $M_s$ | Saturation magnetization |
| $\mu$ | Permeability |
| $\varepsilon$ | Permittivity |
| $\omega_p$ | Plasma frequency |
| $v$ | Electronvelocity |
| $e$ | Electron charge |
| $m_o$ | Electron mass |
| $Y_o$ | Free space admittance |
| $\alpha$ | Attenuation coefficient |

# Publications during the PhD

## i. Publications related to the material presented in the thesis

*[1] M. Shalaby, M. Peccianti, Y. Ozturk, I. Al-Nailb & R. Morandotti, "Terahertz Magnetic Modulation" (to be submitted)*

*[2] M. Shalaby, M. Peccianti & R. Morandotti, "Temporal and Spectral Shaping of Broadband THz Pulses in a Photoexcited Semiconductor" (to be submitted)*

[3] M. Shalaby, F. Vidal, M. Peccianti, R. Morandotti, F. Enderli, T. Feurer & B. D. Patterson, "Terahertz magnetization dynamics," ***Phys. Rev. B Rapid Commun.***, in press (2013)

[4] M. Shalaby, M. Peccianti, Y. Ozturk & R. Morandotti, ***Nat. Commun.*** 4, 1558 (2013)

[5] M. Clerici, Marco Peccianti, Bruno E. Schmidt, L. Caspani, M. Shalaby, M. Giguere, A. Lotti, A. Couairon, F. Legare, T. Ozaki, D. Faccio & R. Morandotti, ***Phys. Rev. Lett.*** 110, 253901 (2013)

[6] M. Shalaby, M. Peccianti, Y. Ozturk, M. Clerici, I. Al-Naib, L. Razzari, T. Ozaki, A. Mazhorova, M. Skorobogatiy & R. Morandotti, ***Appl. Phys. Lett.*** 100, 241107 (2012)

[7] M. Shalaby, H. Merbold, M. Peccianti, L. Razzari, G. Sharma, T. Ozaki, R. Morandotti, T. Feurer, A. Weber, L. Heyderman, B. Patterson & H. Sigg, ***Appl. Phys. Lett.*** 99, 041110 (2011)

## ii. Publications related to the material presented in the thesis

*[8] M. Shalaby, J. fabianska, M. Peccianti, F. Vidal, H. Sigg, B. Patterson, T. Feurer & R. Morandotti, "Crawling terahertz waves in photoexcited nanoslits" (to be submitted)*

*[9] S. P. Ho, M. Shalaby, M. Peccianti, M. Clerici, A. Pasquazi, Y. Ozturk, J. Ali & R. Morandotti, "A Novel Optical Approach for THz Radiation Features Characterization" (to be submitted)*

[10] M. Shalaby, M. Peccianti, Y. Ozturk & R. Morandotti, *submitted to OPN, special issue, year in Optics* (2013)

[11] M. Clerici, D. Faccio, L. Caspani, M. Peccianti, O. Yaakobi, B. E. Schmidt, M. Shalaby, F. Vidal, F. Legare, T. Ozaki, and R. Morandotti, " ***New Journal of Physics***, in press (2013) invited

[12] L. Razzari, A. Toma, M. Clerici, M. Shalaby, G. Das, C. Liberale, M. Chirumamilla, R. P. Zaccaria, F. Angelis, M. Peccianti, R. Morandotti & E. Fabrizio, ***Plasmonics*** 8, 133 (2013)

[13] I. Al-Naib, R. Singh, M. Shalaby, T. Ozaki & R. Morandotti, ***IEEE J. Sel. Top. Quantum Electron***. 19, 8400807 (2013)

[14] L. Razzari, A. Toma, M. Shalaby, M. Clerici, R. P. Zaccaria, C. Liberale, S. Marras, I. Al-Naib, G. Das, F. De Angelis, M. Peccianti, A. Falqui, T. Ozaki, R. Morandotti & E. Di Fabrizio, ***Opt. Express*** 19, 26088 (2011)

[15] B. D. Patterson, R. Abela, U. Flechsig, B. Pedrini, M. Shalaby, M. van Daalen, Th. Feurer, and M. Klaui, "*Preparations for SwissFEL science*," ***Paul Scherrer Institut Scientific Report***, Switzerland (2009)

## Contributions to international conferences during the PhD

[1] M. Shalaby, M. Peccianti, Y. Ozturk, R. Morandotti, "Non-Reciprocal Broadband Terahertz Isolator", CLEO, San Jose, USA (2013)

[2] M. Clerici, M. Peccianti, B.E. Schmidt, L. Caspani, M. Shalaby, M. Giguère, A. Lotti, A. Couairon, F. Légaré, T. Ozaki, D. Faccio, R. Morandotti, "A Scaling Mechanism for Increasing the Terahertz Emission from Ionization of Air", CLEO, San Jose, USA (2013)

[3] S.P Ho, M. Shalaby, M. Peccianti, M. Clerici, A. Pasquazi, Y. Ozturk, J. Ali, R. Morandotti, "Terahertz Characterization via an All-Optical, Ultra-Thin-Knife-Edge Technique", CLEO, San Jose, United State (2013)

[4] M. Clerici, M. Peccianti, B.E. Schmidt, L. Caspani, M. Shalaby, M. Giguère, A. Lotti, A. Couairon, F. Légaré, T. Ozaki, D. Faccio, R. Morandotti",A Scaling of the Terahertz Emission From Laser Induced Plasma with Increasing Pump Wavelength", Photonics North, Ottawa, Canada (2013)

[5] S.P Ho, M. Shalaby, M. Peccianti, M. Clerici, A. Pasquazi, Y. Ozturk, J. Ali, R. Morandotti, "An All-Optical, Zero-Thickness Knife-Edge For Terahertz Characterization", Photonics North, Ottawa, Canada (2013)

[6] M. Clerici, D. Faccio, L. Caspani, M. Shalaby, M. Peccianti, B.E. Schmidt, O. Yaakobi, F. Vidal, F. Légaré, T. Ozaki, R. Morandotti, "THz-Optical Four-Wave Mixing in Air and Coherent N2+ Emission", OTST, Kyoto, Japan (2013)

[7] L. Razzari, A. Toma, M. Clerici, M. Shalaby, G. Das, C. Liberale, M. Chirumamilla, R. Proietti Zaccaria, F. De Angelis, M. Peccianti, R. Morandotti, E. Di Fabrizio, "Resonant Nanoantennas for Terahertz Light", OTST, Kyoto, Japan (2013)

[8] M. Clerici, M. Peccianti, B.E. Schmidt, L. Caspani, M. Shalaby, M. Giguère, A. Lotti, A. Couairon, F. Légaré, T. Ozaki, D. Faccio, R. Morandotti, "Scaling of the Terahertz Field From Two-Color Driven Gas Ionization With Increasing Pump Wavelength", OTST, Kyoto, Japan (2013)

[9] S.P Ho, M. Shalaby, M. Peccianti, M. Clerici, A. Pasquazi, Y. Ozturk, J. Ali, R. Morandotti, "A Novel Optical Approach for THz Radiation Features Characterization", OTST, Kyoto, Japan (2013)

[10] M. Shalaby, M. Peccianti, Y. Ozturk, R. Morandotti, "The Non-Reciprocal THz Isolator", OTST, Kyoto, Japan (2013)

[11] M. Shalaby, M. Peccianti, R. Morandotti, "Temporal and Spectral Shaping of Broadband THz Pulses in a Photoexcited Semiconductor", OTST, Kyoto, Japan (2013)
[12] M. Shalaby, M. Peccianti, Y. Ozturk, I. Al-Naib, R. Morandotti, "Terahertz Magnetic Modulation", OTST, Kyoto, Japan (2013)
[13] M. Shalaby, J. Fabiańska, M. Peccianti, Y. Ozturk, F. Vidal, A. Weber, L.J. Heyderman, H. Sigg, B. Patterson, T. Feurer, R. Morandotti, "THz Field-Enhancement Modulation in Optically Shunted Gold Nanoslit", OTST, Kyoto, Japan (2013)
[14] M. Shalaby, M. Peccianti, Y. Ozturk, M. Clerici, I. Al-Naib, L. Razzari, A. Mazhorova, Maksim Skorobogatly, R. Morandotti, "Broadband THz Faraday Rotation in a Magnetic Liquid", OTST, Kyoto, Japan (2013)
[15] S.P Ho, M. Shalaby, M. Peccianti, M. Clerici, A. Pasquazi, Y. Ozturk, J. Ali, R. Morandotti, "An All-Optical, Ultra-Thin-Knife-Edge Technique for Terahertz Characterization", CAP Congress, Montreal, Canada (2013)
[16] M. Clerici, D. Faccio, L. Caspani, M. Shalaby, M. Giguere, B.E. Schmidt, O. Yaakobi, M. Peccianti, F. Vidal, F. Légaré, T. Ozaki, R. Morandotti, "Envelope and field effects in the nonlinear interaction of broadband terahertz fields and optical pulses in air", SPIE Photonics West-Ultrafast Phenomena and Nanophotonics XVII, San Francisco, US (2013)
[17] M. Clerici, D. Faccio, M. Shalaby, M. Giguère, B.E. Schmidt, M. Peccianti, Yaakobi, A. Lotti, F. Vidal, F. Légaré, T. Ozaki, R. Morandotti, "Envelope and Field Effects in the Nonlinear Interaction of Broadband Terahertz Fields and Optical Pulses in Air", OPTO SPIE Photonic West, San Francisco, USA (2013)
[18] M. Peccianti, M. Clerici, M. Shalaby, L. Caspani, A. Lotti, A. Couairon, D. Cooke, T. Ozaki, D. Faccio, R. Morandotti, "Terahertz Field Detection Boost by Nonlinear Collapse of Normally Dispersed Optical Pulses", Nonlinear Photonics, Colorado Springs, USA (2012)
[19] Y. Ozturk, M. Shalaby, M. Clerici, J.-Y. Hwang, A. Pignolet, R. Morandotti, "Ultrafast laser-induced spin dynamics of cerium and bismuth co-modified iron garnet thin film", Photonics North, Montreal, Canada (2012)
[20] I. Al-Naib, R. Singh, M. Shalaby, T. Ozaki, and R. Morandotti, "Effect of the Spatial Arrangement of Split-Ring Resonators on their Response to Terahertz Fields", Photonics North, Montreal, Canada (2012)

[21] M. Shalaby, M. Peccianti, Y. Ozturk, M. Clerici, I. Al-Naib, L. Razzari, A. Mazhorova, M. Skorobogatiy, T. Ozaki1, and R. Morandotti, "Magnetic field induced switching of terahertz pulses in a liquid", Photonics North, Montreal, Canada (2012)

[22] M. Clerici, M. Peccianti, M. Shalaby, L. Caspani, A. Lotti, A. Couairon, D. Cooke, T. Ozaki, D. Faccio, and R. Morandotti, "Chirp enhanced Broadband THz detection in gas", Photonics North, Montreal, Canada (2012)

[23] M. Clerici, D. Faccio, M. Shalaby, M. Giguère, B.E. Schmidt, M. Peccianti, L. Caspani, F. Légaré, T. Ozaki, and R. Morandotti, "Electric-field characterization of long wavelength, few-cycles pulses by electric field-induced second-harmonic FROG", Photonics North, Montreal, Canada (2012)

[24] M. Shalaby, M. Peccianti, Y. Ozturk, M. Clerici, I. Al-Naib, L. Razzari, A. Mazhorova, M. Skorobogatiy, T. Ozaki, and R. Morandotti, "Broadband THz Faraday rotation in a magnetic liquid", Photonics North, Montreal, Canada (2012). (poster)

[25] Y. Ozturk, M. Hidalgo, M. Shalaby, J.-Y. Hwang, A. Pignolet, R. Morandotti, "Magneto-optical analysis of cerium and bismuth substituted iron garnet thin film", Photonics North, Montreal, Canada (2012)

[26] I. Al-Naib, R. Singh, M. Shalaby, T. Ozaki, R. Morandotti, "Engineering the Response of Terahertz Metasurfaces by Spatial Arrangement of Split-Ring Resonators", CLEO, San Jose, USA (2012)

[27] M. Shalaby, M. Peccianti, Y. Ozturk, L. Razzari, M. Clerici, A. Mazhorova, M. Skorobogatiy, T. Ozaki, R. Morandotti, "Polarization-sensitive Magnetic Field Induced Modulation of Broadband THz Pulses in Liquid", CLEO, San Jose, USA (2012)

[28] M. Clerici, M. Peccianti, M. Shalaby, L. Caspani, A. Lotti, A. Couairon, D.G. Cooke, T. Ozaki, D. Faccio, R. Morandotti, "Enhanced Detection of Broadband Terahertz Fields via the Filamentation of Chirped Optical Pulses", CLEO, San Jose, USA (2012)

[29] M. Clerici, D. Faccio, M. Shalaby, M. Giguere, B.E. Schmidt, M. Peccianti, F. Légaré, T. Ozaki, R. Morandotti, "Electric-Field Induced Second-Harmonic FROG Characterization of Long-Wavelength, Few-Cycle Pulses", CLEO, San Jose, USA (2012)

[30] I. Al-Naib, M. Shalaby, T. Ozaki, and R. Morandotti, "Terahertz Diagonal Macro cell Planar Metamaterials", 36th International Terahertz Conference on Infrared, Millimeter, and Waves, (IRMMW-THz), Houston, USA (2011)

[31] M. Clerici, M. Peccianti, M. Shalaby, T. Ozaki, D. Cooke, D. Faccio, and R. Morandotti, "Enhanced Detection of Broadband Terahertz Field by Filamentation of Chirped Optical Pulses", IEEE Photonic Society Annual Meeting, Arlington, USA (2011)

[32] M. Shalaby, M. Peccianti, G. Sharma, L. Razzari, T. Ozaki, and R. Morandotti, "Terahertz Pulse Slicing into Single- and Half-Cycle Pulses", Photonics North, Ottawa, Canada (2011)

[33] H. Merbold, A. Bitzer, F. Enderli, A. Weber, L. Heyderman, B. Patterson, H. Sigg, M. Shalaby, M. Peccianti, R. Morandotti, T. Ozaki, and T. Feurer, "Nonlinear THz Response Based on Strong Field Enhancement in Nanostructured Metamaterials", the 3rd International Topical Meeting on Nanophotonics and Metamaterials, Seefeld, Austria (2011)

[34] M. Shalaby, H. Merbold, M. Peccianti, L. Razzari, G. Sharma, T. Ozaki, R. Morandotti, T. Feurer, A. Weber, L. Heyderman, H. Sigg, and B. Patterson, "Design of High Transmission High Enhancement Uniform Nano-slit Array: 26 MV/cm of Broadband THz Radiation", Colloque de Plasma-Quebec, Montreal, Canada (2011)

[35] M. Shalaby, M. Peccianti, F. Vidal, T. Ozaki, and R. Morandotti, "Intensity-Dependent THz Polarization Rotation in Ferromagnets", Photonics North, Ottawa, Canada (2011)

[36] M. Shalaby, M. Peccianti, L. Razzari, G. Sharma, T. Ozaki, R. Morandotti, H. Merbold, T. Feurer, A. Weber, L. Heyderman, H. Sigg, and B. Patterson, "Broadband Enhanced 26 MV/cm THz Radiation in Uniform Nano-slit Arrays", Advanced Photonics: OSA Optics & Photonics Congress, Toronto, Canada (2011)

[37] M. Shalaby, M. Peccianti, L. Razzari, G. Sharma, T. Ozaki, R. Morandotti, "Ultrafast THz Pulse Shaping: Generation of Half-cycle Pulse from Multi-cycle THz Pulse", Advanced Photonics: OSA Optics & Photonics Congress, Toronto, Canada (2011)

[38] M. Shalaby, M. Peccianti, L. Razzari, G. Sharma, T. Ozaki, R. Morandotti,H. Merbold, T. Feurer, A. Weber, L. Heyderman, H. Sigg, and B. Patterson, "Very high THz fields in uniform nano-slit arrays: broadband enhancement of intense THz radiation", OTST, Santa Barbara, USA (2011)

[39] G. Sharma, I. Al-Naib, M. Peccianti, M. Shalaby, M. Reid, R. Morandotti, and T. Ozaki, "THz induced nonlinearity in photoexcited ZnTe", OTST, Santa Barbara, USA (2011)

[40] M. Shalaby, M. Kläui, T. Feurer, F. Enderli, L. Heyderman, B. Patterson, F. Vidal, T. Ozaki and R. Morandotti, "Magnetic Switching in Permalloy Thin Films Using THz Pulses", Photonics North 2010, Niagara Falls, Ontario, Canada (2010)

## INTRODUCTION

Will a galloping horse during the gait ever lift all four feet completely off the ground? A question that the bare human eye never managed to answer. At the bottom, it comes to frames (pictures) spread over a time scale, the faster you can "draw" the pictures, the more you stack in a time period. Interesting enough, that small time period can fit a "literally" infinite number of frames; at least this is what science has been teaching us so far.

It all comes to tools. We do nothave "super" eyes to grasp fast motions, but at least we may have a look at them later after recording, if we *could*. Edward Muybridge indeed succeeded in doing exactly this! More than a century ago, he developed a camera shutter that opened and closed *faster* than the galloping horse. People watching the still pictures afterwards got convinced that the four hooves really left the ground for an instant of time (Fig. I.1).

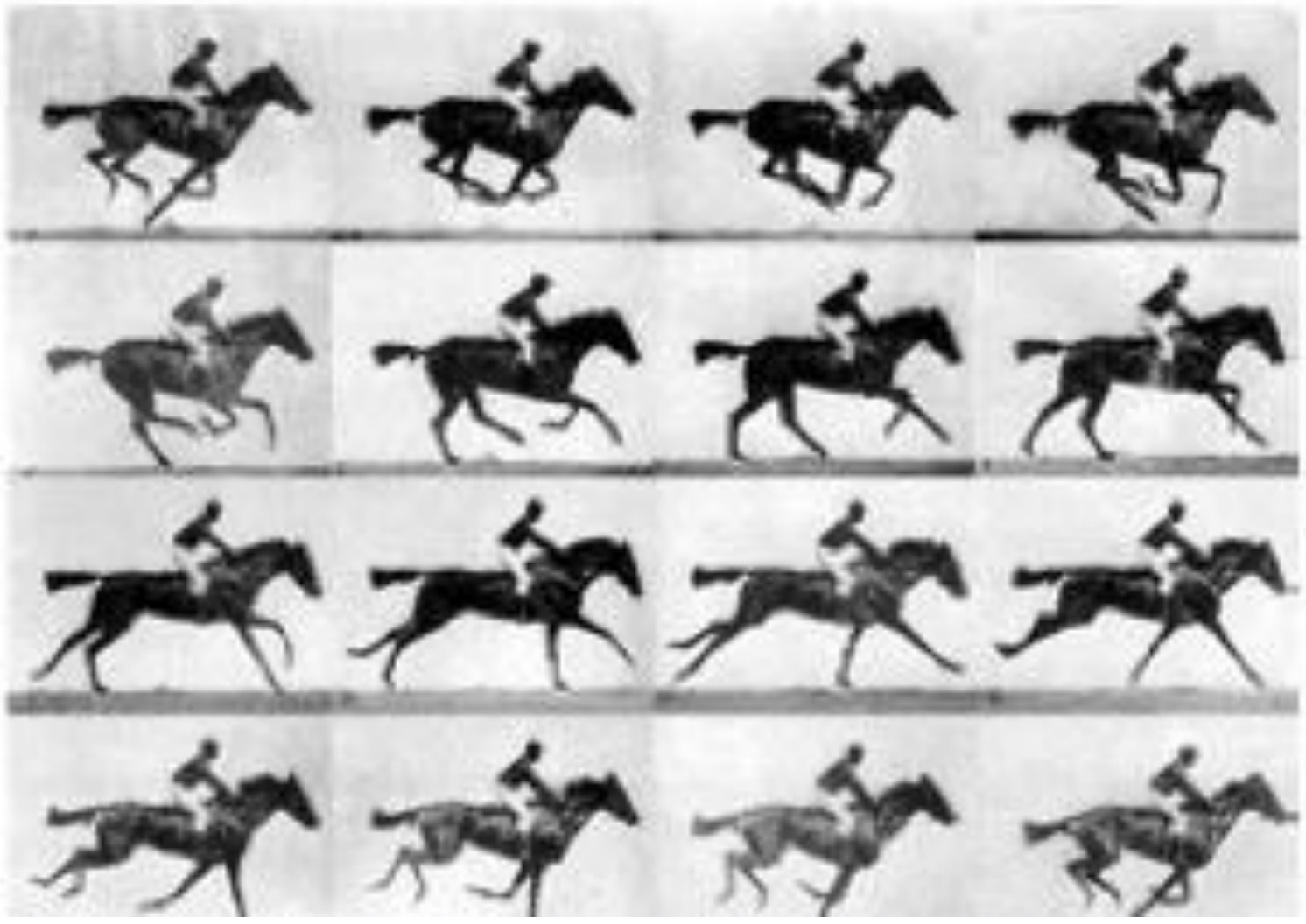

**Figure I.1 Muybridge motion pictures.**

Muybridge's shutter opened and closed every 2 milliseconds and he managed to capture the horse flying through air. Today, we want to look at features smaller than Muybridge's horse, in both time and space. Capture of single atoms frozen in their positions has always been a long-standing dream that is just coming true. On the spatial scale, synchrotron have evolved over the past centuries giving science an indispensible tool to look at spatial features enough to see individual atoms. Unfortunately, it fails when it comes to the temporal resolution.Or in other words, our molecular movie will be a blurred one. Lasers boosted science faster than any other tool with unprecedented temporal resolution; femtosecond (and recently attosecond) is a laboratory reality. However, the spatial resolution is too coarse for our atomic scale. Trials to shrink the spatially-resolved synchrotron pulses down in time to femtosecond scale (*i.e.*femtoslicing) were not better than those targeting shrinking temporally-resolved laser pulses down to the atomic scale (*i.e.* high harmonics generation); in both cases, we have the technology but we don't have enough photons to use them in a real experimental study. It is only recently (2009) thata temporally- and spatially-resolved picture became reality. Fourth generation light sources (X-ray lasers) have enriched science with marvelous results in the past few years and we became very close to watching the atoms *dancing* on their space and time scales.

The application list is long; magnetization dynamics, solution chemistry and surface catalysis, coherent diffraction, ultrafast biochemistry, and correlated electron materials are just a few examples [1].

Magnetism embraces the human civilization; compass is one of the oldest inventions and magnetic memory sustains today's information revolution. However, "one lesson from a study of the history of magnetism is that fundamental understanding of the science may not be a prerequisite for technological progress. Yet fundamental understanding helps [2]". Magnetism is a time-space phenomenon. It is enough to look at its time and length scales (Fig. I.2) in order to get fascinated by various distinct behaviors magnetism shows along them [1]. Each stop hides beautiful science and holds potentials for practical applications. Excitation of a magnetic interaction requires stimulation with energy E. After that stimulus is gone, the magnetic system recovers its equilibrium state in a time $\tau$, given by $\tau \approx h/E$, where $h$ is Planck's constant. Different phenomena on the time scale can then be selectively excited upon the expenditure of their associated energies [1]. Therefore, this requires the stimulus (magnetic field) to have the correct spectral contents ($\nu = 1/\tau$) and amplitude. As $\nu$ increases, the required amplitude increases as well. This magnetic field can simply originate from a current pulse in a wire. Low frequency magnetic field sources (up to gigahertz) are a well-developed technology. As we approach the terahertz (THz) range, most sources start to lose their efficiencies. At the same time, the corresponding magnetic phenomena demand higher and higher fields as the frequency keeps increasing.

In addition to a current in a wire, the switching magnetic field can come from electromagnetic waves. If we consider the first possibility, at THz frequencies, the current should be huge. In a typical experiment, 1 nC were put in a 2-10 ps switching pulse [3]. This corresponds to $10^{10}$ electrons delivered by an electron accelerator. Indeed, the switching was ultrafast on the picosecond time scale setting a record on the fastest magnetic switching realized so far. However, the beam destroyed the central part of the sample and such a unique source is simply not available for the wide scientific community, and there are no means for downscaling.

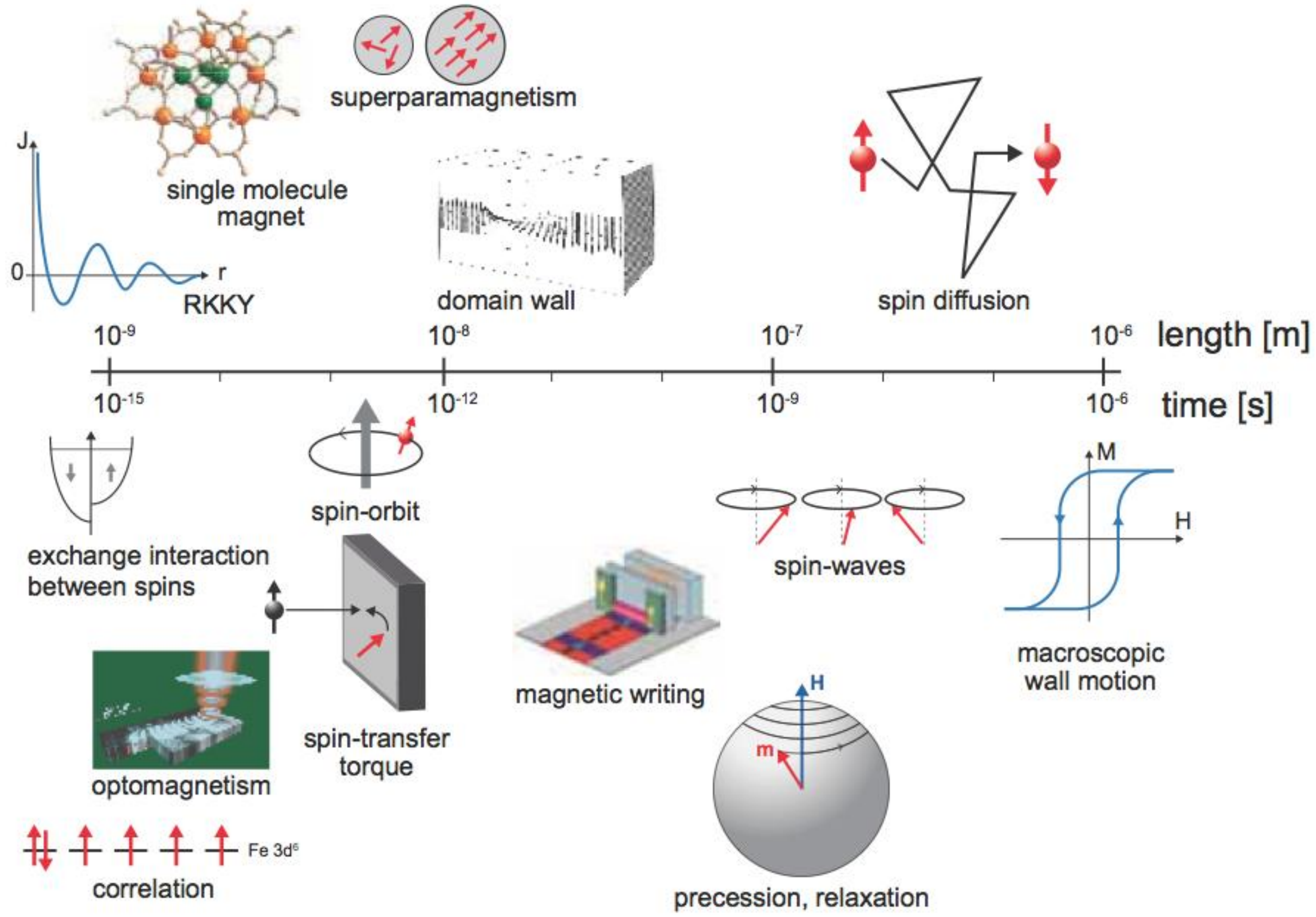

**Figure I.2 Time and length scales in magnetism [1].**

The other option to have such ultrafast magnetic field transients is electromagnetic waves on the picosecond time scale, *i.e.* terahertz waves. Terahertz emerged as a fast growing field in the past decade triggered by the development of intense sources [4-7] and sensitive detectors [8].

Terahertz band or terahertz "gap" is an interesting band between two well-studied bands: microwave and optical. The word terahertz has always been coupled with challenges and promises [9-13]. Terahertz generation is not hard, in a way.Our bodies are terahertz sources, or strictly speaking, anybody with temperature in excess of 10 Kelvin emits terahertz as part of its blackbody emission. However, this terahertz is too weak to be of a real use. Looking for a good source is not an easy job though. Extending the operating bands of microwave and infrared sources was not much of success, as those sources tend to be less efficient as we approach the terahertz band. Detectors did not have better luck. So in brief, terahertz did not have the choice but to evolve as a distinct science with its independent technology. However, the development of femtosecond sources had the most contribution to terahertz science. Most of the well established techniques for terahertz generation and detections are based on femtosecond lasers.

Terahertz finds a wide range of applications extending from triggering nonlinear phenomena [14-18] to linear applications such as imaging [9,19], communications [10], and spectroscopy of chemicals and explosives [20,21].

Magneto-terahertz is a recent term being introduced to science, with almost nothing known about magnetic materials behaviors at Terahertz frequencies. Even the existence of such a thing was arguable until not long time ago.

This thesis work started in the search of an answer to this question: Can an energetic terahertz pulse trigger magnetization dynamics, to be later probed by a delayed pulse from an X-ray laser? [1].

To answer this question, we performedoriginal simulations on terahertz magnetization dynamics (Chapter 2). In photonics, a non-magnetic response of material is almost taken for granted with the material permeability $\mu$ = 1. At low frequencies (*i.e.* up to the gigahertz ones), many materials exhibit magnetic properties and their associated magnetic responses are fast compared to the field oscillations. This slowly varying field can thus be assumed constant with respect to the magnetic response. The electromagnetic and magnetic systems are then considered independently. On one hand, a trace of magnetism is manifested in the electromagnetic system in the form of a nonlinear permeability, *i.e.* hysteresis response. On the other hand, the magnetic system is considered by monitoring the time evolution of the magnetization vector. This is commonly formulated using Landau-Lifshitz-Gilbert (LLG) equation [2]. At terahertz frequencies, the problem becomes more complexbecause the field oscillations are fast enough to occur on a similar time scale to the associated magnetic one. The independent solutions are no longer valid and simultaneous consideration (coupling) of the two systems is therefore mandatory. In Chapter 2, we perform these numerical simulations of the coupled systems (using LLG for the magnetic system and Maxwell for the propagation one).

Numerical results not only proved that terahertz can trigger ultrafast magnetization dynamics, but also brought to attention two important things. First, terahertz pulses have to be tailored in order to suit magnetic experiments. Second, terahertz can also be a suitable wave to effectively probe some magnetic systems. Those results set the motivations for all experiments presented in Chapters 3 and 4, respectively.

In Chapter 2, it will be shown that magnetization dynamics are very sensitive to the time profile of the terahertz pulse and pulse shaping is necessary. For example, assuming an ideal single cycle pulse (with a positive lobe followed by a negative one), if the first part of the pulse reverses the magnetization, the trailing negative one cancels this reversal. In practice and with multiple terahertz oscillations, the switching process can be non-deterministic. Ideally, a half cycle (gaussian-like) pulse should be used. Such a pulse cannotbe readily obtained from intense terahertz sources and there are no means -to the best of our knowledge- to temporally shape (picosecond)terahertz pulse. In section 3.1, we propose a way to shape the terahertz pulse temporally and spectrally. The technique depends on using optical pulses to excite free carriers in semiconductors and thus increasing its conductivity on the ultrafast time scale. With controlling the delay between the optical pulse and another synchronized terahertz pulse, the later experiences significant shaping.

A very intense terahertz field is another requirement (for terahertz-induced magnetization dynamics) found from numerical calculations. Although the technology of terahertz generation has evolved recently, the terahertz field intensities still lags behind the requirement of many proposed nonlinear experiments. This point is experimentally addressed in chapter 3 with two different approaches: field enhancement (section 3.2) and intense generation (section 3.3).

Field enhancement in sub-wavelength metallic structures has drawn a lot of attention since the observation of high optical transmission in sub-wavelength hole arrays [22]. In the terahertz regime, Seo et al. [23] recently reported a very high field enhancement in a single 70 nm-wide slit etched in a thin gold film. A single slit can be depicted as a nanocapacitor loaded with the charge collected by the surrounding metal surface. When terahertz radiation impinges on the metal sheet, light-induced current creates transients of charge imbalance across the gap. This imbalance, in turn, leads to a field enhancement inside the gap that scales with the incident wavelength. Such a result is somehow revolutionary. However, the transmission of a single slit structure is very low, which hinders experiments on noisy nonlinear sources. In our study, we took a step forward looking for a design of the structure exhibiting concurrent high enhancement and high transmission.

Concerning the second approach, we looked into the generation of more intense terahertz radiation. We focused on terahertz generation from laser-induced plasma. The technique depends on focusing short optical pulses in air (gas) where the ionized air acts as a nonlinear source for terahertz radiation. Most of the previous studied focused on using the typical 800 nm-centered optical pulses. However, recently, it has been shown that using longer optical wavelengths leads to more intense plasmas. Such a result was employed in the field of high harmonic and X-ray generation. As the stronger the plasma is, the more intense the terahertz gets, applying this approach (*i.e.* wavelength scaling) to our case should in principle lead to more intense terahertz emission. We investigated the scaling of terahertz generation with the optical wavelength (up to 2000 nm) and obtained dramatic increase in the terahertz generation. This study is the subject of section 3.3.

Numerical calculations in Chapter 2 shed light on another aspect of the terahertz-magnetism: linear applications. The main motivation of the calculations was to prove that terahertz could trigger ultrafast magnetization dynamics and find the experimental conditions to do that. The former isaddressed in Chapter 2 and the latter isthe subject of experiments in Chapter 3. The goal was to ultimately probe those dynamics with an X-ray pulse. However, it was found from numerical calculations that terahertz can act as a probe as well. Monitoring the polarization rotation of the transmitted terahertz pulse can show the magnetic state of amaterial. In a different way, handling the magnetization state of some materials can be used to control the propagation of terahertz waves. At terahertz frequencies, magneto-photonic devices are non-existing. This is because (1) the lack of knowledge on the magnetic properties of materials at terahertz frequencies and (2) the broad bandwidth response of materials required by short terahertz pulses. In Chapter 4, we experimentally introduce three novel experiments (devices) to the terahertz community: (1) broadband non-reciprocal phase retarders, (2) terahertz magnetic modulators, and (3) a terahertz isolator. We used a Ferrofluid (magnetic liquid) to build the former two and a ferrite (permanent magnet) to build the latter.

Ferrofluid consists of magnetic nanoparticles suspended in a carrier liquid. The magnetic nanoparticles are normally randomly distributed but when an external magnetic field is applied, they tend to align along its direction. They are usually ferromagnet (*i.e.* retain the magnetization state when the magnetic field is removed). However, due to this random distribution, they do notexhibit any net magnetization state when the external magnetic field is switched off. Therefore, they behave as paramagnets. Ferrofluids were developed more than a century ago but have not been studied in the terahertz regime. After the first spectroscopic measurements, we found that those liquids exhibit very high transmission in the terahertz range while preserving similar magnetic properties to the ones in the optical regime. This allowed for the usage of longer samples and therefore longer interaction lengths and more pronounced effects. Moreover, the fact that the magnetization curve is very nonlinear (*i.e.* the magnetization builds up very quickly at low applied magnetic fields, then it saturates) makes it useful for low magnetic field applications. We focused in two experimental configurations: out-of-plane and in-plane magnetization. In the out-of-plane geometry, the material is magnetized according to the popular Faraday geometry and thus non-reciprocal phase retardation is expected. In addition, the phase retardation in our case is broadband. This isthe first demonstration of broadband non-reciprocity at terahertz frequencies. If the in-plane configuration is considered, ferrofluidsshow unique characteristic of magnetic field-channel formation. Under an external magnetic field, the nanoparticles line up in channels. If the channel is in-plane, the material becomes non-symmetric with respect to the terahertz polarization plane. If the terahertz electric field is parallel to the channel's direction, high terahertz attenuation is foreseen. Such effect allowed us to realize the first terahertz modulator.

Finally, we demonstrate the terahertz isolator. Isolators are used to shield the electromagnetic sources against unwanted back-reflected radiation. They are based on non-reciprocal phase retarders adjusted to a specific angle of retardation. One of the objectives behind the Ferrofluids study here was to realize a broadband terahertz isolator. However, the associated retardation was not enough to reach that target. We finally moved to another promising class of materials: ferrites. Ferrites are permanent magnet with very low conductivity and strong magnetic properties. At terahertz, they have a big advantage because the associated resonance is in the sub-terahertz band. This means that their response in the terahertz regime is expected to be flat (broadband). Using a ferrite sample, we reached high degrees of phase retardation and experimentally realized the first isolator capable of handling short terahertz pulses.

# 1 EXPERIMENTAL TECHNIQUES FOR TERAHERTZ MAGNETISM

Terahertz science has been lagging behind so long mainly due to technological constraints. Despite falling between two well-investigated spectral regimes, microwave and infrared, a THz experiment has always been hard to set up. The operating principles of sources, devices, and detectors from the surrounding spectral bands cannot be reproduced in the THz band for several reasons. In essence, microwave electronics is not fast enough and optical quantum-mechanical transitions cannot be easily exploited due to the very low THz photon energy.

It is only recently–stimulated by the development of femtosecond lasers- that THz sources and detectors have become widely available. Terahertz evolved rapidly in the past decade into a distinct regime with its particular devices and experimental knowledge.

Terahertz-magnetism, a multidisciplinary research area, is the main focus of this work. In addition to the general experimental THz methods, specific techniques associated with magnetism are required. This chapter covers the main experimental techniques used in the thesis.

Short THz pulses (broadband) can be generated from either linear accelerators [4,24] or laser-based systems. The latter represents the platform used for the investigations presented here. Using lasers, THz can be generated using photoconductive switches [25,26] or laser-induced nonlinear generation from gases [27] and/orcrystals [5-6]. Considering the complex nature of the experimental setups presented here, generation using crystals is advantageous especially when the THz polarization, focusing, and guidance need to be varied during a single experiment. Optical rectification (OR) of femtosecond laser pulses in nonlinear optical crystals was then the generation technique of choice in this work.

## 1.1 Terahertz generation: optical rectification

The rigorous field-matter interaction model of the OR can be casted from Maxwell's equations [28,29]

$$\nabla X\, \boldsymbol{E} + \frac{\partial \boldsymbol{B}}{\partial \boldsymbol{t}} = 0 \qquad (1.1.1)$$

$$\nabla X\, \boldsymbol{H} = \frac{\partial \boldsymbol{D}}{\partial \boldsymbol{t}} + \boldsymbol{J} \qquad (1.1.2)$$

$$\nabla \cdot \boldsymbol{B} = 0 \tag{1.1.3}$$

$$\nabla \cdot \boldsymbol{D} = \varrho \tag{1.1.4}$$

where$\boldsymbol{J}$and $\varrho$ denote the free current density and charge density, respectively. $\boldsymbol{D}$,$\boldsymbol{B}$,$\boldsymbol{E}$, and $\boldsymbol{H}$ are the electric displacement, the magnetic induction, the electric field and the magnetic field, respectively. Those field vectors are related by the permittivity $\varepsilon$and the permeability $\mu$ tensors through$\boldsymbol{D} \equiv \varepsilon\boldsymbol{E} = \varepsilon_0\boldsymbol{E} + \boldsymbol{P}$and $\boldsymbol{B} \equiv \mu\boldsymbol{H}$. $\boldsymbol{P}$is the polarization field and physically represents the field-matter interaction, accounting for the density of the induced electrical dipoles.

Assuming no free charge (*i.e.*$\varrho = 0$), and in consideration of the negligible magnetic properties of dielectrics at optical frequency, *i.e.*$\mu = \mu_0$From Eq. 1.1.1 and 1.1.2, we can relate $\boldsymbol{E}$to $\boldsymbol{P}$through the wave equation

$$\nabla^2\boldsymbol{E} - \frac{1}{c^2}\frac{\partial^2\boldsymbol{E}}{\partial t^2} = \mu\frac{\partial^2\boldsymbol{P}}{\partial t^2} \tag{1.1.5}$$

The second derivative of polarization represents the source of the EM (THz) waves. Following the standard approach to describe the electronic polarizability of media, the polarization can be Taylor-expanded into apower series in terms of the $n^{\text{th}}$-order nonlinear susceptibility tensor $x^{(n)}(\boldsymbol{r},t)$as

$$\boldsymbol{P}(\boldsymbol{r},t) = x^{(1)}(\boldsymbol{r},t)\boldsymbol{E}(\boldsymbol{r},t) + x^{(2)}(\boldsymbol{r},t) : \boldsymbol{E}(\boldsymbol{r},t)\boldsymbol{E}(\boldsymbol{r},t) + x^{(3)}(\boldsymbol{r},t) : \boldsymbol{E}(\boldsymbol{r},t)\boldsymbol{E}(\boldsymbol{r},t)\boldsymbol{E}(\boldsymbol{r},t) + .. \tag{1.1.6}$$

Optical rectification originates from the second order nonlinearity (which exists only in noncentrosymmetric crystals). A basic description of the process can be inferred simply considering the second order source terms of Eq. 1.1.6 for the case of a scalar co-polarized pump and a nonlinear product. In such a case, for a pulsed pump oscillating at a frequency $\omega$ described as $E(\boldsymbol{r},t) = e(\boldsymbol{r},t)e^{-i\omega t}$ with $e(\boldsymbol{r},t)$ being the pulse envelope, the nonlinear contribution to the source term of Eq. 1.1.5 can be written as

$$\frac{\partial^2\boldsymbol{P}(\boldsymbol{r},t)}{\partial t^2} \propto \frac{\partial^2}{\partial t^2}\left[e(\boldsymbol{r},t)e^{-i\omega t} + cc\right]^2 = \frac{\partial^2}{\partial t^2}\left[e(\boldsymbol{r},t)^2e^{-i2\omega t} + cc\right] + 2\frac{\partial^2}{\partial t^2}|e(\boldsymbol{r},t)|^2 \tag{1.1.7}$$

Where *cc* denotes the complex conjugate. The two source terms originate respectively from the optical sum and frequency difference process. The latter is known as optical rectification, as a consequence of the fact that the optical carrier is cancelled and the source term is simply the second derivative of the pulse power envelope. The finite spectrum associated to OR results from the frequency difference process of different frequency components of the pump pulse. This process requires the conservation of energy and momentum between the interacting fields. Hence considering two distinguishable frequency component of the optical pump, the frequency and momentum of the THz wave are simply determined from the following relations

$$\Delta\omega_O = \omega_{O1} - \omega_{O2} = \Omega_{THz} \tag{1.1.8}$$

$$\Delta k_O = k_{O1} - k_{O2} = k_{THz} \tag{1.1.9}$$

where the subscripts O and THz distinguish between optical and THz related quantities and O1 and O2 refer to the two distinct generating waves. Dividing side-by-side Eq. 1.1.8 and. 1.1.9, in the limit of infinitesimal $\Delta k_O$ we obtain the fundamental (phase matching) condition for THz generation from the OR

$$v_{G,O} = \frac{\partial\omega_O}{\partial k_O} = \frac{\Omega_{THz}}{k_{THz}} = v_{Ph,THz} \tag{1.1.10}$$

$v_{G,O}$and$v_{Ph,THz}$ are the group velocity of the generating optical pulse and the phase velocity of the THz wave, both being frequency-dependent. Intuitively the phase mismatch regulates the energy exchange between the pump and the THz optical product. Neglecting other simultaneous nonlinear processes, and focusing our attention onthe direction of propagation, a given THz spectral component periodically receives and transfers power by back-interacting with the optical pump. As the mismatch increases, the length of such transfer decreases, which in turn reduces the maximum achievable energy transfer.

As it generally happens in all nonlinear frequency mixing processes, the phase mismatch defines the *coherence length*, *i.e.* the length scale characterized by the maximum energy transfer towards THz generation in OR.

The low phase-mismatch and the fair-to-good nonlinear optical coefficients are the main reasons for the wide adoption of Zinc Telluride (ZnTe) crystals in optical rectification sources. Zinc Telluride is an isotropic nonlinear crystal exhibiting phase-matching around 1 THz for an optical source at a wavelength of 800 nm,thusconveniently matchingthe typical emission wavelength of the widely available Ti-Sapphire femtosecond laser sources.

## 1.2 Terahertz detection: electro-optical sampling

Electro-optical (EO) sampling is typical nonlinear interaction geometry, were the instantaneous value of the THz electric field modulates the polarization of an optical probing field. It can then be used for THz detection.

In a typical configuration, the short optical pulse gates the THz wave in a nonlinear crystal. The optical-THz delay is changed to scan the total THz waveform. The underlying optical interaction mechanism is the Pockels effect, where the (relatively long) THz field behaves as a static field applied to an EO crystal. A synchronized optical beam thus experiences THz-induced birefringence (phase retardation). The phase retardation in a properly-oriented/ (110) cut ZnTe crystal (given that the optical beam is co-propagating and co-polarized with the THz wave and both are perpendicularly impinging on the crystal) can be expressed as [28,29]

$$\Delta\emptyset = \frac{\omega\, L}{c} n_O^3 r_{41} E_{THz} \tag{1.2.1}$$

where$n_O$ and $r_{41}$ are the refractive index and the EO coefficient of the crystal at the frequency $\omega$ of the optical probe. Figure 1.1 shows an example of time trace and spectrum of a THz pulse generated using OR and detected using EO sampling in ZnTe crystals.

A typical way to measure that birefringence is to use a balanced detection setup, where a $\lambda$/4-plate converts the two circular polarized eigenmodes of optical probe in two linear polarization. Trivially in the absence of the THz, the two components have the same amplitude as they arise from a linearly-polarized probe transmitted through the detection crystal. As the THz induces a birefringence, it changes the probe ellipticity, *i.e.* affecting the amplitude balance between the two circularly-polarized eigenmodes. A Wollaston prism after the $\lambda$/4-plate then separates the p- and s-polarization components to be detected by photodiode (x) and (y).The intensities of the signals reaching them are then

$$I_x = \frac{I_0}{2}(1 - sin\,\Delta\emptyset) \tag{1.2.2}$$

$$I_y = \frac{I_0}{2}(1 + sin\,\Delta\emptyset) \tag{1.2.3}$$

where$I_0$is the intensity of the optical probe. For $\Delta\emptyset \ll 1$ (which is typically the case in the EO sampling), the difference signal is

$$I_s = I_y - I_x \approx \frac{I_0 \omega L}{c} n_O^3 r_{41} E_{THz} \quad (1.2.4)$$

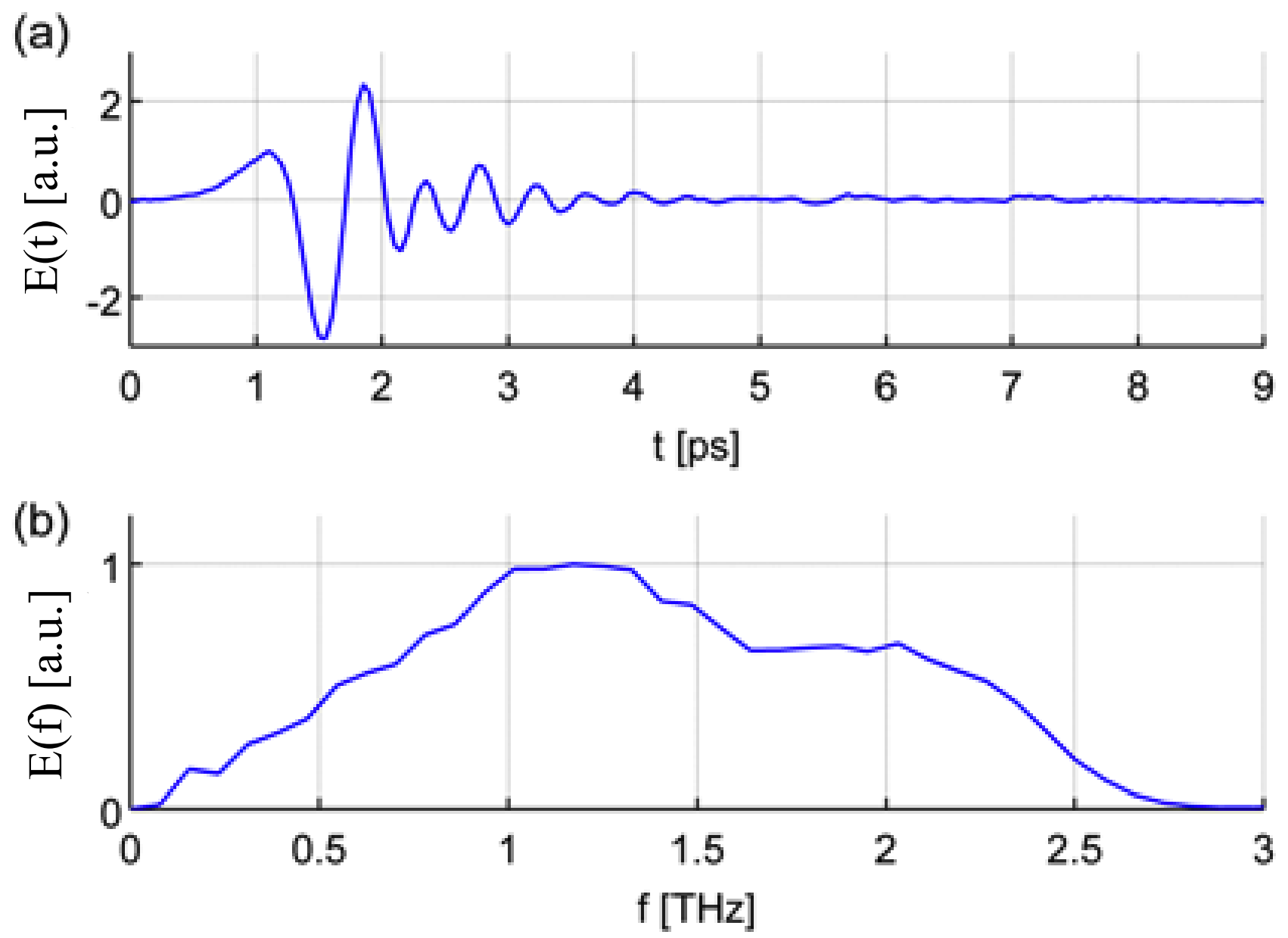

**Figure 1.1** **Terahertz pulse example. (a) Time trace and (b) spectrum of the pulse generated by means of the OR of 800 nm-centered pulse train (of 35 fs duration and 2.5 kHz repetition rate) in a 1 mm-thick ZnTe. Another 0.5 mm-thick ZnTe was used for detection through EO sampling.**

Equation 1.2.4 states that the detected signal is directly proportional to the THz electric field. In addition to the phase matching condition that holds as a basic requirement between the optical and the THz fields, other factors can also affect the detection bandwidth, like the absorption of the medium. Zinc Telluride has a phononic mode at 5.3 THz, which is conveniently far from the phase matching frequency (~ 1 THz). Most of the experimental results presented in this thesis were obtained using THz, generated from OR and detected by EO in ZnTe crystals.

## 1.3 Terahertz time domain spectroscopy (THz-TDS)

Although THz technology developed fast over the past few years, incoherent detection techniques (*i.e.* measuring the average power) are solely reserved to continuous wave and highly energetic pulsed emission. However, THz coherent detection techniques evolved quite fast to become the main detection platform. The EO sampling explained above is just an example. In addition to the high sensitivity, coherent detection methods are inherently phase-sensitive, enabling the access to the temporal field waveform of the detected THz. In other words, the THz complex spectrum (amplitude and phase) can be directly obtained. Such a property gives THz science powerful and fascinating spectroscopic advantages. Linear THz spectroscopy can thus be used to infer the complete dielectric response of materials. Indeed, this is presently the most popular THz application [12,13]. Retrieval of the THz dielectric response is the key step in the characterization of materials, as it will be clearly appreciated in the different parts of this thesis. The procedure can be summarized in two steps. First we experimentally determine the sample transfer function

$$t(\omega) = |t(\omega)|e^{i\Phi(\omega)} = \frac{E_s(\omega)}{E_r(\omega)} \qquad (1.3.1)$$

where$E_s(\omega)$ and $E_r(\omega)$ are the Fourier transforms of the THz field waveforms transmitted through the sample and through the air (reference), respectively. Second, from the transfer function, we mathematically extract the dielectric response of the material. A standard method implies the application of the transfer matrix approach [30-32]. This approach relates the THz fields (*i.e.*the incident, reflected, and transmitted components) at the sample interfaces to the dielectric and geometrical parameters of the sample. Using this approach, Eq. 1.3.1 can be expanded into

$$t(\omega) = t_{a,s}t_{s,a}e^{-i(\tilde{n}_s-1)\omega d/c} \qquad (1.3.2)$$

where$t_{a,s} = 2/(1+\tilde{n}_s)$ and $t_{s,a} = 2\tilde{n}_s/(1+\tilde{n}_s)$ are the transmission Fresnel coefficients at the input (air:sample) and output (sample:air) interface, respectively. $\tilde{n}_s = n + i\left(\frac{\alpha\, c}{2\omega}\right)$is the complex refractive index of the sample. Under the assumption that $\left(\frac{\alpha\, c}{2\omega}\right)^2 \ll n^2$ –which is valid for all the measurements performed in this work– separating the amplitude and phase components of Eq. 1.3.2 yields decoupled equations for the real refractive index $n$ and the absorption coefficient $\alpha$

$$n(\omega) = -\Phi(\omega)\frac{c}{\omega d} + 1 \qquad (1.3.3)$$

$$\alpha(\omega) = \frac{2}{d} ln\left(\frac{(n(\omega)+1)^2|t(\omega)|}{4\, n(\omega)}\right) \qquad (1.3.4)$$

For a given sample thickness *d*, the complete dielectric response can be readily found using Eq. 1.3.3 and 1.3.4.

## 1.4 Time-resolved terahertz spectroscopy

The aforementioned linear spectroscopy is the basic THz-TDS. More complex techniques focus on nonlinear spectroscopy where the response of the sample depends on the intensity of the triggering field (pulse). This can take one of various configurations depending on the phenomenon to be studied. A triggering pulse can drive the transition of a system towards a new state. For example, an electron beam can be used to switch the magnetic domains of a magnetic material [3]. If the material response is fast enough, evidences of the nonlinear interaction can be observed in the pumping pulse itself. The optical rectification technique explained above is an example. If the process is efficient, monitoring the spectrum of the THz-generating optical pulse gives insights on the optical rectification process.

An important class of experiments,*i.e.time-resolved* spectroscopy (TRS), focuses on the delayed effect induced by a pump.

In a typical TRS experiment, two fields are used, a pump and probe. The state of the sample is monitored against the pump-probe delay ($\tau$). The pump and probe fields can be of the same nature (spectrum, temporal profile, polarization etc) or different ones. A THz-pump / X-ray-probe experiment (highlighted in the introduction) is an example. A typical configuration of THz-TRS is concerned with optical-pump / THz-probe of semiconductors. In section 3.1, we use this setup to induce ultrafast temporal and spectral shaping of a THz pulse. Figure 1.2 shows a schematic diagram of the TR-optical-pump / THz-probe setup used in our studies.

An intense optical pump is split between the generation and detection lines. The generation line is further dividedbetween the (OR) generation and optical pump. The optical pump impinges on the sample, placed at the focus of a focusing (parabolic) mirror. If the optical photon energy exceeds the bandgap, semiconductors can be photoexcited through the *single photon absorption process* that induces the local formation of free carriers.The transmitted THz field ($E_t$) through a photoexcited semiconducting layer of thickness $d$, much smaller than the THz wavelength for normal incidence is given by [33]

$$E_t = (2Y_0 E_i - Jd)/(Y_0 + Y_s) \tag{1.4.1}$$

Where $E_i$ is the incident field. $Y_o$and$Y_s$ are the admittance of free space (~1/377 S) and sample, respectively. $J = n\ e\ v$is the electron current density with $n$, $v$, and $e$ being the electron density, charge, and velocity, respectively.An increase in the charge density upon an optical excitation is thus accompanied by an increase in the local conduction and consequently an attenuation of the transmitted THz pulse.  Figure 1.3 shows the result of an optical pump THz probe of a silicon substrate.

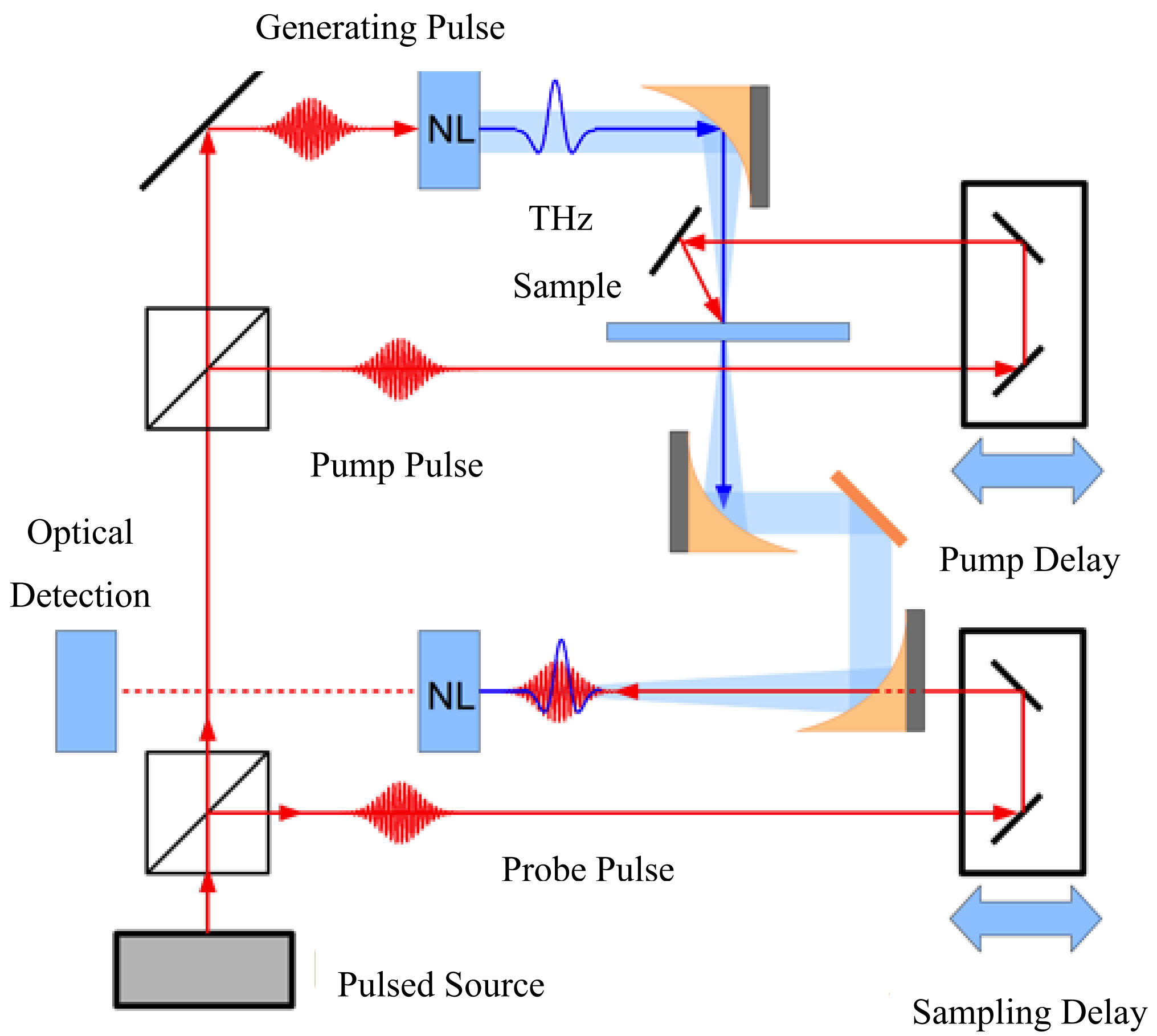

**Figure 1.2 Time-resolved optical-pump / THz-probe characterization setup. "NL" refers to a nonlinear crystal (ZnTe).**

The optical fluence was 695 μJ/cm$^2$ and the peak of the THz pulse (in Fig. 1.1(a)) is scanned against the probe pump delay. As the optical pump starts to overlap in time with the THz pulse, the THz field gets attenuated and even totally shielded. More details will be given in the part of the thesis dedicated to THz pulse shaping (section 3.1).

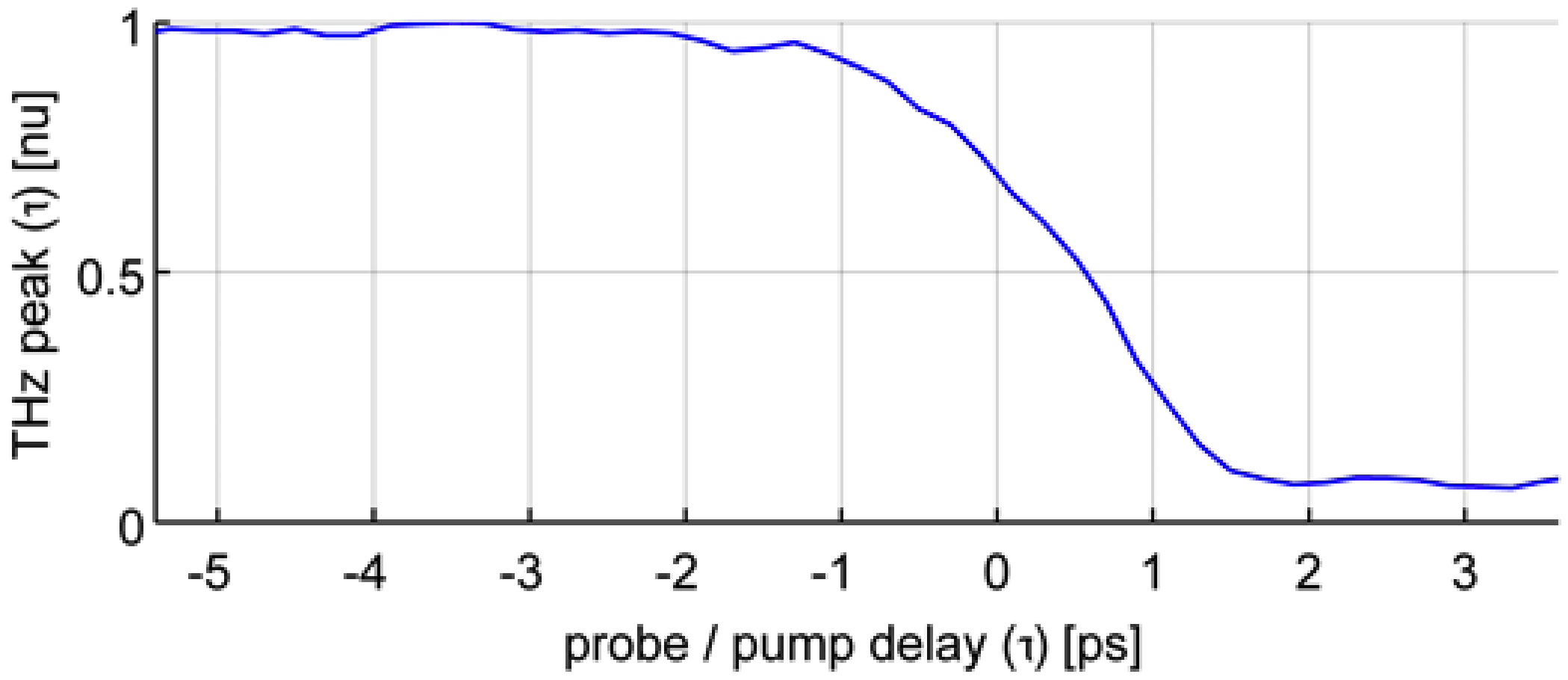

**Figure 1.3** **Terahertz transition during the optical-pump / THz-probe of silicon.**

## 1.5 Terahertz magnetic ellipsometry

The linear THz-TDS technique described above assumes the conservation of the THz linear polarization upon propagation through the sample. However, such hypothesis is in general violated when effects like optical retardation and dichroism occur in the medium. Magneto-photonic systems deal intensively with these specific effects. For example, light experiences polarization rotation if it propagates through a magneto-optical material placed in a magnetic field $\boldsymbol{H}$. Such a rotation significantly depends on the experimental configuration. If the $\boldsymbol{H}$ is applied in parallel to the light polarization plane (Fig. 1.4(a)), the rotation is reciprocal (*i.e.* it does not depend on the sign of $\boldsymbol{H}$) and the rotation angle is proportional to the square root of the magnetization component $\boldsymbol{M}$. Moreover, the rotation depends on the angle $\theta$ between $\boldsymbol{M}$ and $\boldsymbol{H}$, reaching a maximum for $\theta$= 45° and vanishing for $\boldsymbol{M}$ either aligned or orthogonal to $\boldsymbol{H}$. On the contrary, an out-of-plane magnetization (Fig. 1.4(b)) induces a non-reciprocal rotation. The rotation is much stronger in this case as it is directly proportional to $\boldsymbol{M}$. The rotation is also dependent on its sign. This effect is commonly known as Faraday rotation. The non-reciprocity obtained in this configuration sets the basis for an important class of devices called non-reciprocal phase retarders, with the optical *isolator* being the most remarkable example. Large part of the thesis is devoted to this category of devices [2].

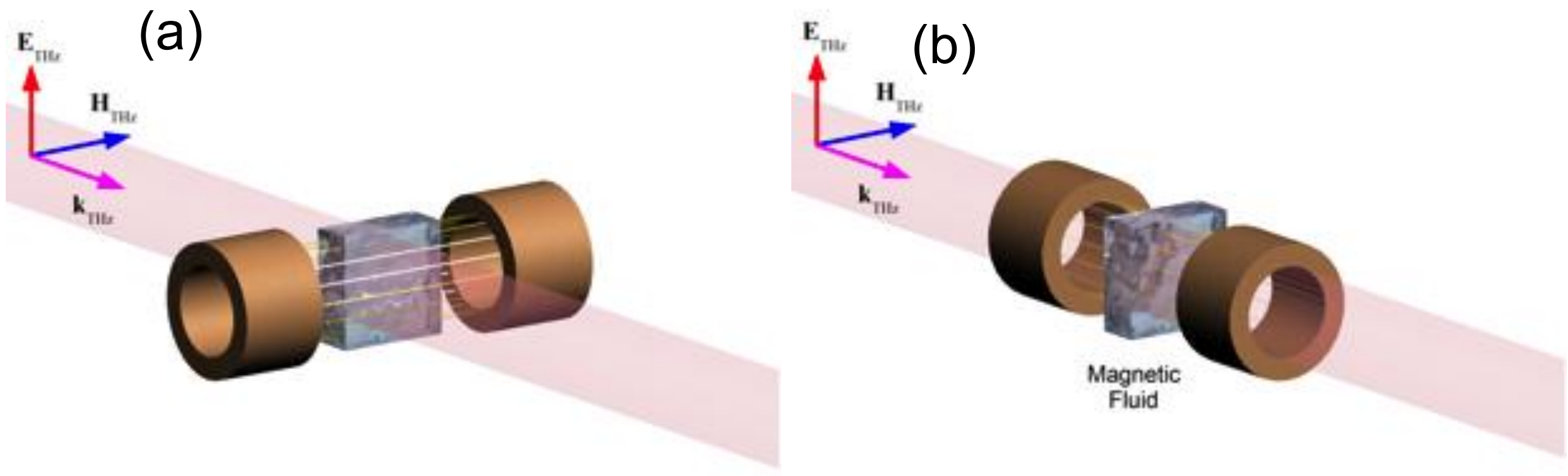

**Figure 1.4 Magneto-photonic experimental configurations.(a) In-plane and (b) out-of-plane setups.**

The origin of both polarization rotation and the related induced change of ellipticity can be understood through the representation of the light polarization in terms of its right- and left-circular eigenmodes. Differences in their refractive indices and absorption coefficients lead to rotation and ellipticity, respectively. In the optical regime, a practical approach to the experimental characterization of those quantities requires optical waveplates, which are not available in the THz domain, due to the associated large bandwidth. A popular alternative solution commonly used in THz setups is then to reconstruct the circular eigenmodes by measuring two orthogonal linear polarization components. This approach is enabled by the field sensitivity of TDS detection. Figure 1.5 shows a schematic diagram of a THz ellipsometry setup where a set of three polarizers is used (WGP1; WGP2; WGP3). WGP1 and WGP3 ensure vertical polarization of both the generated and detected THz pulses. Two measurements are taken for WGPs aligned to ±45° from the angle of maximum transmission: $E_{+45^o}(f)$ and $E_{-45^o}(f)$. From which, we get

$$\begin{pmatrix} E_l(f) \\ E_r(f) \end{pmatrix} = \frac{1}{2}\begin{pmatrix} -1+i & 1+i \\ 1+i & -1+i \end{pmatrix}\begin{pmatrix} E_{+45^o}(f) \\ E_{-45^o}(f) \end{pmatrix} \tag{1.5.1}$$

where $E_l(f)$ and $E_r(f)$ are the corresponding left circularly polarized (LCP) and right circularly polarized (RCP) light, respectively. From which the rotation angle $\emptyset$ and ellipticity $\varphi$ are readily found [34] as:

$$\emptyset(f) = \frac{arg\,[E_r(f)] - arg\,[E_l(f)]}{2} \tag{1.5.2}$$

$$\varphi(f) = \frac{|E_r(f)| - |E_l(f)|}{|E_r(f)| + |E_l(f)|} \tag{1.5.3}$$

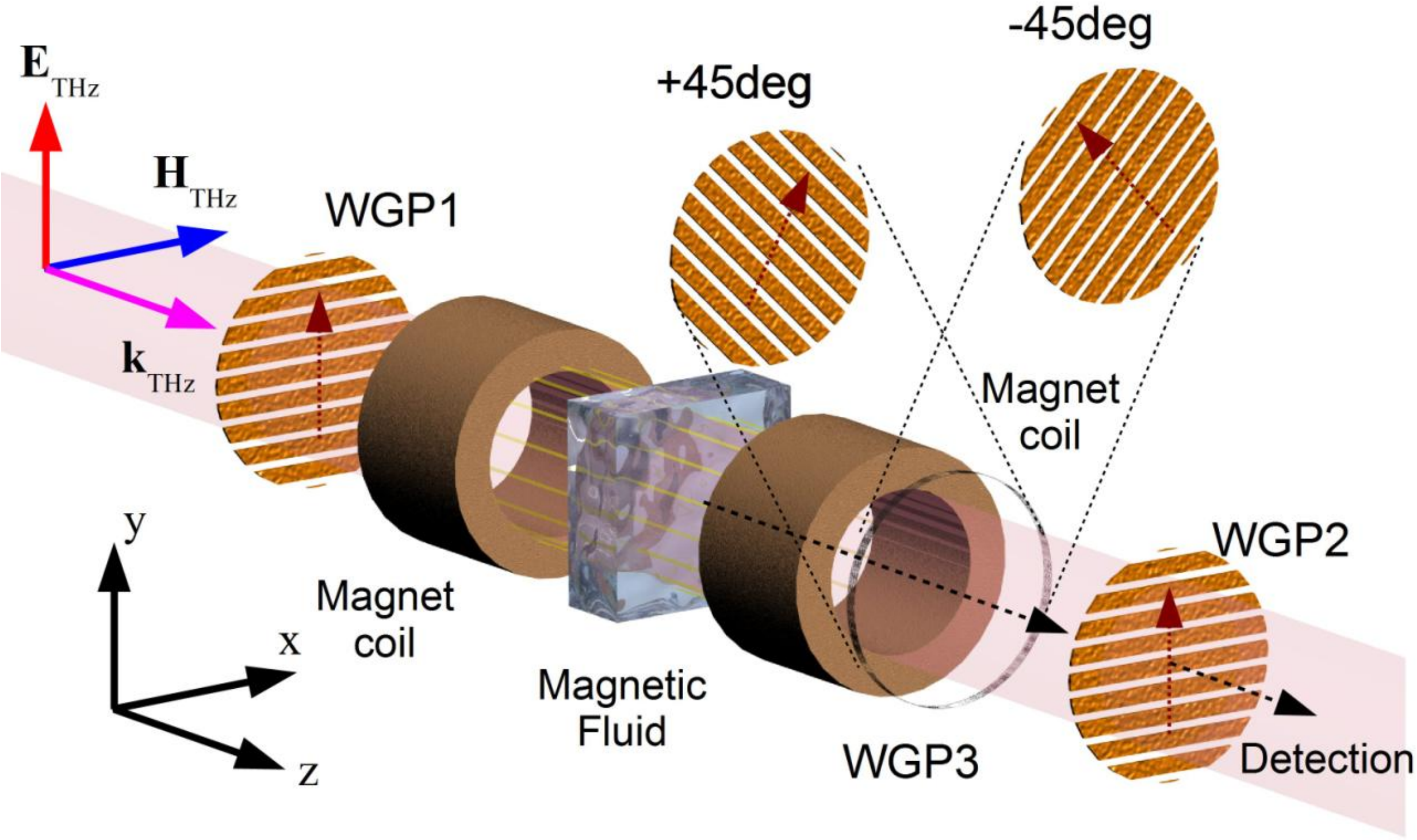

**Figure 1.5 Terahertz ellipsometry.**

The techniques previously described were used to obtain most of the results presented in the thesis where the band of 0.8 – 3 THz was used. This bandwidth was directly obtained using ZnTe crystals for both generation and detection. If other frequency ranges are required, other crystal and possibly other laser wavelengths will be needed. However, if a very large bandwidth becomes necessary, it will be hard to find a crystal that satisfies the phase matching condition over the whole bandwidth. In addition, crystals tend to suffer from frequency-dependent absorption of both THz and optical waves. For this kind of applications, the use of a gas mediumwhere phase matching and absorption problems do not exist will be more suitable. In section 3.3, we will introduce two other experimental schemes where an ionized air (plasma) is our nonlinear medium: THz generation via laser-induced plasma and air-biased coherent detection. They are very specific techniques and were used only in the measurements presented in section 3.3. So, they will be covered in detail as part of the experiment presented there.

# 2 TERAHERTZ MAGNETIZATION DYNAMICS

The THz-driven magnetic phenomenon is a novel science problem bridging the –otherwise rarely overlapping– magnetism and photonics communities. At high frequencies, materials tend to lose their magnetization properties. Magnetization is negligible at optical frequencies in natural media, *i.e.* themagnetic relative permeability is approximately equal to unity, $\mu = 1$. Up to the Gigahertz frequencies, the magnetic field-matter interaction is usually associated with static or slowly varying fields, *i.e.* the magnetic response is very fast with respect to field variations. The formalization of the interaction is then pursued in the assumption of constant forcing fields, and such approach accurately predicts the response up to microwave frequencies.

At THz frequencies, the problem takes a differentperspective. Field oscillations are slower than the optical ones. Magnetic response of some materials is foreseen at THz frequencies and experiments have reported magnetic responses induced by short (ps-scale) magnetic field pulses generated using electron accelerators [3]. Most importantly, this magnetic response occurs at the same time-scale as THz field oscillations. This means that the static field hypothesis is not generally valid in this regime. This makes THz-magnetism interaction a peculiar problemin whichboth the electromagnetic propagation and the magnetic response mechanics should be considered simultaneously. This chapter presents the first numerical calculations of THz-magnetic interactions. The obtained results set the basis and motivations for all the experimental work presented in the thesis, as it will be shown below.

## 2.1 Magnetism at low frequency

In general terms, the magnetic response of a material is described by its hysterisis curve, which shows the magnetic state $\boldsymbol{M}$of the material in a given magnetic field$\boldsymbol{H}$. It is useful to define arbitrary material and field time constants:$\tau_M = 1/f_M$ and $\tau_H = 1/f_H$, respectively. If we spectrally analyze the magnetic response of the material and the applied magnetic field, $f_M$and $f_H$ represent the lowest and highest frequencies in the respective spectra which are significant enough to affect our problem.

The hysterisis curve is a static curve, *i.e.* it assumes an instantly steady state of magnetization, *i.e.*$\tau_M \ll \tau_H$. In such a case, the time response of the material is not a critical parameter and frequency domain (*i.e.* hysterisis) is enough to formulate the problem. The applied magnetic field $\boldsymbol{H}$ can originate from a static source or can be the one associated to the EM wave.Ahysterisis-based picture is ideally descriptive if $\boldsymbol{H}$is static, *i.e.*$\tau_H \to \infty$.As $\boldsymbol{H} \propto 1 / \tau_M$, the higher $\boldsymbol{H}$ is, the lower $\tau_M$will be and the more interesting the phenomena observed on the magnetic time scale will be.If$\boldsymbol{H}$ comes from an EM source, the hysterisis curve still maps the material response with the instantaneous $\boldsymbol{H}$as long as $\tau_M \ll \tau_H$.

The problem complicates when the field is no longer static on the scale of the magnetization dynamics. This can be modeled, like any nonlinear EM propagation problem, using Maxwell's equations where the magnetic reponse of the material is given by the$H$-dependent magnetization $M(H,f) = \mu_o \mu(H,f)$ where $f$ is the frequency. Figure 2.1 shows a typical example where a 1 $kHz$-centered oscillating wave propagates through a Permalloy film. Calculations were reproduced after Lubber's work [35]. At a low magnetic field, the transmitted wave shows almost a linear response forthe sample. At a high field (which is the case shown in the figure), a high nonlinearity is shown due to the nonlinear magnetic response of the material, *i.e.* $\mu(H,f)$.

Although certainstatic and oscillating $\boldsymbol{H}$ can lead to the same $\tau_M$, there are fundamental differences between the two cases. On one hand, the presence of constant $\boldsymbol{H}$ induces precession of the magnetic moments with a constant frequency defined by the magnitude of $\boldsymbol{H}$. On the other hand, an oscillating $\boldsymbol{H}$*drives* the magnetization dynamics with that frequency, revealing properties of the magnetization mechanics at that frequency.

When the characteristic time of the field evolution is much larger than $\gg \tau_M$, increasing the frequency of the oscillating field allows for a faster dynamics of the magnetic dipoles. Note that as the EM frequency increases, higher peak fields are required to drive the corresponding dynamics.

The hysteresis curve provides a good description of the magnetization dynamics up to the MHz-GHz regime. As the frequency increases further this interaction model failsto predict the effect of the "inertia" in the magnetic system, which affects the magnetization dynamics. The assumption of a static stimulus is no longer satisfied. The time evolution of magnetization is ruled by the Landau-Lifshitz-Gilbert (LLG) model as will be shown later on. The calculation of its temporal solutions requires a detailed look at the magnetic system with the different field components. The first step is to evaluate the effective magnetic field inside the magnetic system.

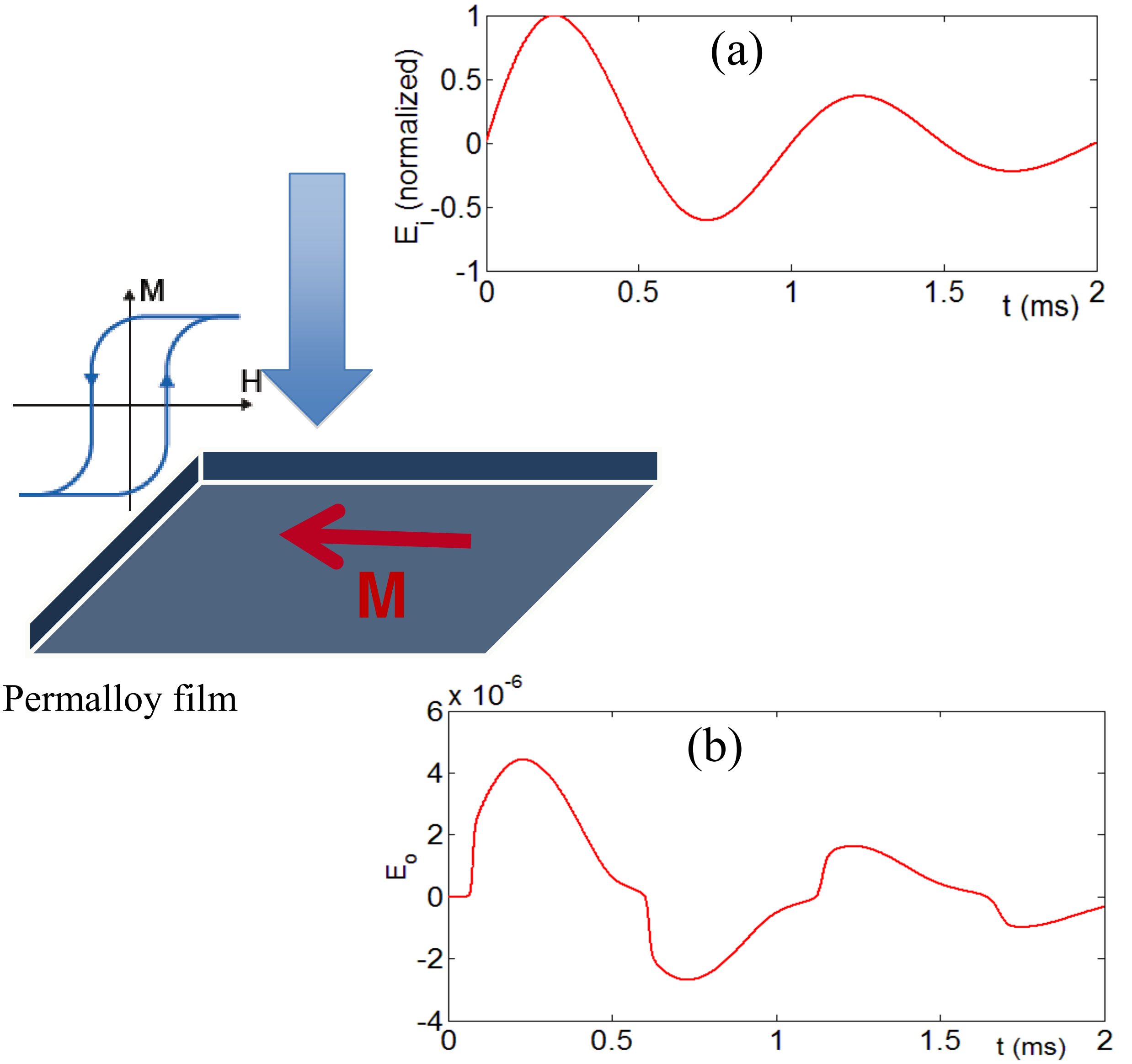

**Figure 2.1** **Low frequency magnetization dynamics. (a) An incident (damping 1 kHz) electromagnetic wave on a Permalloy film. (b) The transmitted wave showing the material nonlinear response. Results are reproduced after Lubber's [35].**

## 2.2 Effective field in a magneticsystem

The vector $\boldsymbol{M}$, as mentioned above, can describe the magnetization state inside the material. Its magnitude -called saturation magnetization $M_s$-is a characteristic of the material. Under the application of an external $\boldsymbol{H}$, the direction of $\boldsymbol{M}$ changes, but $M_s$ stays invariant.So it is always convenient to use the representation:$\boldsymbol{M} = M_s\boldsymbol{m} = M_s(m_x\tilde{x} + m_y\tilde{y} + m_z\tilde{z})$ with $m_x$, $m_y$, and $m_z$ being the Cartesian components of the unity vector $\boldsymbol{m}$. In the steady state, $\boldsymbol{m}$generally tends to align to the applied magnetic field. However, in a magnetic medium, there are other internal field components that also affect the direction of $\boldsymbol{m}$. So, an effective magnetic field $\mathrm{H}_{eff}$ should be used instead. The effective magnetic field is defined through a variational approach [36]

$$\boldsymbol{H}_{eff}(\boldsymbol{r},t) = -\frac{1}{\mu_o}\frac{\delta U}{\delta \boldsymbol{M}(\boldsymbol{r},t)} \tag{2.2.1}$$

with$U$being the total energy, given by

$$U = U_{app} + U_{ex} + U_d + U_{an} \tag{2.2.2}$$

where$U_{app}$, $U_{ex}$, $U_d$, and $U_{an}$are the applied, exchange, demagnetization, and anisotropy components, respectively. The energy components and the corresponding field factors can be written as [36]

$$U_{app} = -\int \boldsymbol{M}.\boldsymbol{H}_{app}\,d\boldsymbol{r} \,\&\, \boldsymbol{H}_{app} = -\delta U_{app}/\delta\boldsymbol{M} \tag{2.2.3}$$

$$U_{ex} = A\int((\nabla m_x)^2 + (\nabla m_y)^2 + (\nabla m_z)^2)d\boldsymbol{r} \,\&\, \boldsymbol{H}_{ex} = (2A)\nabla^2\boldsymbol{m} \tag{2.2.4}$$

$$U_d = -\frac{1}{2}\int \boldsymbol{M}.\boldsymbol{H}_d d\boldsymbol{r} \,\&\, \boldsymbol{H}_d = -\delta U_d/\delta\boldsymbol{M} \tag{2.2.5}$$

$$U_{an} = -(K/2)\int(m_x^4 + m_y^4 m_z^4)d\boldsymbol{r} \,\&\, \boldsymbol{H}_{an} = (2K/M_s)(m_x^3\hat{x} + m_y^3\hat{y} + m_z^3\hat{z}) \tag{2.2.6}$$

where$A$ and $K$ are the exchange and anisotropy constants. The applied field $\boldsymbol{H}_{app}$ can come from static and/or oscillating sources. However, in all the calculations presented in this chapter, we will assume that no static fields are applied.The exchange field $\boldsymbol{H}_{ex}$ tends to align neighboring magnetic moments to maintain uniform magnetization. This field is effective over a material-dependent length scale, given by the exchange length [2]

$$l_{ex} = \sqrt{2A/\mu_o}\,/\,M_s. \tag{2.2.7}$$

This parameter represents the length scale on which the magnetization is uniform. The demagnetization field$\boldsymbol{H}_d$ can be casted in terms of a demagnetization tensor $N$ as$\boldsymbol{H}_d = -N\,\boldsymbol{M}$. For a flat ellipsoidal geometry [2] -of which the thin films considered here are typical examples- the demagnetization tensor $N$ is a diagonal matrix with diagonal components $(N_x, N_y, N_z)$fulfilling the relation $N_x + N_y + N_z = 1$ [2]. For a thin film in the x-z plane, the in-plane components $N_x$ and $N_z$are equal to zero and therefore$\boldsymbol{H}_d = -\boldsymbol{M}_y$. This magnetization component attempts to keep $\boldsymbol{M}$ in plane. Finally, the crystal anisotropy field $\boldsymbol{H}_a$ accounts for the dependence of the potential energy on the orientation of the magnetization relative to the crystal axes, as well as for its contribution to the overall free energy [2]. The potential energy will be at its minimum when $\boldsymbol{M}$ is aligned with a particular crystal axis. In this work, we consider two ferromagnets: Permalloy and Cobalt. While the former has minimal crystal anisotropy, the latter exhibits strong uniaxial anisotropy.

## 2.3 Precessional and damping mechanisms

In a steady state, the direction of $\boldsymbol{H}_{eff}$ coincides with that of $\boldsymbol{m}$. If an external magnetic field is applied such that $\boldsymbol{H}_{eff}$starts to change direction formingan angle $\Phi_{\boldsymbol{M}-\boldsymbol{H}}$ with $\boldsymbol{m}$, a non-zero torque component is generated. Such a torque results in the formation of a new $\boldsymbol{H}_{eff} \,//\, \boldsymbol{m}$equilibrium state. This is perceived ina two-step process: precessional and damping (Fig. 2.2). First, the magnetic moment$\boldsymbol{\mu}$starts to precess around the direction of $\boldsymbol{H}_{eff}$. Second, $\boldsymbol{\mu}$damps to the direction of $\boldsymbol{H}_{eff}$.

Formally, the mechanics of the two processes are described by the LLG equation [2]

$$\frac{\partial \boldsymbol{M}}{\partial t} = \gamma\left(\boldsymbol{M} \times \boldsymbol{H}_{eff}\right) - \frac{\alpha}{M_s}\left[\boldsymbol{M} \times \frac{\partial \boldsymbol{M}}{\partial t}\right] \tag{2.3.1}$$

Where$\alpha$ is the Gilbert's damping constant, the first term being the precessional term and the second is the damping one.

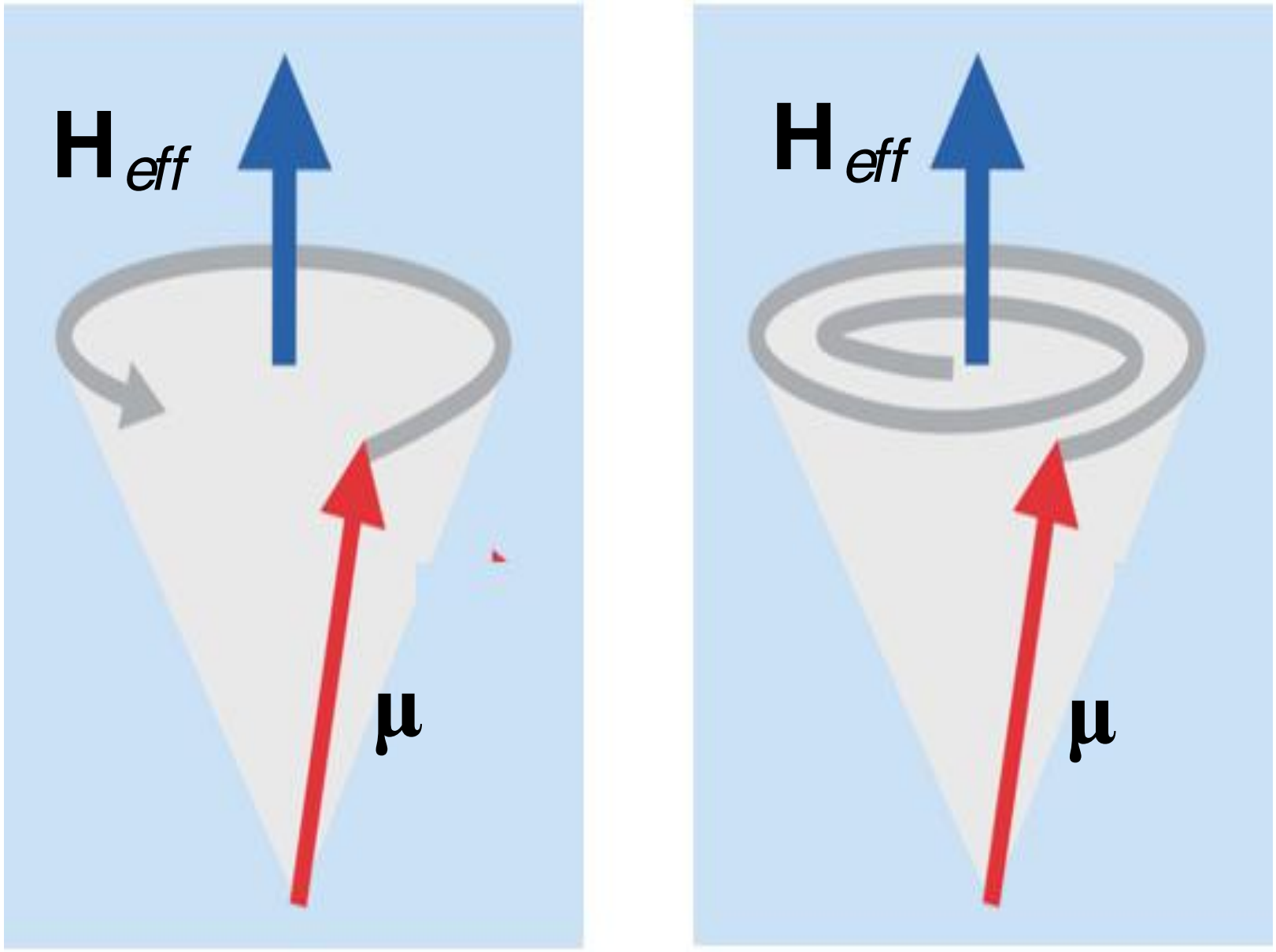

**Figure 2.2 Magnetization dynamics. (a) Precessional and (b) damping mechanisms [1].**

## 2.4 Model for terahertz-magnetization dynamics

The temporal dynamics of $\boldsymbol{M}$ under the application of $\boldsymbol{H}$ is governed by the LLG equation as shown above. The LLG assumes that the fields are spatially and temporally uniform in the sample region. Triggering magnetization dynamics with THz pulses violates this assumption because THz is simply a propagating wave experiencing spatio-temporal field dynamics inside the sample. This evolution originates fromboth linear and nonlinear contributions to the field-matter interaction. On the linear side, the field experiences Fresnel reflections at the sample interfaces and is generally attenuated with the propagation distance. These linear phenomena can be directly taken into account by solving,in time and space, Maxwell's equations for the propagating THz wave. The magnetic problem can then be solved using a completely decoupled and propagation-independent LLG model.

However, THz-triggered magnetization dynamics involve a nonlinear field-matter relation that cannot be directly incorporated with such a model consisting of decoupled LLG and Maxwell's equations. At a given depth inside a film, the precessions of magnetic moments depend on both the local magnetic field and on the precessions of the neighboring magnetic moments. In the case of a THz excitation, the local magnetic field $\boldsymbol{H}$ depends nonlinearly on the magnetic precessions. Therefore, the complete description of the nonlinear interactions between a THz pulse and the ferromagnetic film requires the coupling between the LLG equationand the Maxwell's equations [2]. The solution of this system strongly depends on both material properties and pulse parameters. The coupled LLG-Maxwell model is thus fundamental for such studies.

## 2.5 Coupled LLG-Maxwell

Terahertz propagation can be described by solving Maxwell's equations for the field vectors: $\boldsymbol{M}$-the magnetization, $\boldsymbol{D}$-the field displacement, $\boldsymbol{B}$-the magnetic induction, and $\boldsymbol{E}$-the electric field. Furthermore, the Landau-Lifshitz Gilbert equation links the applied (THz) field $\boldsymbol{H}_{app}$ to the magnetization $\boldsymbol{M}$. The latter is the common quantity between the magnetic (LLG) and thepropagation (Maxwell) systems and the simultaneous solution is thus pivoted around it and $\boldsymbol{B}$ (Fig. 2.3).

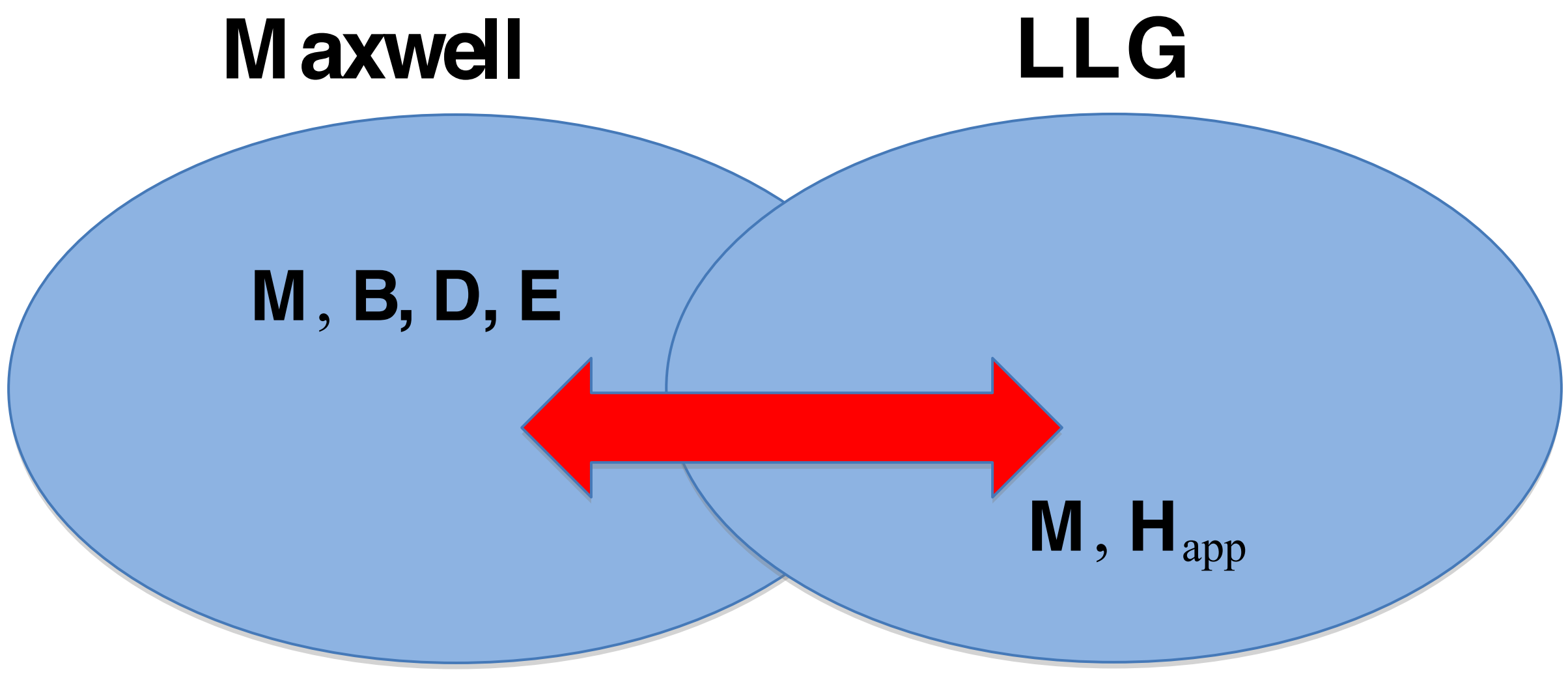

$$\boldsymbol{H}_{app} = \boldsymbol{B} / \mu o - \boldsymbol{M}$$

**Figure 2.3** **A coupled LLG-Maxwell model. The terahertz-magnetization model depends on the simultaneous solution ofthe LLG and Maxwell equations, where $M$ is the common parameters in the coupled model.**

In a typical integration scheme, Maxwell's equations are solved for the spatial distribution of the THz magnetic field at a certain temporal point. This value of magnetic field is plugged in the solution of the LLG equation, which gives a new spatial map of the vector$\boldsymbol{M}$. $\boldsymbol{M}$is used in the calculation of the next temporal point of Maxwell's equations. This temporal coordinateupdates the magnetic field to solve the LLG equation at the following time step.

Here,the finite difference time domain (FDTD) technique is used for the numerical simulations [37]. This technique solves both LLG and Maxwell equationsin a stepwise time domain. The material dielectric response is described in the frequency domain using the Drude model:

$$\varepsilon(\omega) \ = \ \varepsilon_{\infty}\left(\frac{\omega_p^2}{\omega(iv_c + \omega)}\right) \tag{2.5.1}$$

where$\varepsilon_{\infty} = 1$ and the plasma frequency $\omega_p$is calculated using [38]$\omega_p \ = \ (N\, e^2/\varepsilon_o m_o)^{1/2}$where $N$, $e$, and $m_o$ are the free electron density, the electron charge, and electron mass, respectively.

This frequency-dependent dielectric response is implemented in the time domain simulations via *z*-transformation with a time step $\Delta t$ [37]

$$\boldsymbol{E}(z) \ = \left(\frac{\boldsymbol{D}(z)}{\varepsilon_o}\right)\left(1 \ + \left(\frac{\omega_p^2 \Delta t}{v_c}\right)\left(\frac{1}{1 - z^{-1}} - \frac{1}{1 - e^{-v_c \Delta t} z^{-1}}\right)\right)^{-1}. \tag{2.5.2}$$

The coupled model can then be written as

$$\frac{\partial \boldsymbol{B}}{\partial t} \ = \ -\nabla \times \boldsymbol{E} \tag{2.5.3}$$

$$\boldsymbol{H} \ = \frac{\boldsymbol{B}}{\mu_o} - \boldsymbol{M} \tag{2.5.4}$$

$$\frac{\partial \boldsymbol{D}}{\partial t} \ = \ \nabla \times \boldsymbol{H} \tag{2.5.5}$$

$$\boldsymbol{H}_{eff} \ = \boldsymbol{H} + \boldsymbol{H}_d + \boldsymbol{H}_{ex} + \boldsymbol{H}_{an} \tag{2.5.6}$$

$$\frac{\partial \boldsymbol{M}}{\partial t} \ = \ \gamma\left(\boldsymbol{M} \times \boldsymbol{H}_{eff}\right) - \frac{\alpha}{M_s}\left[\boldsymbol{M} \times \frac{\partial \boldsymbol{M}}{\partial t}\right]. \tag{2.5.7}$$

Now after the numerical model is established, the last step before running the simulation is to estimate the THz field intensity required to start magnetization dynamics. A useful approach is to relate the applied fields to the induced magnetic precessions, *i.e.* ferromagnetic resonance (FMR).

## 2.6 Ferromagnetic resonance

In a FMR experiment, a magnetic sample is placed in a static magnetic field, which induces precessions of magnetic moments. Then, the frequency of a propagating EM wave through the sample is swept looking for a resonance between the EM wave and the magnetic precession frequencies. At such a frequency the EM wave is absorbed in the sample. The resonance frequency depends on both the internal magnetization and the static field and can be tuned by changing the static field. Those quantities are related by the well-knownKittel formula [39]. In this typical configuration, the static field is the driving one. On the contrary, our problem is concerned with starting magnetization dynamics with the EM wave (THz field) in the absence of a static field. Kittel formula is not applicable to our case, but we could derive a new resonance formulalooking at the evolution of magnetization dynamic described by the equation of motion (the precession term in LLG, Eq. 2.3.1):

$$\frac{\partial \boldsymbol{M}}{\partial t} = \gamma(\boldsymbol{M} \times \boldsymbol{H}_{eff}) \tag{2.6.1}$$

Under the assumption of an x-polarized THz magnetic field and a z-magnetized sample, $\boldsymbol{H}_{eff}$ can be decomposed into

$$\boldsymbol{H}_{eff} = (H^x + H_{ex}^x)\hat{x} + (H_d^y + H_{ex}^y)\hat{y} + (H_{an}^x + H_{ex}^z)\hat{\mathrm{z}}. \tag{2.6.2}$$

The exchange field isalways parallel to $\boldsymbol{M}$ and, therefore, they vanish in the solution of the equation of motion. The anisotropy field is negligible for the polycrystalline Permalloy film considered here. At the end, the derivation will be generalized to materials with crystal anisotropy as well, such as Cobalt. The components of the equation of motion can then be written as:

$$\frac{\partial M_x}{\partial t} = \gamma(-M_z H_d^y) \tag{2.6.3}$$

$$\frac{\partial M_y}{\partial t} = \gamma(M_z H^x) \tag{2.6.4}$$

$$\frac{\partial M_z}{\partial t} = \gamma(M_x H_d^y - M_y H^x) \tag{2.6.5}$$

In order to maximize the torque exerted on the spins, we set $\Phi_{\boldsymbol{M}-\boldsymbol{H}} = 90°$ where $\Phi_{\boldsymbol{M}-\boldsymbol{H}}$ is the initial angle between $\boldsymbol{M}$ and $\boldsymbol{H}$. Consequently, the required peak $\boldsymbol{H}$ and $\omega$ are given by the initial conditions. Initially the demagnetization field is zero. Assuming quasi-monochromatic dynamics timedependence $e^{i\omega t}$ and by simplifying Eq. 2.6.3; 2.64; 2.6.5, the instantaneous frequency is found to be $\omega = \gamma H^{x}$. Moreover, under no externally applied DC field,$\boldsymbol{H}_{an}$ initially aligns with $\boldsymbol{M}$and therefore does not contribute to $\omega$. This statement generalizes our solution to materials other than Permalloy with high crystal anisotropy,such asCobalt. After accounting for the boundary conditions at the air-film interface, we estimate that an incident THz peak magnetic field of 6 T is sufficient to initiate magnetic interactions in the lower part of the THz band considered ~ 0.3 THz.

## 2.7 Terahertz-triggered magnetization dynamics

Using the model described above and the THz peak value estimated from FMR, we simulated the nonlinear interaction between a propagating THz wave and a thin Permalloy film. Figure 2.4(a) shows the time profile of asingle cycle THz pulse (STP) and its spectrum. We assume that $\boldsymbol{M}$is initially z-oriented and the sample is illuminated by a linearly polarized THz wave having $\boldsymbol{H} = H\hat{x}$, as shown in Fig. 2.4(b).

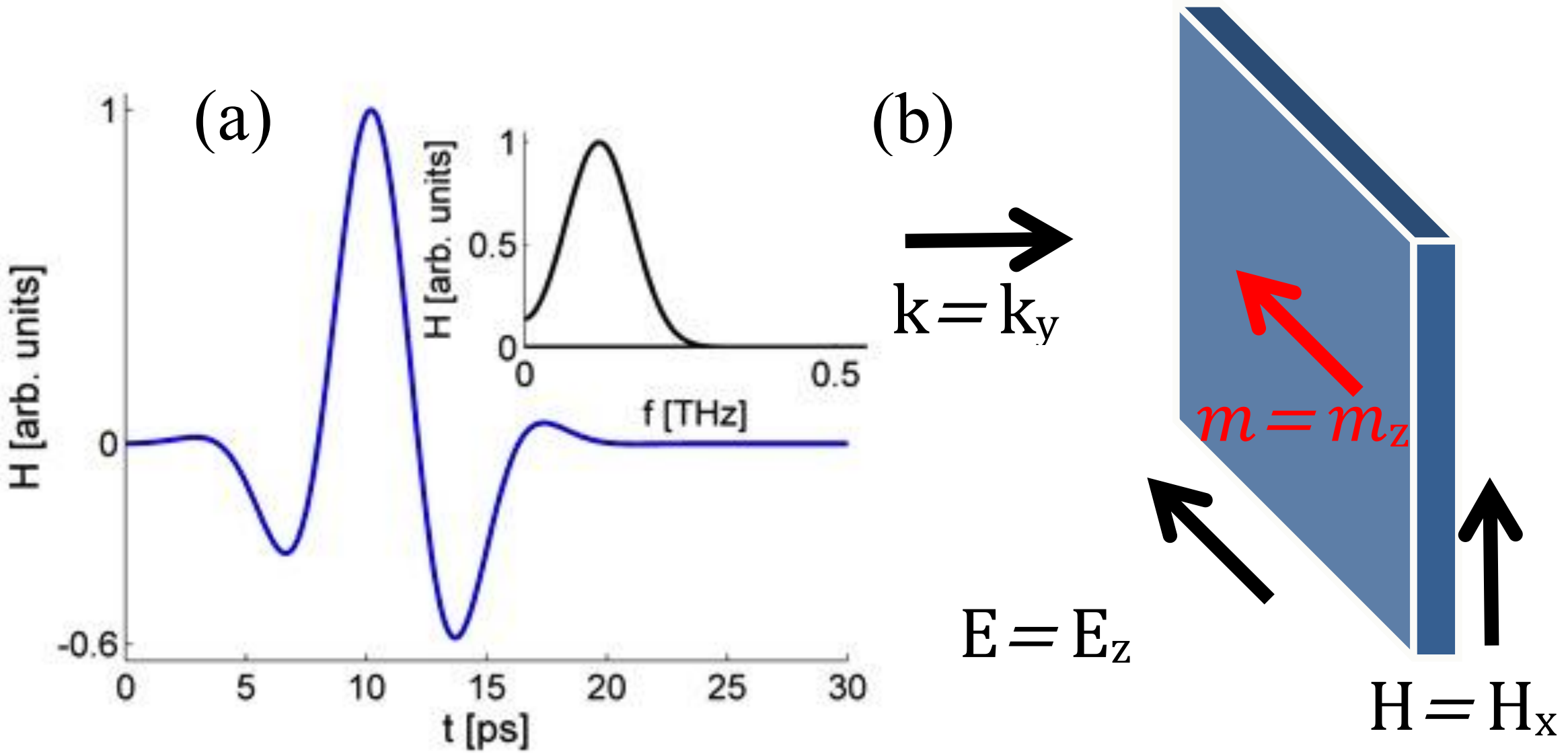

**Figure 2.4** **Magnetic-interaction geometry. (a) Temporal waveform and spectrum of aTHz pulse. (b) Sketch of the magnetic film along with the initial directions of both the magnetization and the THz field vectors.**

Figure 2.5shows the temporal dynamics of the THz magnetic field and of the magnetization components along the film depth. At the front surface of the film, $\boldsymbol{M}_z \times \boldsymbol{H}$initially generates a torque, which results in the precessions of the magnetic moments with significant out-of-plane component$\boldsymbol{M}_y$. The shape anisotropy of the film induces a strong demagnetization effect due to the field contribution $\boldsymbol{H}_d = -\boldsymbol{M}_y$which counteracts the out-of-plane magnetization state. Those two opposing effects (out-of-plane magnetization and demagnetization) lead to damping mechanism acting towards the film plane. Since polycrystalline Permalloy films in general do not exhibit crystal anisotropy, there is no preferred direction for the damping away from the film edges. Hence, the final orientation of $\boldsymbol{M}$ can be arbitrary in the film plane depending on the direction of the effective magnetic field $\boldsymbol{H}_{eff}$. During this interaction process, total switching of the magnetization from $+\hat{z}$ to $-\hat{z}$ is realized at some instants.

Considering now the depth-dependence of the interaction, the $\hat{y}$ component of $\boldsymbol{H}_{eff}$ and therefore the magnetization dynamics depend nonlinearly on the local propertiesof the magnetic system and on the THz pulse.

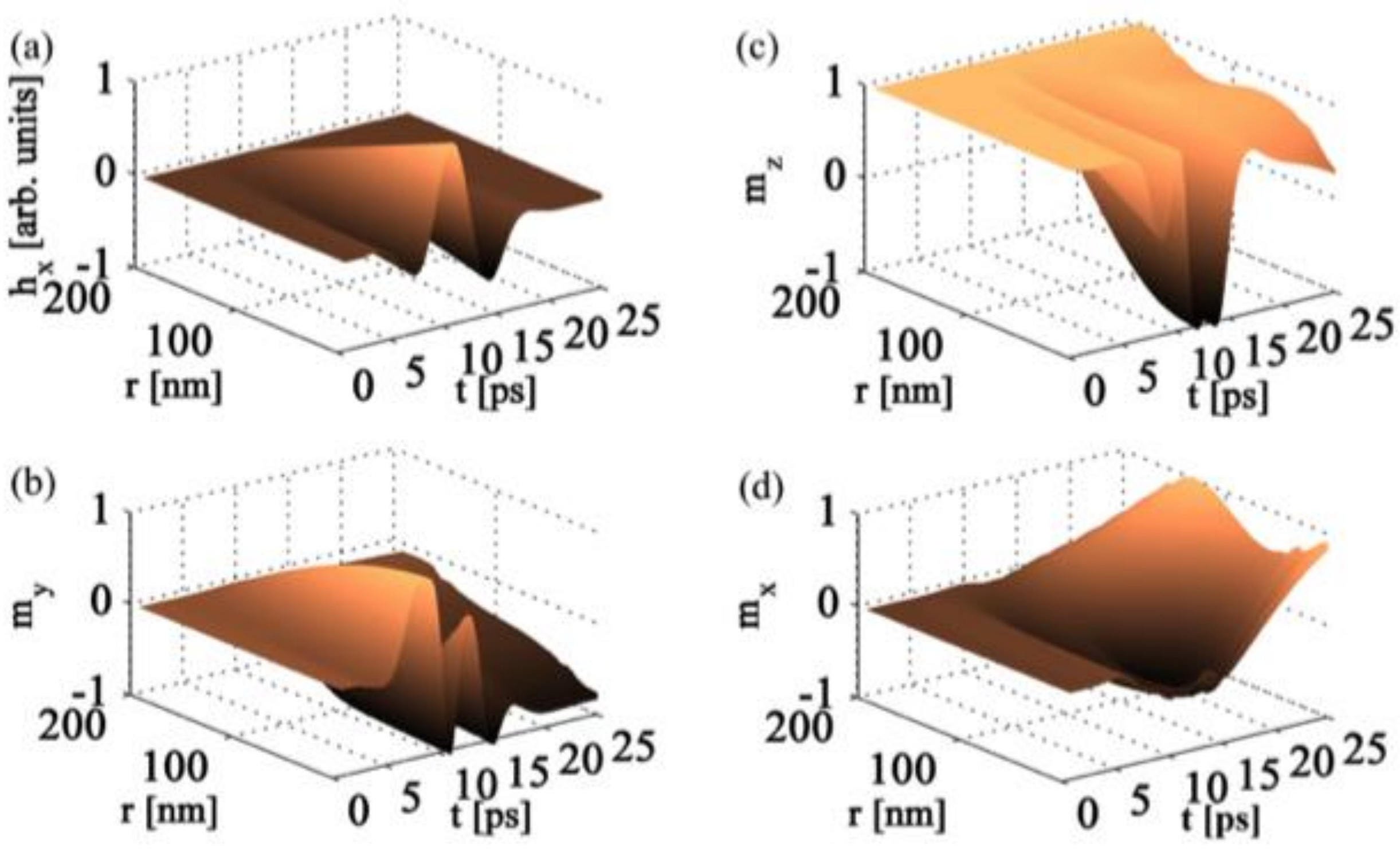

**Figure 2.5** **Spatio-temporal THz magnetization dynamics. (a) The magnetic field of the triggering THz pulse. (b) The out-of-plane magnetization components. (c) and (d) are the in-plane magnetization components.“r” represents the depth along the y-axis.**

At a given depth $r_o$, $\boldsymbol{M}(r_o)$ is affected by the components of$\boldsymbol{M}(r)$ within the range $r_o - l_{ex} < r < r_o + l_{ex}$ where $l_{ex}$ is the exchange length. Due to this nonlocal effects on the orientation of $\boldsymbol{M}$, the preceding part of the magnetic film$(r < r_o)$-where the reorientation has already been triggered- enhances the switching in $r = r_o$, whereas the successive film sections ($r > r_o$) induce a restoring force that tends to maintain the magnetization inits original state. Those two opposing mechanisms lead to a smooth magnetization transition between neighboring magnetic layers. Figure 2.5(a) shows that the magnetic field of the THz pulse decays (almost exponentially) along the film depth. The rate of decay depends nonlinearly on the magnetization dynamics along the propagation direction. This, in turn, leads to a corresponding decay in the induced torque $\boldsymbol{M}_z \times \boldsymbol{H}$. The dependence of the nonlinear interactions on the depth is demonstrated in Fig. 2.5(b), (c), and (d)by considering both time and depth evolution of the THz-magnetization dynamics.

The above-mentioned dynamics lacks control over the magnetization evolution and the final state of magnetization is hard to precisely control. A Terahertz-switched magnetic memory requires the interaction length to be comparable to $l_{ex}$. This ensures coherent –single domain– precessions of magnetic moments.Moreover, in magnetic switching both pulse shape and crystal anisotropy play significant roles.

## 2.8 Magnetization dynamics: effect of crystal anisotropy

UsingPermalloy, as shown in Fig. 2.5, leads to a final state of magnetization anywhere in the film plane with no preference on the final direction. This problem can be overcome by using a material with a strong crystal anisotropy, such as uniaxial Cobalt, which limits the final $\boldsymbol{M}$ state to $|\pm M_z|\hat{z}$ where $\hat{z}$ is assumed to be the crystal anisotropy axis. Figure 2.6 shows the interaction between the same STP we considered above but here with a 5 nm-thick Cobalt film. The film thickness is comparable to Cobalt $l_{ex} = 3.4$ nm[2]. Hence the film is assumed to be a single magnetic domain.As the magnetic field of the THz pulse increases, the generated torque tips the magnetization vector out-of-plane, *i.e.*a significant $\boldsymbol{M}_y$component rises. At the same time, and because of the vectorial nature of the problem, $\boldsymbol{M}_z$ changes as well, as part of the 3-dimentional dynamics. With the triggering field (THz pulse) vanishing, the magnetization dynamics relaxes to an equilibrium state. In the present case, the crystal anisotropy forces the magnetization vector to pointtowards the $\hat{z}$ axis. In comparison, this behavior is strikingly different in the Permalloy film case, where the magnetization could end in any direction. Figure 2.7 shows the time-extended dynamics where the trajectory is shown in a 100 ps window.

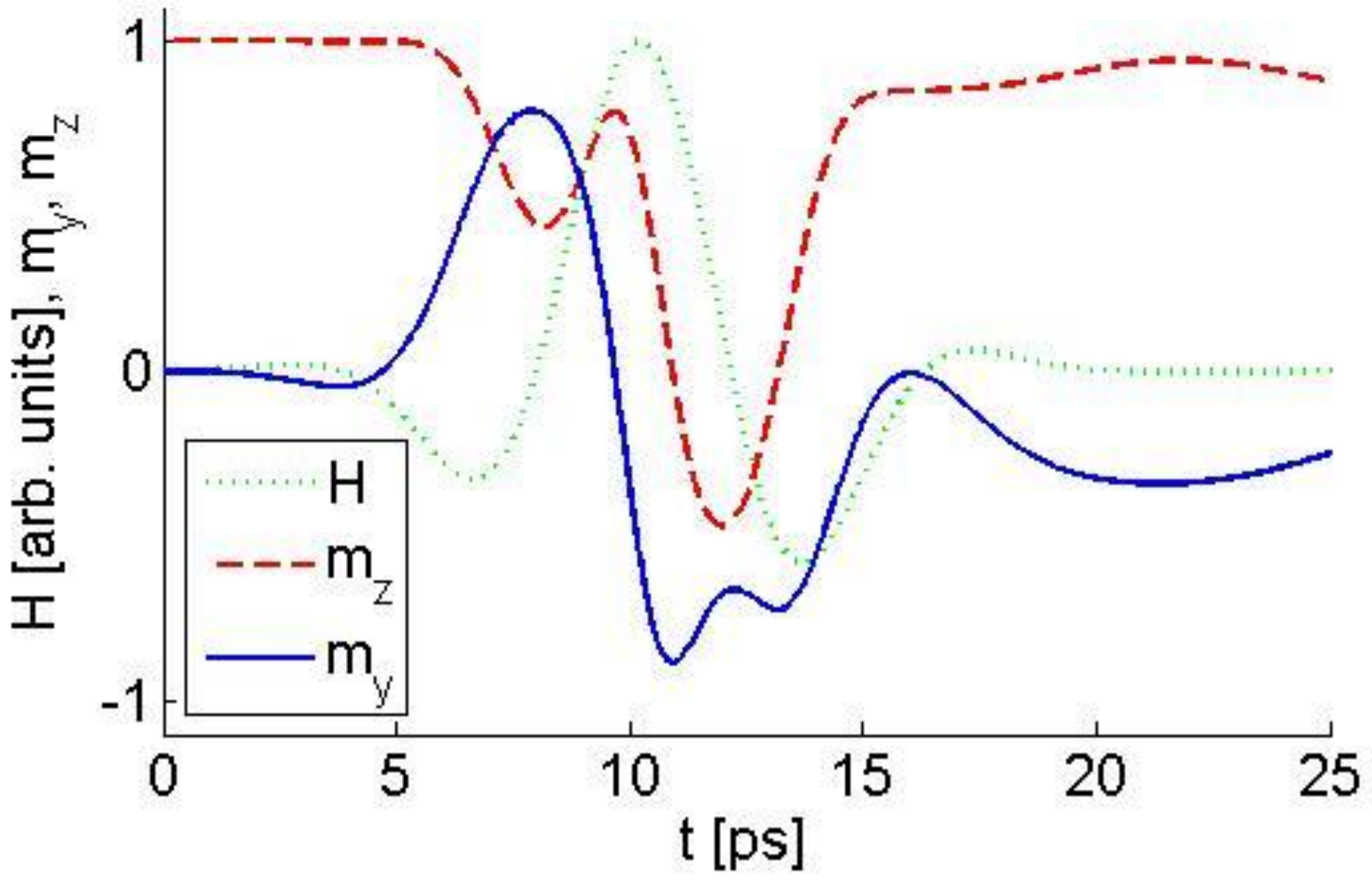

**Figure 2.6** **Effect of the crystal anisotropy on the dynamics. The triggering magnetic field of a STP along with the magnetization components is shown in a time window of 25 ps. A 5 nm-thick Cobalt film with strong crystal anisotropy was used.**

## 2.9 Magnetization dynamics: effect of pulse shape

Crystal anisotropy has been shown above to limit the final magnetization state to its axis. However, the magnetization trajectory, depicted in Fig. 2.7, shows a double magnetic switching. This double switching arises from the vectorial nature of the problem and the oscillating fields of the STP. While the positive part of the THz field switches $\boldsymbol{M}$ from $\boldsymbol{M}_z > 0$ to$\boldsymbol{M}_z < 0$, the negative part switches it backto$\boldsymbol{M}_z > 0$. Therefore, the time profile of the THz pulse becomes a critical switching parameter.

In this respect, half cycle THz pulses (HTPs) can be advantageous over STPs especially when magnetic switching applications are considered [40,41]. If we define$A^+$and $A^-$to be the positive and negative field peaks, respectively, a HTP is defined as a pulse with a high aspect ratio $|A^+/A^-|$.The propagation of electromagnetic waves enforces them to have a zero time average.

So, the difference in peaks is compensated by a corresponding difference in their semi-periods.Figure 2.8 shows a HTP where the negative tail (from t = 5 ps to t = 15 ps) is extended in time to ensure azero temporal average. To enable a consistent comparison, an HTP having a spectral content comparable to the previously introduced STP is considered in this work. The calculations in section 2.8 were repeated using an HTP instead of the original STP. The weak interaction induced by this (the temporally long and low peak field) part induces dynamics far below THz speed scales, and therefore it does not contribute to the ultrafast magnetization reversal process. The corresponding dynamics isshown in Fig. 2.9 and 2.10 where the double switching problem is then overcome. The switching was also obtained in less than 5 ps, a value that agrees with the FMR prediction introduced in a previous paragraph.

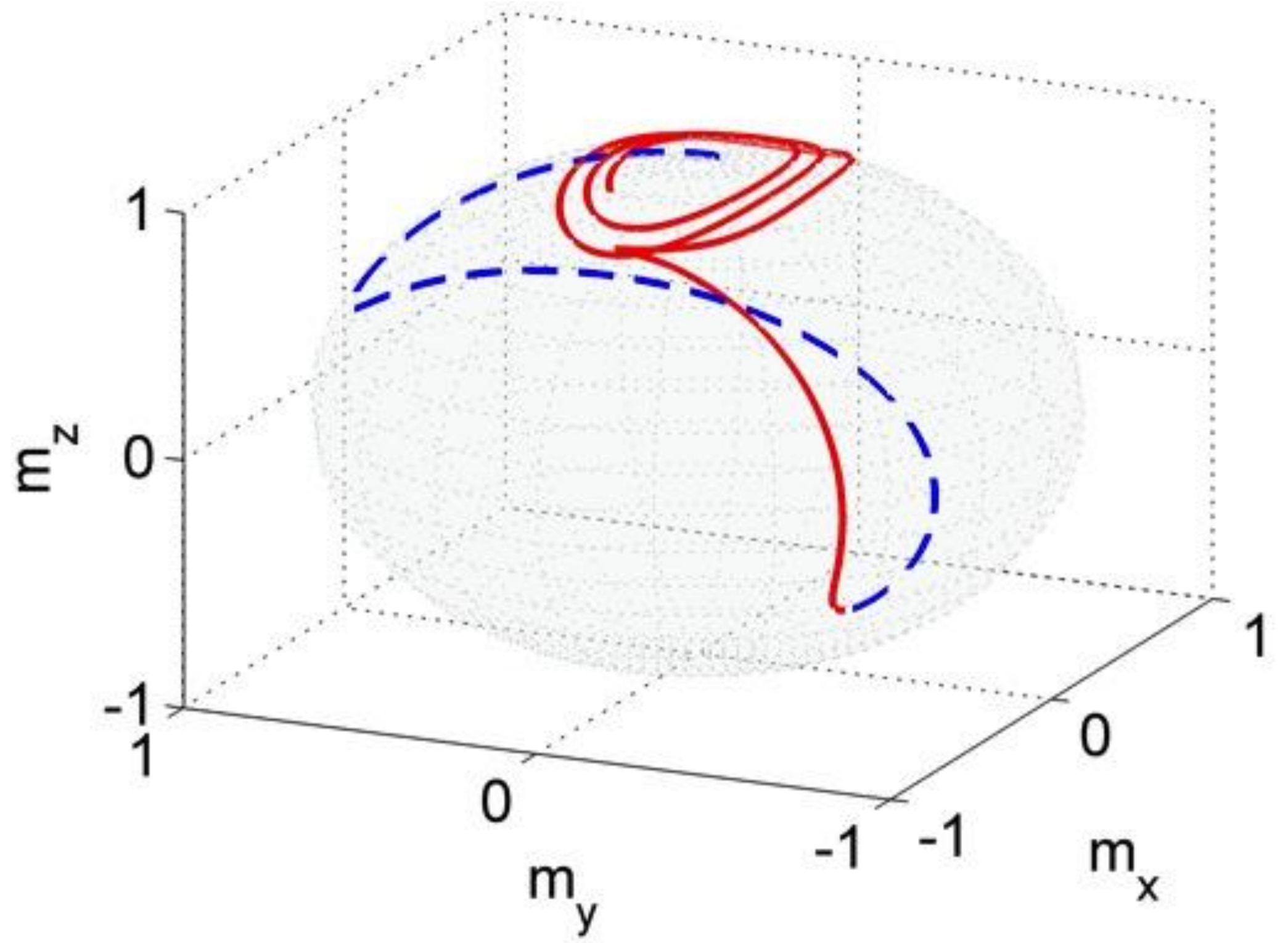

**Figure 2.7** **Effect of the crystal anisotropy on the dynamics: Magnetization trajectory. The dashed blue line shows the initial magnetic switching, which is cancelled afterwards (red solid line) because of the oscillating nature of the THz pulse.**

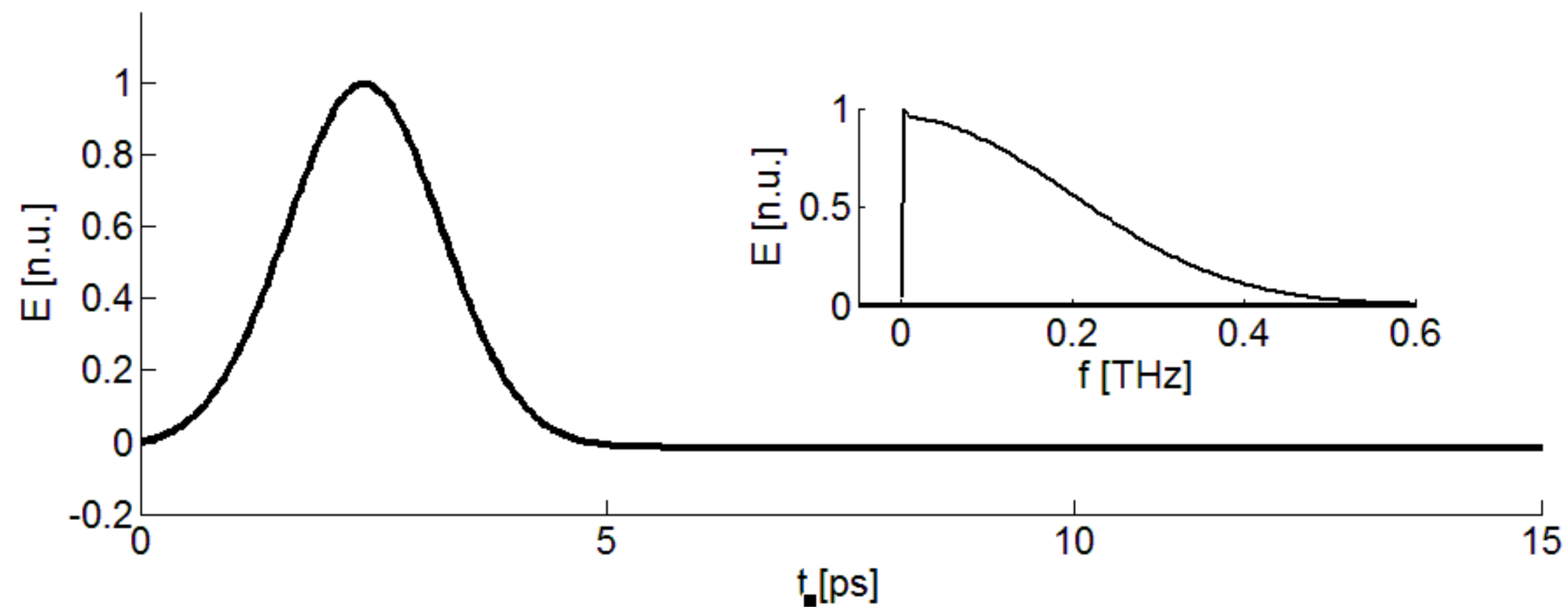

**Figure 2.8** A half cycle THz pulse. The corresponding spectrum is shown as an inset

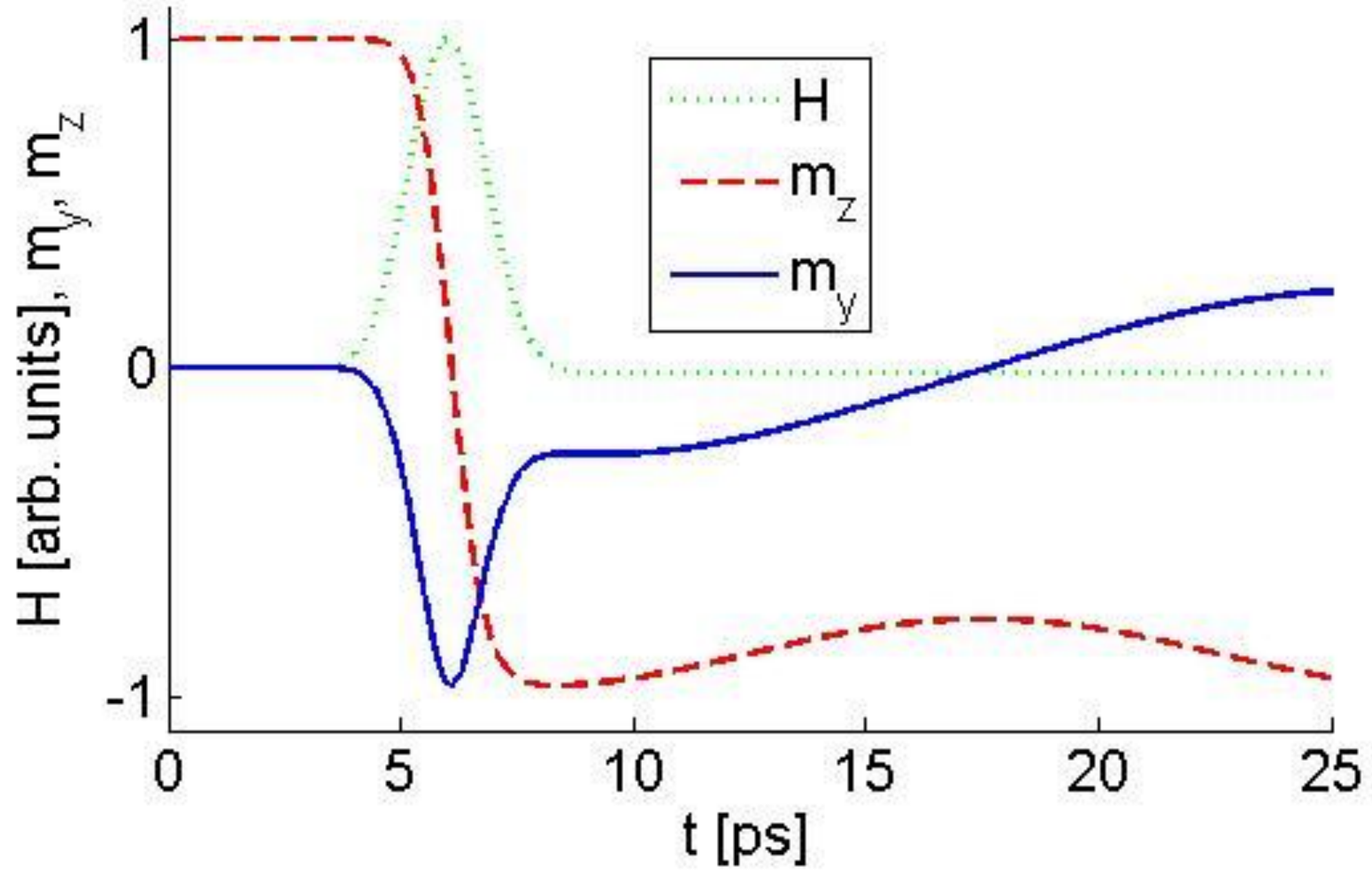

**Figure 2.9** Effect of the pulseshape on the dynamics. The triggering magnetic field of an HTP along with the magnetization components is shown in a time window of 25 ps. A 5 nm-thick Cobalt film with strong crystal anisotropy was used.

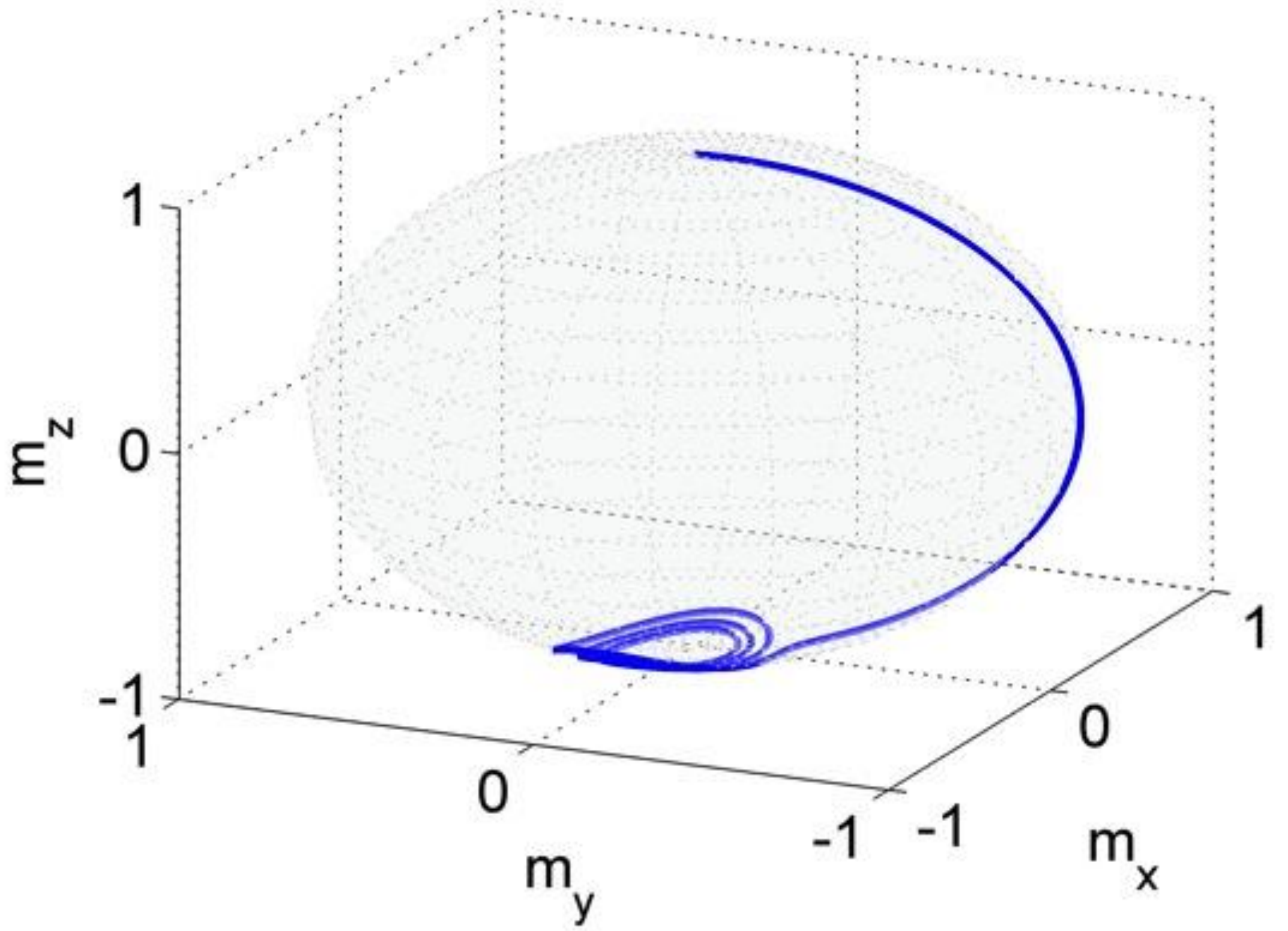

**Figure 2.10** **Effect of the pulse shape on the dynamics: Magnetization trajectory. The double switching problem observed with STP is overcome here.**

## 2.10 Terahertz probe: magnetization dynamics-induced polarization rotation

So far, the results presented here focused on the change in the magnetization state. But what can we say about propagation? In section 2.1, Luebber's results have shown that evidence of nonlinear interactions can be inferred by monitoring the transmitted THz pulse. Those nonlinear contributions are easily distinguishable at low frequencies where the temporal oscillations of the fields are much slower than the associated interaction dynamics. Moreover, at lower frequencies, the skin depth of the material is high (as it is inversely proportional to the square root of the frequency) and hence thick films (*i.e.* having significant nonlinear contribution) can be used. Those two factors suggest that looking for a similar effect, if any, on the transmitted THz pulse is not trivial. However, the magnetic nature of the problem opens up another way to probe the dynamics. Propagation of EM waves through ferromagnets leads to a frequency-dependent rotation of the plane of polarization. This rotation depends on both material properties and EM waves parameters. As mentioned in Chapter 1, two main configurations can be thought of bothin-plane and out-of-plane magnetization. The latter is a strong one. At ahigh magnetic field level (6 T), the THz pulse triggers a strong out-of-plane magnetization. In the condition of maximum THz-induced torque on the magnetic vector, $\Phi_{\boldsymbol{M-H}}$=90°, the field experiences the maximum rotation. On the contrary, at a low field intensity, the generated torque is too weak to induce significant out-of-plane dynamics. Field rotation is then associated to a mainly in-plane dynamics of $\boldsymbol{M}$. In this regime, the problem configuration is Cotton Mouton and the maximum rotation is obtained for $\Phi_{\boldsymbol{M-H}}$=45° [2].

The effect of the nonlinear interactions studied here on this rotation is investigated for the STP considered above with field intensities 60 mTand 6T, respectively. Figure 2.11 shows the induced rotation at 0.2 THz. For a 6 T pulse, the THz field is strong enough to overcome the shape anisotropy and start the magnetization dynamics. As in a typical Faraday configuration, the angle of rotation is maximized when $\boldsymbol{M}$ is parallel to the direction of propagation. As shown above, at 6 T the interaction is mediated by a significant $\boldsymbol{M}_y$ and is maximum when $\Phi_{\boldsymbol{M-H}}$=90°. In contrast, the excitation field in the latter case of 60 mT is too low to start significant magnetic interactions in the THz regime and therefore $\boldsymbol{M}_y(\boldsymbol{r}, t)$is negligible. The rotation in this case is small and is given by the in-plane magnetization components with a maximum rotation obtained when $\Phi_{\boldsymbol{M-H}}$=45°.

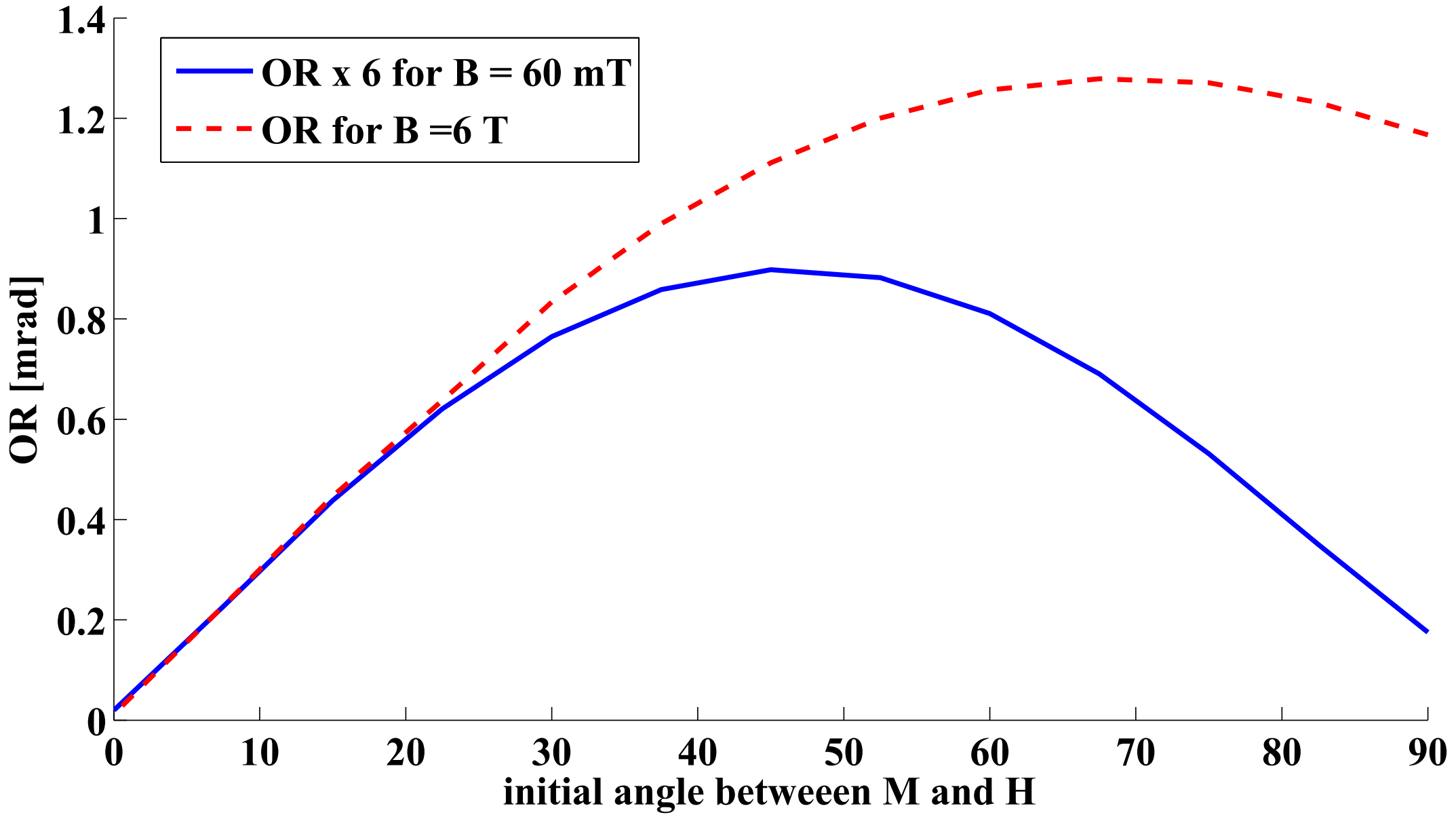

**Figure 2.11** **Magnetization-induced THz polarization rotation (OR).**

We would like to point out here that in some materials with exceptionally high crystal anisotropy [42], the coherent rotation of $\boldsymbol{M}$ can lead to the emission of EM radiation in the THz regime by way of a free induction decay mechanism of the spin system in ferromagnets which adds up to the incident THz radiation. This contribution is negligible here because polycrystalline Permalloy shows minimal crystal anisotropy.

The general description of nonlinear interactions given above can be controlled for both magnetism and THz applications. This is done by tailoring both material properties and THz parameters depending on the specific application. For example, employing the rotation of the THz polarization plane in building THz isolators requires long interaction length to obtain large angles of rotation. This can be implemented by using insulating ferromagnets where the decay of THz fields with propagation is dramatically reduced. Moreover, the choice of material with specific crystal and shape anisotropies depends on the THz power level. Assuming a normal incidence, in the case of low THz power applications, the interaction is minimal and so is the induced THz polarization plane rotation. Using a material with out-of-plane crystal anisotropy, gives $M_y = M_s$and hence, such rotationis significantlyenhanced. In contrast, this is not suitable for high power applications because the generated strong torque $\boldsymbol{M}_y \times \boldsymbol{H}$will initiate in-plane interactions, which in turn reduce$\boldsymbol{M}_y$ and therefore the polarization plane rotation.

In conclusion, we addressed in this chapter the problem of THz magnetization dynamics. We used theoretical predictions along with time domain simulations to prove that THz waves, under certain conditions can be used to nonlinearly trigger magnetization dynamics. However, we found that this requires intense and properly shaped THz pulses. In chapter 3, we experimentally address this. Finally, we found that THz can probe magnetic systems. This also suggests that magnetic materials can be used to control THz waves. This finding was the motivation for the developments of magneto-THz devices, as it will be experimentally shown in Chapter 4.

# 3 HARNESSING TERAHERTZ PULSES FOR NONLINEAR MAGNETIC EXPERIMENTS

The calculations of THz magnetization dynamics reported in Chapter 2 drew the attention to the main requirements for this kind of experiments: properly shaped pulses and high fields. In this chapter, we experimentally address them. The first section is concerned with a novel technique for THz pulse shaping, whereas in the following parts the development of suitable approaches towards intense THz radiation will be described in details. Specifically the results of the investigation in plasmonic THz field enhancement (section 3.2) and the successful up-scaling of the generation efficiency of plasma-based THz sources (section 3.3) will be presented.

## 3.1 Temporal and spectral shaping of terahertz pulses

Several classes of nonlinear field-matter interactions depend in general on the envelope and spectral contents of the electromagnetic field involved [15,42,43]. However, in some important cases at THz frequencies, the nonlinear interaction process is field dependent, *i.e.* it is affected by the temporal field waveform. Hence, the output field shape from the THz sources needs to be tailored to suit the experimental requirements. As introduced in Chapter 2, field induced magnetization dynamics is a relevant example where a half cycle THz pulse (a pulse with a significant asymmetry between the amplitudes of the positive and negative parts of the oscillating fields) is required to gain control on the process. In addition, properly shaped THz pulses were shown to coherently guide ions over a collective microscopic path in Ferroelectric materials [44].

In this section, a novel technique to shape the temporal and spectral profiles of the THz pulseis presented. This technique depends on nonlinear photo-excitation of free carriers in semiconductors by means of the optical pump–THz probe spectroscopy introduced in Chapter 1.

### 3.1.1 Terahertz pulse shaping techniques

A number of techniques have been proposed for THz pulse shaping through (i) the manipulation of the generation process [45-48] or (ii) the linear filtering of a freely propagating THz beam in masks and waveguides [49-51].The former has been mainly addressed viathe generation with photoconductive antennas and (narrow band) periodically poled Lithium Niobate. However, these sources are typically weak and not suitable for nonlinear THz experiments. On the other hand, linear filtering mainly affects the spectrum amplitude with no direct control on the THz field waveform.

### 3.1.2 Photo-excitation of semiconductors

As introduced in chapter 1, a fundamental quantum-mechanical process in semiconductor is the single-photon absorption that depends on the creation of free carriers promoted by the absorption of a photon having energy larger than the band gap. Under illumination with an intense optical field, a significant number of free carriers can be created. It is important to recall here, thatthe transmitted THz field ($E_t$) through a photo-excited semiconducting layer of thickness $d$ much smaller than the THz wavelength is given by$E_t = (2Y_0E_i - Jd)/(Y_0 + Y_s)$where $E_i$ is the incident field. $Y_o$and$Y_s$ are the admittance of free space and sample respectively. $J = nev$is the electron current density,while$n$, $v$, and $e$arethe electron density, charge, and velocity,respectively [33]. An increase in the charge density upon optical excitation is thus accompanied by a corresponding increase in the current density and by the consequent attenuation of the transmitted THz pulse. If the optical intensity is strong enough, the THz is totally shielded. However, if the pump consists of a short optical pulse and the probe/pump delay $\tau$ (when $\tau > 0$ the pump impinges before the THz probe transmission) is carefully selected, significant shaping of the THz time profile can be obtained as shown here. In other words, the induced-carriers transition curve temporally attenuates the broadband THz pulse.

### 3.1.3 Gabor-limit and the speed of temporal shaping

A 1 THz-centered pulse has time oscillations on the scale ofpicoseconds. Let usimagine now that we want to isolate a half cycle of the THz waveform via a fast photo-excitation process. Towards this goal, short (femtoseconds) optical pulses induce a seemly instantaneous photo-excitation. We then assume that the only relevant processes involved are the reflection and absorption induced by the free carriers' layer. It is noteworthy to mention that although the probe pulse is the main mean to trigger the phenomenon, the observedtemporal resolution is inherently limited by the THz probe bandwidth. It is not possible to observe clipping dynamics faster than the THz wave period. This observation (stemming from information theory) is known as Gabor limit and it is a fundamental consequence of the relation between time and spectrum in causal systems.

As a result, in a transform-limited pulse (*i.e.* single cycle), no significant reshaping or change of the spectral contents should be perceived for a given charge-induced attenuation.To produce a significant reshaping of the pulse spectrum, we deliberately induce significant chirp in the THz pulse. The pulse is then stretched in time and different spectral contents of the pulse are localized around different points on the temporal axis. In this way, the time-synchronized carrier-induced modulation presented here can affect the spectral distribution.

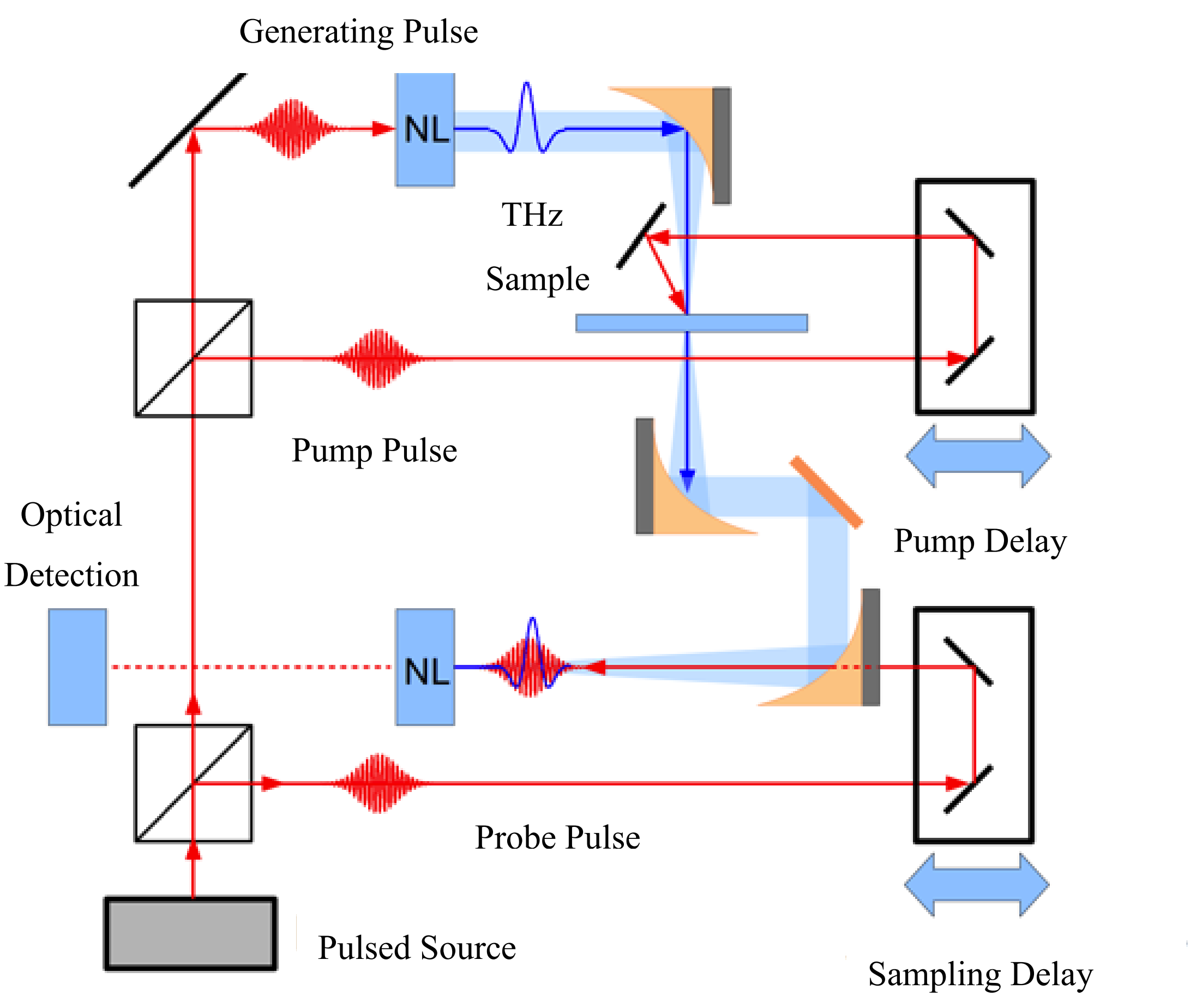

**Figure 3.1** **Optical-pump / THz-probe setup.Characterization setup for THz temporal pulse shaping. "NL" refers to the nonlinear crystal (ZnTe).**

### 3.1.4 Temporal shaping

Measurements in this partwere performed using a time resolved optical pump-THz probe technique (see Chapter 1 for details). Figure 3.1 shows a schematic diagram of the setup where the energy of a 35 fs pulse train (centered at a wavelength of$\lambda$=800 nm and with a repetition rate of 2.5 kHz) is split between the optical pump, the THz generation, and the THz detection beamlines. Generation and detection were performed using optical rectification and electro-optical sampling, both performed in a ZnTe crystal.

Specifically, we induced a photo-excited layer on the surface of a 2 mm-thick high resistivity silicon wafer. The waveform and the spectrum of the THz pulse transmitted through the silicon waferare shown in Fig. 3.2(a).The three main peak amplitudes are designated as $A_1$, $A_2$, and $A_3$. $\tau$is the delay between the THz probing signal and the optical pump where $\tau = 0$ corresponds to $A_2$being reduced by a factor of $\sqrt{2}$. In Fig. 3.2 (b), the sample is shined with a 695 μJ/cm$^2$ optical pulse arriving significantly later than the THz pulse. As we operate in a strong carrier excitation regime, a weak THz pulse attenuation is observed because the carrier lifetime is comparable to the period of the pump pulse train.

The waveform shown here is the typically generated THz pulse from optical rectification of femtosecond pulses in a ZnTe crystal. A highly asymmetric pulse, *i.e.*with $|A_1/A_2| >> 1$ and $|A_1/A_3| >> 1$ is hard to achieve by simply shaping the generation. However, as introduced above, it is fundamental in an important class of THz-nonlinear interactions [51,52]. For example, when an oscillating symmetric pulse is to be used to trigger the magnetization switching in a magnetic material, the effective torque on the magnetic moment depends on the sign of the applied magnetic field. As a result, a field-symmetric pulse produces a weak net switching effect, and often leads to a cancelation of the magnetic switching or a non-deterministic magnetizationdynamics. Such field asymmetry is obtained here by varying $\tau$in such a way that $A_2$ and $A_3$ are gradually reduced.

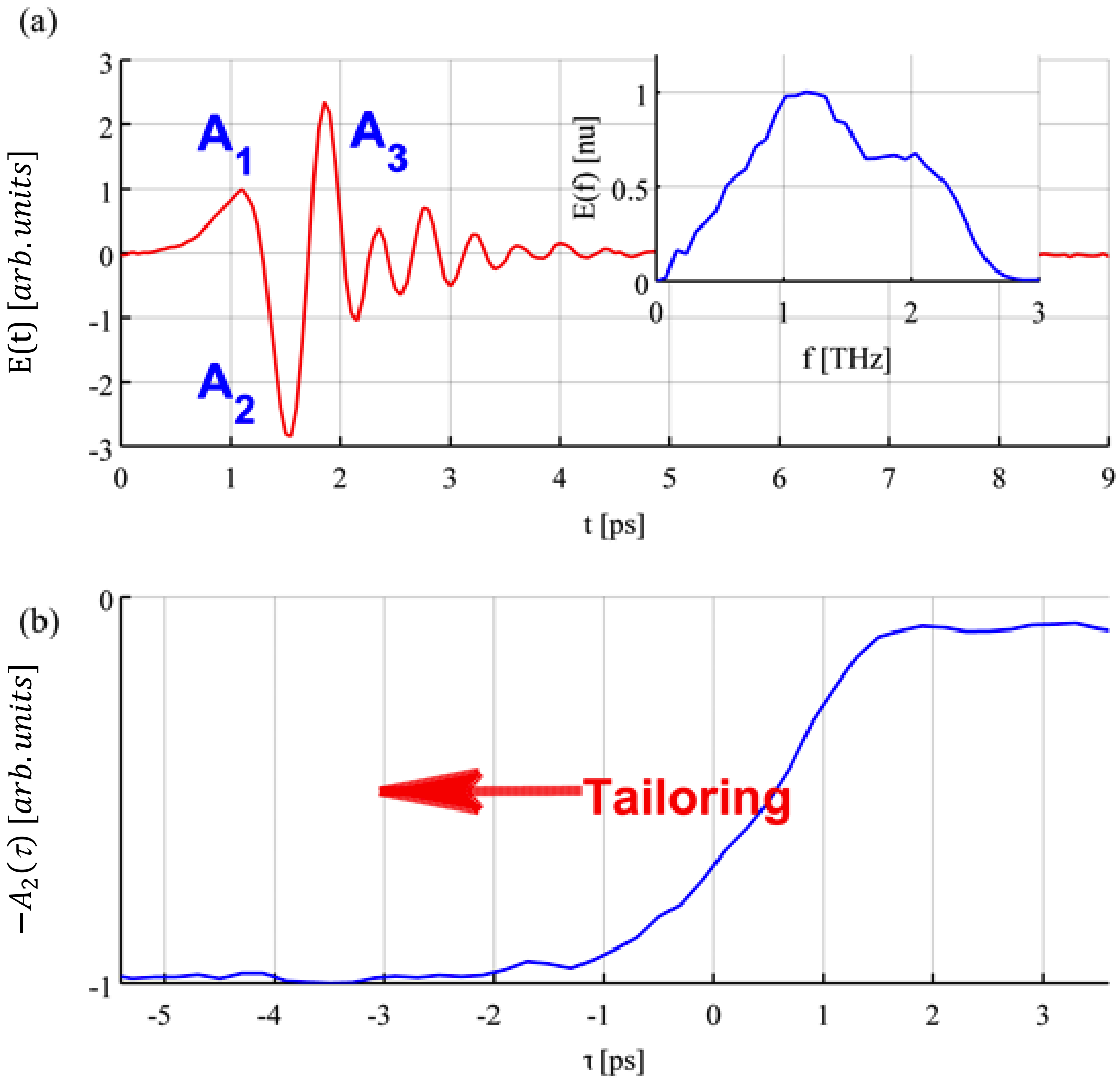

**Figure 3.2** **(a) Temporal profile and spectrum of the transmitted THz pulse through a silicon wafer. The THz arrives before the optical pump. (b) Optically-induced THz transition in silicon. The THz negative field peak ($A_2$) is scanned against the THz probe-optical pump delay $\tau$.**

Figure 3.2(b) shows the amplitude transition curve where the peak $A_2$ is shown against the delay $\tau$ over a period of 9 ps. $A_2$ is attenuated by > 90% when the THz arrives right after the optical pump. Pulse asymmetry is thus expected to change as $\tau$ is varied close to the temporal overlap between the probe and the pump. The asymmetry build-up process is illustrated in Fig. 3.3 where the original and temporally shaped pulses are shown (in normalized units) for different τ points.

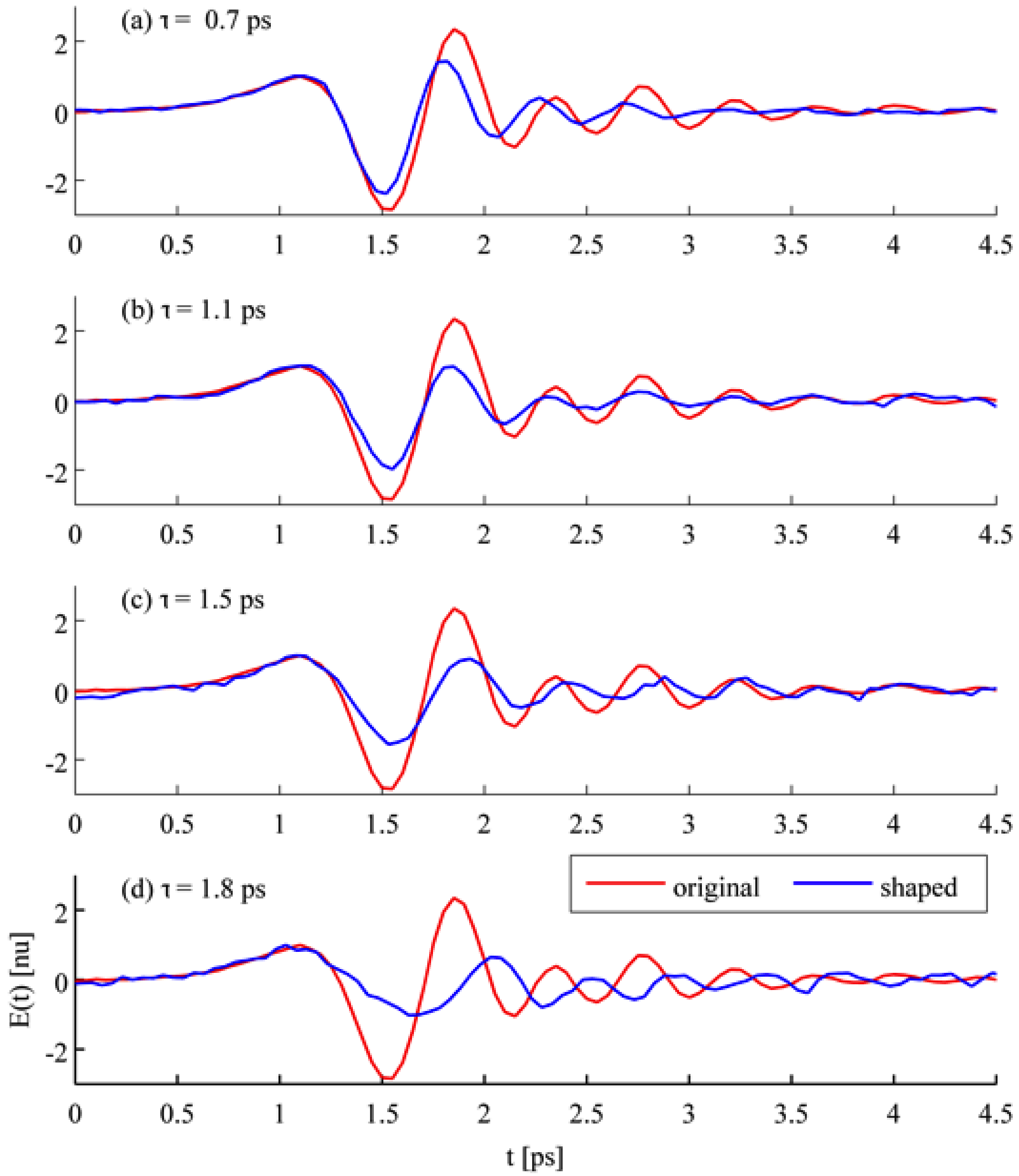

**Figure 3.3** **The transmitted THz amplitude with (blue) and without (red) optical excitationfor different probe/pump 0.7; 1.1; 1.5; 1.8 ps. The plots are normalized to the amplitude of the first peak.**

As the delay of the pump is reduced, it starts to overlap with the THz pulse. Consistently, $A_3$ is attenuated firstand then $A_2$ follows. In switching experiments, there is usually a threshold field, below which the field becomes too weak to overcome the switching inertia and reverse the process. Therefore, a high contrast between $A_1$ and $A_{2/3}$ is targeted. To evaluate the efficiency of the temporal shaping process, we define the peak field modulation (shaping) depth as

$$M_{i1} = \frac{(\mathring{A}_{i1})^2 - (A_{i1})^2}{(\mathring{A}_{i1})^2}, \quad i = 2,3 \tag{3.1.2}$$

where $A_{21}$ and $A_{31}$ are the asymmetry factors given by $A_{21} = A_2/A_1$ and $A_{31} = A_3/A_1$, respectively. Circle superscripts denote the quantities calculated on the original unmodulated pulse. The reason we chose the square of the field in this calculation is that in magnetization dynamics for example, the field-induced torque depends on the perturbation of the total energy with respect to the magnetization vector (Eq. 2.2.1). The waveform reshape, shown in Fig. 3.3 for different $\tau$, is mapped using Eq. (3.1.2) into $M_{21}$ and $M_{31}$ in Fig. 3.4, where modulation depths as high as 87% and 92%, are shown. Such high modulation corresponds to a strong asymmetry being introduced in the THz pulse.

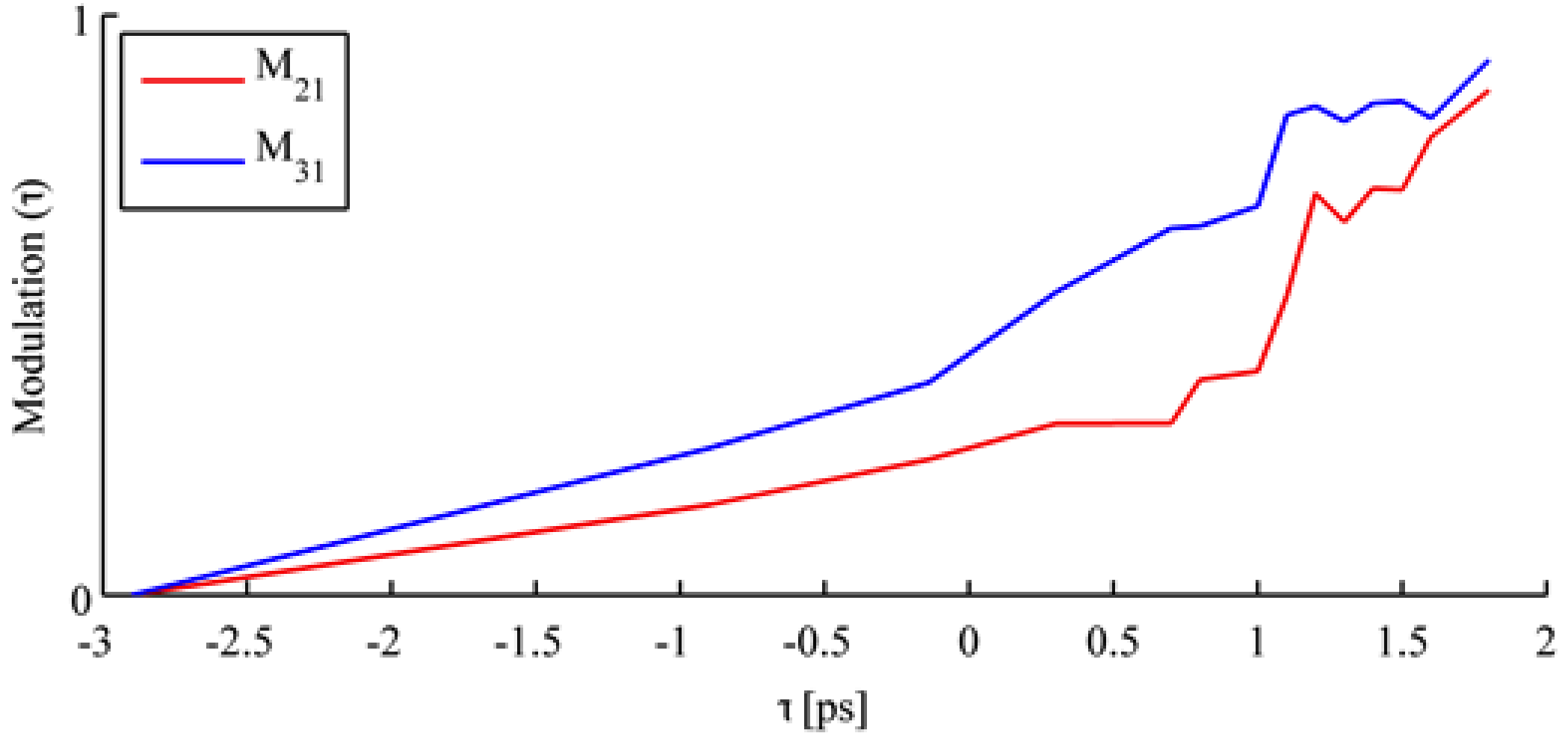

**Figure 3.4** **Temporal modulation depth. The photo-induced asymmetries in $A_{21}$ and $A_{31}$ are shown as a function of the delay $\tau$.**

The pulse temporal asymmetry depends also on the impinging fluence that leads to different electron current densities and different THz attenuations. Figure 3.5 shows the transition curve of $A_2$ for different optical fluences. The latter can be combined with the delay to achieve a marked manipulation of pulse asymmetry.

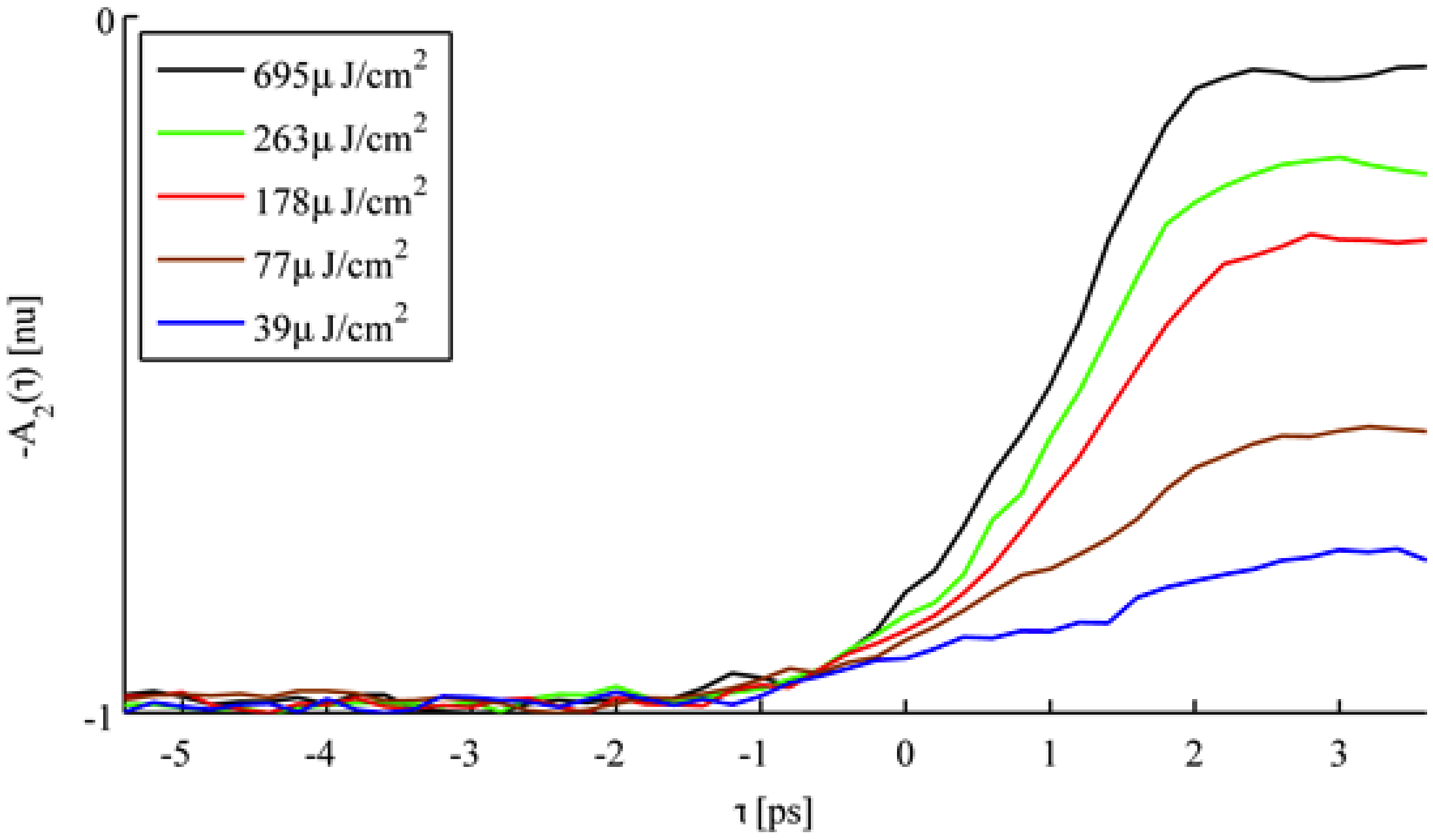

**Figure 3.5** **The effect of optical fluence at the semiconductor surface.The THz transition curves are shown for different intensity levels.**

The temporal shaping of a close-to-transform-limited pulse using the technique shown aboveintroduces very marginal changes in the spectral contents of the pulse. THz pulses, generated using optical rectification in ZnTe, are in general slightly chirped due to the chromatic dispersion of the phase mismatch between the THz pulse and the optical pump in the rectification process. During the temporal shaping process,measurable but small spectral changes were introduced.

### 3.1.5 Spectral shaping

As highlighted before, exploiting a similar temporal technique on highly chirped THz pulse allows for significant shaping of the spectrum. The spectral shaping process consists of two steps (i) the pulse chirping and (ii) time-resolved optical-pump / THz-probe of silicon. We chirped the pulse exploiting its propagation in a dispersive waveguide consisting of a copper tube, 278.3 mm-long and with an inner diameter of 27.1 mm (Fig. 3.6). We estimated its group velocity dispersion to be anomalous at 1 THz, $\beta_2$ =-1.8 ps$^2$/m (we neglect higher order dispersion contributions).

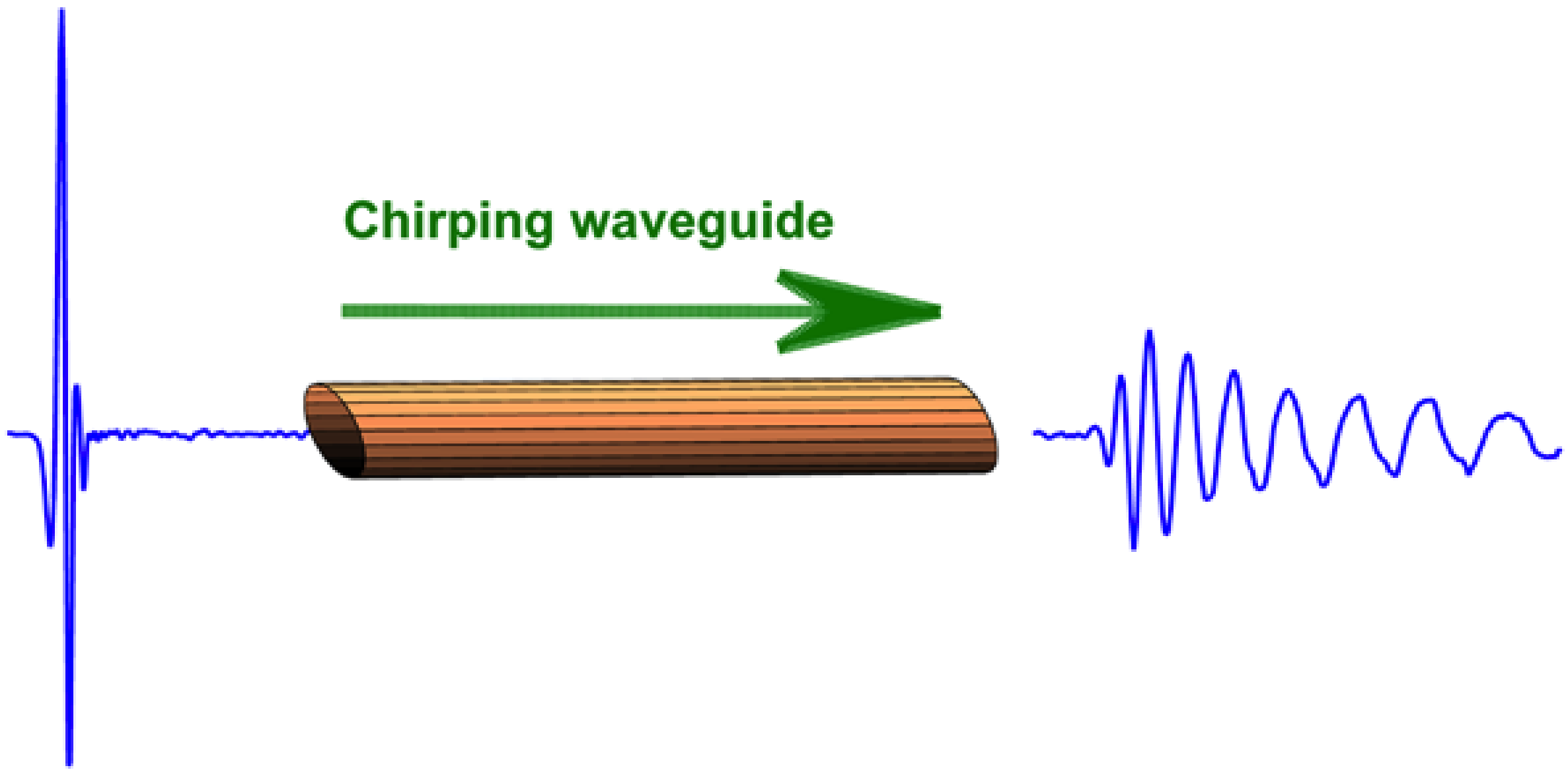

**Figure 3.6** **Chirping of the THz pulse. A hollow copper waveguide is used to introduce anomalous THz pulse chirping.**

Because of the negative chirp of the THz pulse, low frequency componentsare delayed with respect to the higher frequency components. As $\tau$ increases, the output spectrum loses its low frequency components. Figure 3.7 shows a map of the transmitted pulse as a function of the probe-pump delay $\tau$. The reference $\tau = 0$ was arbitrarily selected. These spectral changes are shown in the corresponding spectral map of Fig. 3.8.Finally, a comparison is shown by plotting the time traces and spectra of the THz pulses before and after the excitation, at four delay points in Fig. 3.9.

As shown in Fig. 3.8 and clearly pointed out in Fig. 3.9, the THz spectrum progressively undergoes shaping as the delay increases. The change in the spectrum is clearly appreciated in Fig. 3.1.10(a) where both the spectrum center ($f_2$) and the FWHM edges' frequencies ($f_1$ and $f_3$) are shown. All the spectra shown in this figure are in normalized units. As shown, this process is accompanied by (i) narrowing of the FWHM bandwidth and (ii) pushing the spectrum center up in frequency. Finally, the spectral modulation parameter $M_f = (f_2 - f_2^{\circ})/f_2^{\circ}$ s shown in Fig. 3.10(b) where modulations as high as 52% were obtained.

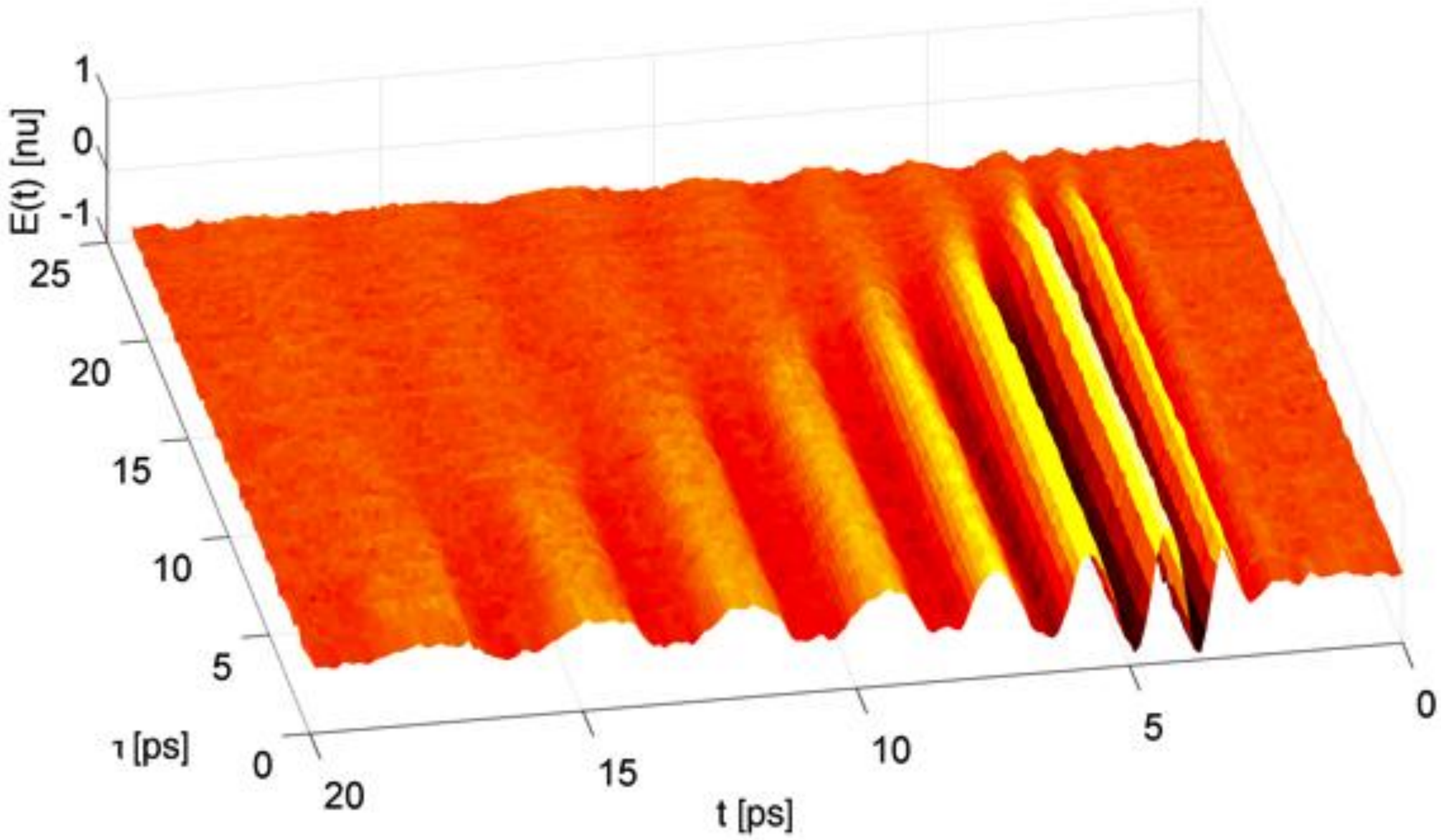

**Figure 3.7 Terahertz spectral shaping shown in the time domain.**

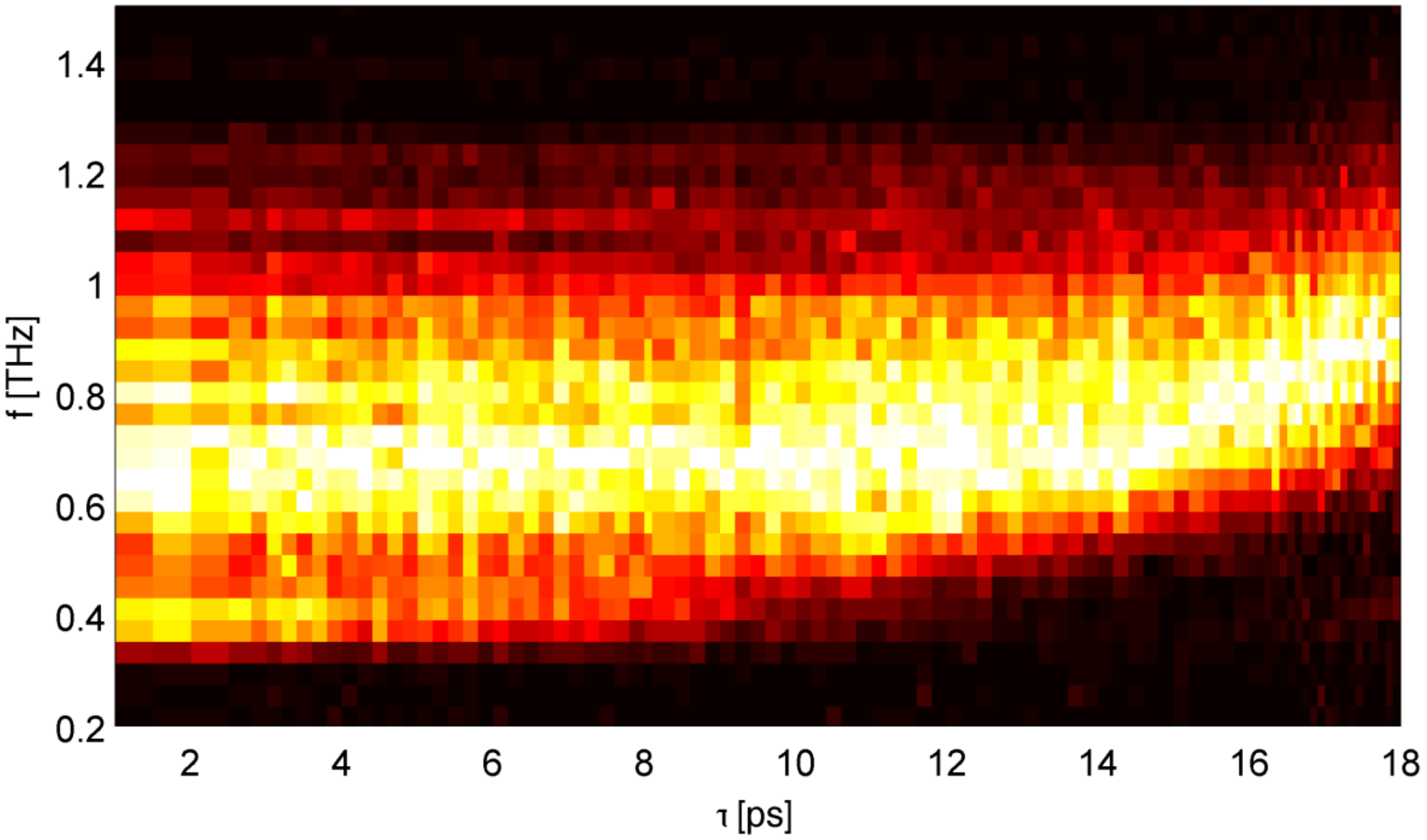

**Figure 3.8 Terahertz spectral shaping vs the probe / pump delay.**

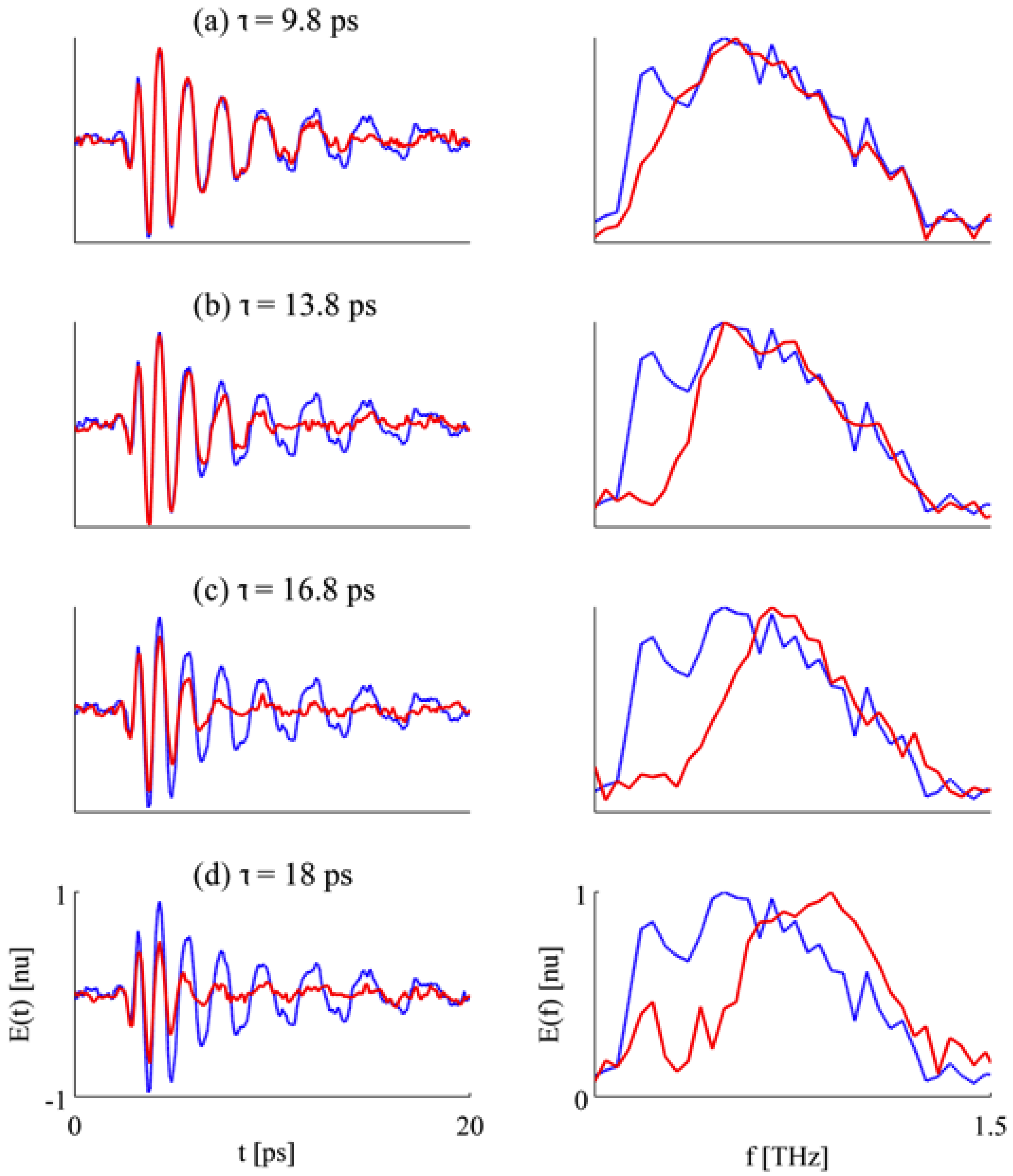

**Figure 3.9** **Spectral shaping of the THz pulse shown at delays of 9.8 ; 13.8 ; 16.8; 18 ps in the time and frequency domains, respectively.**

It should be noted that for a p-polarized THz impingingon the sample ata Brewster angle, the optically-inducedreflection exhibits the complementary effect, *i.e.* the spectrum median is downshifted. Alternatively, low-pass filtering can be achieved by positively chirping [53] the input pulse.

To conclude this part, we have shown that optical pump-THz probe of carriers in semiconductors can be used to shape the THz pulse. Although the main objective was to introduce significant asymmetry in the temporal profile of the THz pulse, the same method was extended for THz spectral shaping after chirping the THz pulse.

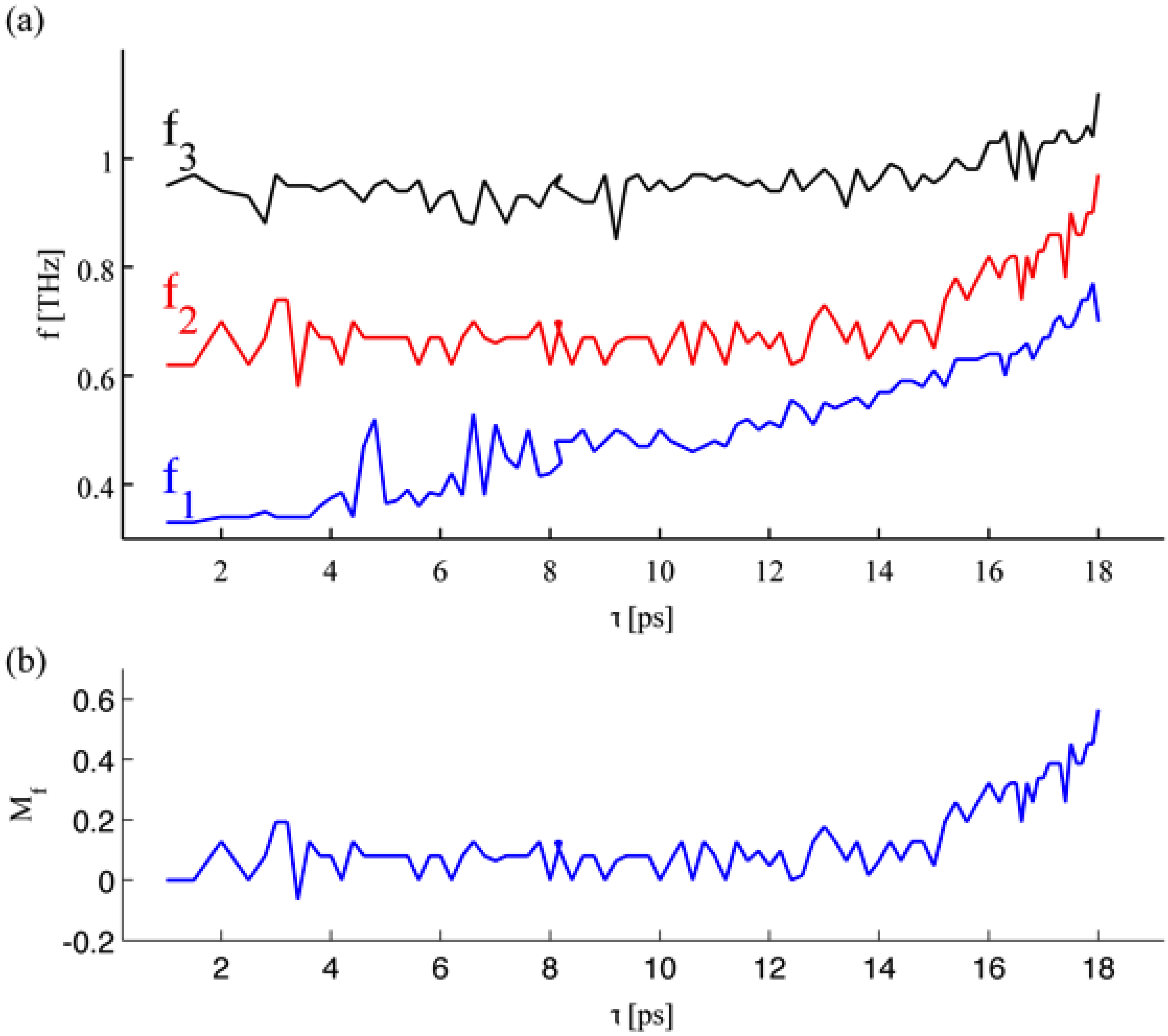

**Figure 3.10** **Spectral modulation. (a) Trace lines of FWHM frequency edges and the spectrum median. (b) Spectral modulation parameter shown for different delay points.**

## 3.2 Concurrent high terahertz field enhancement and transmission in nanoslits arrays

Transmission and enhancement properties of electromagnetic waves in sub-wavelength apertures have drawn much interest since the first report on the extraordinary optical transmission in sub-wavelength hole arrays [22,54]. In the THz regime, field enhancement is of a particular importance due to the technological limitations on the peak THz fields practically achievable from current sources which limit the realization of nonlinear THz experiments as highlighted in Chapter 2. Seo*et al.* reported a field enhancement factor $F$ approaching 780 at 0.1 THz using a single 70 nm-wide slit etched in a freestanding 70 nm-thick gold sheet [23]. In their work, a slit is described as a nanocapacitor being loaded with the charge collected by the surrounding metal surface. An impinging THz radiation on the metal sheet induces currents, which create transients of charge imbalance across the nanogap. This imbalance leads to a field enhancement inside the gap that scales as $1/f$ where $f$ is the light frequency. Despite the high field enhancement obtainable from a single slit, such a structure suffers from a low field transmission $T$. Nonlinear THz sources havegenerally low repetition rates and the associated TDS detection typicallyoffers low signal to noise ratios. Hence, the use of low transmission structures in many experimental configurations is unpractical.

In the following part, we will show the design and the experimental demonstration of1D arrays of nanoslits featuring high $T$ while preserving high F in the frequency range of 0.2–2.7 THz.

### 3.2.1 The challenge of highly transmissive highly enhancing nanoslits structures

A straightforward approach towards high transmission is to increase the number of nanoslits[55,56]. However, in general this may come at the expense of field enhancement as this action physically alters the geometry that supports plasmonic enhancement. As mentioned above, the accumulated charge across the nanogap depends on the induced current over the metal surface, which suggests a finite charge collection area around the gap. As the spacing between the slits decreases beyond a certain limit, neighboring slits start to share charges,which leads to a decrease in the enhancement. It is then important to tune the geometrical parameters to obtain reasonable values of both $F$ and $T$ by optimizing the array parameters.

### 3.2.2 Structure and design approaches

Figure 3.11 shows a schematic diagram of a nanoslits array with the four geometrical parameters l, a, h, and $d$ representing the slits length, width, spacing and film thickness, respectively.

We assume infinitely long slits (l= 2 mm) when compared to our THz spot size (1.2 × 1.2 mm at the focus: full width at $1/e^2$ value).The response of the structure can be described by a non-resonant $1/f$ response following the approach of Sao *et al.*[23]. Two design approaches are considered here: numerical and analytical. For the numerical one, both the time-dependent and the frequency-dependent simulations [57] of the wave propagation are used to calculate the near field (inside the nanoslits) and far field (at the detector) for different values of the design parameters. The analytical approach relies on the calculation of the far field transmission from the nanoslits arrays following the formalism introduced in [55]:

$$T = \frac{i2\mathrm{a}}{d\,sin(k\mathrm{h})[1+(\mathrm{a}/d)^2]+i2\mathrm{a}\,cos(k\mathrm{h})} \quad (3.2.1)$$

where k is the wavevector of the electromagnetic wave. The transmission is related to the enhancement F by the geometrical coverage ratio (filling factor) through

$$F = T/\beta \quad (3.2.2)$$

Where $\beta = \mathrm{a}/d$ is the filling factor.

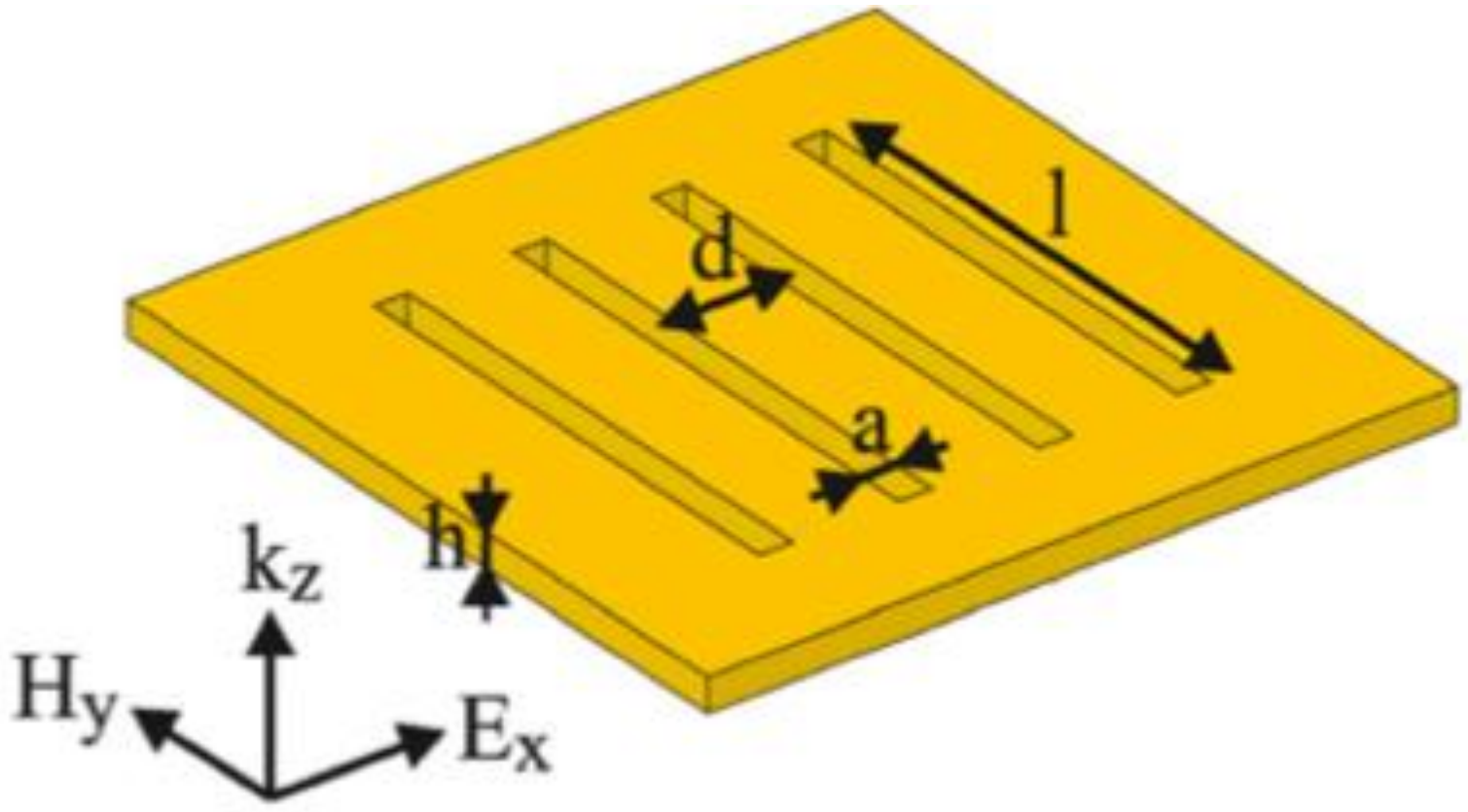

**Figure 3.11 Schematic diagram of a nanoslits array showing the geometrical design parameters.**

### 3.2.3 Parametric dependence

In general the field enhancement increases as the slitwidth a decreases, down-limited by the charge-screening length (below 1 nm) [23]. In practice, the effective threshold is given by the minimum feature size that can be achieved in the fabrication process. Figure 3.12 shows the enhancement factor for different slit widths in a freestanding gold film (at $d$ = 100 μm and h = 60 nm). We managed to fabricate a slitwidth of 40 nm using electron beam lithography as will be shown later. For a slits length of 2 mm, this resulted in rectangular apertures with extreme ratios of 50,000.Decreasing the thickness has also the effect of enhancing the transmission due to the reduced absorption experienced by the plasmonic wave.

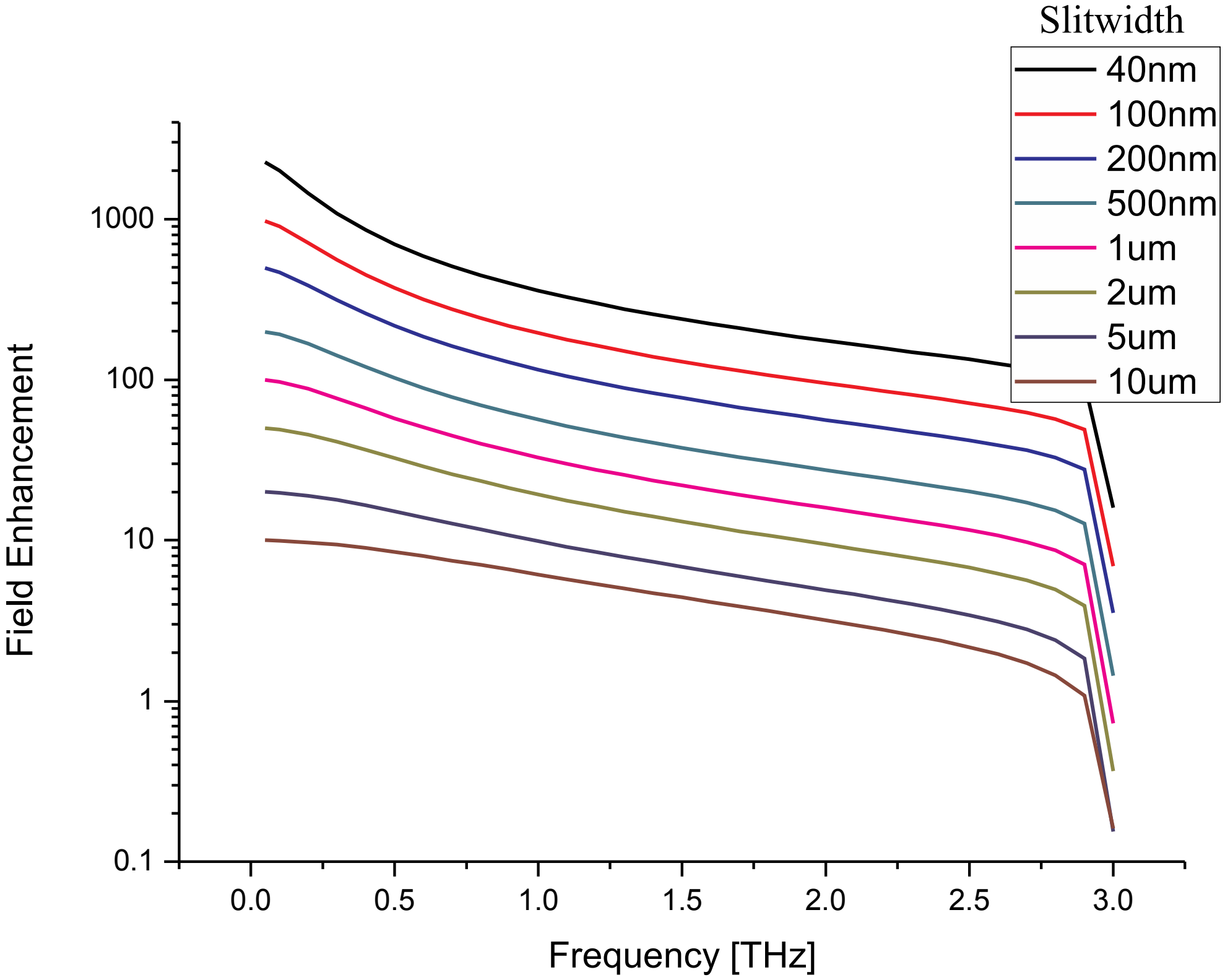

**Figure 3.12** **Effect of slit width on the enhancement factor. The slits periodicity and film thickness were fixed to 100 μm and 60 nm, respectively.**

Figure 3.13 shows the enhancement factor for different film thickness in a freestanding gold film (at $d$ = 100 μm and a = 40 nm.) We choose h= 60 nm which is comparable to the gold skin depth at 1 THz (70 nm) to reduce the direct transmission of the THz through the unpatterned gold.

Based on the results presented above, we chose a = 40 nm and h =60 nm. Those two numbers were selectedby solely looking for maximum enhancement. Deviation from those numbers will change the transmission, but will dramatically decrease the enhancement factor as it is clear from Fig. 3.12 and 3.13. The critical design parameter is the spacing $d$. In this part, we used both the analytical model and Eq. 3.2.1 and 3.2.2 in comparison with the numerical simulations [58]. The calculated $F$ and $T$ are shown in Fig. 3.14 for different periodicities.

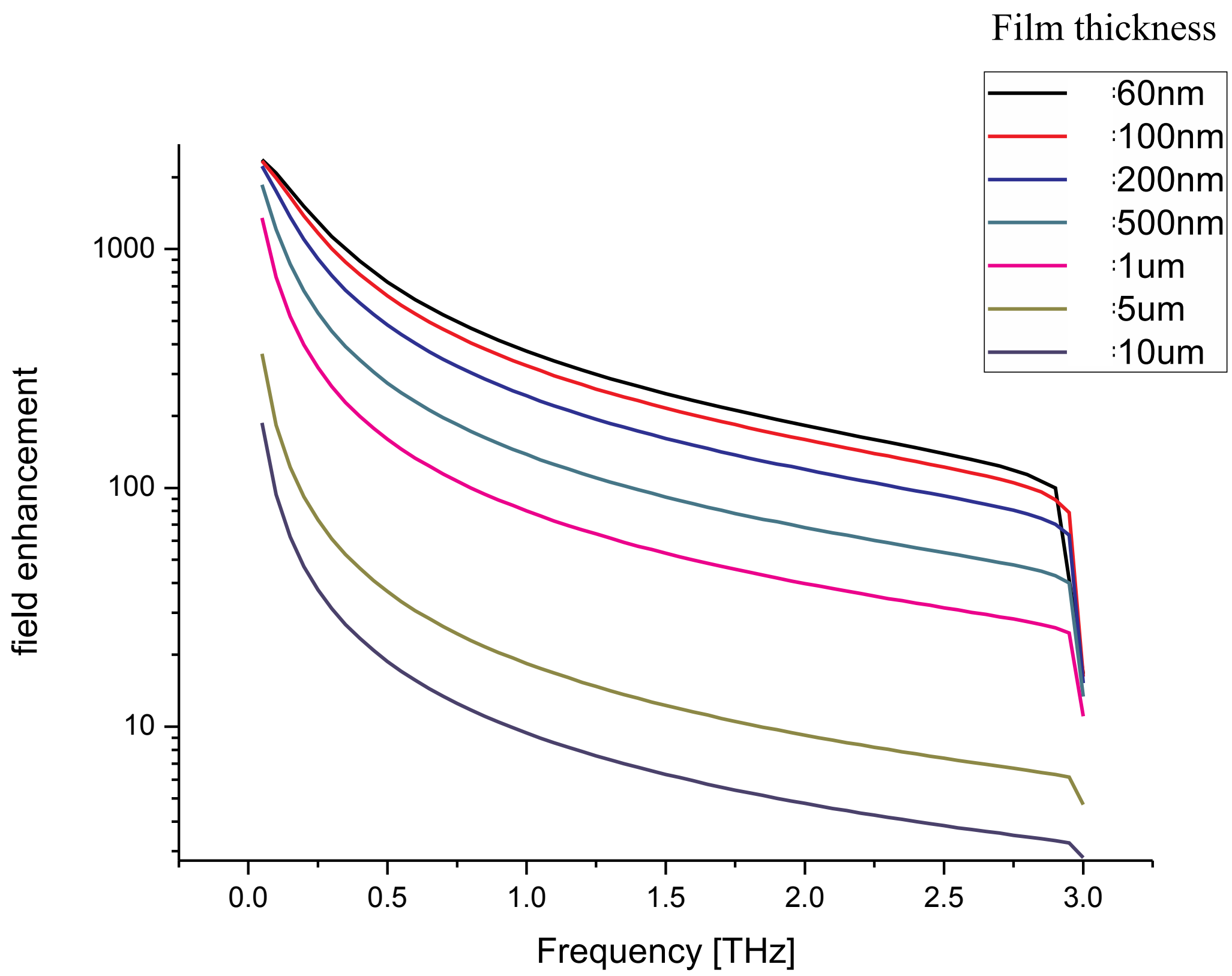

**Figure 3.13 Effect of film thickness on the enhancement factor. The slits periodicity and width were fixed to 100 μm and 40 nm, respectively.**

Direct comparison between the predictions of the analytical and the numerical modelsshows a very good agreement. However, we depend mainly on the numerical calculations here because (1) it give slightly more accurate results given that we are able to model the dielectric response of the gold film using the Drude model.

The Drude model provides more accurate predictions than the simple assumption of aperfect conductor, made in the analytic model (3.2.1) and(3.2.2). In addition the substrate effect can be easily incorporated in the numerical model.

Fig. 3.14 clearly shows the intuitive picture of $T$ versus $F$ presented in section 3.2.1. As the slit periodicity decreases, $F$ decreases but $T$ increases, with a more pronounced variation at high frequencies. Two frequency lines are taken down to Fig. 3.15(a) and (b) where the $T-F$ plot is shown for 0.5 and 1 THz, respectively. Above a certain periodicity, the enhancement does not seem to change much with the geometry. Considering the case of 0.5 THz, for example, a periodicity of 20-30 μm is therefore a suitable choice.

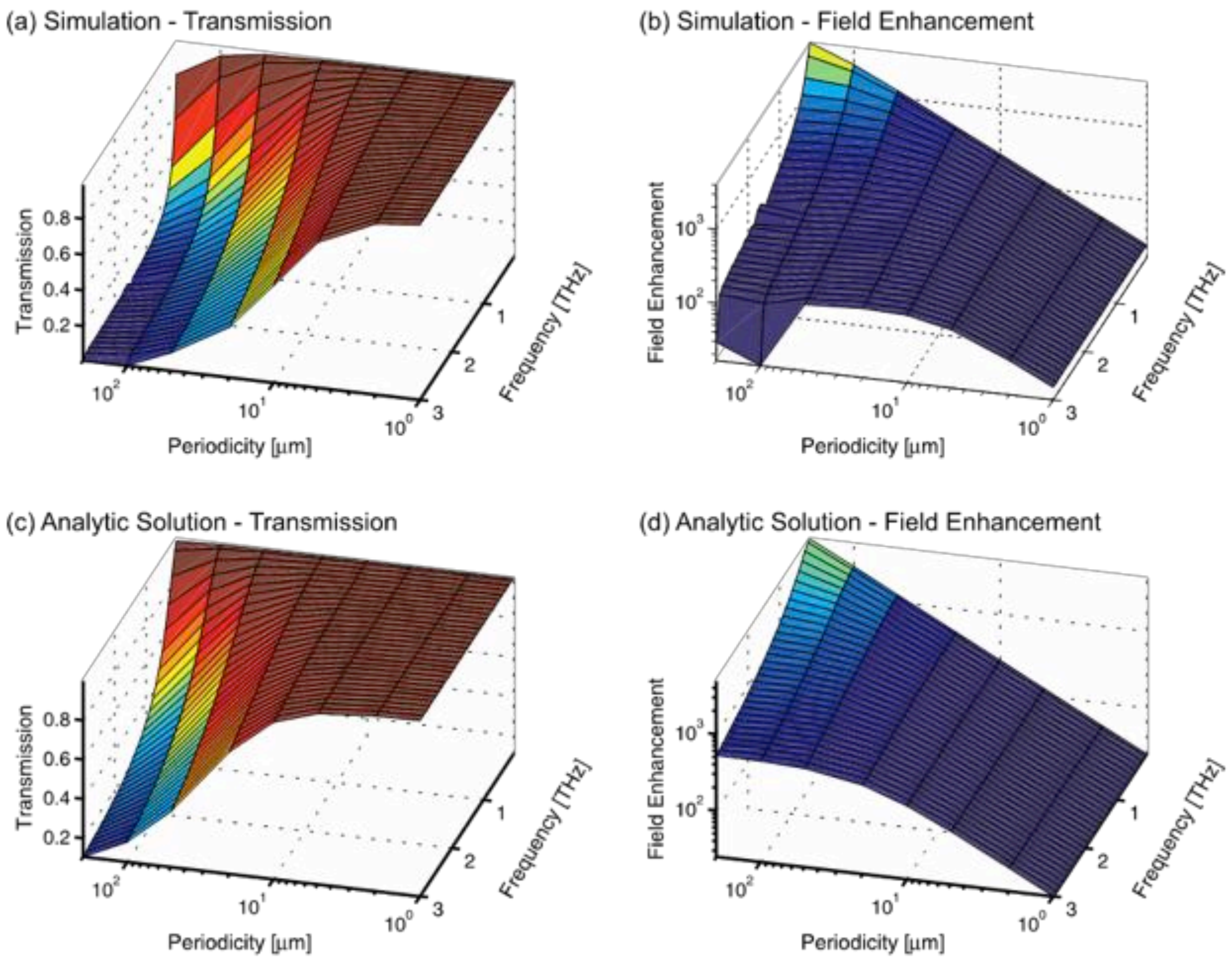

**Figure 3.14** **Effect of slit periodicity on the transmission and enhancement factor. (a) The simulated far-field transmission T and (b) thecorresponding field enhancement F recorded in the mid-gap. (c) Analytical transmission T according to Eq. 3.2.1 and (d) the corresponding field enhancement calculated using Eq. 3.2.2. The slit width and film thickness were fixed to be 40 nm and 60 nm, respectively.**

As we used a broadband source, it is good to consider different frequency components. Figure 3.16 shows the broadband enhancement, plotted against the array periodicity. As it can be clearly appreciated, no significantdifference is expected for $d$ ranging between 100 μm, 200 μm, and the ideal single slit case. We therefore chose $d$ = 100 μm as the design parameter. In comparison to the single slit case, while both are expected to have similar enhancement factors, the former should show much higher transmission due to the multiple scattering slits.

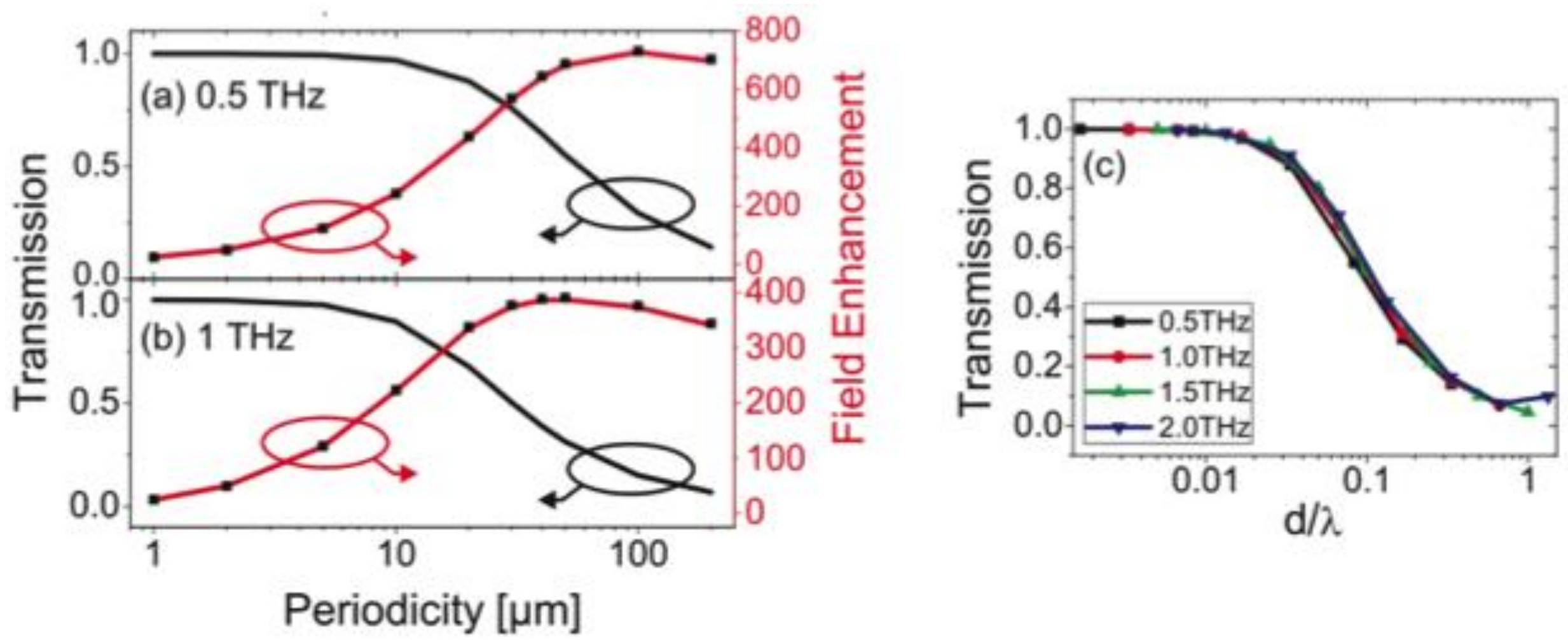

**Figure 3.15 Effect of periodicity on transmission and enhancement**

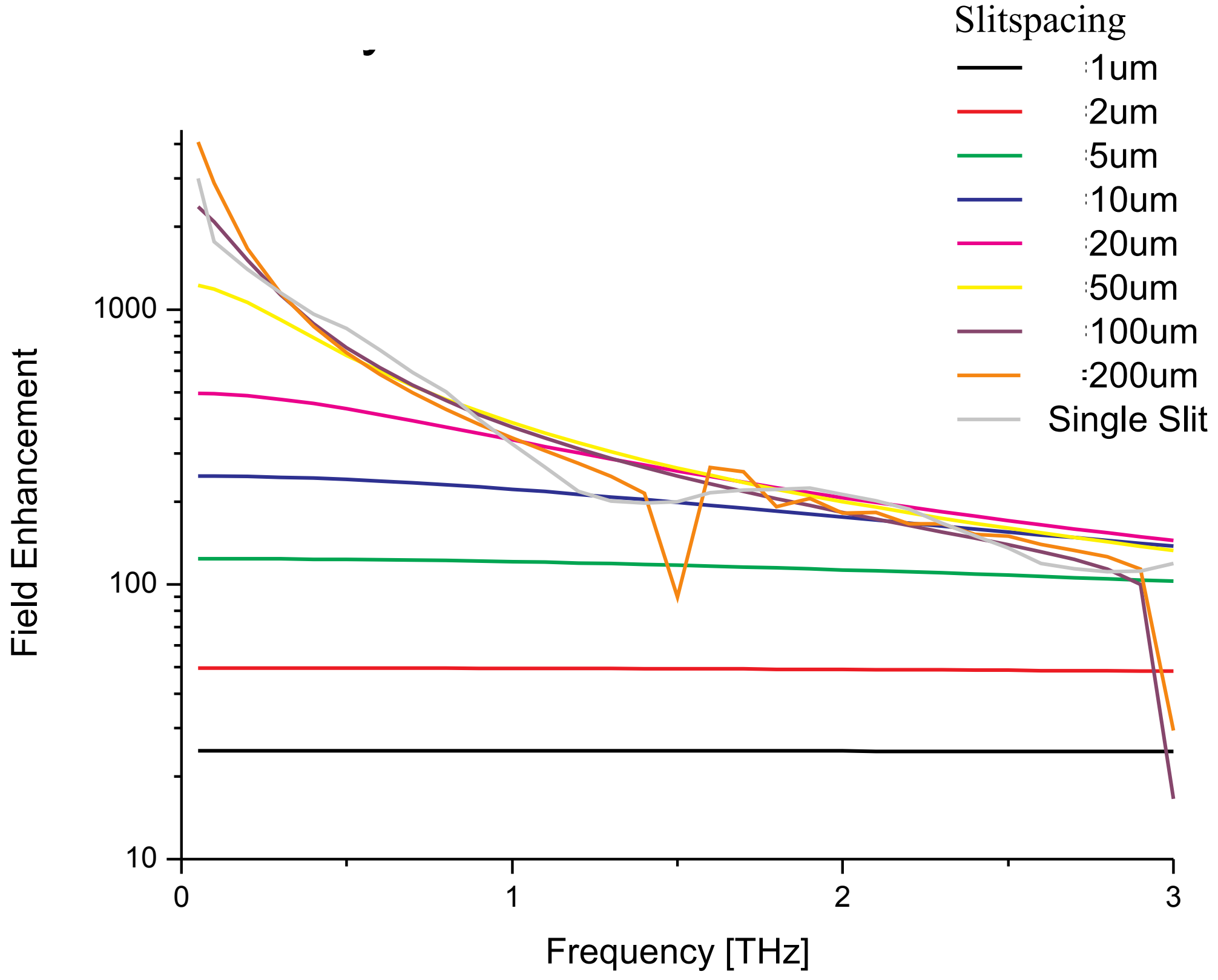

**Figure 3.16 Effect of slit periodicity on the enhancement factor. The slits width and film thickness were fixed to be 40 nm and 60 nm, respectively.**

To summarize, the chosen design parameters values are a = 40 nm, h = 60 nm, $d$ = 100 µm. In addition, we fabricated another sample with $d$ = 1 µm, for comparison purposes, as it will be shown in the near field characterization section. All our samples were fabricated on a 500 µm-thick high-resistivity Silicon substrate.

### 3.2.4 Sample preparation and fabrication imperfection

The samples presented here were fabricated using e-beam lithography. The fabrication steps (shown in Fig. 3.17) can be summarized in: (a) coating the silicon substrate with an e-beam positive tone resist (PMMA) (b) exposure of an area of 3 mm x 3 mm which leaves behind resist rectangular areas of dimensions 2 mm x 40 nm, (c) evaporation of metal layers: 5 nm of Cr (for adhesion) + 60 nm of Au, and (d) lift off process where an acetone solvent is used to remove any PMMA (and the superimposed metal layers) on the silicon substrate. This process leaves behind a patterned structure of gold on the top of the silicon substrate. Then, the unpatterned region is covered once moreby gold through the following steps: (e) the photoresist is applied again to the sample, (f) Au is evaporated on the top of the photoresist, and finally (g) lift off is performed. As a result, the whole sample area of the substrate will be covered with Au and the central 2 mm x 2 mmportion will be formed byan array of nanoslits, 40 nm-wide and 2mm-long.

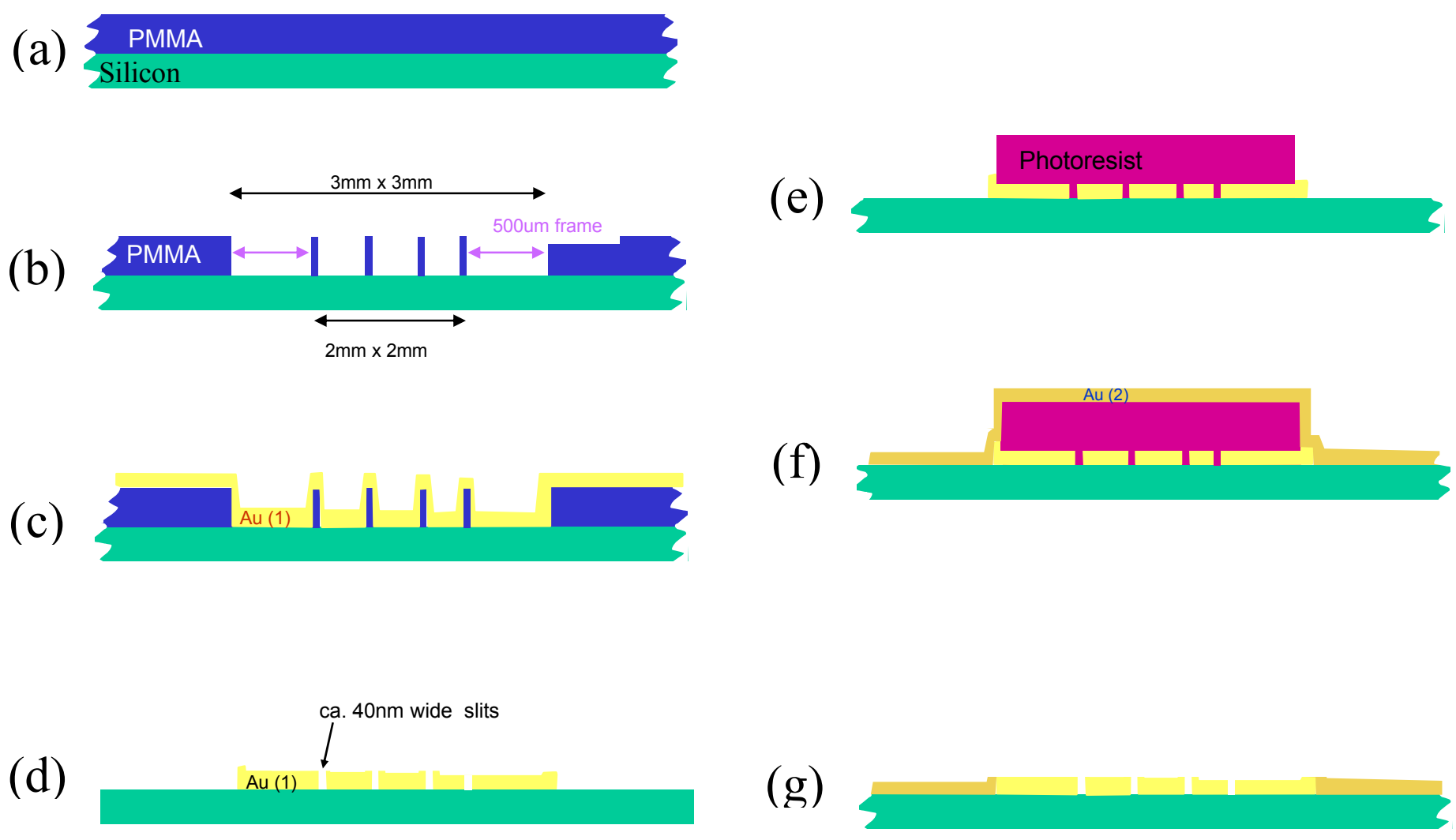

**Figure 3.17** **Sample fabrication. A positive resist e-beam technique was used to pattern arrays of 40 nm-wide, 2 mm-long nanoslits in a 60 nm-thick gold films**

Samples of different periodicity were fabricated. In particular, a 1 μm and 100 μm-spaced slits samples were used in our experimental campaign. The characterization was performed using Scanning Electron Microscope (SEM) as shown in Fig. 3.18. Approximately 5% of the slits were found closed due to incomplete Lift off. In addition, the slit width gets narrower at one end (around 20 nm-wide). This defect affected approximately 15 μm of the 2 mm length of the slits.

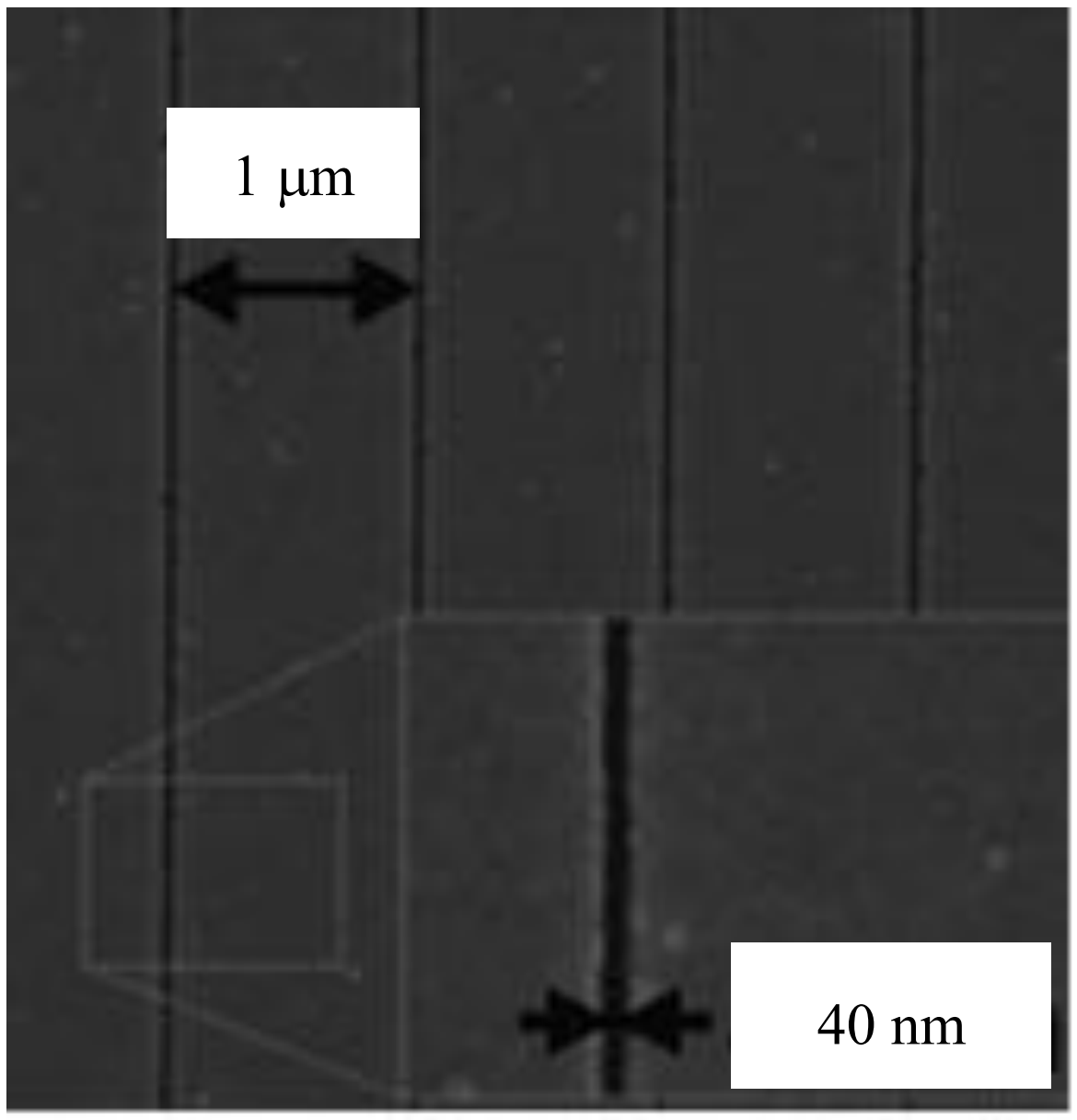

**Figure 3.18** **Samples characterization under SEM.**

### 3.2.5 Sample characterization: spatiotemporal field evolution

There wereno direct experimental tool to probe the enhanced field inside a 40 nm-wide slit. Hence, the experimental characterizations were performed monitoring the spatiotemporal evolution of the THz waves right after the slits and in the far field. The spatiotemporal electric field distribution was measured using a THz polaritonics platform [59-61]. The setup relies on placing the sample between two Lithium Niobate (LN) crystals. The THz is generated in the first one and propagates through the sample reaching the second (detecting) crystal. Scanning the (optical) probe pulse spatially over the detecting LN crystal and temporally (with varying delay) gives a spatiotemporal map of the THz electric field. Three temporal snapshots are shown in Fig. 3.19of (a) the substrate and (b) the 1 μm- as well as(c) the 100 μm-periodic structures on substrates.

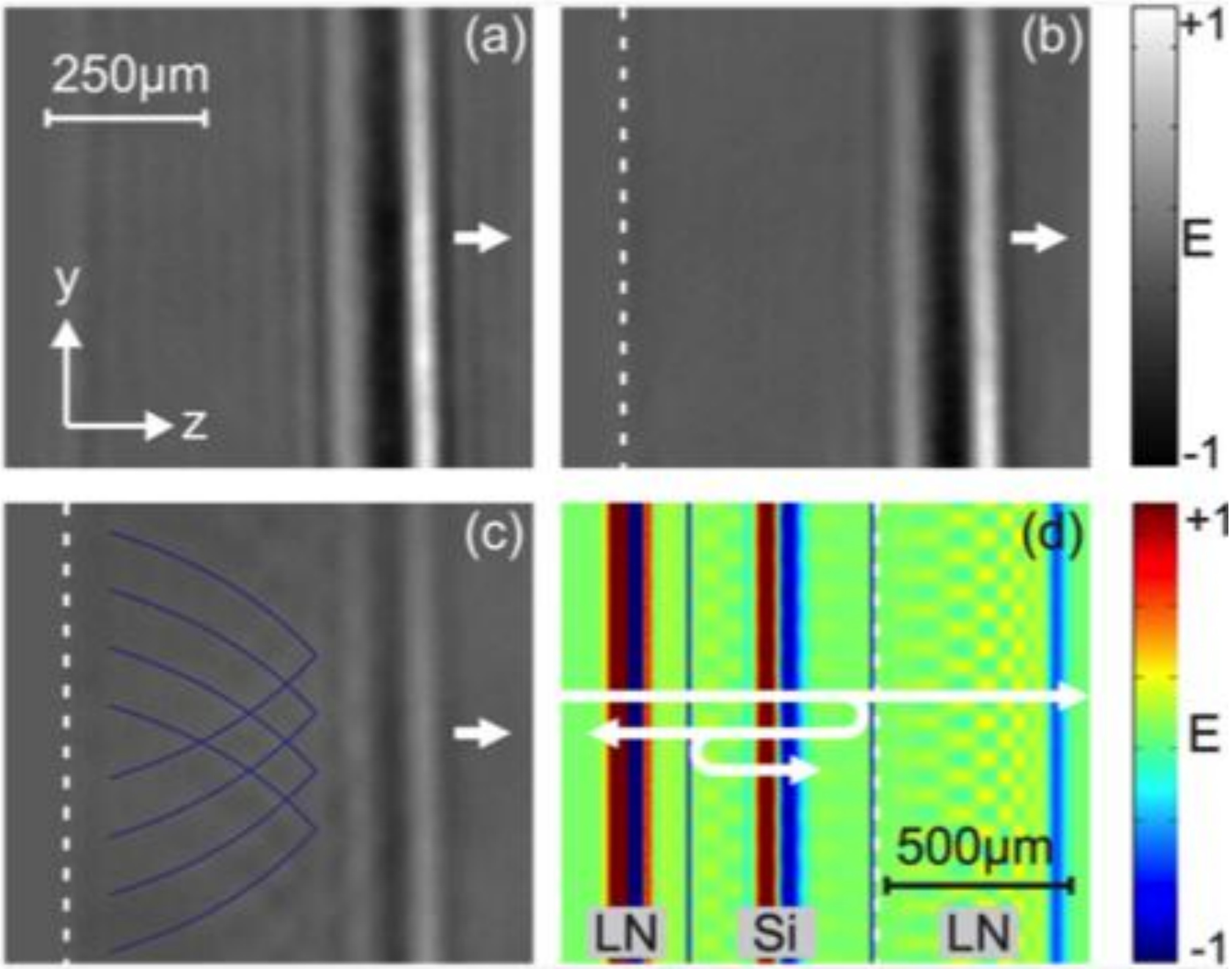

**Figure 3.19** **Snapshots of the spatial field distribution measured approximately 10 ps after the THz pulse has passed through a substrate (a) without slits, and with slits having (b) 1 µm, and (c) 100 µm spacing. The images are plotted on the same absolute gray scale. (d) Simulation corresponding to (c) showing the entire sandwich geometry.**

The white arrows mark the propagation direction. On one hand, in (b), the THz is of high intensity, comparable to that of the substrate and its waveform does not show any peculiar features. Numerical simulations predicted almost unity transmission from this sample although its coverage is only 4%, *i.e.* the enhancement factor is ~25. On the other hand, the waveform from the100 µm-spaced slits sample (c) shows an interference pattern trailing the zero$^{th}$ order waveform as marked by the blue curves in the figure. This interference pattern results from the coherent superposition of the fields transmitted through the individual slits. Figure 3.19(d) shows a time dependent simulation of the employed LN sandwich geometry. The waveform transmitted into the right LN crystal reproduces all details observed in the measurement, *i.e.* the interference features as well as the weak replica following the main THzwaveform, which originates from the multiple reflections in the silicon substrate.

### 3.2.6 Sample characterization: far field measurement

A quantitative estimation of the field enhancement in our experiments was obtained from far field measurement using THz-TDS (more detail in Chapter 1). In particular, we present here the characterization of the 100 μm-spaced array, where high enhancement is expected. The THz field was generated by means of optical rectification of 70 mJ laser pulses in a large aperture (2” in diameter) ZnTe crystal [62]. Figure 3.20(a) and (b) show the transmitted THz field through the sampleand its spectrum compared to the measurement in a silicon substrate (reference). The field enhancement is estimated with respect tothe total transmitted energy and the slitfilling factor (Eq. 3.2.2). It is shown in Fig. 3.20(c) over the spectral range of our source (0.2–2.7 THz) along with numerical simulations for the freestanding slit array and the slit array on the substrate.

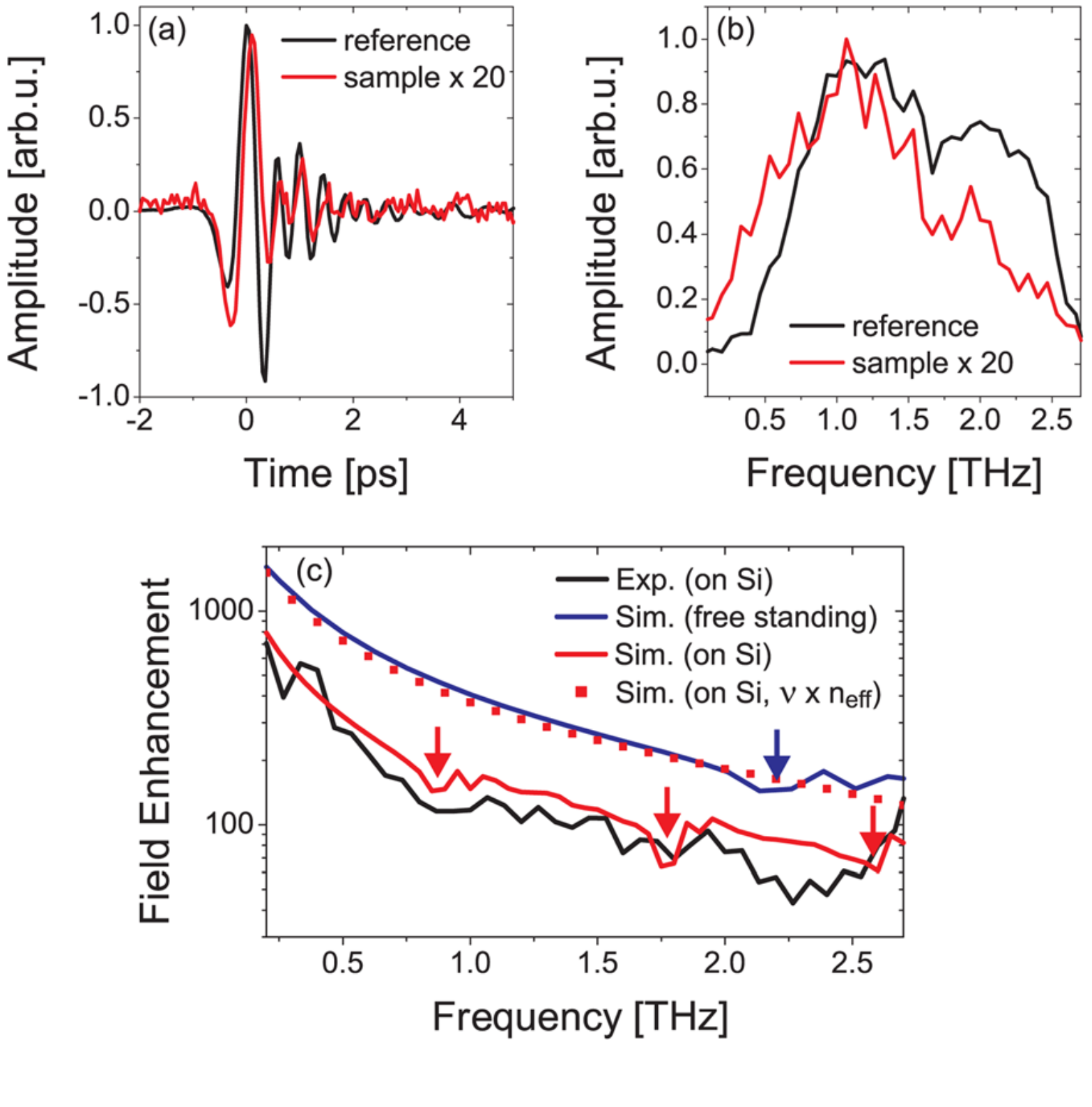

**Figure 3.20** **Far field characterization. (a) Time profiles and (b) spectra of the transmitted fields through the reference substrate and the sample. (c) Field enhancement.**

Good agreement is obtained between measurements and simulations for the slits on the Si substrate. It is clear that the substrate reduces the field enhancement. This effect is explained by considering the structure responding to an effectively smaller wavelength $\lambda_{eff} = \lambda_0/n_{eff}$ where $\lambda_0$ is the free space wavelength and $n_{eff} = \sqrt{(n_{Si}^2 + n_{Air}^2)/2}$. In the same figure, we confirm this effect by scaling the simulation of the substrate by this factor to reproduce the freestanding case. The arrows indicate the effect of the grating modes (where the slit array diffracts the THz radiation into the sample plane) [63]. Those arrows are shifted towards lower frequencies when the substrate is considered, as $\lambda_{\text{eff}}$ becomes the relevant wavelength. Quantitatively, the sample has a coverage $\beta = 1/2500$. The transmission T reaches values as high as 30% at 0.2 THz, which corresponds to a field enhancement $F = 760$. By scaling the detected peak of 10.6 *kV/cm*transmitted through the sample, we estimate an in-slit field of 26 MV/cm.

To conclude, we investigated the design of a THz nanoslits array for concurrent high transmission and enhancement. We managed to increase the transmission by a factor of 20 (in comparison with a single-slit structure) while maintaining the high enhancement exhibited by the single slit structure over the 0.2–2.7 THz frequency range. Combining our results with an intense THz source, we experimentally obtained THz field strengths as high as 26 MV/cm in the near field. Such intense radiation is of particular interest for THz nonlinear experiments. It is noteworthy to mention here that the THz enhancement involves only the electric field with no significant enhancement predicted for the magnetic field. We expect here that the complementary structure (array of nanowires) could be used to obtain significant enhancement of the magnetic field component.Unfortunately, relating the far field measurement of the transmitted field in such a structure with the enhancement is trickier in this case. For example for very sparse wires distribution, most of the impinging field does not interact with the structure and reaches the detector. Nonlinear phenomena locally dependent on the enhanced magnetic field could be used to study the phenomena.

## 3.3 Intense generation: wavelength scaling of the terahertz generated from laser-induced plasmas

Numerical results from Chapter 1 show the necessity of having intense THz radiation to trigger the nonlinear magnetization dynamics. Although we demonstrated how plasmonic structure could be used to locally enhance the field strength, such enhancement is not enough to reach the experimental requirements. It is thus useful to combine this approach with intense THz generation.

In this part, after a brief comparison between the main techniques to generate intense THz pulses, we present our advances towards the up-scaling of the energy of the THz generated from laser-induced plasmas.

### 3.3.1 Intense terahertz sources: a comparative look

The generation of intense THz pulses is a challenging topic. The reason is two-fold: first, the generation mechanisms are generallyinefficient. Second, the relatively long wavelengths associated with THz waves hinder the tight focusing of the field, especially when compared to the optical counterparts. Intense radiation can be obtained mainly from electron accelerators or laser sources. The former is realized when an accelerated electron beam from a linear accelerator crosses the boundary between two media with different dielectric constants. The electric fields associated with those charged particles in the two media are different. At the crossing interface, the charged particles have to “shake off” that difference, which is usually in the form of electromagnetic radiation [64,65]. This radiation is known as the transition radiation. The characteristics of the radiation depend on the acceleration of the particles and their time scale. Careful control over those parameters can lead to the generation of arbitrarily-tuned intense THz. Using this technique, 100 μJ of ~1 THz-centered pulses have been demonstrated [24]. More energetic pulses (140 μJ) have also been experimentally shown [4] but, in comparison, the spectrum is boosted up in frequency to be centered around 10 THz. Although the two energy levels are within a factor of 1.4, the corresponding difference in the peak field is very high, (1 MV/cm and 20 MV/cm, respectively). This arises from the difference in the spectral contents (and hence the tighter focus obtained in the latter case) and the temporal duration.

Intense laser pulses can be used to generate THz exploiting nonlinear optical processes, particularly in nonlinear crystals and gases. Second order nonlinear optical process andlaser-induced plasma are the typical respective generation mechanisms. In Chapter 1, optical rectification of femtosecond pulses has been shown to generate THz pulses. Phase matching and nonlinear coefficients are the main factors determining the efficiency of generation. For example, considering the optical rectification in ZnTe, 800 nm-centered optical pulses (commonly available from Ti-Sapphire lasers) are phase-matched with approximately 1 THzcentral frequency. This makes ZnTe a very popular crystal for THz generation. Nevertheless, its nonlinear coefficients are small and thick crystals have poor performances in terms of the generated THz bandwidth.

For example, in one of the sources used in this thesis research, 50 mJ of optical pulses in a large aperture ZnTe crystal have been shown to generate 1.5 μJ of THz radiation, corresponding to 200 kV/cm of peak electric field [62].

Other but more complex approaches towards higher conversion efficiencies have been demonstrated. A popular technique relies on selecting a crystal with much higher nonlinear coefficient thusachieving the phase matching using the known tilted-pump-pulse-front technique [66] to compensate for the phase mismatch. This configuration however requires a very peculiar pumping geometry in order to tilt the pump pulse group front to correctly overlap with the phase front of the generated THz. This condition is achieved using a set of optical gratings and a spatio-temporal image system. Lithium Niobate ($LiNbO_3$) has been extensively exploited in this design but setup complexity and imaging imperfections make reaching high field quite challenging. High conversion efficiencies have been shown in literature. However 1.2 MV/cm is the highest field ever measured using this technique [5]. Compared to ZnTe, this result was achieved using 4 mJ pulses. In fact, this demonstrates the strong effect of the nonlinear coefficient on the generation efficiency. The second approach depends on achieving phase matching using optical pulses centered around wavelengths other than 800 nm. This technique became recently accessible thanks to the availability of high-energy optical parametric amplifiers, capable of efficiently converting 800 nm pulses into others, centered around different wavelengths. For example, phase matching in highly nonlinear organic crystals was achieved with wavelengths in the NIR. Using1.5μm-centered pulses with energy of 950μJ (generated from 10 mJ, 800 nm pulses), 1.5 MV/cmTHz field was demonstrated in a Dast crystal, setting a recent record on the most intense THz radiation obtained from nonlinear crystals [67]. Provided the availability of the phase-matching wavelength, this technique is way simpler and more efficient than the tilted-pump-pulse-front one shown above. However, its main limitation lies in the relatively low damage-threshold of the available organic crystals and in the difficulty of fabricating large crystal for up-scaling the generation process.

Although the three laser techniques mentioned above have contributed to the science of nonlinear THz spectroscopy [14-18,68,69], they are generally limited by the crystal properties with a bandwidth around few THz. Even with the phase matching makeup techniques mentioned above, the simultaneous satisfaction of phase matching over bands wider than few THz is generally difficult to achieve. This is to be added to optical and THz absorption in crystals.

An alternative approach is to useagas medium for THz generation. This technique depends on focusing intense enough optical pulses to ionize the gas molecules, which become our new nonlinear medium. Terahertz bandwidths extending over tens of THz can be directly obtained. However -and in comparison with the second order nonlinear optical rectification- this technique is somehow less efficient, but at the same time, does not require phase-matching.

In this part, we investigate suitable mechanisms to increase the conversion efficiency of plasma-based THz sources. Our approach focuses on the wavelength dependence of the THzgeneration mechanisms from photo-excited molecules in a gas. In brief, our results demonstratethat by increasing the pump wavelength toward the Mid-IR band, the conversion efficiencydramatically scales up. Similar interaction geometries havebeen employed in the research field of HighHarmonics Generation.

### 3.3.2 Mechanism for terahertz generation from laser-induced plasmas

In the simplest description, by focusingfew tens μJ femtosecond laser pulses in air, the laser intensity can surpass the ionization threshold, generating a plasma.The THz emission mechanism is essentially related to the acceleration of the ionizedelectrons due to the ponderomotive force induced by the transverse intensity gradient of theimpinging optical beam [70-75]. The emitted THz in this configuration is usually at an angle from the direction of propagation (Fig. 3.21(a)). In addition to this pondermotive-based generation, other configurations have been demonstrated and were shown to be more efficient. For example, applying an externalDC bias across the laser-induced plasma (Fig. 3.21(b)) creates a transverse polarization thatsignificantly enhances the THz generation. Mixing two colors (fundamental and second harmonic fields, Fig. 3.21(c)) has also been shown to be an efficient one [75]. This technique is also referred to as AC bias generation and it is the main focus of our work.

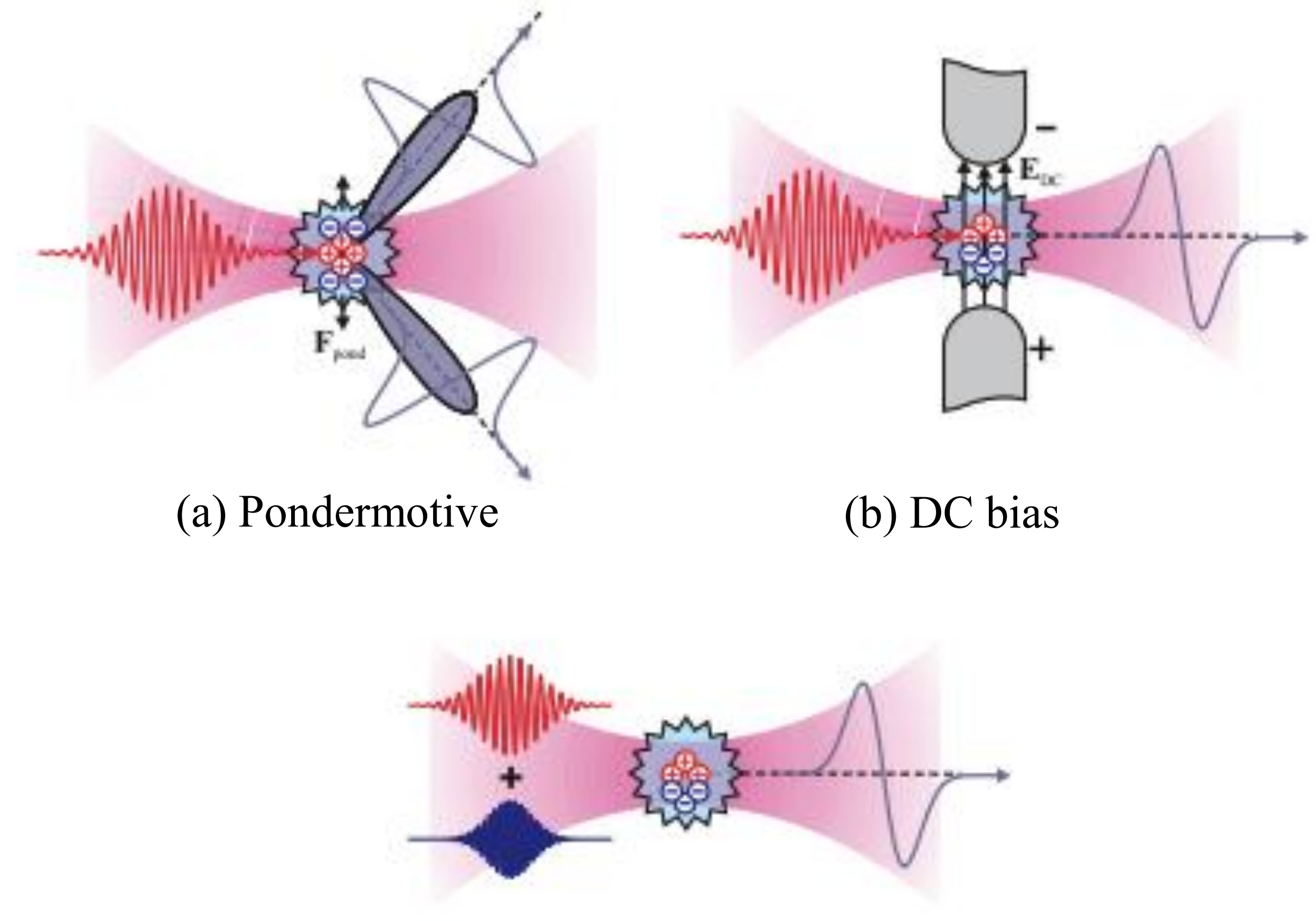

(a) Pondermotive (b) DC bias

(c) AC bias (ω-2ω)

**Figure 3.21 Different mechanisms for THz generation from laser-induced plasmas. (a) Pondermotive. (b) DC bias. (c) AC bias ($\omega - 2\omega$) [75].**

The mathematical analysis is similar to the one presented in section 1.1 for the optical rectification except that the problem here is a third order nonlinear interaction rather than second order. The input fundamental and second harmonic waves can be represented by $\omega$ and $2\omega$. For no phase difference, the emitted THz can be expressed as

$$E_{THz} \propto x^{(3)}\sqrt{I_{2\omega}}I_{\omega} \quad (3.3.1)$$

where $x^{(3)}$ is the plasmathird order susceptibility. $I_{\omega}$ and $I_{2\omega}$ are the intensities of the fundamental and second harmonic, respectively. This description predicts reasonably well thepolarization of the emitted THz.However, it lacks consistency in describing the dependence on the delay between pump and second harmonic. In addition, recent more rigorousdescriptions in terms of transient photo-induced currents correctly predict the largeconversion coefficient observed in the AC biased THz generation [76].

### 3.3.3 Experimental setup

Figure 3.22 shows the experimental setup of THz generation and detection. The pump pulse (provided from a 100 Hz, 65±5 fs Ti-Sapphire system) was focused using a 4" equivalent focal length parabolic mirror. A 100 µm-thick Beta Barium Borate (BBO) crystal was used to generate the second harmonic field. The input beam-waist was 8 mmwide. The generated THz, resulting from mixing the fundamental and second harmonic in the plasma, passes through a filter thatblocks the mid-infrared radiation products (>20 THz), usually emitted by photo-excited plasmas.

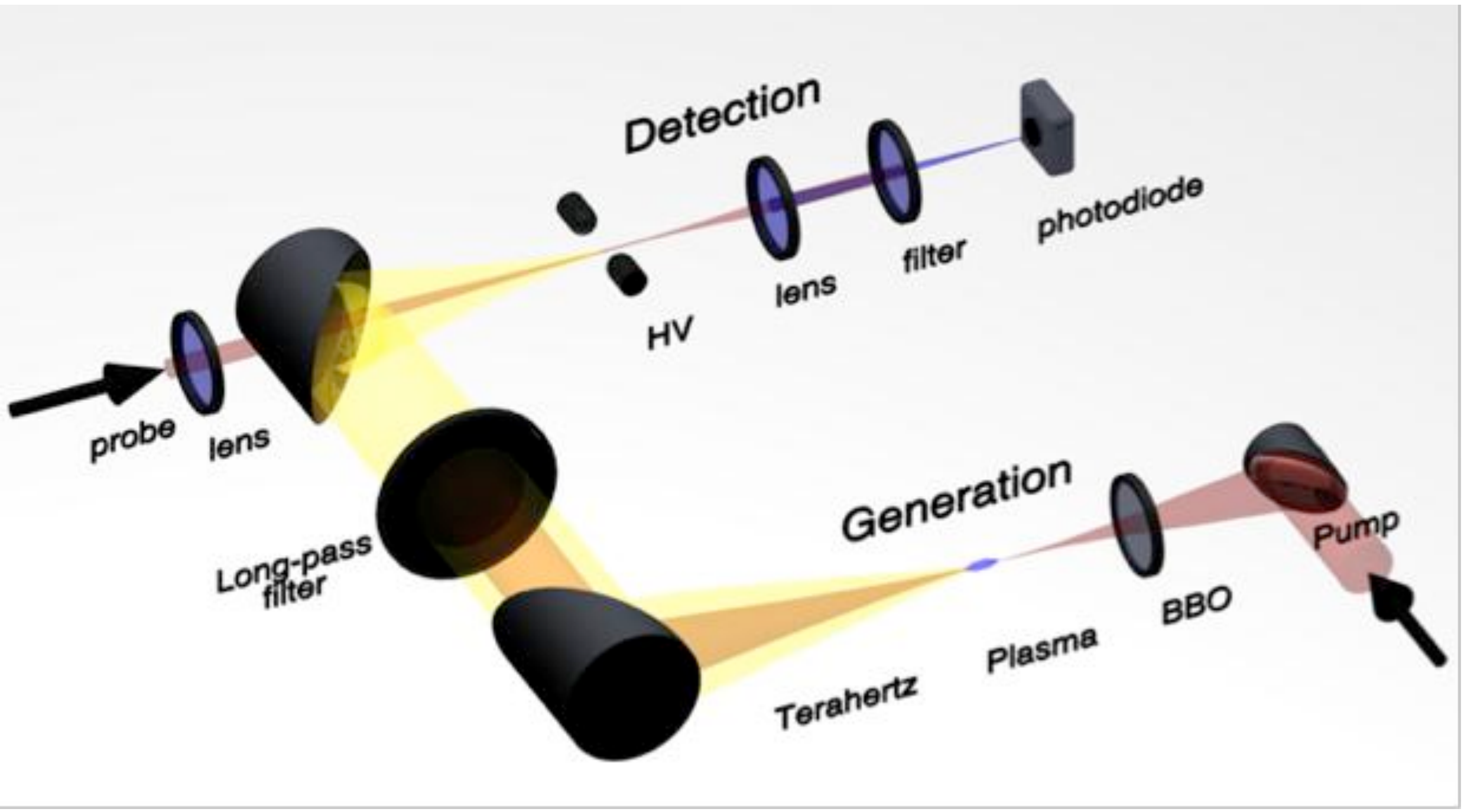

**Figure 3.22 Experimental setup for THz generation from laser-induced plasma and detection usingthe ABCD technique.**

Terahertz characterization was performed using energy measurement and time domain spectroscopy.

### 3.3.4 Energy measurement

Broadband THz pulses have only been obtained recently. So, calibrated instruments for this wide band are still to come. In order to measure the THz energy, we used the following procedure. We measured the energy using a Pyroelectric detector (Molectron, Coherent). The optical pulses were shielded with three filters. According to the specifications provided by the company, this device exhibits a flat spectral response for frequencies > 3 THz, where most of the energy in our pulses lies. In addition, we performed the same measurementusing another detector (Microtech Instruments), calibrated between 1 THz and 3THz. Wewere then able to calibrate the energy readings of the first wideband detector against a secondnarrowband.

### 3.3.5 Broadband Detection: Air Biased Coherent Detection

Energy measurement gives access to the efficiency of the generation process, but does not show the spectral distribution of the energy components. We employed time domain spectroscopy (TDS) to measure the time profile of the THz pulse and then calculate the spectrum. In Chapter 1, we explained the electro-sampling technique as a means for TDS. This technique is widely used throughout experimental measurements presented inthis thesis. However the bandwidth of this technique is limited by the phase matching between THz and probe pulses and by the detection crystal absorptionresonances.Air Biased Coherent Detection (ABCD) emerged as a suitable alternative for wideband detection [8,77]. The interaction medium is the air (gas) and hence the phase-matching and absorption problems disappear. The process can be formalized as electric field induced second harmonic generation (EFISH)in a medium having a third order nonlinearity. The generated intensity of the second harmonic is thusproportional to the THz probe field intensity, according to the field relation $E_{2\omega} \propto \chi^{(3)} E_{THz} I_{\omega}$.

In order toretrieve the information on the field, an external field$E_{ext}$ is applied through apair of electrodes. Additional contributions arise in the second harmonic intensity: *i.e.* $I_{2\omega} \propto (E_{THz})^2 + E_{ext} E_{THz} + (E_{ext})^2$. The second term is linearly dependent on the THzfield. In the measurement, the sign of $E_{ext}$ is switched periodically and $E_{THz}$is detected as theonly oscillating contribution to $I_{2\omega}$. In our experiment (Fig. 3.22), we used a 45 fs-wide (full width at half maximum), 800 nm, 20 μJ probe pulse. The optical and THz beams were collinearly superimposed (by letting the optical beam pass through a 1.5 mm-wide hole in the THz-focusing parabolic mirror). The optical beam was focused to a ~10 μm-wide spot in the center of the THz focus. The generated second harmonic signal (400 nm-centered) was detected using an amplified photodiode after filtering the residual 800 nm light.

### 3.3.6 Wavelength scaling of the radiated energy

The photo-induced plasma characteristics depend strongly on the pump wavelength. As the wavelength increases, the laser fields impose a stronger Pondermotive force on the electrons leading to higher acceleration. This property was previously used to extend the cut-off frequency of the high-orderharmonic generation [77,78]. Terahertz generation should be enhanced in a similar way. Motivated by this prediction, we investigated the wavelength scaling of THz generation from laser plasmas. We used a photocurrent model [80] to predict the gain in energy conversion associated to the wavelength change. Under the assumption of a wavelength-independent excitation (plasma) volume, we expect the resulting THz emission to scale quadratically with the wavelength. Based on that, we performed measurement sweeping the fundamental pump wavelength from 0.8 to 2 μm. Figure 3.23 shows the experimental measurement along with numerical fitting.

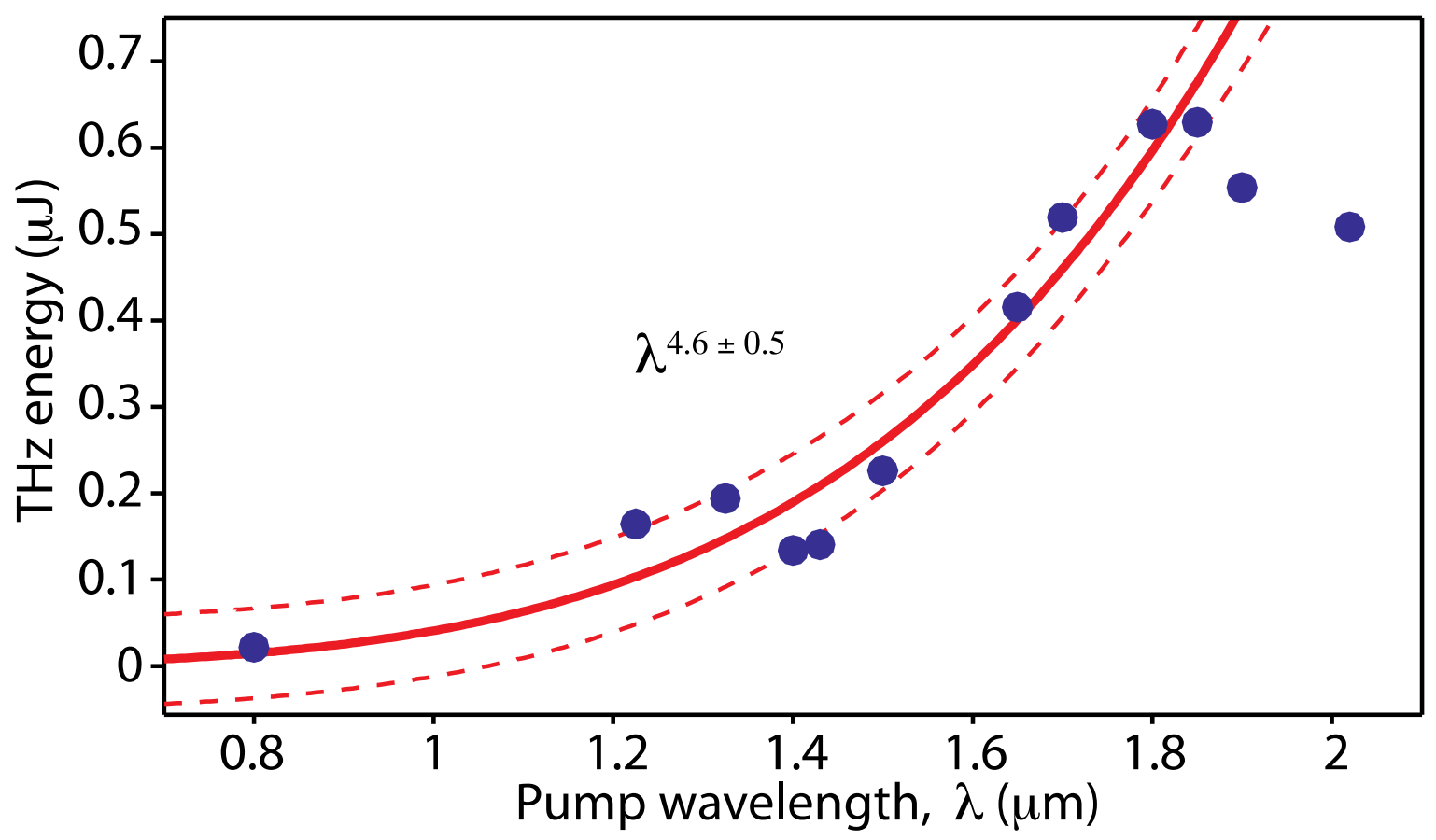

**Figure 3.23** **Wavelength scaling of the THz energy. Measurement of the THz energy at different pump wavelengths (blue dots). Red line shows a numerical fit: $\lambda^{4.6\pm0.5}$. More details are given in the supplementary materials of *M. Clerici et al., Phys. Rev. Lett. 110, 253901 (2013).***

As shown the measurement fits well with a $\lambda^{4.6\pm0.5}$ scaling up to a pump wavelength of 1.8 μm, after which, the efficiency of the THz generation drops. Such drop is essentially due to the focusing geometry in the experimental setup. In general the interaction volume and peak field intensity vary with the wavelength as we used a fixed focal length (numerical aperture) parabolic mirror. Hence, for a Gaussian beam with a given beam diameter, the focused light –and so is the induced plasma volume- scales as $\lambda$, while the peak intensity goes as $\lambda^{-2}$. Including those corrections, numerical simulations clearly highlight a deviation from the quadratic scaling law that relates to a reduced efficiency above 1.8 μm. In Fig. 3.24, the complete numerical simulations are compared with the experimental results, with a remarkably good agreement.

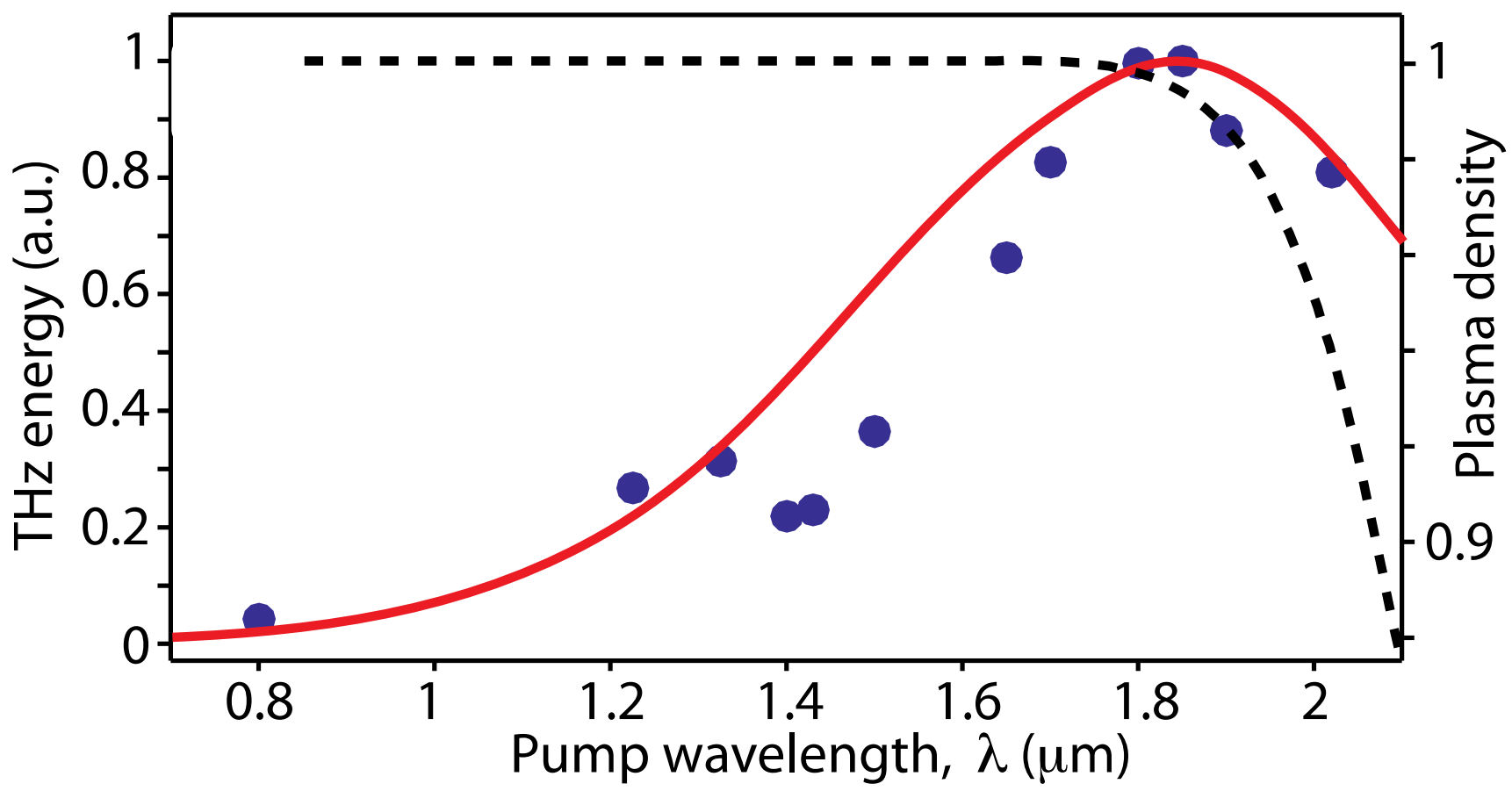

**Figure 3.24** **Wavelength scaling of the THz energy. Measurement of the THz energy at different pump wavelengths (blue dots).The red line shows the numerical simulations using a photocurrent model after including the geometrical parameters associated to the focusing.**

Finally, to investigate the distribution of the spectral contents as the wavelength is varied, we used the ABCD technique to record three THz waveforms for pumping wavelengths of 800, 1450 and 1850 nm, respectively. Figure 3.25 shows the time profile and the corresponding spectrum. For the first two waveforms, the spectrum is mainly centered around 3-4 THz, where as the pump wavelength increases towards 1850 nm, the spectrum center shifts to ~ 6 THz. An image of the THz at the focus recorded using a thermal camera is shown in Fig. 3.25(c). Using this spatial profile along with the energy measurement and the time trace, we estimate the peak electric field to be around 4.4±0.4 MV/cm. This represents the highest THz electric field ever measured from a tabletop source.

In conclusion, we demonstrated a significant enhancement at longer wavelengths of the THz emitted from photo-induced plasma due to the inherent scaling of the field-plasma interaction law. It is important to stress here that the real significance of this results lies on the fact that no reliable increase in the generation intensity of a plasma source (by simply increasing the pump power) has been ever demonstrated before. This is mainly due to the associated saturation and nonlinear effects. The presented technique suggests further increase in the total energy as longer pump wavelength are used, provided that proper focusing is adopted.

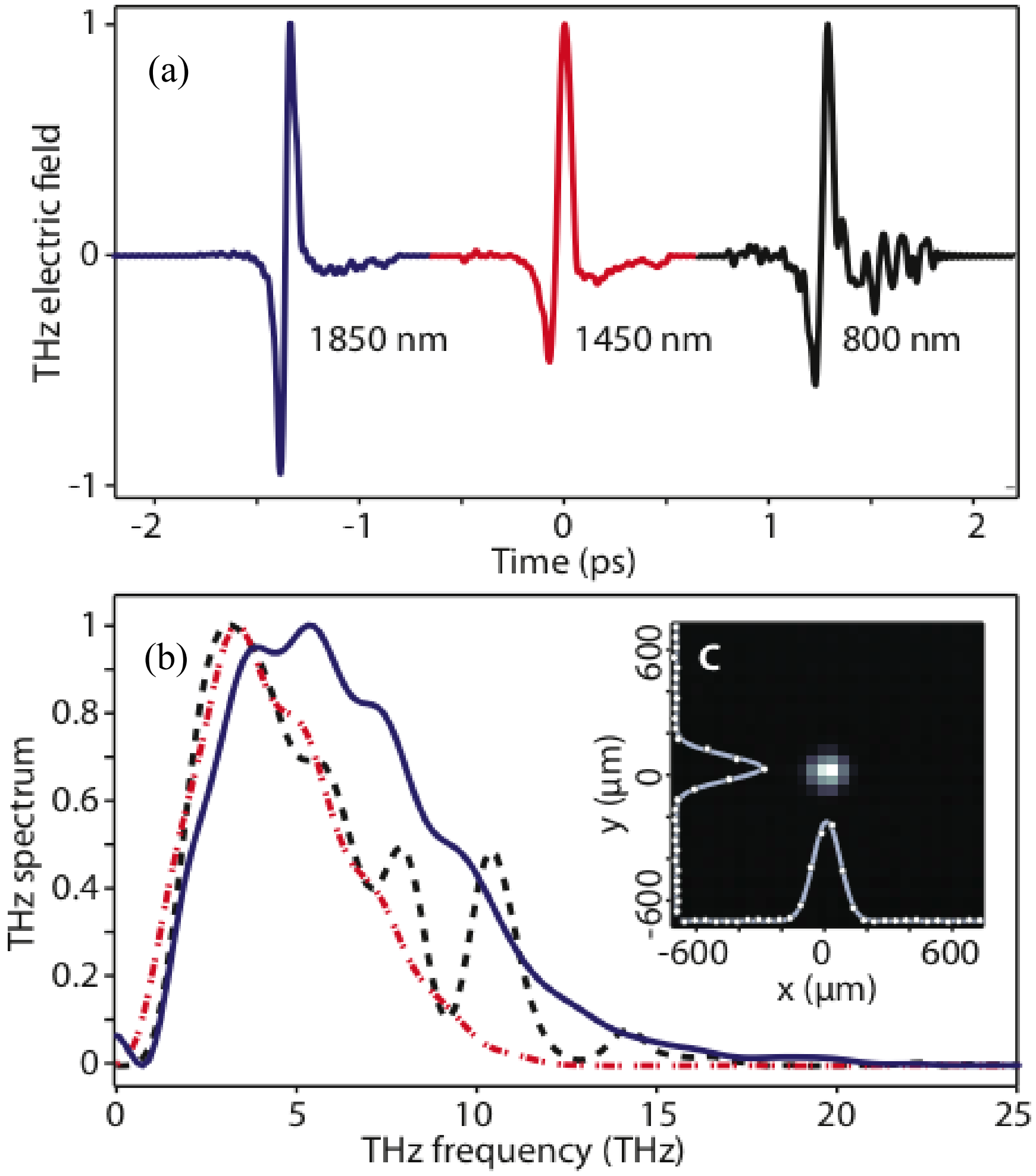

**Figure 3.25** **Terahertz spectroscopy of the generated THz. Three traces were measured in (a) the time domain. (b) The corresponding spectra. (c) An image of the THz spot at the focus. The pulse duration of the optical pump pulses were 60±5 fs. The energies of the fundamental and second harmonic pulses were 400 µJ and 20 µJ, respectively. The pulse repetition rate was 100 Hz.**

# 4 MAGNETO-TERAHERTZ (M-THZ) DEVICES

Studies of THz-magnetization dynamics brought to the surface the importance of linear THz-magnetic applications. On the magnetic "side", the results presented in Chapter 2 particularly suggest that the THz pulses could reveal the magnetic state of certain materials. Generally speaking, EM waves can be used to study magnetic materials whether in the reflection or the transmission geometry. In that sense, high frequency waves are advantageous because of their small wavelengths and, consequently, of their high spatial resolutions. Unfortunately, as we move towards optical frequencies, most materials either lose their magnetic response (with a relative permeability of one) or absorb/scatter light.

Then, what about longer EM oscillations, *i.e.* THz waves? Can they be suitable magnetic diagnostic tools? This is still an open question, especially in light of the very few studies on magnetic materials using THz waves performed so far.

On the photonic side, the THz science simply lacks "THz-devices". Generation and detection technologies grew rapidly over the past few years. Tunable narrowband and ultra broadband sources as well as detectors, extending over more than a spectral decade, are now available at the THz spectrum. However, the supporting devices (such as modulators, isolators, filters, etc.) remain the actual bandwidth bottleneck. The reason is two fold. First, the study of material properties at THz frequencies is still at its beginning. Required THz sources and detectors just become available. Second, THz imposes rigid bandwidth constraints. Many significant THz applications (such as imaging, spectroscopy, and sensing) require short (broadband) pulses. Even in a potential application of THz in communications, the frequency of (narrowband) THz sources will need to be "swept over" wide spectral ranges to sustain more channels. So the dilemma is not only finding materials with the proper response at THz frequencies (which is already challenging) but also such properties should be maintained within a broadband spectrum.

Magneto-photonic devices are potentially important but presently non-existent in the THz band. Motivated by this, the remaining part of my thesis work has been devoted to the practical realization of such devices. In this chapter, three novel magneto-THz devices are presented: (i) broadband non-reciprocal phase retarders, (ii) a magnetic modulator, and (iii) the non-reciprocal isolator. To the best of our knowledge, they are the first of their kind at the THz frequencies.

## 4.1 Ferrofluid-based Broadband Non-reciprocal phase retarder

Material response to EM waves can generally be described by a complex dielectric function, which contains the refractive (delay) and absorptive properties of the material. Any material is a phase retarder (PR), *i.e.* it induces a certain delay in the propagating wave. Phase retarders come at the core of fundamental parts in communications and signal processing systems. Modulators, isolators, switches, circulators, delay lines, and filters, for example, are based on PR.

Any natural material is, therefore, a PR by definition. In addition, if it is birefringent in the spectral range of the propagating wave, the latter experiences a phase shift between its two distinct linear field eigen-components (the *ordinary* and the *extraordinary* wave), that alter the field polarization state. However, such retardationis reversed if the wave is back-reflected to propagate through the same material again and the polarization transformation is canceled out. In other words, the phase retardation has the same sign when the electromagnetic wave vector is reversed and the system is called *reciprocal*. Such reciprocity is broken in a specific class of materials in response to externally applied magnetic fields or internal remanent magnetization states. In such a case, if the EM wave is back-reflected, the newly induced retardation over the reflected pass adds up to the original forward retardation. Such a directional retardation finds many important applications. For example, an electromagnetic isolator is a popular shielding device in laser laboratories and photonic circuits. It is based on a non-reciprocal phase retarder (NRPR) operating at a specific retardation as will be shown in detail in another part of the thesis. In this chapter, the first broadband NRPR operating with THz pulses is demonstrated using Ferrofluids. This study also represents the first study of Ferrofluids properties in the THz regime. Ferrofluid will turn out to have broadband rotation and (relatively) low absorption as shown.

### 4.1.1 Properties of Ferrofluids

Ferrofluids are suspensions of magnetic nanoparticles in a carrier fluid. The nanoparticles are usually ferromagnetic ones, such as ferrites ($Fe_3O_4$). Each nanoparticle is considered a separate magnetic domain [81]. Brownian motion of the nanoparticles gives them the free motion as a suspension in the form of a colloid against the gravitational forces. In addition, they are usually coated with a surfactant material, which provides electrostatic resistance against the agglomeration. The carrier fluid can vary from water to some organic liquids. However, due to the highly absorbing nature of water with respect to THz waves, all the studies performed here are based on an organic carrier liquid. In terms of magnetic properties, although Ferrofluids are based on ferromagnetic nanoparticles, they behave as paramagnets, *i.e.* they do not exhibit anymagnetic properties in the absence of external magnetic fields. This originates from the random distribution of the freely suspended nanoparticles. Thisis illustrated in Fig.4.1, where magnetic moments (domains) normally point in different directions. Under the application of external magnetic biases, those moments tend to align to the magnetic field direction, giving rise to a net magnetization. Moreover, in a microscope scan, those nanoparticles are seen to form clusters. Cluster formation is a fundamental property of Ferrofluids in the presence of magnetic fields and is responsible for some interesting and unique properties as will discuss in detail in section 4.2.

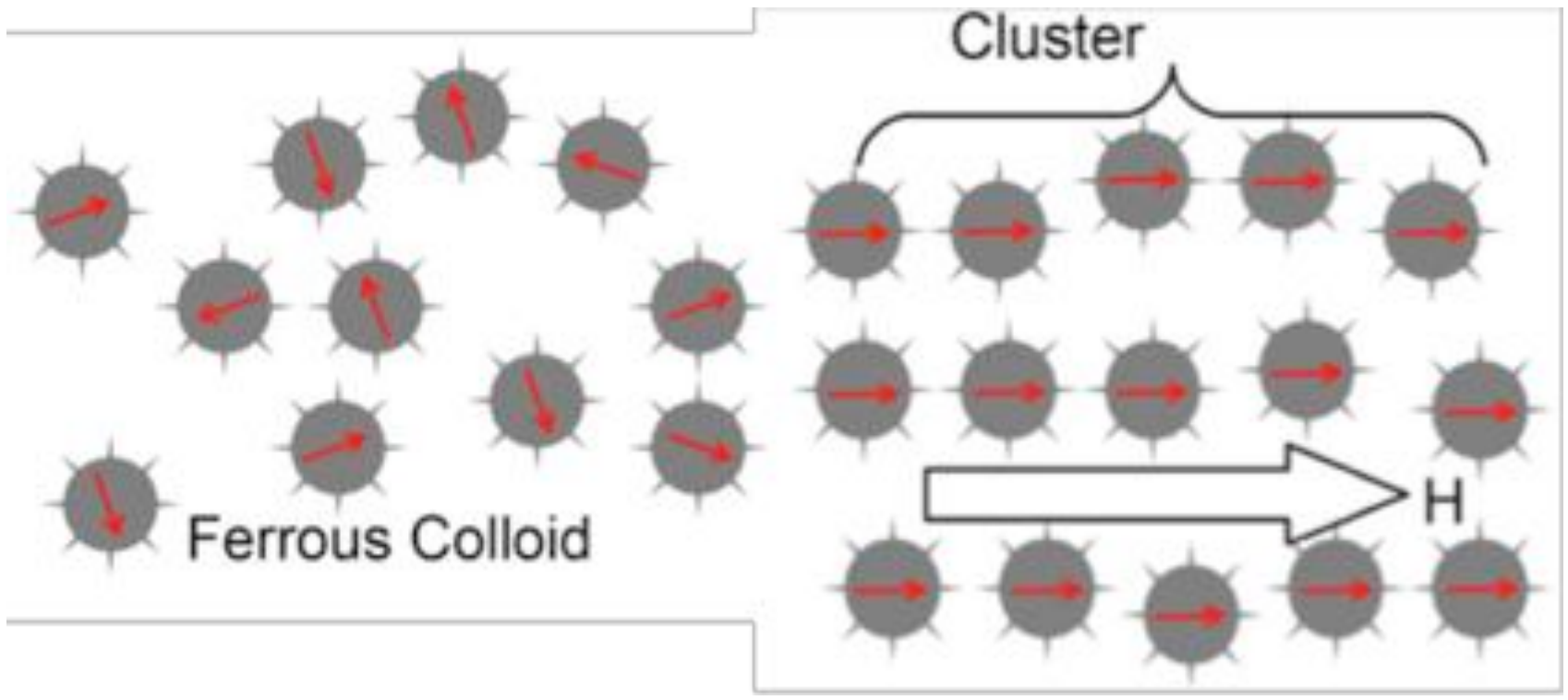

**Figure 4.1** **Cluster formation of magnetic nanoparticle. In the absence of external magnetic fields, the nanoparticles are randomly distributed with their magnetic moments oriented in different directions and no net magnetization is perceived. Under an external magnetic bias, the nanoparticles tend to align to the field axis forming clusters.**

### 4.1.2 Magnetic properties of Ferrofluids

The magnetization curve of a material describes its response to an externally applied magnetic field. In the case of a Ferrofluid, at low particles concentrations (as is the case for all studies here), the particles are considered independent, non-interacting and the overall magnetization can be described as a statistical average. Magnetic nanoparticles are very sensitive to field

gradients,along which they can easily be swept. In response to a uniform magnetic field, individual nanoparticles tend to align along the field axis. The alignment process can be described in terms of two motions: Neel and Brownian [81]. While the former tends to physically rotate the particle and align its magnetic moment to the applied field, the latter tends to reorient the magnetic moment without any real physical movement. The two mechanisms occur at different time scales in a fraction of the second. In all the studies presented here, DC magnetic fields are used and thus a steady state condition is assumed. In a given sample, the Ferrofluid behavior can be described on average by a Langevin function [81]

$$L(\propto) = \frac{M}{M_s} = coth(\propto) - \frac{1}{\propto}, \qquad \propto = (m\mu_0 H)/k_B T \tag{4.1.1}$$

Where $M$ and $M_s$ are the magnetization level and the saturation magnetization of the sample, respectively. Further,$m$, $\mu_o$,$H$, $k_B$, and $T$ are the magnetic moment, the free space permeability, the applied magnetic field, the Boltzmann constant, and the temperature, respectively. Figure 4.2 shows this Langevin behavior plotted at room temperature with two interesting properties depicted.

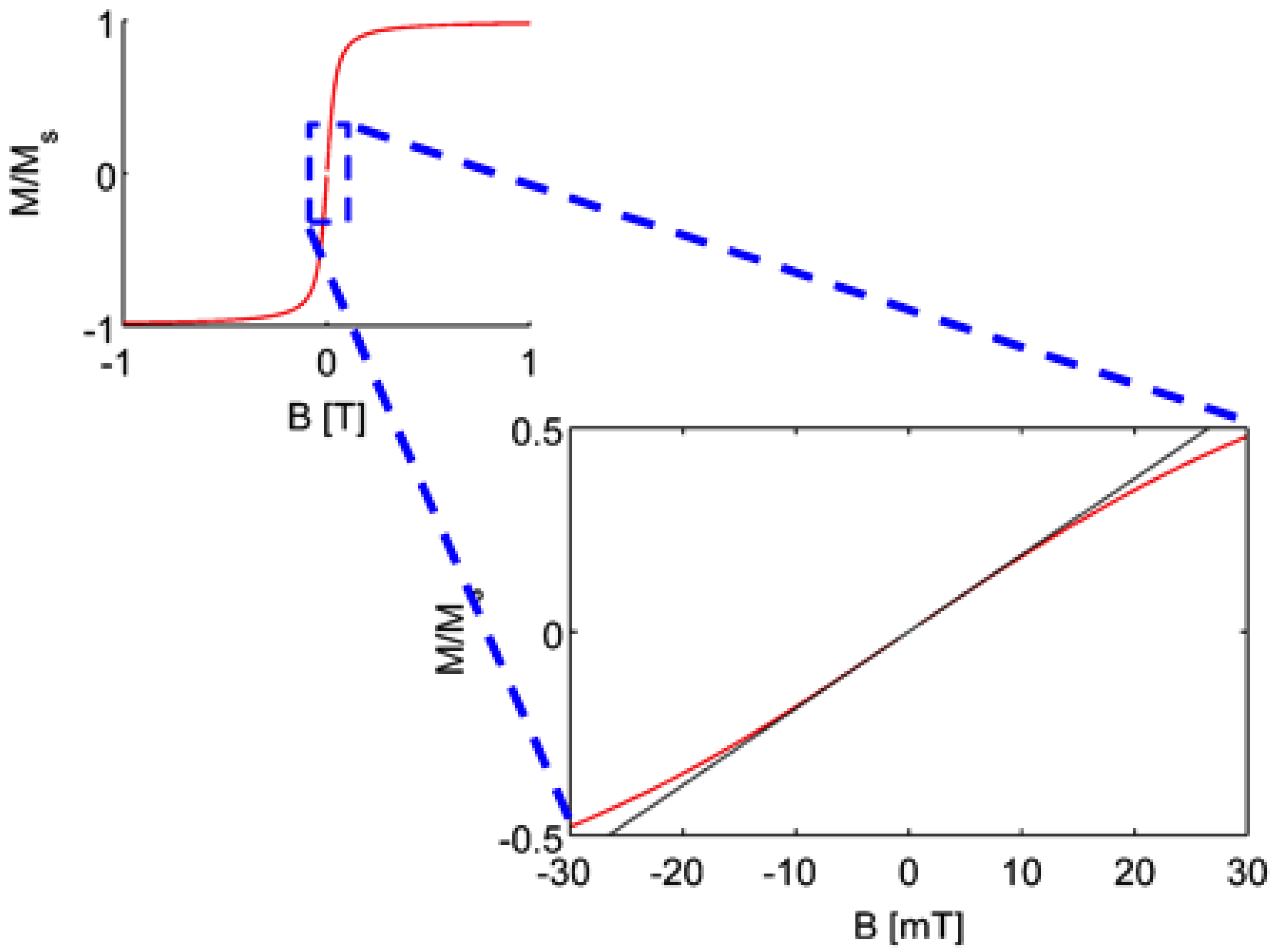

**Figure 4.2** **Magnetization state of Ferrofluids showing the Langevin response under the application of external magnetic fields at room temperature. The zoom-out of the initial part of the curve shows the sharp increase building up at half the saturation level for a small field of ~30mT (shown in red.) Linear approximation ($k_B H/3$) of the Langevin function is shown in black.**

First, the paramagnetic response: as highlighted in the introductory part, Ferrofluids do not exhibit any net magnetization in the absence of external magnetic fields. This is confirmed by the hysteresis curve collapsing to a zero point-crossing curve. Second, the nonlinear response to external magnetic fields: Ferrofluids magnetization curve is characterized by a strong nonlinearity, starting with a sharp rise of the magnetization followed by a very slow asymptotic saturation. Such a property gives them a big advantage when it comes to practical applications as it suggests that significant amounts of magnetization can be obtained at low values of the applied magnetic fields.

A close look (zoom-out of the central part, Fig. 4.2) at this shows that half of the saturation magnetization can be reached at a very small magnetic field of ~30 mT. Moreover, in thisregion the Langevin function can be approximated by a linear response described by $L(\propto) = k_B H/3$[81].The linear response represents an advantage in terms of tunability of the magnetization.

### 4.1.3 Dielectric properties of Ferrofluids

Ferrofluid studies presented in this part are based on a commercial Ferrofluid, EFH1 provided by Ferrotec (USA). The magnetic nanoparticle is~10 nm-sized magnetite ($Fe_3O_4$.) The carrier fluid is a hydrocarbon and the particle concentration is 7.3% in weight. As a sample, we used a 10 mm-thick THz-transparent cuvette. To calculate the dielectric function, we employed the THz-TDS (detailsare given in chapter 1) to measure the transmitted THz in three cases: (1) the air reference, (2) one side of the cuvette, and (3) the Ferrofluid-filled cuvette (sample.) Then, we calculated the transmission functions of the side of the cuvette and the full sample with respect to the air reference. We used the transfer matrix approach to retrieve the complex dielectric response of, *i.e.* the refractive index and the absorption coefficient. From the first transfer function we obtained the dielectric response of the cuvette side. We then used it along with the second transfer function to calculate the dielectric function of the EFH1. Details of the transfer matrix approach and the index retrieval procedure are given in Chapter 1. Figure 4.3 shows the refractive index and the absorption coefficient of the EFH1 over the broad spectrum of 0.2-0.9 THz, limited by the signal to noise ratio of the characterization setup. The plots show some of the interesting properties of the EFH1 at THz frequencies. With respect to the refractive index, the low values (~1.54) lead to low reflection losses when it comes to practical applications. It is noteworthy mentioning that most of the materials show much higher refractive index at THz (as compared to optical frequencies) and thus suffer from high losses at THz frequencies. In addition, the EFH1 shows negligible dispersion over the wide band shown in the figure. This property is very favorable in many applications where long propagation distances are required, such as e.g., in waveguides. Also, as we will show later, this liquid can be used to obtain a phaseretardation, therefore a certain length of propagation is required to accumulate the required phase retardation. The absorption coefficient is relatively low as well. In principle, this is a very interesting property that allowed us to use a long sample (1 cm) in the experiments presented here.

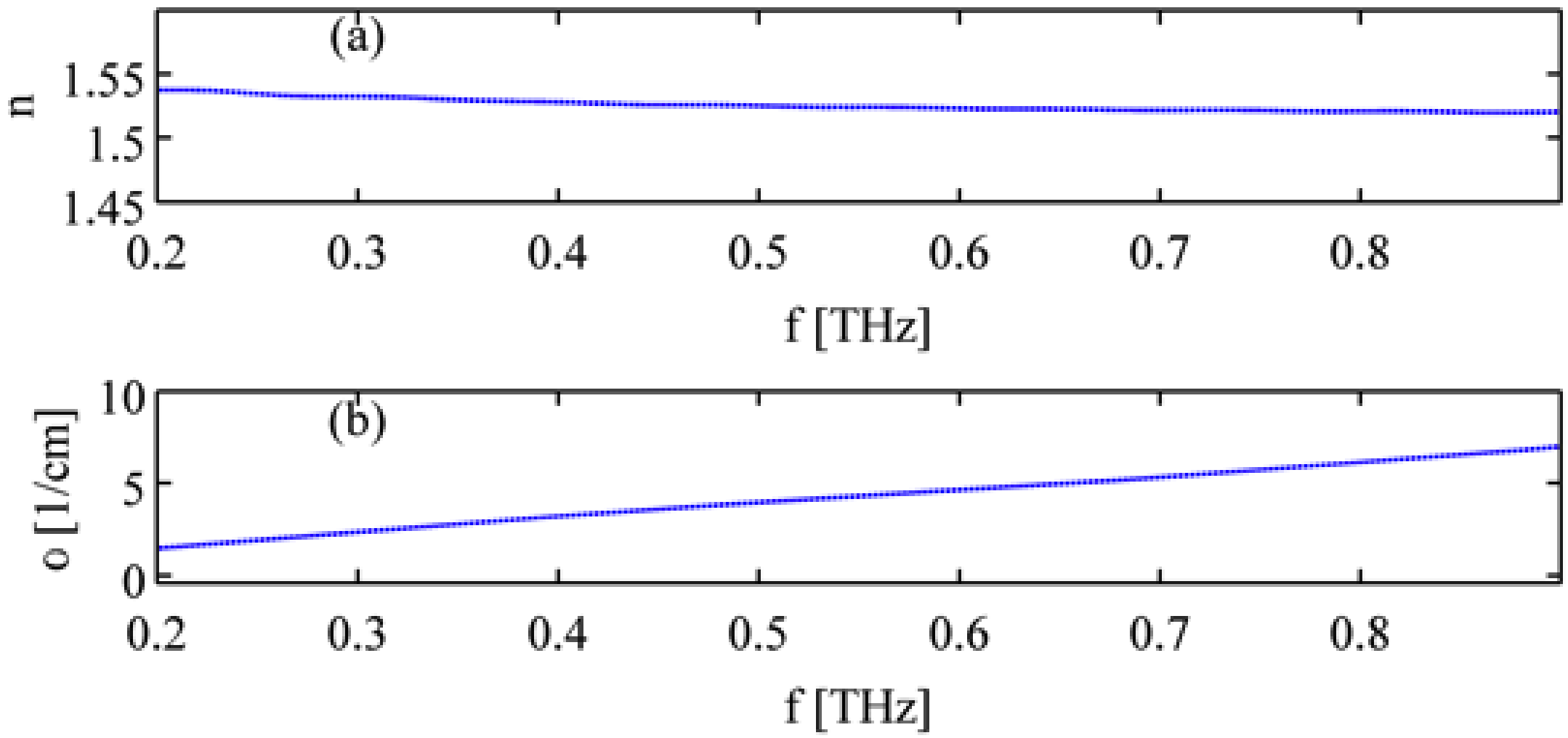

**Figure 4.3** **Dielectric response of a commercial Ferrofluid (EFH1) at room temperature. (a) Refractive index. (b) Absorption coefficient.**

### 4.1.4 Measurement of non-reciprocal phase retardation and ellipticity

Non-reciprocal phase retardation is manifested in the difference ofthe propagation velocity associated to the right- and left-circularly polarized eigenmodes, whereas the*ellipticity*measures the difference in their relative absorption coefficient. Thosequantities can be addressed by characterizing the polarization state changes upon propagation in the sample. This measurement is usually performed using an *ellipsometer*that probes the polarization state using waveplates, polarizers and power detectors. Waveplates are, generally speaking, spectrally dependent and thus hard to use with broadband THz pulses. However, as the complex THz spectrum can be retrieved from TDS detection, another spectrum-insensitive characterization technique can be used.

The polarization state is fully reconstructed by using two crossed states of an analyzing polarizer as shown in Fig.4.4. Both the polarization rotation (half the value of retardation) and ellipticity can be readily calculated from those two orthogonal states. The input (WGP1) and output (WGP3) polarizers are placed to ensure a horizontal polarization of both the generated and the detected THz pulses. Details of the calculations are given in Chapter 1.

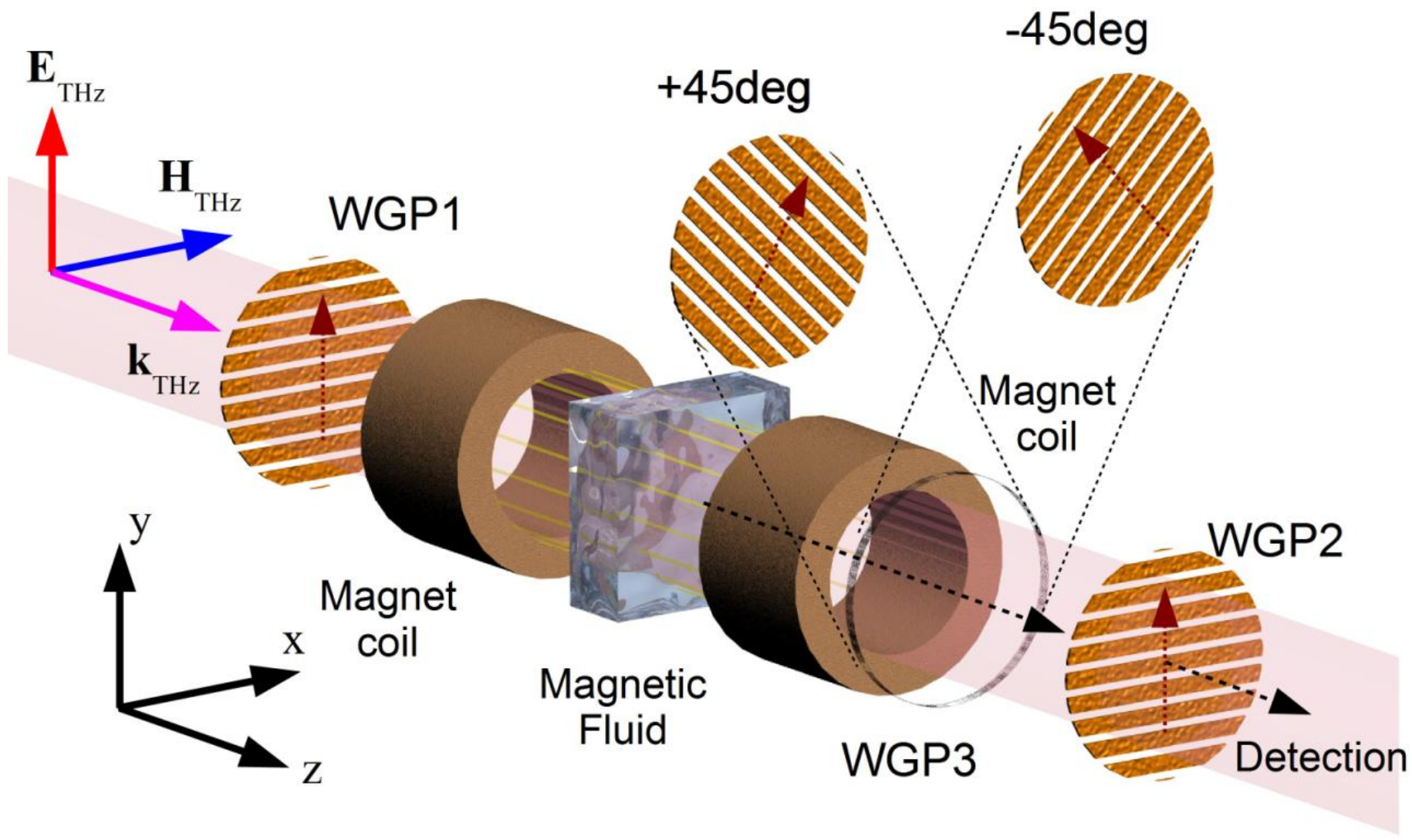

**Figure 4.4** **Terahertz ellipsometry setup to measure both the NRPR and the ellipticity. WGP1 and WGP3 ensure a horizontal polarization of both the generated and the detected THz pulses. Measurements of two crossed states of WGP2 (45° and -45°) give access to the circular components of the propagation eigenmodes and the calculations of both the NRPR and the ellipticity.**

For THz generation and detection, we used optical rectification and electro-optical sampling in two ZnTe crystals. We used a relatively thick (2 mm) crystal for detection in order to obtain a long enough time window (without multiple reflections inside the detecting crystal) for the data analysis. Parameters and specifications of the system are given in Chapter 1. Measurements have been performed in a closed box purged with Nitrogen to avoid THz absorption by water molecules. The magnetic field was varied from 0 to 30 mT. The measured rotation is shown in Fig.4.5 over the spectral range of 0.2-0.9 THz. A rotation as high as 11 mrad/mm was measured using a very small magnetic bias of 30 mT.

In Fig. 4.2, we showed that the built-up magnetization can be approximated by a linear dependence with the applied magnetic field. In addition, the rotation angle is generally linearly proportional to the internal magnetization. This suggests that the rotation angle should be a linear function of the applied magnetic field. This is confirmed by the experimental results presented in Fig. 4.6, where the rotation at 0.2, 0.4, 0.6, 0.8 and 0.9 THz is plotted vs the magnetic field.

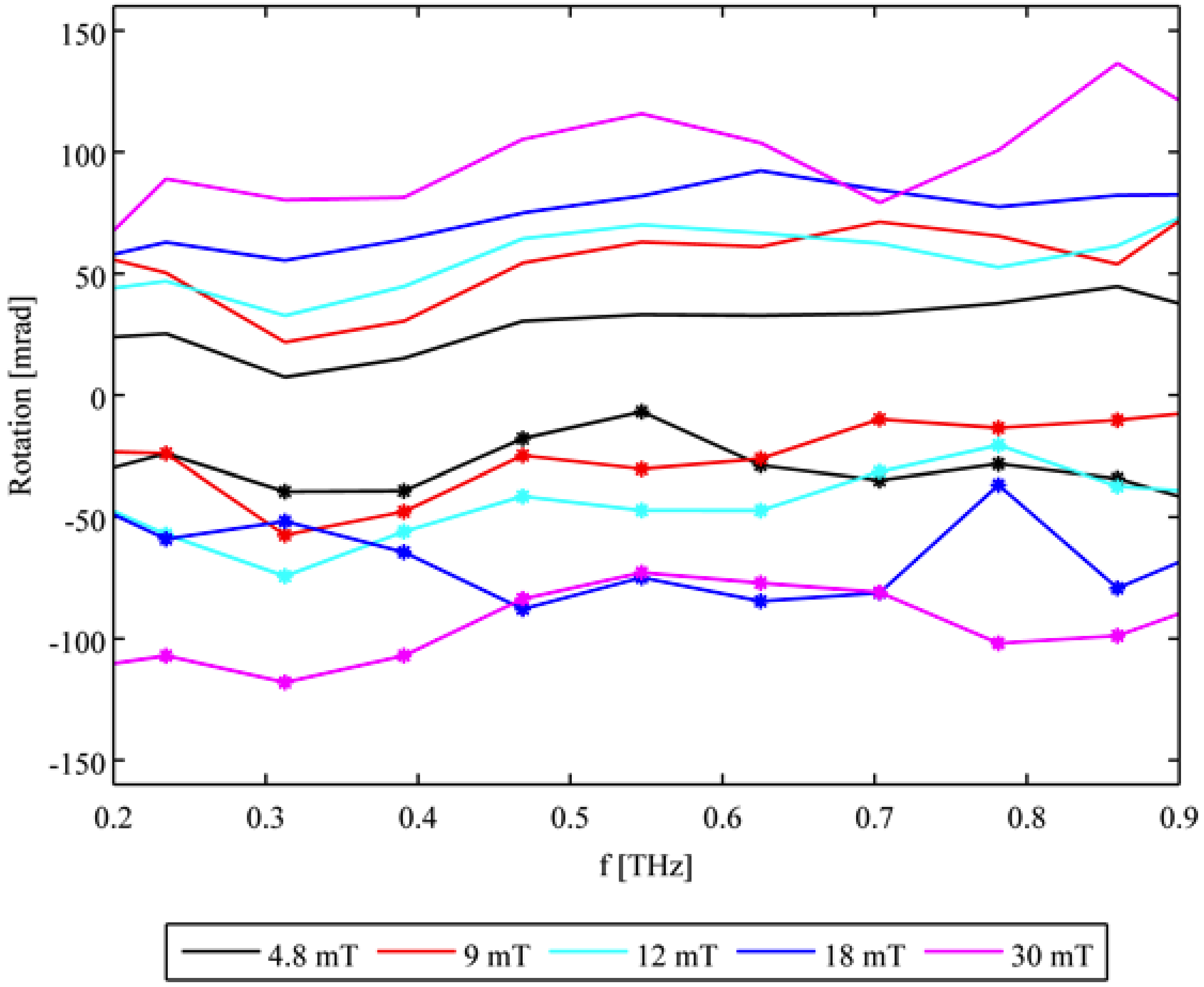

**Figure 4.5** **The spectral dependenceof the EFH1 rotation. Rotation is flat over the spectral range of 0.2-0.9 THz. The dotted lines show the rotation when the magnetic field direction is reversed. The reversal of the rotation signs, in the latter case, confirms the non-reciprocity of the rotation.**

### 4.1.5 Verdet constant and figure of merit of EFH1

We finally compare the performance of our rotator with a relevant case from literature. As shown above, rotation is calculated per unit length. At a given magnetic field bias, changing the propagation length changes the rotation angle. However as the length increases, terahertz absorption increases too thus limiting the maximum achievable rotation. An important factor to evaluate the performance of a rotator is the figure of merit (FOM = rotation x absorption coefficient) that relates the maximum available rotation to the absorption.At room temperature, few studiesof terahertz Faraday rotation are found in literature [42,82,81]. Figure 4.7 compares the Verdet constant of EFH1 with that of doped silicon [82].

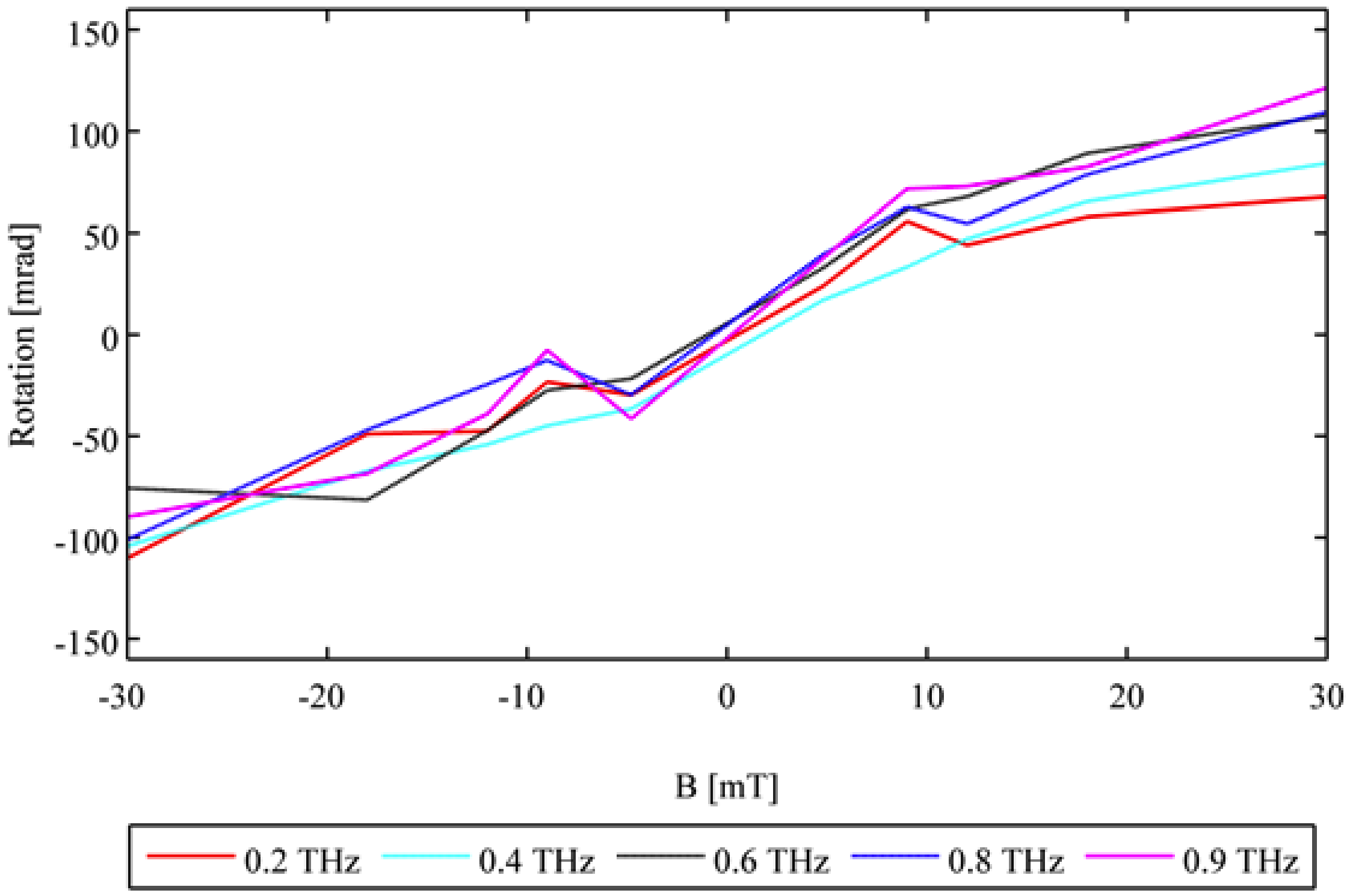

**Figure 4.6** **The magnetic field dependence of the EFH1 rotation. Rotation is shown for fiveTHz frequency lines [0.2 0.4 0.6 0.8 0.9] THz where the linearity of the rotation dependence on theapplied magnetic field is confirmed.**

Although doped silicon has a higher Verdet constant than EFH1, it is strongly frequency dependent. In applications where very narrowband operations are considered, doped silicon can represent a convenient solution. However, for broadband operation, doped silicon does not guarantee a constant rotation over the input pulse spectrum. Considering the FOM, Fig.4.8 shows that the lower losses of EFH1 lead to a strong FOM exceeding that of the doped silicon.

In conclusion, we investigated NRPRs at the THz frequencies. Working with broadband pulses requires broadband retarders. Targeting broadband operation, we realized the first broadband NRPRs operating over a bandwidth 0.2-0.9 THz. Using a Ferrofluid, we obtained significant rotation reaching 11 mrad/mmusing ~30 mTof magnetic field. In addition, its high transparency allows for a high figure of merit. These advantages add to the flexibility of the liquid form for integration in other reciprocal or non-reciprocal devices like waveguides, switches, etc. This work also represents the first study of the properties of Ferrofluids at the THz frequencies.

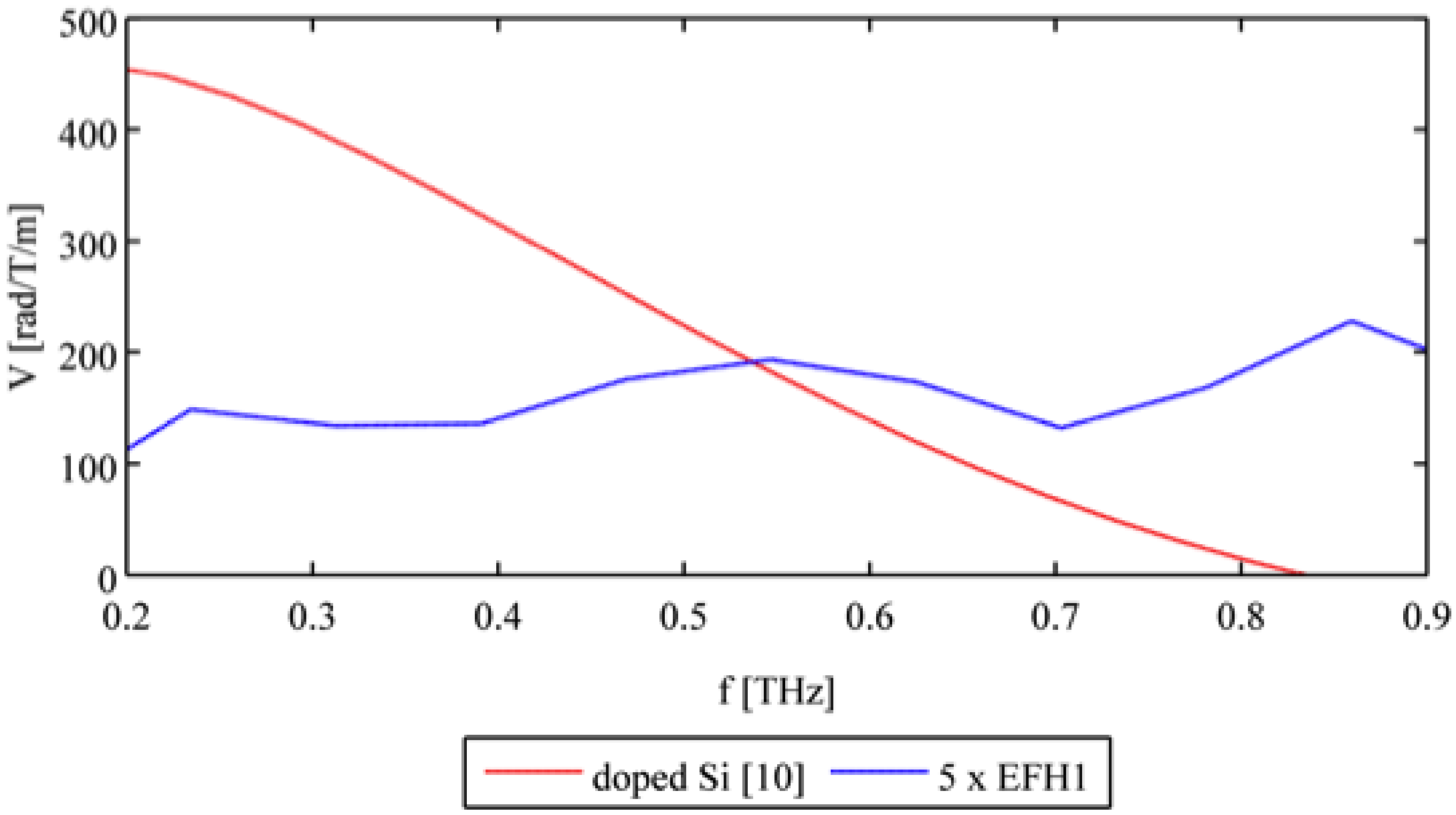

**Figure 4.7A comparison between EFH1 and doped silicon in terms of Verdet constant. Doped silicon shows higher but strongly frequency dependent Verdet constant. EFH1 shows a broad Verdet constant.**

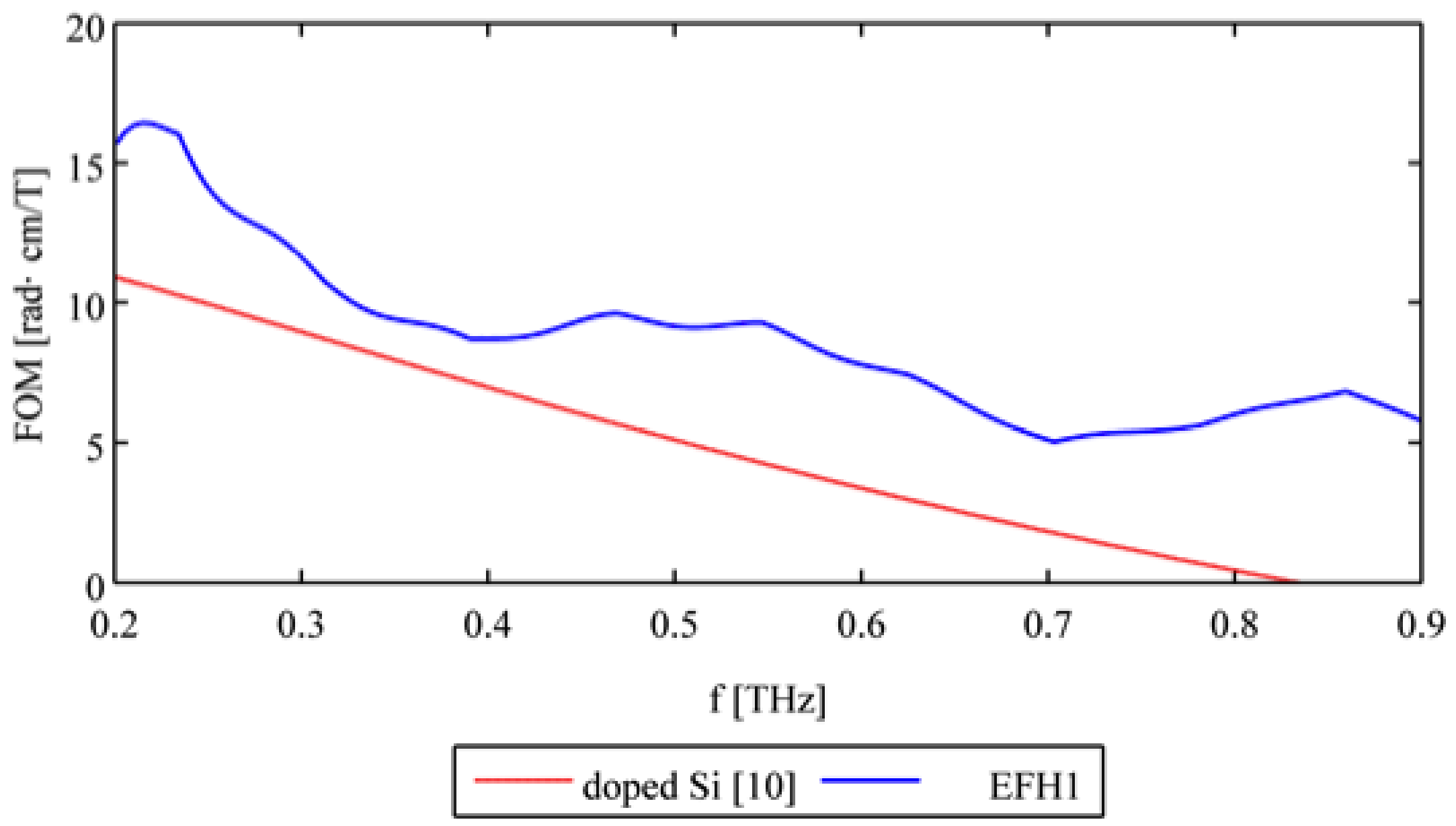

**Figure 4.8** **A comparison between doped silicon and EFH1 in terms of FOM. The comparatively low losses of EFH1 lead to a significant increase in the FOM, exceeding that of the doped silicon.**

## 4.2 Terahertz Magnetic modulation

Terahertz signal processing recently rose to prominence in the THz community with endless potential applications. Increasing demand for high bandwidthsand data rates in the wireless communication systems keeps pushing up the frequency limit, presently reaching the edge of the THz band. The terahertz radiation is still hard to manipulate, mainly because of the lack of suitable materials as well as efficient modulation geometries. Terahertz signal processing has been demonstrated using optical [84-89], electronic [90-92], and thermal [93-95] techniques exploiting materials' dielectric and thermal properties. Different techniques can vary in bandwidth, structure complexity, speed, flexibility, or modulation depth with the application being the final selection criterion. The photoexcitation of a semiconductor can shield a THz wave propagating through it and modulate the THz transmission as it has been shown in section 3.1. This optical modulation occurs on the ultra fast time scale, but it requires an intense laser, which hinders its application beyond the laser laboratories. Thermal excitation of semiconductors changes their THz conductivity and consequently their absorption. Strong thermal modulation can be obtained exploiting the insulator-to-metal transition in $VO_2$that takes place at relatively moderate temperature changes. However, the modulation speed occurs on the time scale of few tens of milliseconds.Finally, the electrical excitation of a Schottky diode built on the top of a semiconducting layer can modulate a propagating THz wave. The speed of this technique falls between that of the thermal and the optical ones but the structural complexity and the compatibility with specific (and generally narrowband) structures are big disadvantages. A conclusion is drawn that there is no optimum technique. In the end, it comes to application.

The magnetic field is an important tool to control material response to EM waves. However, most magnetic materials tend to lose such response as the frequency increases. Extreme conditions of high magnetic field and/or low temperatures are thus crucial to change the material properties. No wonder, then, that magnetic modulation has been absent so far at THz frequencies. Following the big advantages (*i.e.* low losses and low magnetic field requirements) of the Ferrofluids we demonstrated at THz frequencies in section 4.1, we investigated possible magnetic modulation of THz pulses using Ferrofluids. In this part, we show that magnetic field-induced clustering of nanoparticles can efficiently modulate the THz pulses at very low magnetic field intensities, a goal that is difficult to achieve in solid materials even at much higher magnetic fields. We chose Ferrofluids because of the low absorption and the high achievable modulation as will be shown next.

### 4.2.1 Clustering of magnetic nanoparticles and absorption mechanisms

At the beginning of this chapter, the basic mechanism of clustering of magnetic nanoparticles was briefly described. The axis of the cluster is always parallel to magnetic field direction. If the cluster axis is parallel to the THz propagation direction k, (as is the case for non-reciprocal phase retardation studies in section 4.1) THz electric polarization does not experience any significant asymmetry induced by the cluster and therefore the input polarization angle is irrelevant. However, if the cluster is formed in the plane of the field polarization, the THz wave experiences a cluster-induced dichroism, *i.e.* polarization-dependent absorption [96]. This absorption arises from the reorientation of the particles. In this respect, several absorption mechanisms are perceived, e.g.: (a) Rayleigh scattering by the clusters, which can be neglected because the THz wavelength is orders of magnitude larger than the nm scale of the cluster size. (b) Absorption by the field-induced imaginary electric and magnetic polarizations. The latter is the eddy currents losses in the particles and can be ignored here because of the low conductivity of the insulating magnetic nanoparticles. Absorption by the field-induced imaginary electric polarization (*i.e.*

current generated within the colloidal nanoparticles) is thus the main mechanism responsible for light absorption presented here [97].

Under the application of an external magnetic field, if the angle between the cluster axis and the THz electric polarization is $\theta$, it is conventionally useful to describe two waves: ordinary and extraordinary. They refer to the components of the THz polarization at $\theta = 0°$ and $\theta = 90°$, respectively. In the case of ordinary waves, field-induced current and, thus, light absorption reach maximum values. Conversely, light transmission increases in the latter ($\theta = 90°$) case in comparison with the rest (zero) field case (Fig.4.9). This increase is attributed to the alignment of light-absorbing nanoparticles orthogonal to the THz electric field and therefore the loss of the otherwise polarization-insensitive average induced currents.

### 4.2.2 Terahertz linear dichroism in Ferrofluids

The absorption coefficient of a Ferrofluid[94,95] is given by

$$\alpha(0) = (36\pi\phi_M/c)\sqrt{\varepsilon_1}\sigma/(\varepsilon + 2)^2 \tag{4.2.1}$$

Where $\phi_M$, $\varepsilon_1$, and $\sigma$are the concentration, the relative dielectric constant, and the conductivity of the magnetic nanoparticles, respectively. $\varepsilon$is the relative dielectric constant of the carrier liquid.

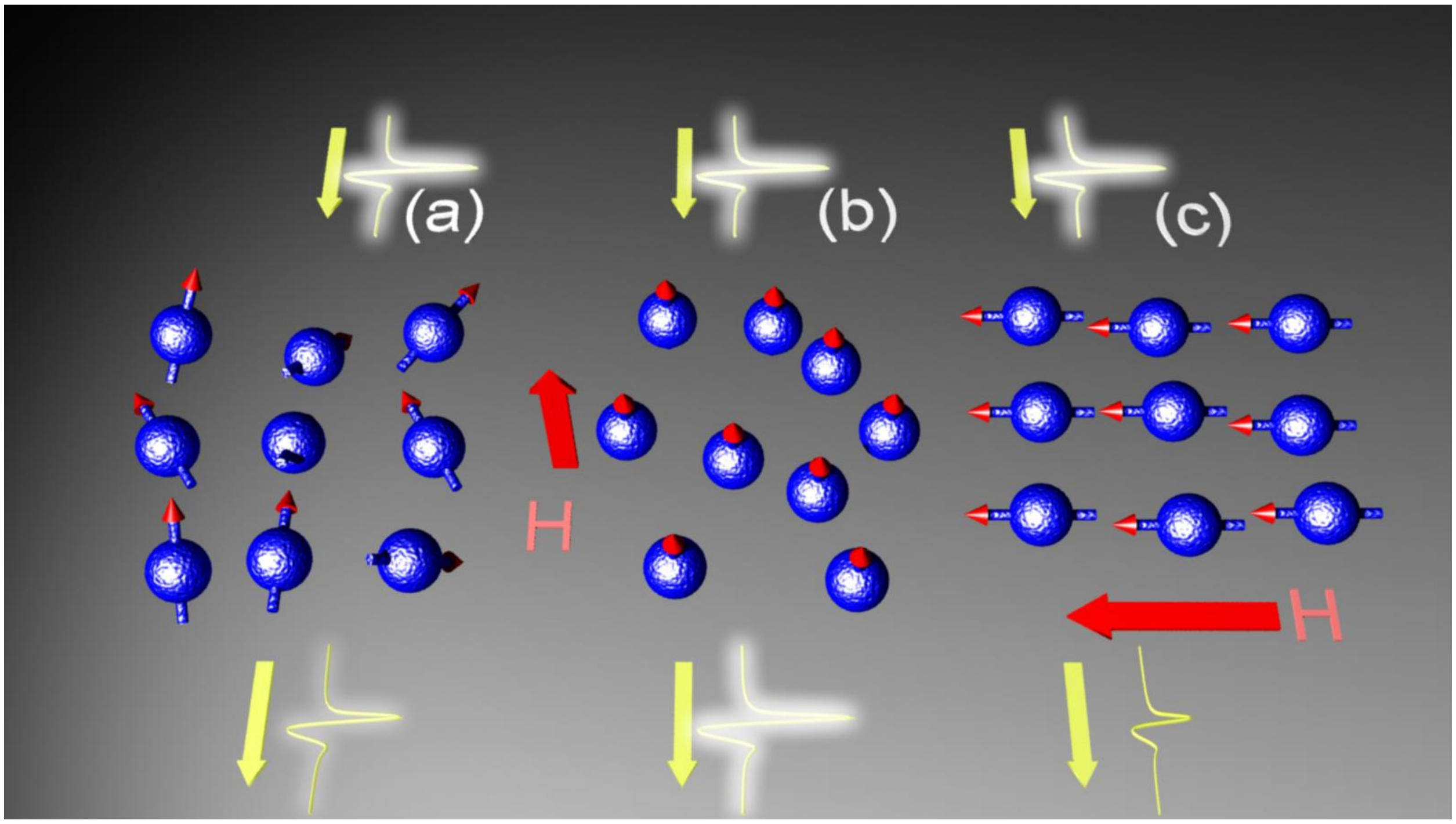

**Figure 4.9** **Cluster formation of the magnetic nanoparticles. In the absence of external magnetic fields, the nanoparticles are randomly distributed with their magnetic moments oriented in different directions and no net magnetization is perceived (a). External magnetic field tends to align the nanoparticles to its axis forming clusters. If the cluster is createdin the direction of electric field polarization (*i.e.* ordinary wave) high attenuation is induced (c). An extraordinary wave experiences a corresponding increase in the transmission (b).**

The ordinary and extraordinary waves experience totally different absorptions. Quantitatively and in a respective order, their absorption coefficients [96,97] are given by

$$\alpha_{//} = 4\pi\sigma\sqrt{\varepsilon_1}\phi_M/c[1 + \langle N\rangle(\varepsilon - 1)]^2 \quad (4.2.2)$$

$$\alpha_{\perp} = 16\pi\sigma\sqrt{\varepsilon_1}\phi_M/c[(\varepsilon + 1) - \langle N\rangle(\varepsilon - 1)]^2 \quad (4.2.3)$$

Where $\langle N\rangle$is the average electric depolarization factor along the cluster axis. At low magnetic fields (as is the case of the measurements presented here,) the variation in $\langle N\rangle$ from that of a sphere is taken to be small and it can be expanded to $\langle N\rangle = 1/3 - \Delta N$where $\Delta N$is the deviation factor. The induced change in $\Delta\alpha_{//} = \alpha_{//} - \alpha(0)$and $\Delta\alpha_{\perp} = \alpha_{\perp} - \alpha(0)$can then be written as

$$\Delta\alpha_{//} = 6[(\varepsilon - 1)/(\varepsilon + 1)]\alpha(0)\Delta N \quad (4.2.4)$$

$$\Delta\alpha_{\perp} = -3[(\varepsilon - 1) - (\varepsilon + 2)]\alpha(0)\Delta N \quad (4.2.5)$$

And the two absorption coefficients can thus be simply related [96,97] by

$$\Delta\alpha_{//} = -2\Delta\alpha_{\perp} \quad (4.2.6)$$

Phenomenologically, this is manifested in an induced linear dichroism in a traveling wave. Although the last relation found in the studies related to near IR narrowband experiments, interestingly enough, it does not depend on the frequency as long as the wavelength is much longer than the nm-sized nanoparticles chains. This gives it a big advantage for broadband THz applications where the spectrum can easily extend over a frequency decade from a broadband source.

### 4.2.3 Measurement of terahertz linear dichroism

Our samples were 10 mm-long cuvettes filled with magnetite ($Fe_3O_4$)-based Ferrofluids EFH1 and EFH3 with the particle concentrations of 7.8% and 12.4%, respectively. The nanoparticles are around 10 nm-sized. Two experimental configurations were used corresponding to the ordinary and extraordinary waves. The direction of the applied field and the polarization convention are shown in Fig.4.9. For the ordinary wave, the THz electric and magnetic fields are assumed to be (x-) horizontally and (y-) vertically polarized, respectively. Two wire grid polarizers were placed before and after the sample and vertically aligned to ensure a horizontal polarization of the generated and the detected THz fields. In the case of the extraordinary wave measurement, the THz generation crystal, the detection crystal, and the wire-grid polarizers were rotated 90°. THz was generated using the optical rectification nonlinear technique in a ZnTe crystal and detected using the electro-optical sampling in a second ZnTe crystal. The magnetic field was supplied by an electromagnet (GMW-3470) and was in-plane and horizontally aligned during all the measurements in this section. The transmission measurements of the ordinary and the extraordinary waves through EFH1 are shown in Fig.4.10.

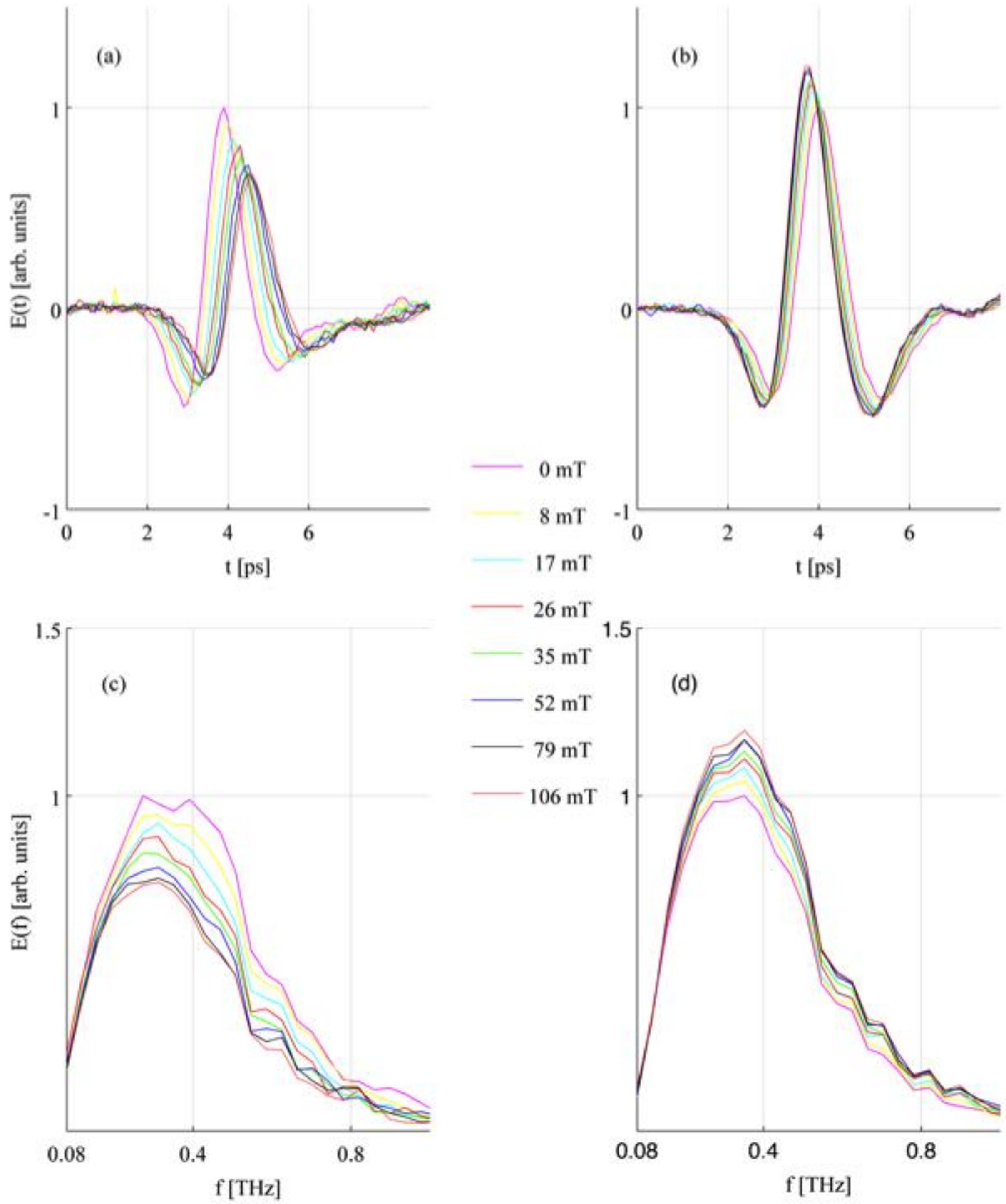

**Figure 4.10** **Cluster-induced effects in the THz transmission. (a) and (c) show the attenuation at different magnetization levels in both the time profile and spectrum of the ordinary wave. (b) and (d) show the corresponding increase in the transmitted extraordinary wave.**

### 4.2.4 Estimation of terahertz modulation depth

To calculate the induced absorption coefficients from experimental measurements, we first write the THz field as $E(\omega)_\circ = E_\circ(\omega)e^{-\alpha d}$ where $\alpha$ and $d$ are the absorption coefficient and the sample thickness, respectively. Perturbation in the sample refractive index was found to be less than 3% for the levels of magnetic field used in our experiments. Therefore, we assume that Fresnel losses are constant and the induced losses are totally contained in the attenuation factor $e^{-\alpha d}$. We then define the logarithmic transmission

$$t_i = ln\frac{E(\omega)_i}{E(\omega)_\circ} = \Delta\alpha_i d \quad i \in (\perp,//) \tag{4.2.7}$$

Equation (4.2.7) shows that the transmission amplitude is a thickness-scaled induced absorption. The latter is shown in Fig. 4.11(a) for the ordinary and the extraordinary waves in comparison with the theoretical predictions from Eq.4.2.6. For the two levels presented in the figure (17mT and 35mT,), a very good agreement is obtained between the theory and the experiment. Most importantly, it is manifested over broadband frequency range shown here up to a frequency decade.

Following the literature, we evaluate the efficiency of the modulation process by calculating the intensity of the modulation depth, which we define as

$$I_m = \frac{|E^\circ|^2 - |E|^2}{|E^\circ|^2} \tag{4.2.8}$$

Circle superscripts denote the measurement in the original unmodulated pulse. Figure 4.12(b) shows the intensity modulation depth of the ordinary waves for the two levels of the magnetic field presentedin Fig.4.11. A modulation depth, as high as 66%, is shown at 0.8 THz using only 26 mT of magnetic field. Such a field is very low and can be obtained simply bycirculating a small current in a wire. The modulation increases with frequency. That originates from the linear dependence of the induced absorption on the rest absorption coefficients (Eq.4.2.4 and 4.2.5) as well as fromthe linear dependence of the latter on frequency.

### 4.2.5 Nonlinear regime and saturation effect

In Fig. 4.13, we show that the modulation increases with the applied magnetic. However, the induced magnetization in Ferrofluid grows rapidly with the applied field first, then the process of magnetization build-up and goes nonlinearly (*i.e.*saturates). This behavior is described by the Langevin function $L\,(KH)$ of the magnetization $M = coth\,(KH) - 1/KH$ where $H$ is the applied field, and $K$ is a temperature dependent parameter.

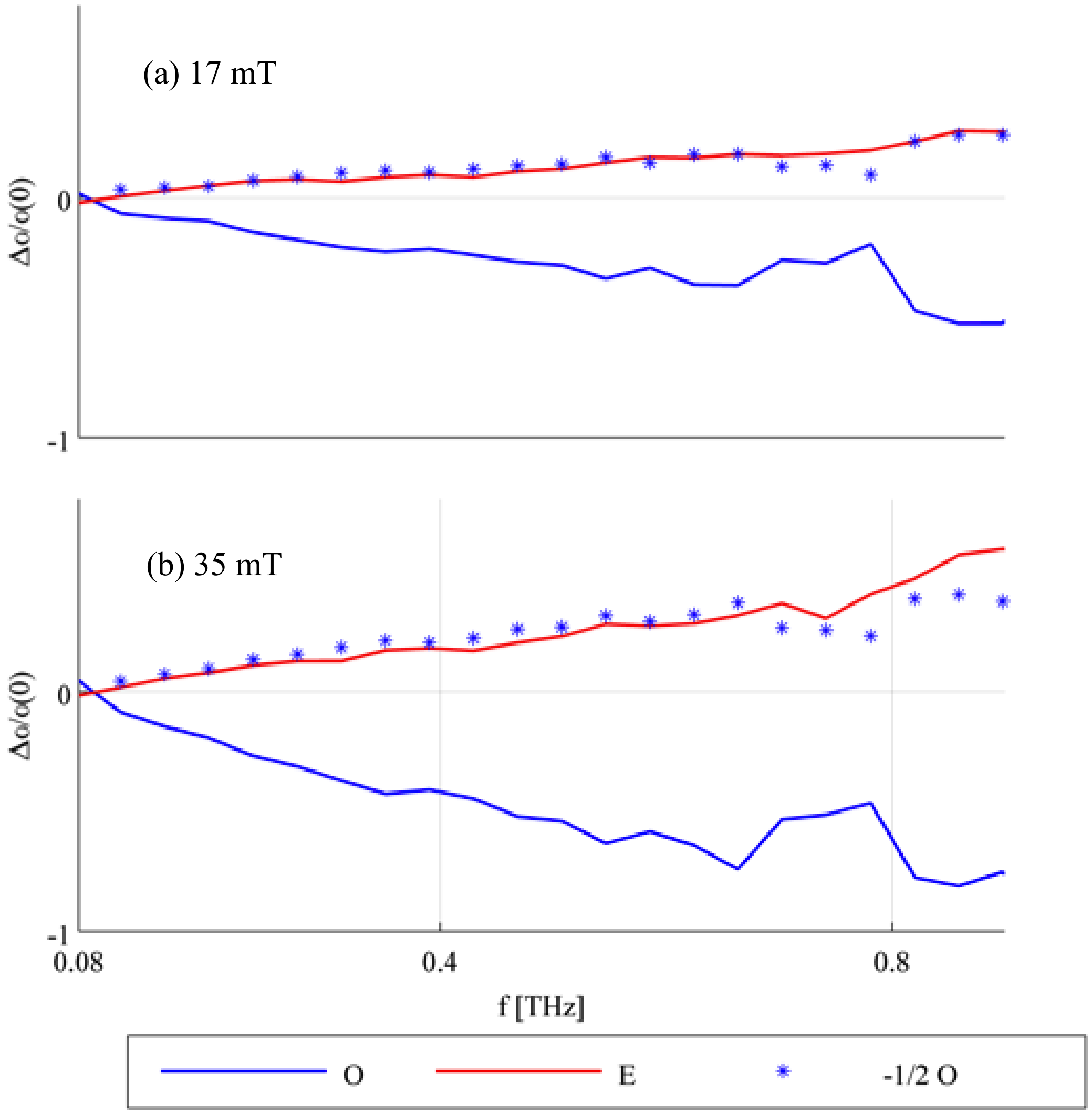

**Figure 4.11** **Terahertz linear dichroism. The induced absorptions in the ordinary and extraordinary waves are experimentally shown. Also the theoretically estimated absorption of the extraordinary wave is compared with experiments under the application of (a) 17mT and (b)35mT magnetic fields, respectively.**

At the same time this nonlinear behavior sets an upper limit on the applicability of the technique, the rapid increase in $M$ at low magnetic fields gives Ferrofluids an advantage when it comes to its integration and flexibility of use in low magnetic field applications. At the same time EFH1 requires 1 T to reach the saturation magnetization level of 40 mT, while only 30 mTof applied field is required to reach half the saturation level. In addition, in this region, the magnetization process, and the dependent effects, follow linear behavior in function ofthe applied field. In these calculations, we neglected the demagnetization factor, which is very small in our case, given the low saturation magnetization. As the induced absorption is proportional to $M$, the former is thus expected to follow the applied field witha similar Langevin-like behavior. This is experimentally demonstrated here in Fig. 4.13 where a magnetic field up to 600 mTwas considered.

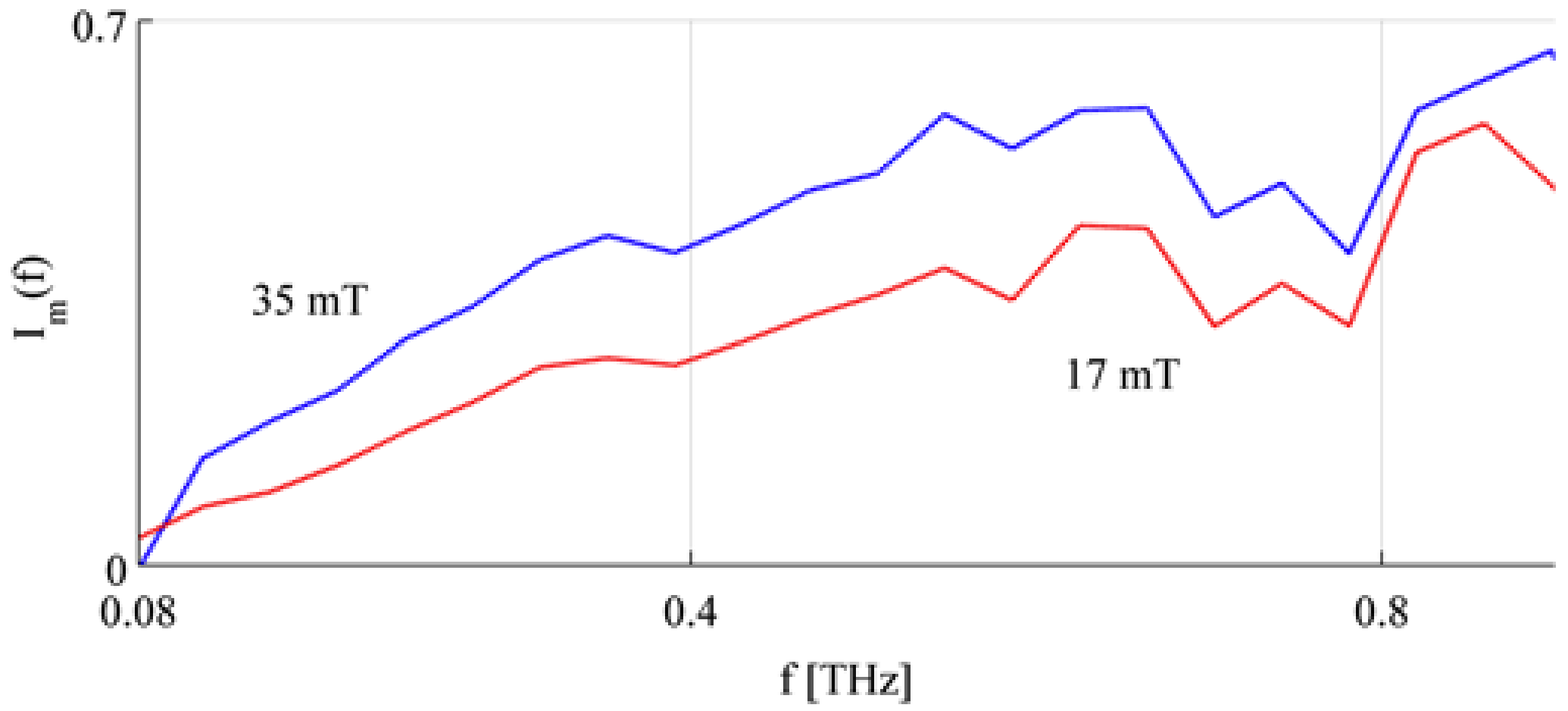

**Figure 4.12** **Magnetic modulation depth in EFH1. Broadband measurement of the magnetic modulation depth in the frequency range (0.08-0.8) THz is shown for two levels of magnetic field: 17 mT and 35 mT. Modulation increases with the THz frequency and the applied magnetic field.**

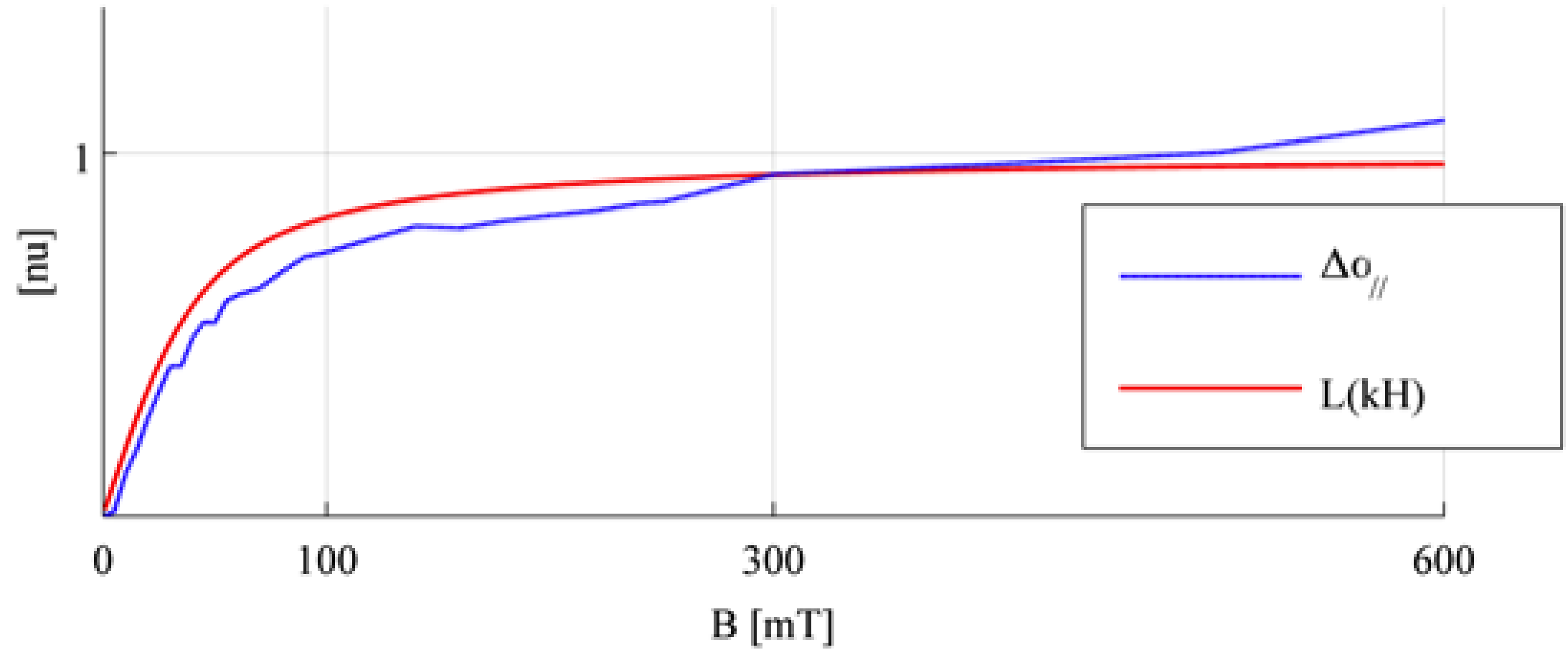

**Figure 4.13** **Nonlinear behavior of the magnetic modulation. The Langevin response of a magnetic liquid is shown in the magnetic field range going up to 600 mT. The induced absorption of the ordinary wave is shown to have a similar behavior.**

### 4.2.6 Field polarity and concentration effect

In addition to the nonlinear behavior described above, several factors can affect the modulation process such as the field polarity, the concentration, the viscosity of the carrier fluid, the magnetic nanoparticle dimensions, and the temperature. The attenuation process is mediated by an increase in the electrical conductivity as the number of particles aligned with the magnetic field increases. The modulation process is thus expected to be independent of the THz polarity. This is confirmed in Fig. 4.14 where the THz pulse is shown under the application of two oppositely polarized magnetic fields ±44 mT. As shown, the two pulses undergo similar attenuation thus confirming the insensitivity of the process to the magnetic field polarity.

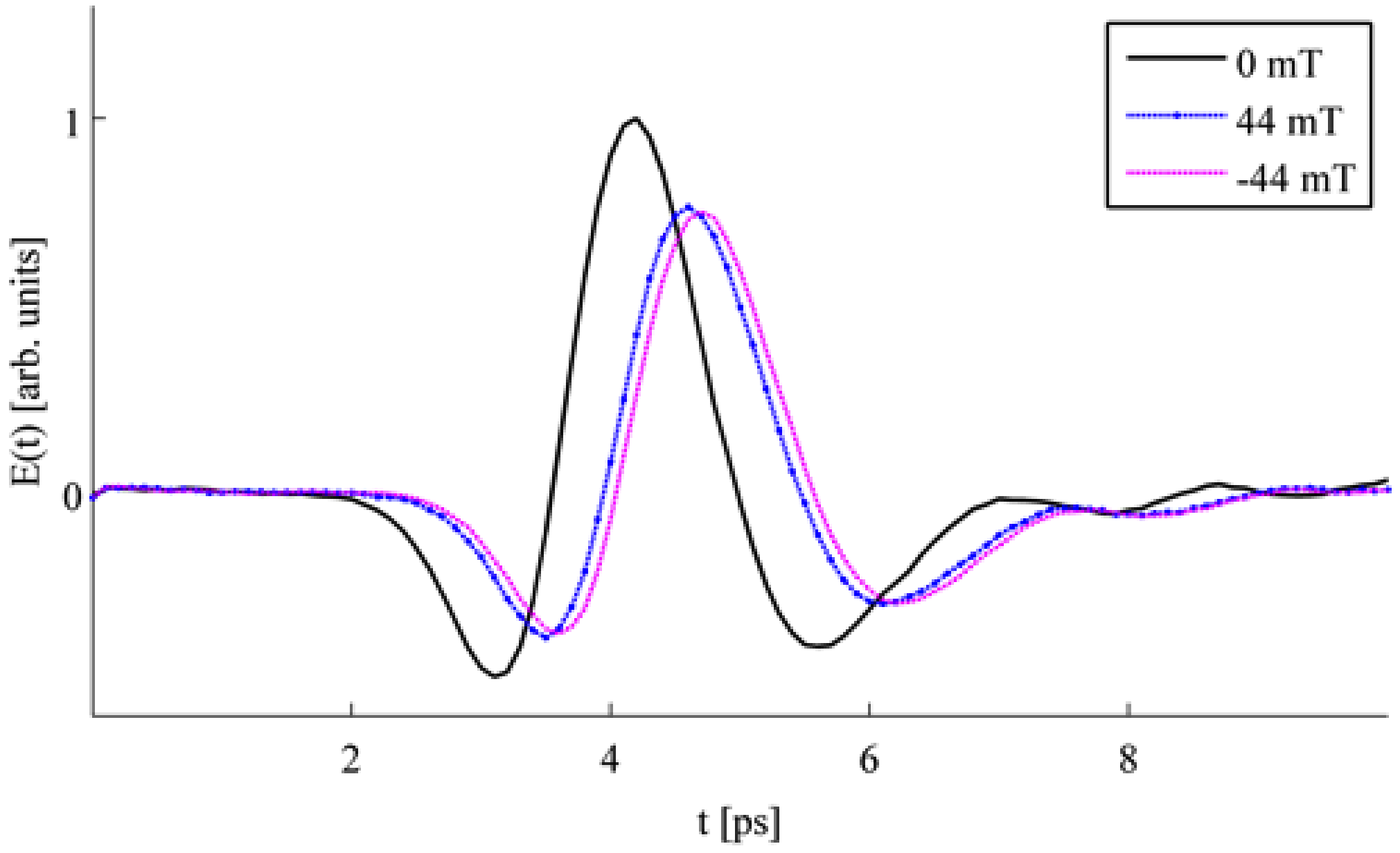

**Figure 4.14 Magnetic field Polarity and THz modulation. The measurements of the THz pulse in the absence of an external magnetic field and with two oppositely polarized fields. The induced absorption is shown to be insensitive to the sign of the magnetic field.**

The effect of the concentration is considered here by comparing two Ferrofluids from the same series (EFH1 and EFH3) with their concentrations related by $\phi_{M_EFH3} = 1.5\phi_{M_EFH1}$. As $\Delta\alpha$ is directly proportional to the concentration (Eq. 4.2.1; 4.2.2; 4.2.3,) we therefore expect $\Delta\alpha_{EFH3} = 1.5\Delta\alpha_{EFH1}$. This last relation is experimentally verified and shown in Fig. 4.15 where the measurement of $\Delta\alpha_{EFH3}$ and $1.5\Delta\alpha_{EFH1}$are shown with excellent agreement at magnetic field levels of 13 mT,26mT; and70mT, respectively. This implies that a higher absorption modulation can be obtained by increasing the concentration of the sample. However this comes at the expense of an increased absorption. The advantage of using higher concentration liquids over longer sample length can be seen if this liquid is coupled with other structures where the thickness can't be arbitrarily varied (like metamaterials) or higher thickness induces more losses associated with the structure itself (like waveguides.)

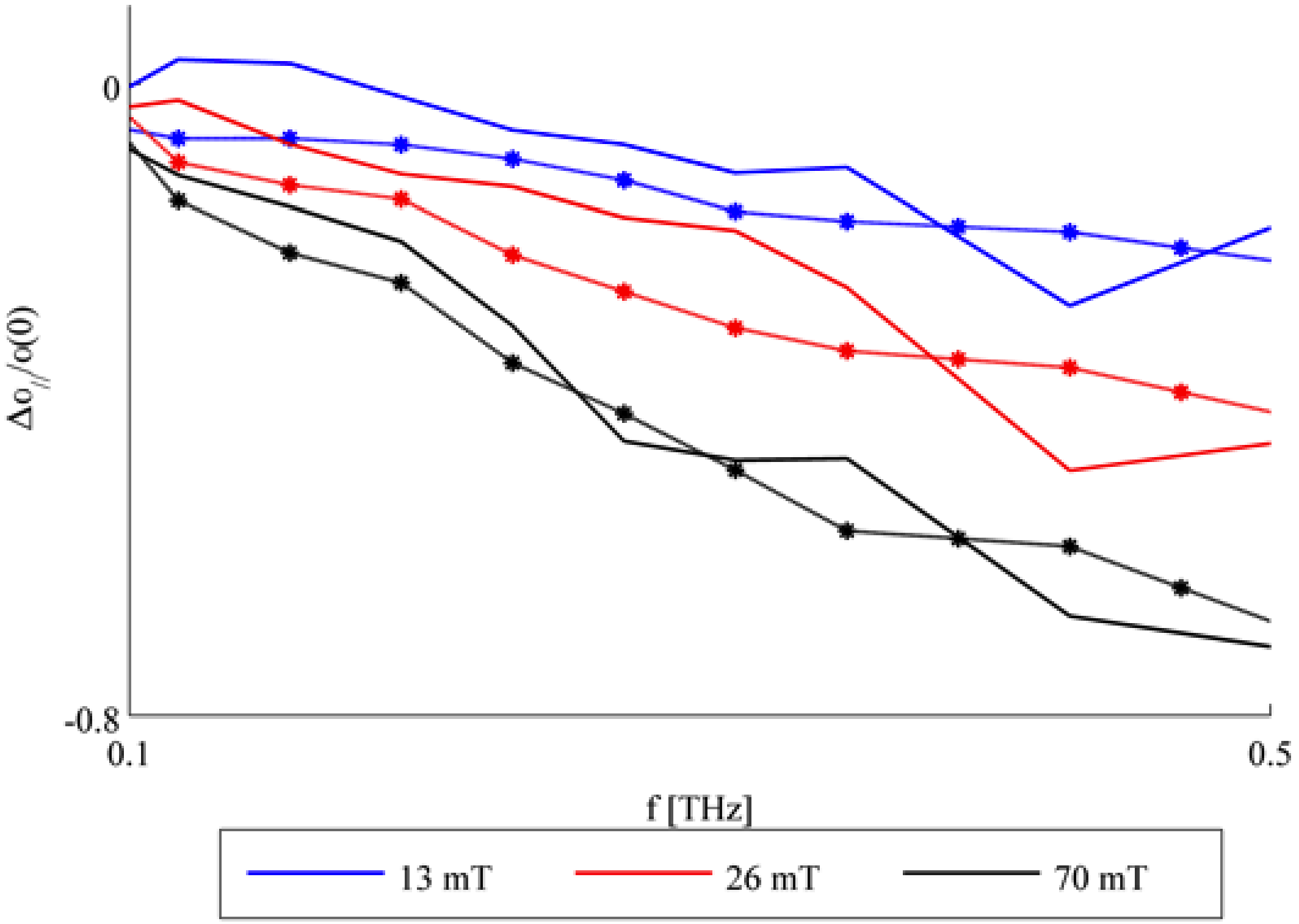

**Figure 4.15** **The effect of nanoparticles concentration on the THz modulation. The induced absorption in EFH3 (marked with asterisks) and EFH1 (after scaling it by a factor of 1.5 to account for the ratio of nanoparticles concentrations.) Agreement between the two measurements confirms the scalability of the induced absorption with the nanoparticles concentration.**

In conclusion, we used magnetic field-induced clustering of nanoparticles in Ferrofluids to modulate THz transmission. The demonstrated technique combines high modulation depth and low magnetic field requirements while preserving the flexibility of using liquids. We believe that our results will open up a new class of THz modulators that can be integrated in other magnetic/nonmagnetic systems.

## 4.3 The non-reciprocal terahertz isolator

A Faraday isolator is an electromagnetic non-reciprocal device, a key element in photonics. The high importance of the isolator stems from its ability to shield electromagnetic sources against the effect of unwanted radiation such as back-reflected light and back-propagating spontaneous emissions. The circulator, a common isolator variant, is widely used to obtain a complete separation between the forward and backward propagating waves. This enables the realization of a desired transfer function in reflection only. In this part, the first isolator operating with THz pulses is demonstrated.

### 4.3.1 Principle of operation of the electromagnetic isolator and the terahertz challenge

In section 4.1, we demonstrated the principle of non-reciprocal phase retarders. An isolator is simply a phase retarder operating at the specific angle of 45º. Over a century ago, Lord Rayleigh described a one-way transmission system based on Faraday rotation [98]. In Faraday rotators, the phase shift between the two propagating circular eigenmodes reverse its sign as the wavevector is reversed. As this is accompanied by a reversal of the propagation coordinates, both forward and backward propagation induce the same phase retardation, *i.e.* the reflected wave possesses a polarization state different from the input one. If the rotation is adjusted to 45º and the rotating medium is placed between two specifically aligned polarizers, light propagates in only one direction.

Such non-reciprocity finds its application in many fundamental systems. In the microwave regime, isolators, gyrators, and circulators are basic examples of non-reciprocal elements that have been vital parts of microwave systems over the past half-century [99-102]. Looking at optical frequencies, isolators are extremely popular in laser devices and photonic circuits. Unidirectional propagation is a typical requirement in fiber laser sources or in amplification chains [103-106].

At the upper end of the GHz band, isolators have already been crucial parts in the operation and testing of important systems like the free electron lasers (240GHz) [107] and in the Planck telescope (320GHz) [108]. The available isolation frequency lines and highest achievable isolation frequency set a limit on the range of systems to which this isolatorcan be applied [109].

As highlighted in the introductory part of the chapter, broadband operation imposes itself as an important requirement in most THz applications. Many experiments require broadband or tunable sources, which can easily cover more than a spectral decade. Not surprisingly, due to these severe constraints, and in spite of its enormous importance, a THz isolator *has not been realized so far*.

### 4.3.2 Selection of the sample

We chose a Strontium Iron Oxide magnet ($SrFe_{12}O_{19}$), a commercially available Ferrite as a sample for the isolator. Faraday rotation at THz frequencies has been demonstrated at room temperature in both solid [82,83] and liquid samples [110]. However, in comparison, our phase retarder has three main advantages towards the realization of an isolation device. First the induced Faraday rotation is weakly sensitive to frequency in the THz band. This originates mainly from the fact that Strontium Ferrites exhibit ferromagnetic resonance around 50-60 GHz (depending on the applied magnetic field), thus far below the THz regime [111]. This allows for a very low-dispersive operation. Second, although Ferrites generally exhibit magnetic properties similar to those of conducting ferromagnets, their conductivities are in general very low. Isolators require a phase retarder where significant polarization rotation is obtained upon propagation. This directly correlates the maximum achievable rotation to the inherent losses. Terahertz low-loss media are, therefore, fundamental for practical devices. Finally, $SrFe_{12}O_{19}$ falls under the general class of permanent (hard) magnets, *i.e.* it retains its magnetic state when the external magnetizing field is removed, hence with respect to the optical counterparts, external magnets are not required to sustain the operation of the isolator.

### 4.3.3 Structural characterization

The sample phase was confirmed using the X-ray diffraction (XRD) technique. XRD is an analytical technique that can be used to reveal the chemical and physical composition of materials. Because X-ray wavelengths are comparable to the interatomic distances, measurement of the diffracted X-ray gives information on the crystal structure. A typical diffractometer records the intensity of the diffracted waves at different angles (XRD $\theta - 2\theta$ spectrum). The angular spectrum is characteristic to a specific crystal structure and is used to identify and confirm the material phase. We carried out the characterization using a Rigaku (D/MAX-2200/PC) X-ray diffractometer with Cu K-alpha radiation line using the JCPDS file no. 33-1340. Figure 4.16 shows the XRD $\theta - 2\theta$ spectrum, which confirms the $SrFe_{12}O_{19}$ crystalline phase.

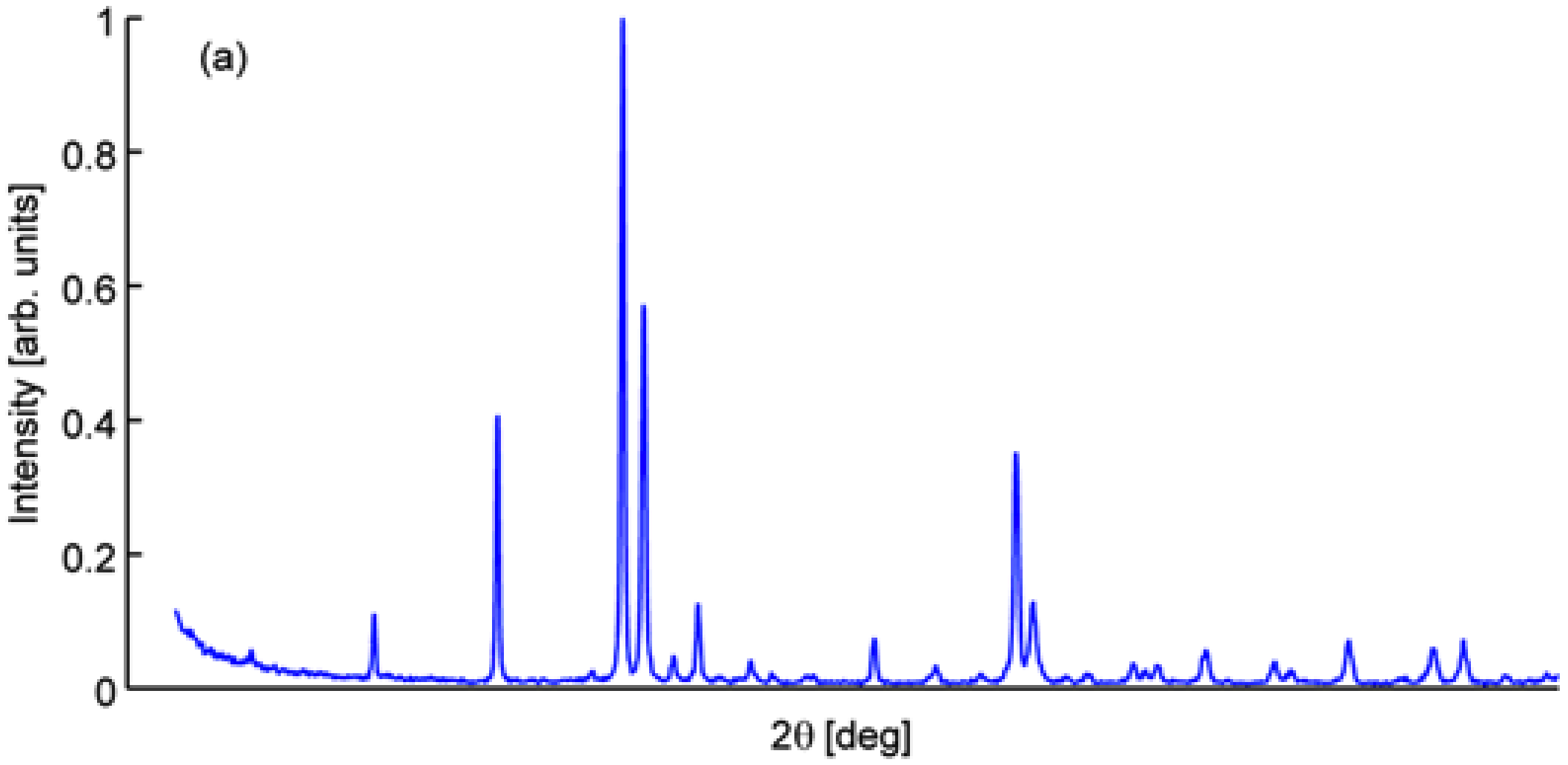

**Figure 4.16** **Structural characterization. The measured $\theta - 2\theta$ XRD spectrum depicting the crystalline phase of $SrFe_{12}O_{19}$**

### 4.3.4 Magnetic medium characterization

The magnetic medium was characterized via the measurement of the hysteresis curve (magnetization state M versus the applied induction field B) using a Lakeshore vibrating sample magnetometer (VSM) (model 7400) at room temperature. Figure 4.17(a) shows the hysteresis behavior of the sample where the saturation magnetization was found to be 360 kA/m. In order to estimate the remanence magnetization after each magnetization stage, we measured the induction field intensity ($B_0$) at specific distances from the sample. Then, we calibrated it against the remanence at saturation, which was obtained from the hysteresis curve. The decay of magnetic induction with the distance from the sample is shown in Fig. 4.17(b).

### 4.3.5 Dielectric properties

Measurements of the complex dielectric function as well as all the spectroscopy measurements in this part have been performed, once more, using a standard THz time-domain spectroscopy (THz-TDS.) THz pulses were generated via optical rectification in a ZnTe crystal illuminated by femtosecond Ti:Sapphire laser pulses (130 fs-wide pulses at a repetition rate of 1 kHz) having a wavelength centered around 800 nm. The detection has been performed by means of the electro-optical sampling technique using a second ZnTe crystal. Details on THz-TDS characterization are given in Chapter 1. We used a relatively thick ferrite sample (3 mm) and the isolation characterization requires the field to pass twice through the rotator. This limits our transparency window to 0.08-0.8 THz. However, as shown in the next paragraphs, the device shows high rotation and the isolation can be achieved with a 1 mm-thick sample.

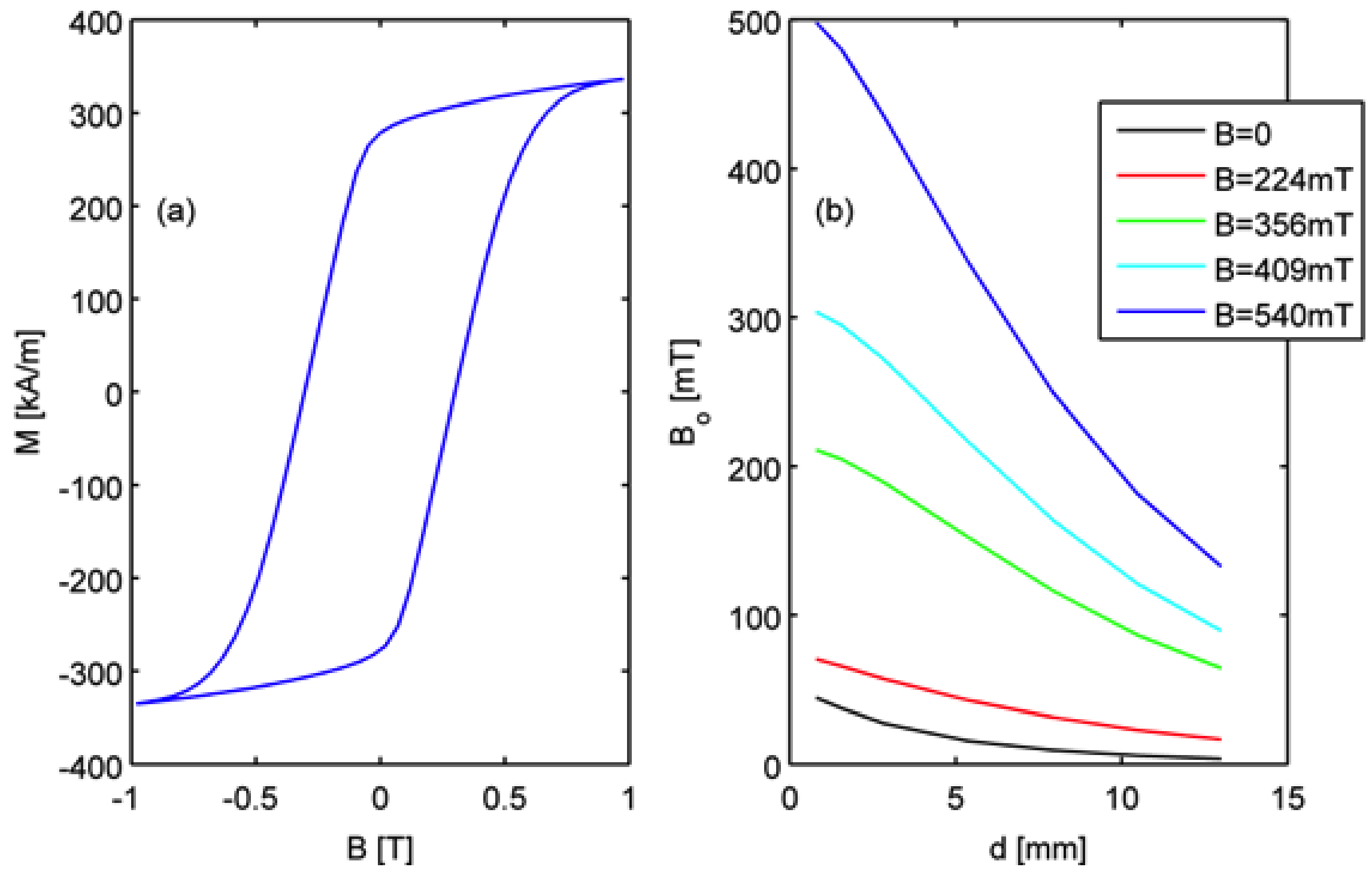

**Figure 4.17** **Magnetic medium characterization.(a) Hysteresis curve of Strontium Iron Oxide obtained from the VSM measurements. (b) Induction field at specific distances from the sample surface measured at different magnetization stages.**

The THz electric and magnetic fields are assumed to oscillate along the y- and x-axis, respectively, whereas the THz wavevector lies in the out-of-plane (sample normal) z-direction.

In order to retrieve the complex refractive index of the sample, we used THz-TDS and index retrieval procedure as explained in chapter 1 and used in section 4.1. The calculated refractive index and absorption coefficient are shown in Fig. 4.18(a) and (b) for the case of an unmagnetized sample. Under such a condition, the medium does not exhibit any significant anisotropy and the residual medium magnetization does not contribute to the phase delay of the THz wave used to calculate the dielectric function. To demagnetize the sample, we applied a reverse magnetic field in order to remove the net out-of-plane magnetization in the entire region illuminated by the THz field. It should be noted, however, that part of the absorption shown in Fig. 4.18 comes from domain wall losses that are reduced upon the magnetization of the sample, *i.e.* following the alignment of the magnetic domains.

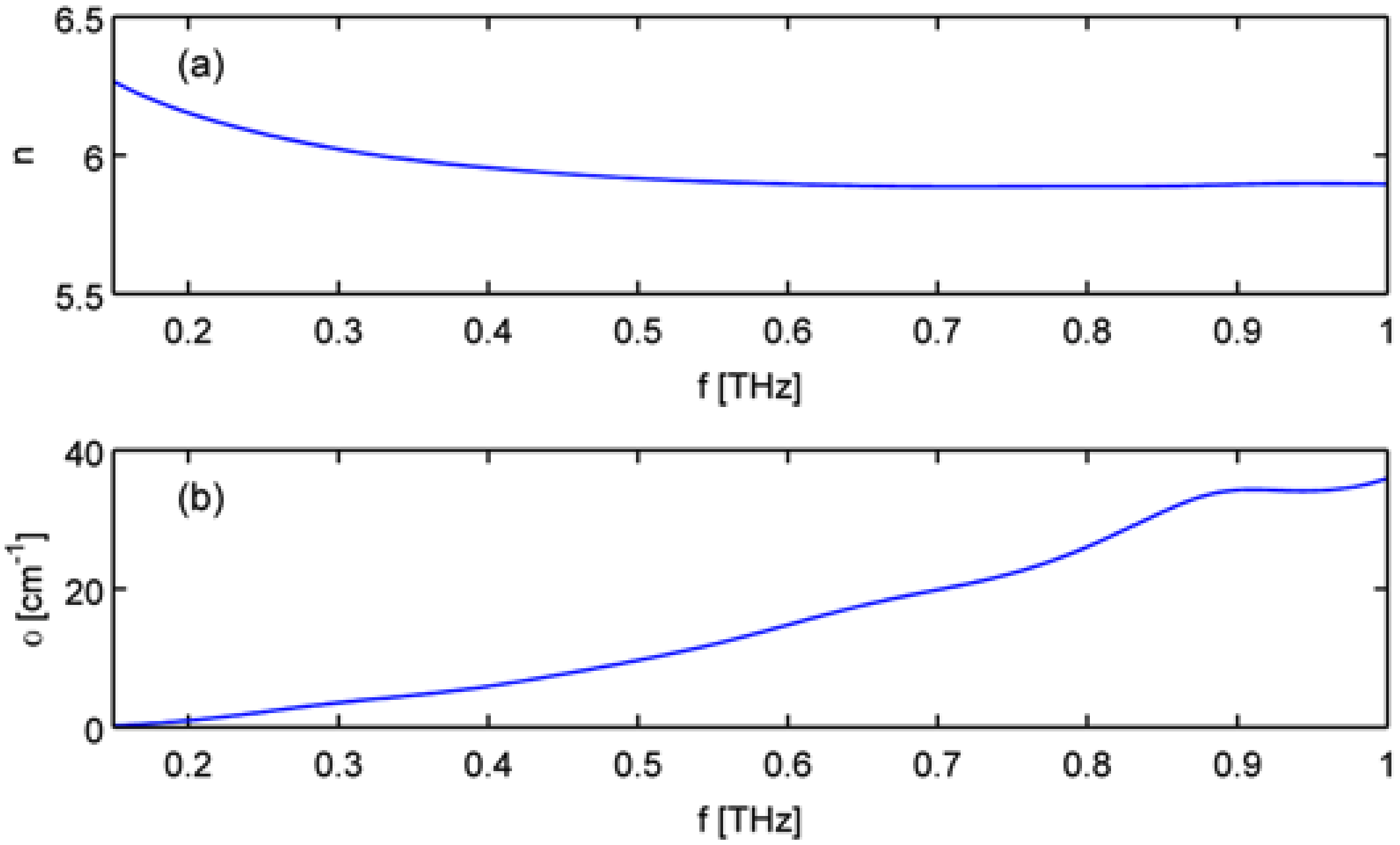

**Figure 4.18** **Dielectric characterization. (a) Refractive index and (b) absorption coefficient of the sample calculated using a transfer matrix technique after the measurement with a THz-TDS system.**

### 4.3.6 Measurement of the phase retardation

We measured the phase retardation using a typical ellipsometry setup as explained in detail in Chapter 1. The setup is shown (again here) as part of Fig. 4.19 where the polarization state was probed by means of a three wire-grid polarizers (WGP1;WGP2;WGP3) configuration.We assume that at 0º, the wire-grid polarizers are transmissive for the vertically polarized THz electric field, *i.e.* their wires are horizontally aligned. This condition corresponds to the maximum transmitted signal.Starting with an unmagnetized state, the sample was permanently magnetized along the THz propagation axis by applying gradually and incrementally an external magnetizing field before each measurement. The out-of-plane remanence magnetization at each measurement was estimated by probing the magnetic induction at a specific distance from the sample surface, and then by calibrating it against the saturation magnetization as explained in 4.19. The magnetization state was also found to be stable over time after the applied magnetic field was removed.

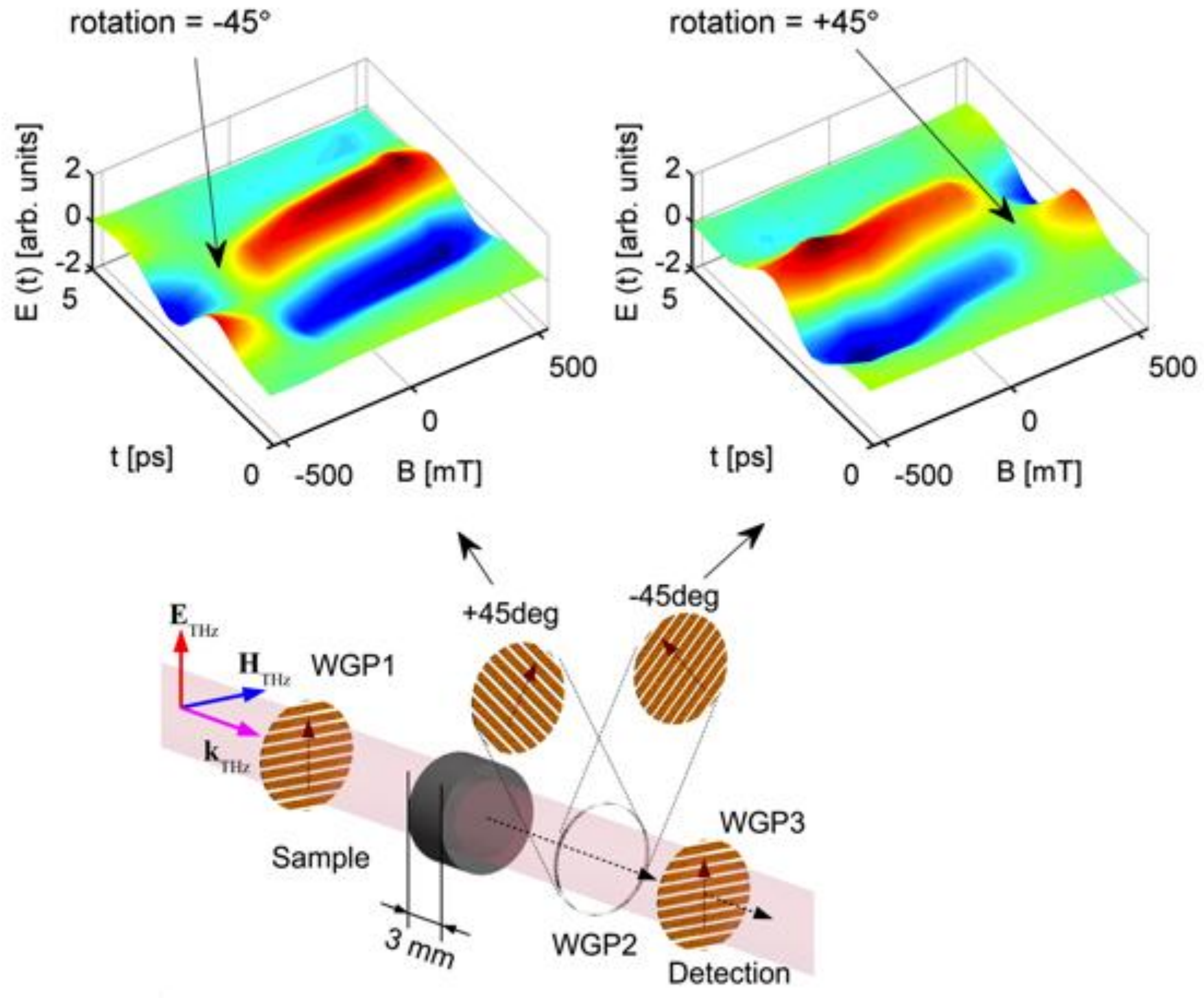

**Figure 4.19** **Forward phase retardation and Mapping. Transmitted THz wave when WGP2 is aligned to 45° and -45°, respectively, is shown as a function of the magnetic field. The -45° and 45° rotations correspond to zero transmittance in the two respective branches.**

WGP1 and WGP3 are set to 0º to ensure the vertical linear polarization of both the generated and detected signals. By adjusting the rotation of WGP2 to 45º and -45º from the maximum transmission position, respectively, we probed the two orthogonal polarization states. As shown in Fig. 4.19, with the increase in the magnetic filed, the THz starts to rotate. Reversal of the direction of the field leads to inversion of the sign of rotation. That is clearly shown by looking at the zero points on the 45º and the -45º measurements of WGP2. Those points correspond to the condition of crossed polarizations, *i.e.* the THz polarization isaligned at -45º and 45º, respectively. Figure 4.20 shows an in-depth look at the rotation where four time traces are plotted. First, when the sample is unmagnetized, a maximum THz transmission is found. Second, when the sample is magnetized to -45º and WGP2 is set to 45º, no signal is detected (because the THz polarization is crossed with the wire grid transmission.) Finally, by keeping the same condition of magnetization, aligning WGP2 to 0º gives $1/\sqrt{2}$ of the peak transmission. A similar value is obtained when WGP2 is aligned even to -45º because the detection polarizer (WGP3) is 0º-aligned and so, sees only $1/\sqrt{2}$ of the peak transmission.

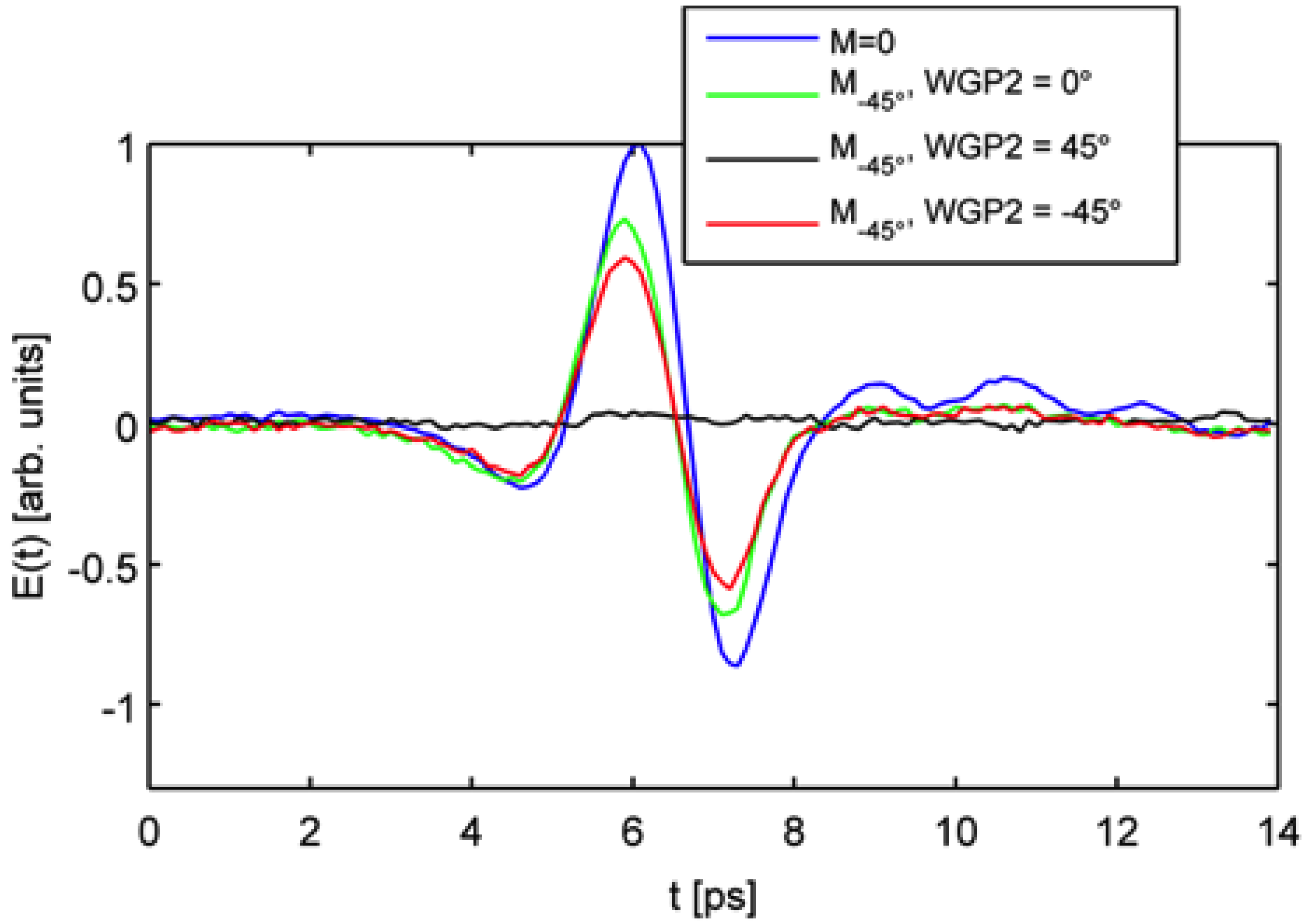

**Figure 4.20** **Rotation effect on the THz transmission. The transmitted THz pulse for certain specific cases: unmagnetized sample and -45º-magnetized samples for three orientations of the analyzer WGP2.**

The rotation picture depicted above can be quantified by spectral analysis of the transmitted pulses. The probed 45º and -45º states of WGP2 can be directly mapped to the circular radiation eigen-modesas detailed in Chapter 1. Both Faraday polarization rotation (equivalent to half the phase retardation value) and ellipticity were readily found for different magnetization levels. While the transmitted field exhibited a negligible ellipticity (which ensures the conservation of the linear polarization of the input THz wave,) a significant retardation reaching 210° was measured for a magnetizing induction of 540 mT. Such a field corresponds to a remenant magnetization of 318 kA/m as obtained from the VSM measurements. Figure 4.21 shows the phase retardation at various magnetic field levels. When the direction of the field is reversed, the rotation changes direction. This confirms the non-reciprocity of the rotation and distinguishes it from regular non-magnetic devices.

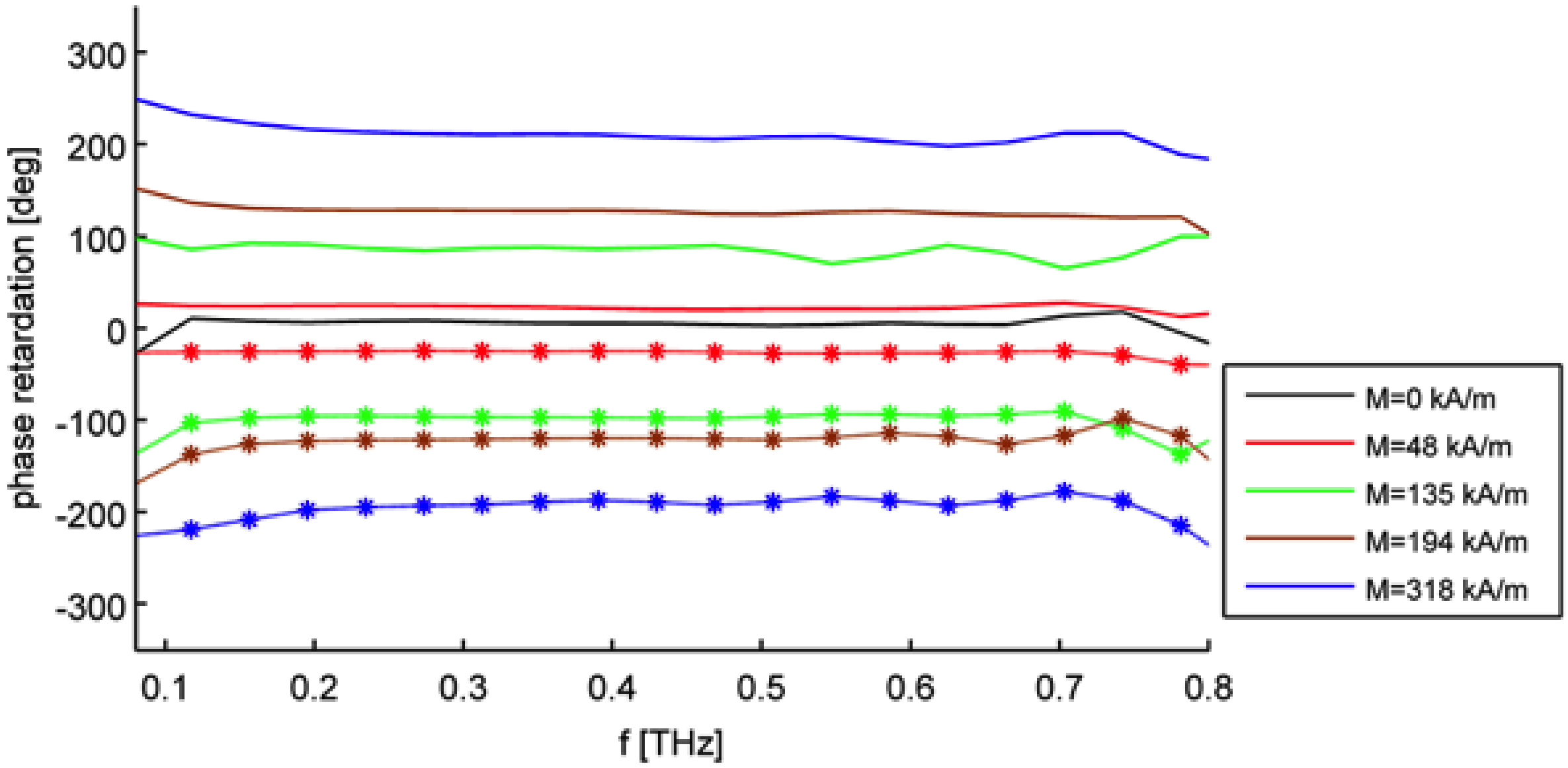

**Figure 4.21** **Magnetic field dependence of phase retardation. Broadband measurement of the phase retardation for different magnetic field levels. The lines with asterisks show the rotation when the direction of the magnetic field is reversed, which confirms the non-reciprocity.**

In the case of the unmagnetized sample, a negligible rotation (< 6°) was measured. This rotation is consistent with the accuracy of the ellipsometry measurement and with the small residual magnetization always induced by the edges of the sample. The retardation was also found to be flat over the considered frequency range. To check the magnetic field dependence of retardation, we show the phase retardation at 0.35 THz for different magnetization ($M$) levels, see Fig. 4.22. As expected, within the experimental accuracy, the retardation is linearly proportional to the internal magnetization.

### 4.3.7 Constructing the isolator

Data provided in Fig. 4.21 show that the sample can induce broadband non-reciprocal adjustable phase retardation, up to 210° for a propagation length of 3 mm. Such a high rotation can be used to build a broadband isolator. The required phase retardation for isolation purposes can be obtained by simply magnetizing the sample to obtain a polarization rotation of 45° upon propagation through it. Due to its non-reciprocity, a reflected wave back-propagating through the isolator encounters a similar phase shift that adds up to a total polarization rotationof 90º, *i.e.* the reflected polarization is crossed with respect to the original wave. If a 0º-aligned polarizer is placed before the isolator, such a back-propagating crossed wave is eliminated and does not reach the source or the other preceding parts in the system. Using the data obtained from our FR measurements, we applied a magnetizing field corresponding to a remanence magnetization ($M$) of 135 kA/m to induce the required 45º rotation. The functionality of the proposed THz isolator has been tested using the backward-waves characterization setup shown in Fig. 4.23, where a flat mirror is normally placed after the sample to allow for the THz wave to back-propagate through the same sample. Two polarizers are used in this configuration following the typical Faraday isolator design: WGP4 is set to 0º and WGP5 is set to 45º.

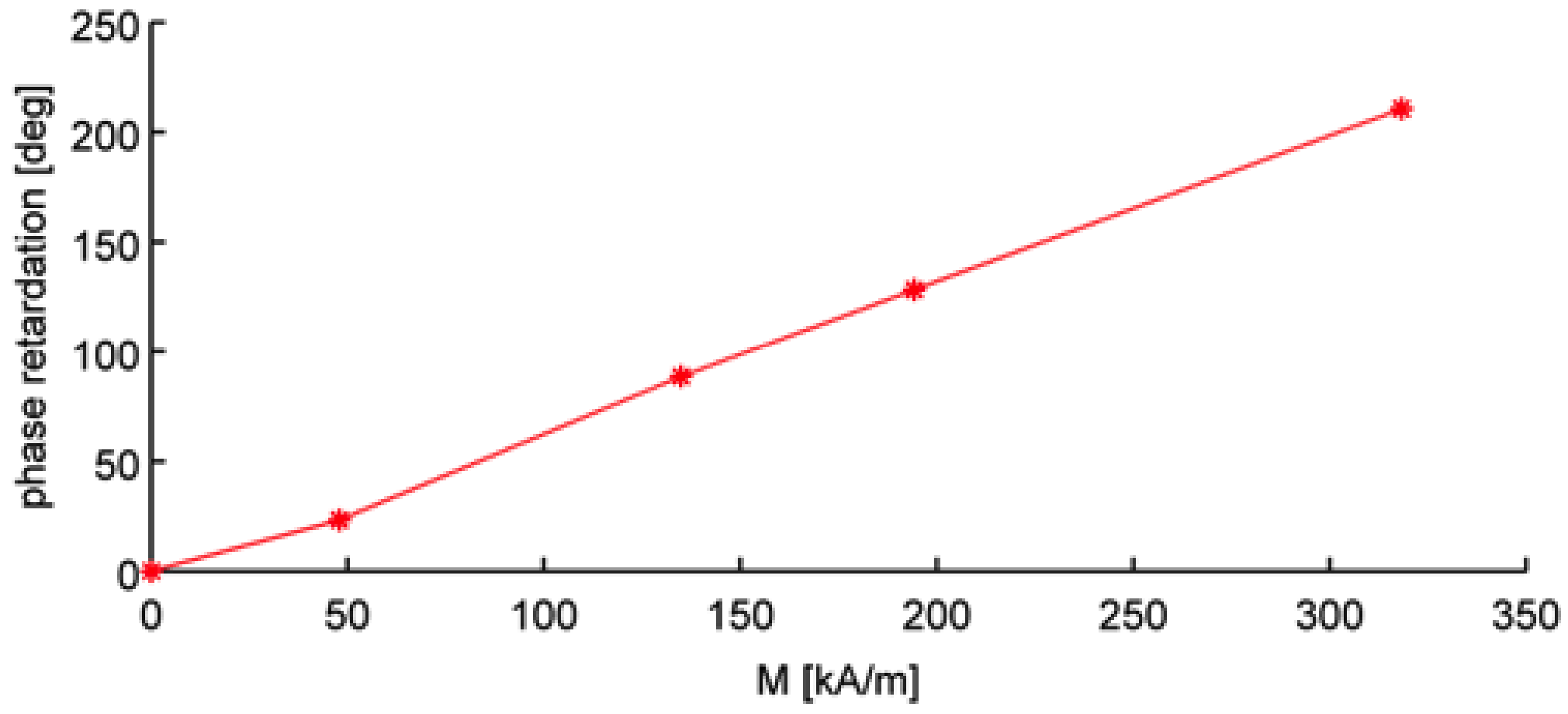

**Figure 4.22** **Linearity of the retardation.Remanence magnetization dependence of the phase retardation measured at 0.35 THz shows the expected linearity of operation.**

The full characterization of the phase retardation of the back-reflected wave is shown in Fig. 4.24 for different magnetization levels. We used the setup shown in Fig. 4.23 after removing WGP5 (because different degrees of retardation were considered.) Figure 4.24 simply maps the forward pass rotation into a two passes rotation. The isolation points correspond to a 90º rotation in the back-reflected wave. It is noteworthy to point out that as the magnetization is further increased, the back-reflected wave can be completely phase reversed as demonstrated in the figure.

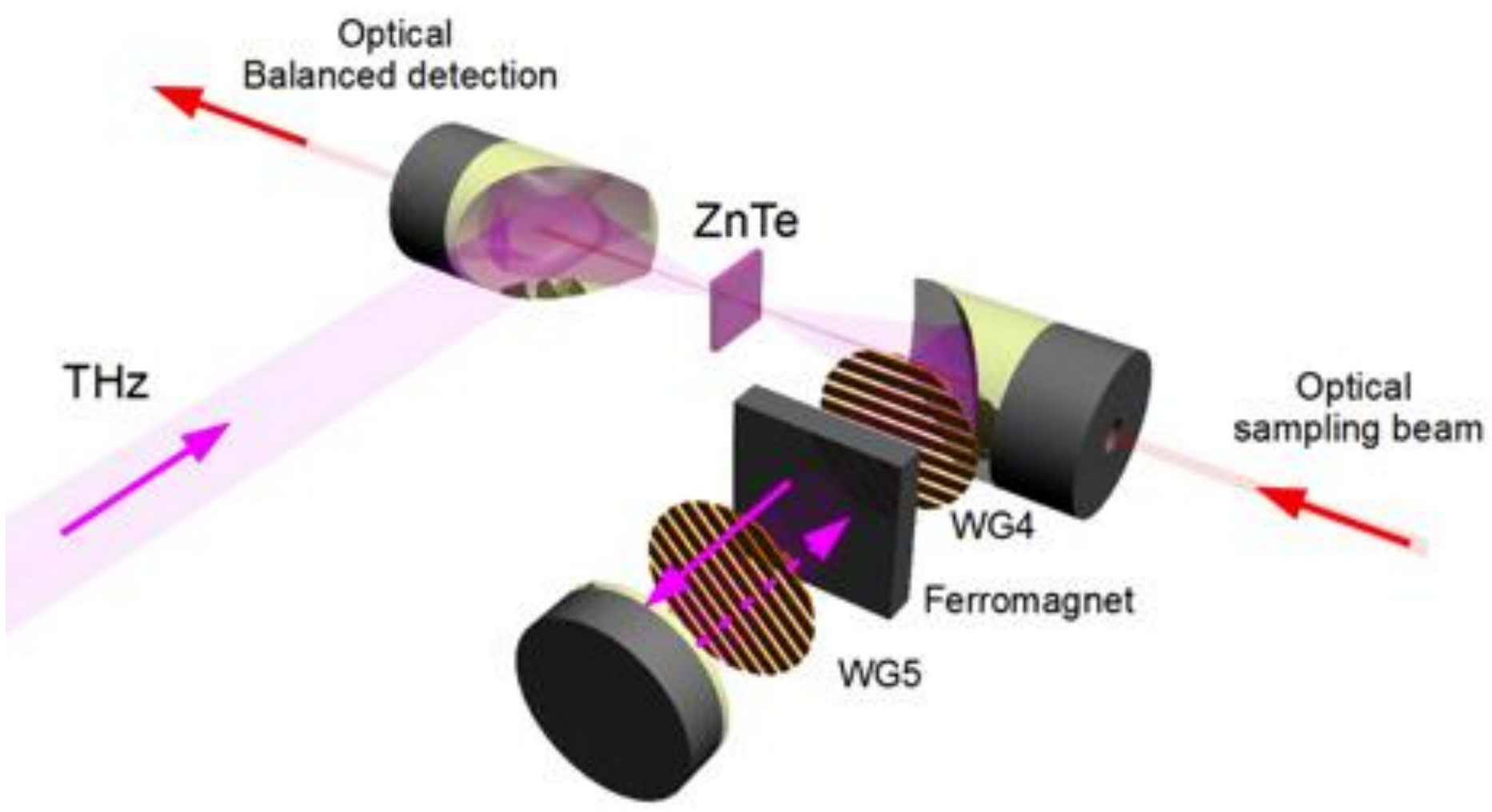

**Figure 4.23** **The isolator setup.Two polarizers (WGP4 and WGP5) were used. WGP4 is set to 0° to ensure a vertical polarization of both the generated and detected signals. WGP5 eliminates any deviation of rotation from the 45° geometry.**

Due to the non-reciprocity, the whole retardation process is flipped when the sign of the magnetic field is reversed, as previously mentioned. To estimate the isolation depth, two time traces extracted from Fig. 4.24 at 0 and 135 kA/m of remanence magnetization are shown in Fig. 4.25. When the sample is not magnetized, the back-reflected field is completely transmitted without placing WGP5 or by placing it oriented to 0º. This demonstrates that no polarization rotation occurs. The polarizer induces a small delay as it can be readily deduced by simply comparing the two plots. Conversely, when the medium is magnetized, no THz radiation is detected and complete isolation (within the sensitivity of our detection) is obtained without placing WGP5 or by placing it oriented at 45º. As the isolation is not achieved for any other orientation of WGP5, this confirms that the rotator exhibits a 45º polarization rotation within the experimental accuracy [112].

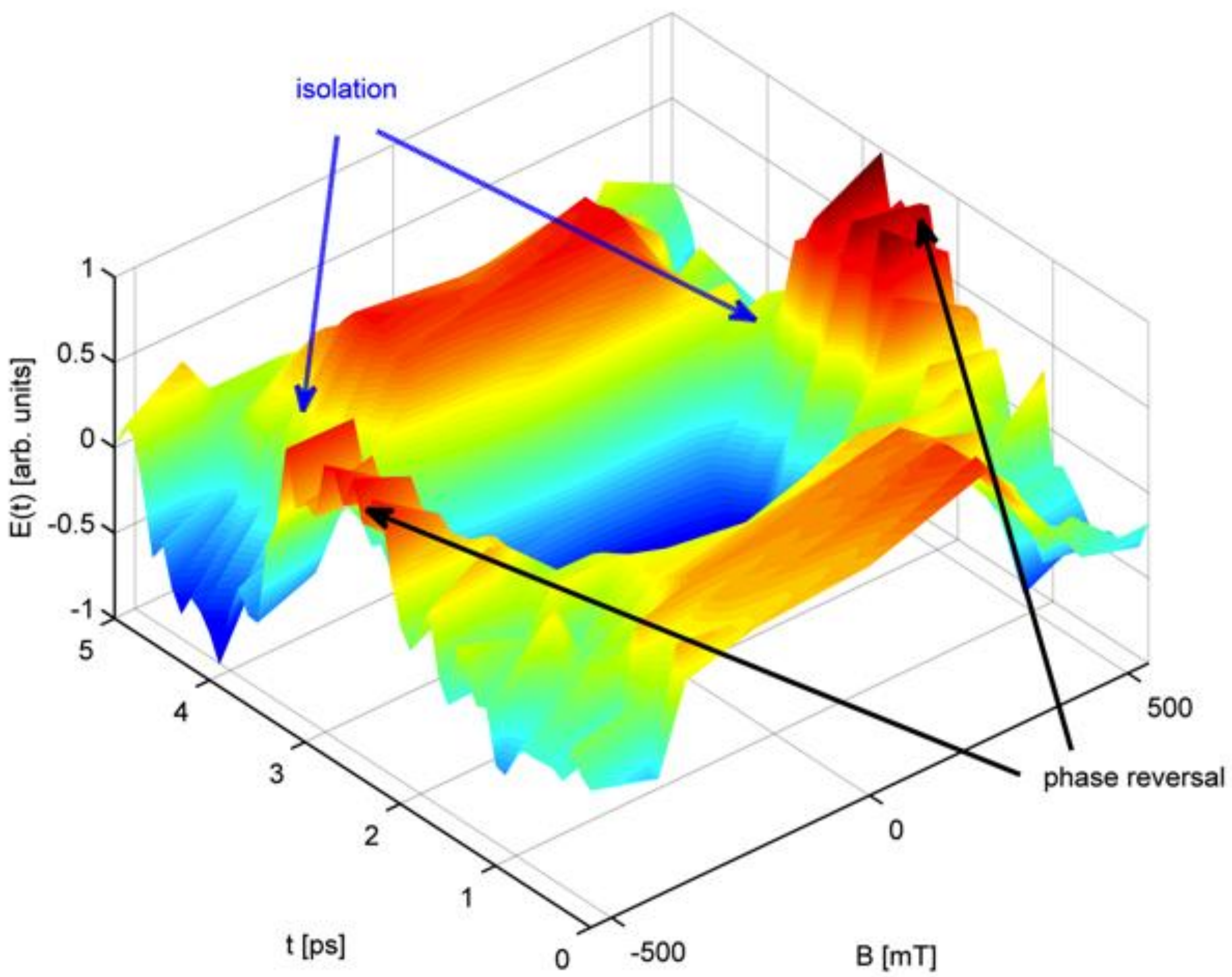

**Figure 4.24** **Reflection phase retardation mapping. Measurement of the back-reflected wave for different magnetizing fields. The isolation points and phase reversal points (where ±90° and ±180° rotations are induced in the reflected wave, respectively) are shown in the plot.**

### 4.3.8 Figure of merit

We determined the magneto-optical Verdet constant, defined as the Faraday rotation normalized by the magnetization strength and the sample thickness $d$, to have a broadband value of $\upsilon \approx 1.53 \times 10^3$ radT$^{-1}$m$^{-1}$, that leads to a figure of merit – defined as FOM=rotation angle x $e^{-\propto d}$ of 3.44, 1.55, and 0.24 radT$^{-1}$ at 0.2, 0.3, and 0.5 THz, respectively.

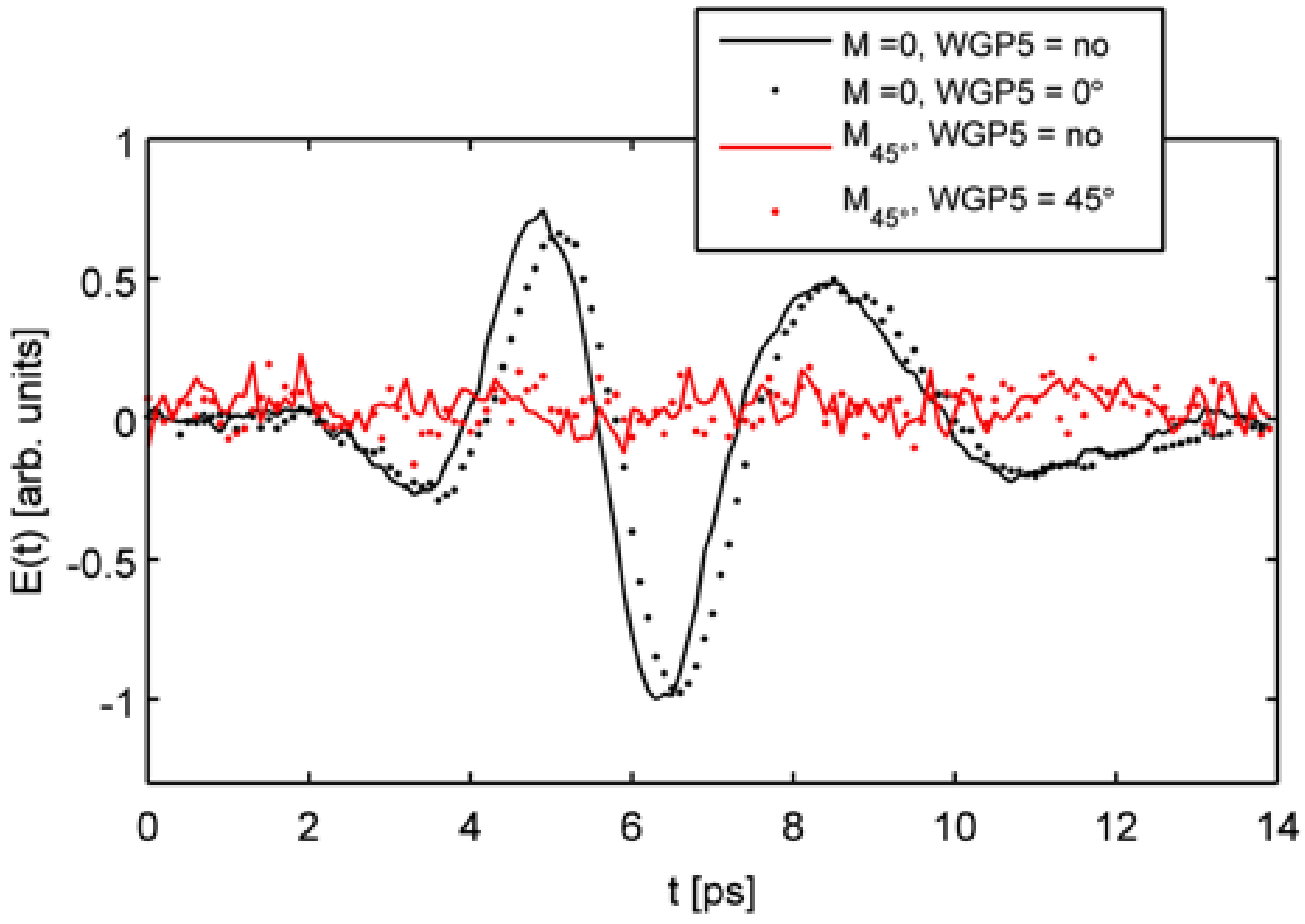

**Figure 4.25** **Isolator testing. The back-reflected THz pulse measured with the isolator setup. The comparison is made between a 45°-magnetized sample (isolator case) and an unmagnetized one ('no' in the top label refers to the absence of WPG5). The amplitude signal to noise ratio was ~ 22.**

### 4.3.9 Losses and frequency dependences

In the practical implementation of the proposed isolator, some factors should be taken into account. Losses and frequency dependences are considered the ones affecting its functionality the most. In unmagnetized ferrites, domain walls induce transmission losses. The latter are reduced when the sample gets magnetized and the domain walls vanish. For example, although Fig. 4.18(b) shows specific (and relatively high) losses associated toan unmagnetized sample, the power transmission of the sample increases by 22% as we magnetize it to obtain a rotation of 45º (required by the isolator) as demonstrated in Fig. 4.26. Most importantly, In addition to $SrFe_{12}O_{19}$ being relatively transparent in the THz regime, it does not exhibit significant circular dichroism when magnetized along the direction of propagation. It means that the right and left circularly polarized modes in the retarder undergo the same attenuation. In other words, the rotation -and hence the isolation properties- are unaffected by the losses. In the light of this, we would like to stress that critical criterion in the choice of a material for an isolator is the independence from the frequency of the phase retardation that should beaffected only by circular dichroic losses.

Although we focused on measurements between 0.08-0.8 THz, extending the operating range to higher frequencies is feasible. The frequency-independent rotation is, in general, expected above the material magnetic resonance. The latter is, conveniently, in the sub-THz regime for many ferrites. In addition, ferrites exhibit low group velocity dispersion in the THz domain, a property that is always considered an advantage when dealing with the propagation of short (broadband) pulses. We could not characterize the retardation above 1 THz due to the limited signal to noise ratio of our detection system and our relatively large sample thickness. However, we expect that by using the required thickness,we could in principleget a45° polarization rotation at the magnetization saturation (~ 1mm) where the isolator transparency window enlarges consistently.

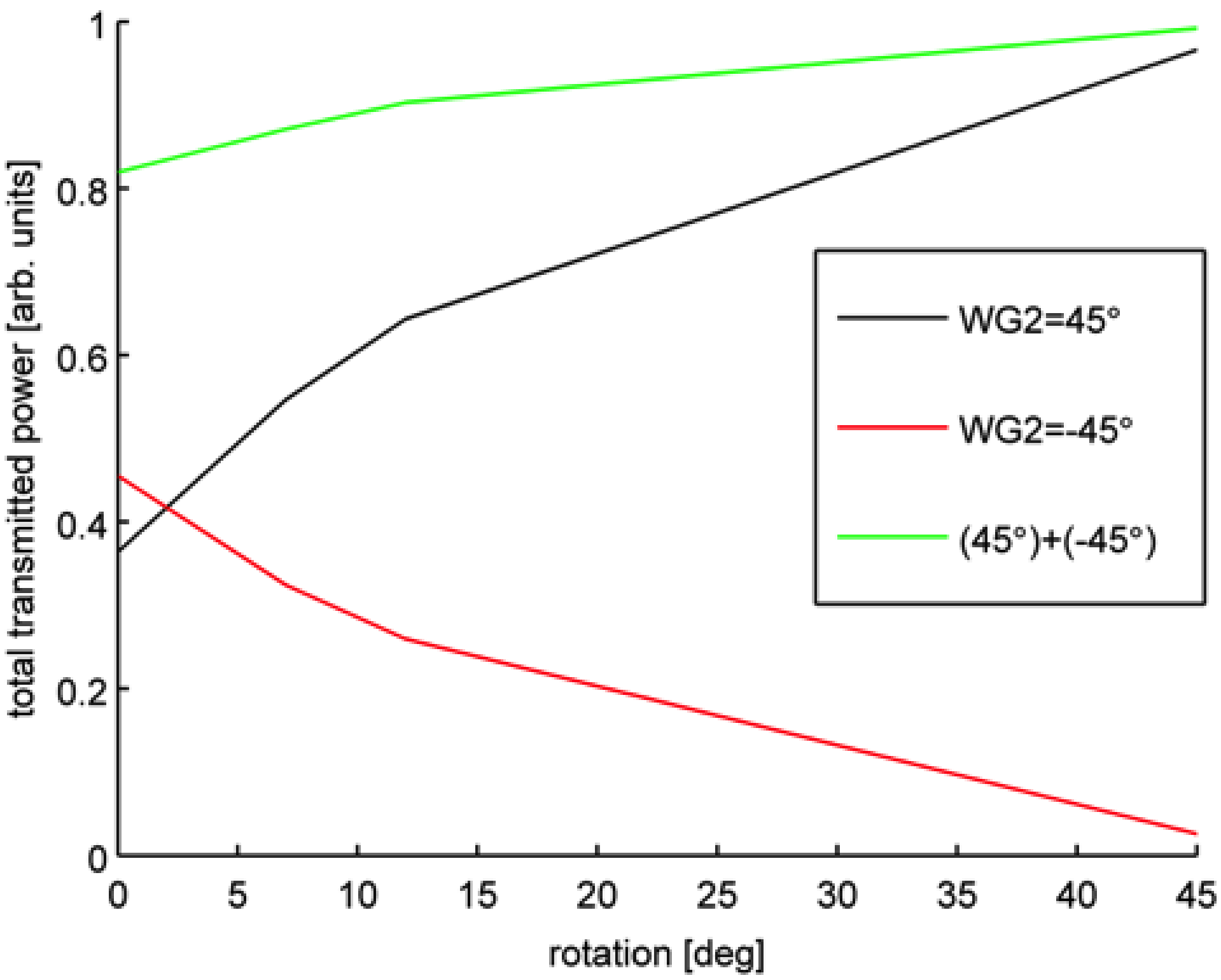

**Figure 4.26 The total transmitted power versus the induced rotation.**

In conclusion, this part shows an adjustable non-reciprocal phase retarder operating at the THz frequencies. By controlling the retardation, we designed and experimentally showed a broadband THz isolator extending over a frequency decade. The general principle of non-reciprocal phase retardation presented here can be applied for non-reciprocal field displacement, coupling and rotation. The presented results will pave the way for a new class of THz devices exploiting magnetic field-induced non-reciprocity, either stand-alone or integrated with other reciprocal systems, as required for the realization of THz lasers and amplifiers.

## CONCLUSIONS AND FUTURE PERSPECTIVES

The thesis work is focused on a variety of magneto-THz phenomena. Our research started with modeling the interactions between a magnetic system and a THz pulse. We, first, found the required THz field intensity for picoseconds magnetization dynamics through the solution of the relevant equation of motion. This gave us an estimate of 6 T(18 MV/cm) of peak THz field, required to initiate magnetization interactions at the lower THz frequencies (~0.2-0.3 THz). Second, we solved the magnetism (Landau-Lifshitz-Gilbert) and propagation (Maxwell) equations in a coupled time domain model. Results showed that an intense enough THz pulse could initiate magnetization dynamics on the ultrafast picoseconds scale. In addition to the high field requirements, calculations have also highlighted the strong sensitivity of the dynamics to the THz pulse shape, with a highly asymmetric pulse being much desired.

Those requirements were then experimentally addressed. First, we looked for a way to temporally shape short THz pulses. Our approach depended on the optical-pump/THz-probe characterization of free carriers in semiconductors. By adjusting the pump/probe delay, we managed to significantly shape the THz pulse. The main object was to induce temporal shaping, but we have also extended the technique to spectral shaping as well. Second, we investigated possible ways to increase the THz peak fields. We took two approaches; one depends on enhancing the THz radiation in sub-wavelength structures. In particular, we focused on nanoslits due to the possibility of high enhancement of broadband pulses from such nonresonant structures. We based our studies on a recent work of high field enhancement in a single slit structure and took a step towards nonlinear THz experiments. Single slit structures suffer from very low transmission that hinders the noisy detection of nonlinear sources. Also, adding more slits to enhance the transmission is typically associatedtoa decrease in field enhancement. We designed a nanoslits array that exhibits simultaneous high enhancement and transmission. Experimental characterization showed increase of the transmission by a factor of 20 compared to a single slit, with a negligible effect on the high enhancement of single slits. The other approach towards more intense radiation was based on generating stronger THz radiation. We investigated a way to improve the generation efficiency from laser-induced plasmas. Previous studies on the subject used broadly available 800 nm-centered laser pulses. We adopted a technique inspired by high harmonics experiments where longer laser wavelengths were found to produce a stronger pondermotive force. We applied this technique to the THz generation and we found a dramatic increase in the generated THz intensity as the pump wavelength increases. For example, using 1800 nm-centered pump pulses, we demonstrated the most intense THz radiation from tabletop source demonstrated to date (and peaking at 4.4 MV/cm).

Returning back to THz magnetism, numerical calculations brought to attention the potentials of linear interactions. Having shown that an intense THz pulse is capable of driving nonlinear magnetization dynamics, the change of the magnetic state was found to beclearly reflected in the polarization rotation of the transmitted THz pulse. This suggested the ability to efficiently harness THz waves with magnetic materials. We targeted three basic magnetic devices that did notexist at the THz frequencies: non-reciprocal phase retarders, modulators, and isolators. In terms of magnetic devices, THz imposes two main challenges. First, most of the materials used at lower frequencies (such as for microwave) lose their responses as the frequency increases towards the THz regime. Second, materials' responses are generally frequency-dependent and short THz pulses are by definition spectrally broadband. Our research aimedtowards materials that overcome the aforementioned limitations, i.e.Ferrofluids and Ferrites. We first exploited the out-of-plane properties of a Ferrofluid, where applying a small external magnetic field bias (30 mT) in the direction of THz propagation led to non-reciprocal rotation (phase retardation/2) of 11 mrad/mm. Then, we moved to the in-plane magnetic effects to construct a magnetic modulator. Ferrofluids have a unique property of magnetic field-induced channel formation. This channel leads to significant absorption of THz radiationwhen the THz electric field is parallel to the channel direction. Using this result, we came out with a THz magnetic modulator. Finally, we investigated the non-reciprocal properties of Ferrites at THz frequencies. We aimed at reaching high non-reciprocal rotation to build an isolator. Our sample showed a very high rotation reaching 194°/T. We magnetized the sample to the 45° rotation typically required for the realization of an isolator. Using THz back-reflection characterization, we showed a fully functional broadband THz isolator.

As for future continuation of these studies, experimental realization of THz-triggered magnetization dynamics is the most important step. The numerical calculations will be incorporated with the results from the supporting (pulse shaping, field enhancement, and intense generation) experiments. Those dynamics will be probed by a delayed X-ray pulse from an X-ray laser. Regarding the linear results, the experiments presented here were just initialsteps. Further improvements of efficiencies by studying different materials and reducing the propagation loses are required. Moreover, our research on Ferrofluids was the first at THz frequencies and this liquid seems a promising material to be further studied and practically applied.

## REFERENCES

[1] B. D. Patterson, *et al.*, *Ultrafast Phenomena at the Nanoscale: Science opportunities at the SwissFEL X-ray Laser* (PSI, Switzerland, 2009)

[2] J. M. D. Coey, *Magnetism and Magnetic Materials* (Cambridge University Press, 2010)

[3] C. H. Back, *et al.*, *Science* **6**, 864 (1999)

[4] D. Daranciang, *et al.*, *Appl. Phys. Lett.* **99**, 141117 (2011)

[5] H. Hirori, A. Doi, F. Blanchard, and K. Tanaka, *Appl. Phys. Lett.* **98**, 091106 (2011)

[6] C. Ruchert, C. Vicario, and C.P. Hauri*, Opt. Lett.* **37**, 899 (2012)

[7] W. Shi, Y. Ding, N. Fernelius, and K. Vodopyanov, *Opt. Lett.* **27**, 1454 (2002)

[8] J. Dai, X. Xie, and X. C. Zhang, *Phys. Rev. Lett.* **97**, 103903 (2006)

[9] T. Kleine-Ostmann and T. Nagatsuma, *J Infrared Milli Terahz Waves* **32**, 143 (2011)

[10] W. L. J. Deibel, and D. M. Mittleman, *Rep. Prog. Phys.* **70**, 1325 (2007)

[11]B. Hu, *et al.*, *Opt. Lett.* **20**, 1716 (1995)

[12] X-C. Zhang, *Phys. Med. Biol.* **47**, 3667 (2002)

[13] M. R. Leahy-Hoppa, M. J. Fitch, X. Zheng, and L. M. Hayden, *Chem. Phys. Lett.* **434**, 227 (2007)

[14] K. Tanaka, H. Hirori, and M. Nagai, *IEEE Trans. Terahertz Sci. Technol.* **1**, 301 (2011)

[15] T. Kampfrath, *et al.*, *Nat. Photonics* **5**, 31 (2011)

[16] L. Razzari, *et al.*, *Phys. Rev. B* **79**, 193204 (2009)

[17] M. C. Hoffmann, *et al.*, *Appl. Phys. Lett.* **95**, 231105 (2009)

[18]M. Liu, *et al.*, *Nature* **487**, 345 (2012)

[19]B. B. Hu and M. C. Nuss, *Opt. Lett.* **20**, 1716 (1995)

[20] X. C. Zhang, *Phys. Med. Biol.* **47**, 3667 (2002)

[21] M. R. Leahy-Hoppa, M. J., Fitch, X. Zheng, and L. M. Hayden, *Chem. Phys. Lett.* **434**, 227 (2007)

[22] T. Ebbesen, H. Lezec, H. Ghaemi, T. Thio, and P. Wolff, *Nature* **391**, 667 (1998)

[23] M. Seo, *et al.*, *Nat. Photonics* **3**, 152 (2009)

[24] Y. Shen, *et al.*, *Phys. Rev. Lett.* **99**, 043901 (2007)

[25] C. H. Lee, *Appl. Phys. Lett.* **30**, 84 (1977)

[26] C. Fattinger and D. Grischkowsky, *Appl. Phys. Lett.* **54**, 490 (1989)

[27] H.G. Roskos, M.D. Thomson, M. Kress, T. Loffler, *Laser & Photonics Reviews* **1**, 349 (2007)

[28] X. C. Zhang and J. Xu, *Introduction to THz Wave Photonics* (Springer, 2010)

[29] Y. S. Lee, *Principles of Terahertz Science and Technology* (Springer, 2009)

[30] M. Skorobogatiy, and J. Yang, *Fundamentals of photonic crystal guiding* (Cambridge University Press, 2009)

[31] M. Born, and E. Wolf, *Principles of optics* (Cambridge University Press, 1999)

[32] C. C. Katsidis and D. I. Siapkas, *Appl. Opt.* **41**, 3978 (2002)

[33] M. C. Nuss and J. Orenstein, *Terahertz time-domain spectroscopy* in Millimeter and Submillimeter Wave Spectroscopy of Solids(Springer-Verlag, 1998)

[34] E. D. Palik and J. K. Furdyna, *Rep. Prog. Phys.* **33**, 1193 (1970)

[35] R. Luebbers, K. Kumagai, S. Adachi, and T. Uno, *IEEE Trans. Electromagn. Compat.* **35**, 90 (1993)

[36] T. L. Gilbert, *IEEE Trans. Magn.* **40**, 3443 (2004)

[37] D. M. Sullivan, *Electromagnetic Simulation Using the FDTD Method* (IEEE Press, 2000)

[38] M. Fox, *Optical Properties of Solids* (Oxford University Press, 2001)

[39] C. Kittel, *Phys. Rev.* **71**, 270 (1947)

[40] D. You, R. R. Jones, P. H. Bucksbaum, and D. R.Dykaar, *Opt. Lett.* **18**, 290 (1993)

[41] Y. Gao, T. Drake, Z. Chen, and M. F. DeCamp, *Opt. Lett.* **33**, 2776 (2008)

[42] M. Nakajima, A. Namai, S. Ohkoshi, and T. Suemoto, *Opt. Express* **18**, 18260 (2010)

[43] R. Zhou, *et al.*, *Appl. Phys. Lett.* **100**, 061102 (2012)

[44] T. Qi, *et al.*, *Phys. Rev. Lett.* **102**, 247603 (2009)

[45] C. Ludwig and J. Kuhl, *Appl. Phys. Lett.* **69**, 1194 (1996)

[46] S. Vidal, *et al.*, *J. Opt. Soc. Am. B* **27**, 1044 (2010)

[47] W. C. Hurlbut, *et al.*, *J. Opt. Soc. Am. B* **23**, 90 (2006)

[48] A. Stepanov, J. Hebling, and J. Kuhl, *Opt. Express* **12**, 4650 (2004)

[49] E. S. Lee, *et al.*, *Opt. Express* **19**, 14852 (2011)

[50] J. Bromage, *et al.*, *Opt. Lett.* **22**, 627 (1997)

[51] M. Machholm and N. Henriksen, *Phys. Rev. Lett.* **87**, 193001 (2001)

[52] K. Yamaguchi, M. Nakajima, and T. Suemoto, *Phys. Rev. Lett.* **105**, 237201 (2010)

[53] R. Mendis and D. Grischkowsky, *J. Appl. Phys.* **88**, 4449 (2000)

[54] F. J. Garcia-Vidal, L. Martin-Moreno, T. W. Ebbesen, and L. Kuipers, *Rev. Mod. Phys.* **82**, 729 (2010)

[55] J. H. Kang, *et al.*, *J. Kor. Phys. Soc.* **49**, 881 (2006)

[56] H. Park, *et al.*, *Appl. Phys. Lett.* **96**, 121106 (2010)

[57] H. Merbold, A. Bitzer, and T. Feurer, *Opt. Express* **19**, 7262 (2011)

[58] D. Kim, *et al.*, *Proc. SPIE* **7214**, 72140H (2009)

[59] T. Feurer, *et al.*, *Annu. Rev. Mater. Res.* **37**, 317 (2007)

[60] H. Merbold and T. Feurer, *J. Appl. Phys.* **107**, 033504 (2010)

[61] H. Merbold, A. Bitzer, F. Enderli, and T. Feurer, *J Infrared Milli Terahz Waves* **32**, 570 (2011)

[62] F. Blanchard, *et al.*, *Opt. Express* **15**, 13212 (2007)

[63] A. Bitzer, *et al.*, *Opt. Express* **17**, 22108 (2009)

[64] V. L. Ginzburg and I. M. Frank, *Sov. Phys. JETP* **16**, 15 (1946)

[65] V. Ginzburg and V. Tsytovich, *Transition Radiation and Transition Scattering* (Adam Hilger, 1990)

[66] A. G. Stepanov, L. Bonacina, S. V. Chekalin, and J. P. Wolf, *Opt. Lett.* **33**, 2497 (2008)

[67] C. P. Hauri, C. Ruchert, C. Vicario, and F. Ardana, *Appl. Phys. Lett.* **99**, 161116 (2011)

[68] F. J. Su, *et al.*,*Opt. Express* **17**, 9620 (2009)

[69] F. Blanchard, *et al.*, *Phys. Rev. Lett.* **107**, 107401 (2011)

[70] D. Cook and R. Hochstrasser, *Opt. Lett.* **25**, 1210 (2000)

[71] K. Y. Kim, a. J. Taylor, J. H. Glownia, and G. Rodriguez, *Nat. Photonics* **2**, 605 (2008)

[72] M. D. Thomson, V. Blank, and H. G. Roskos, *Opt. Express* **18**, 23173 (2010)

[73] T. Bartel, *et al.*, *Opt. Lett.* **30**, 2805 (2005)

[74] K. Y. Kim, J. H. Glownia, A. J. Taylor, and G. Rodriguez, *Opt. Express* **15**, 4577 (2007)

[75] H.G. Roskos, M.D. Thomson, M. Kress, T. Loffler, *Laser & Photonics Reviews* **1**, 349 (2007)

[76] A. V. Balakin, *et al.*, *JOSA B* **27**, 16 (2010)

[77] N. Karpowicz, *et al.*, *Appl. Phys. Lett.* **92**, 011131 (2008)

[78] J. Tate, *et al.*, *Phys. Rev. Lett.* **98**, 013901 (2007)

[79] P. Colosimo, *et al.*, *Nat. Phys.* **4**, 386 (2008)

[80] M. Clerici, *et al.*, ArXiv 1207.4754 (2012)

[81] T. A. Franklin, *Ferrofluid flow phenomena*, Master's thesis (MassachusettsInstitute of Technology, 2003)

[82] O. Morikawa, *et al.*,*J. Appl. Phys.* **100**, 033105 (2006)

[83] A. M. Shuvaev, *et al.*, *Phys. Rev. Lett.* **106**, 107404 (2011)

[84] C. Janke, J. G. Rivas, P. H. Bolivar, and H. Kurz, *Opt. Lett.* **30**, 2357 (2005)

[85] E. Hendry, *et al.*, *Phys. Rev. B* **75**, 235305 (2007)

[86] W. J. Padilla, *et al.*, *Phys. Rev. Lett.* **96**, 107401 (2006)

[87] D. R. Chowdhury, *et al.*, *Appl. Phys. Lett.* **99**, 231101 (2011)

[88] A. E. Nikolaenko, *et al.*, *Opt. Express* **20**, 6068 (2012)

[89] T. Kleine-Ostmann, P. Dawson, K. Pierz, G. Hein, and M. Koch, *Appl. Phys. Lett.* **84**, 3555 (2004)

[90] H. T. Chen *et al.,Nature* **444**, 597 (2006)

[91] W. L. C., *et al.,Appl. Phys. Lett.* **94**, 213511 (2009)

[92] H. T. Chen, *et al.,Nat. Photonics* **3**, 148 (2009)

[93] T. Driscoll, *et al., Appl. Phys. Lett.* **93**, 024101 (2008)

[94] J. Hang and A. Lakhtakiab, *J. Mod. Opt.* **56**, 554 (2009)

[95] R. Singh, *et al., Opt. Lett.* **36**, 1230 (2011)

[96] N. Inaba, *et al.,IEEE Trans. Magn.* **25**, 3866 (1989)

[97] S. Taketomi, *et al.,IEEE Translation J. on Magn. in Japan* **4**, 384 (1989)

[98] L. Rayleigh, *Nature* **64**, 577 (1901).

[99] C. L. Hogan, *Rev. Mod. Phys.* **25**, 253 (1953)

[100] B. D. H. Tellegen, *Philips Res. Rep.* **3**, 81 (1948)

[101] E. A. Ohm, *IRE Trans. Microwave Theor. Tech.* **4**, 210 (1956)

[102] C. L. Hogan, *The Bell System Technical Journal* **31**, 1 (1952)

[103] L. J. Aplet and J. W. Carson, *Appl. Opt.* **3**, 544 (1964)

[104] Z. Yu and S. Fan, *Nat. Photonics* **3**, 91 (2009)

[105] H. Dötsch, *et al.*, *Review. J. Opt. Soc. Am.* B**22**, 240 (2005)

[106] L. Bi, *et al.*, *Nat. Photonics* **5**, 758 (2011)

[107] S. Takahashi, *et al.*, *Appl. Phys. Lett.* **91**, 174102 (2007)

[108] M. Paquay, *et al.,Proc of the 38$^{th}$ European Microwave Conference*, 765 (2008)

[109] D. H. Martin and R. J. Wylde, *IEEE Trans. Microwave Theory Tech.* **57**, 99 (2009)

[110] M. Shalaby, *et al.*, *Appl. Phys. Lett.* **100**, 241107 (2012)

[111] K. A. Korolev, L. Subramanian, and M. N. Afsar, *J. Appl. Phys.* **99**, 08F504 (2006)

[112] M. Shalaby, M. Peccianti, Y. Ozturk, and R. Morandotti, *Nat. Commun.* **4**, 1558 (2013)

Université du Québec
Institut National de la Recherche Scientifique
Centre Énergie, Matériaux et Télécommunications

# Phénomènesmagnétophotoniques aux fréquencestérahertz

Par

Mostafa Shalaby

Thèse présentée
pour l'obtention du grade de Philosophiedoctoral (Ph.D.)

Jury d'évaluation

| | |
|---|---|
| Président du jury et examinateur interne | Prof Luca Razzari, INRS-EMT |
| Examinateur externe #1 | Prof Denis Morris, U Sherbrooke |
| Examinateur externe #2 | Prof Mo Mojahedi, U Toronto |
| Directeur de recherche | Prof RoertoMorandotti, INRS-EMT |
| Codirecteur | Dr MarcoPeccianti, U Sussex |

## Introduction

Un cheval au galop peut-il avoir simultanément les quatre fers en l’air? Aucun observateur n’a jamais pu répondre à cette question grâce à l’œil nu. L’image ci-dessous décompose le mouvement d'un cheval au galop à l'aide de prises de vue étalées sur une période de temps. De fait, plus les prises de vue sont rapides plus il y en aura dans une dite période. Théoriquement, cette période de capture, aussi petite qu'elle soit, pourrait contenir une infinité de prises de vue à condition que les performances de capture le permettent.

Tout est dans l'instrumentation. Notre vision ne nous permet pas de suivre naturellement les mouvements au-delà d'une certaine rapidité, mais nous pouvons les enregistrer et les visualiser par la suite. Et c'est ce qu'Edward Muybridge a fait. Il y a plus d'un siècle, il a développé un obturateur dont la vitesse était plus rapide que celle d'un cheval au galop. C'est ainsi que des images convaincurent les gens que les quatre fers quittaient vraiment le sol simultanément pendant un instant (figure 1).

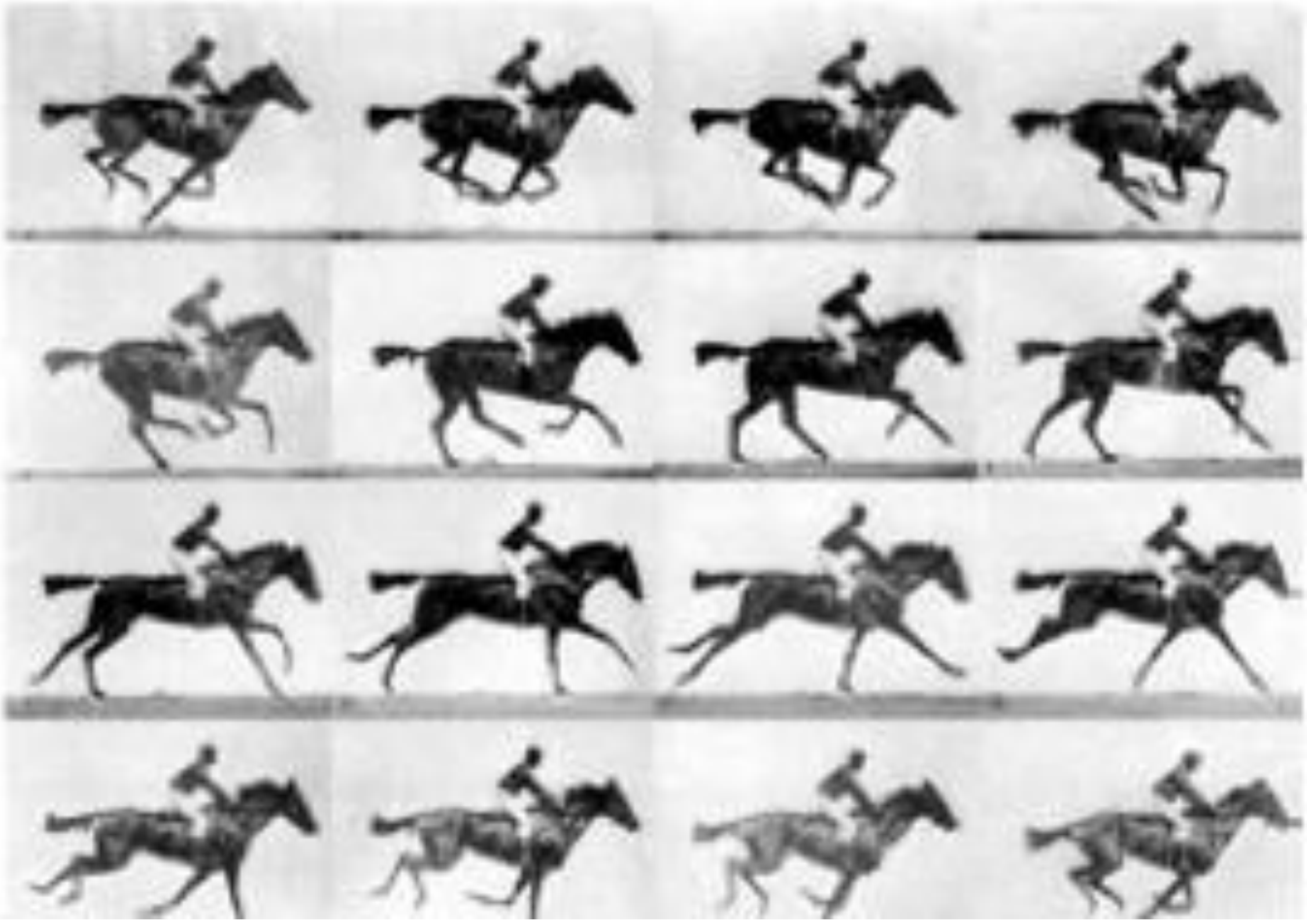

**Figure 1** **Série d'images de Muybridge décrivant le galop d'un cheval.**

La période de l'obturateur de Muybridge était de 2ms pour lui permettre de photographier les mouvements aériens d'un cheval. Aujourd'hui, nous voulons regarder à bien plus petite échelle que le cheval de Muybridge, aussi bien dans le temps que dans l'espace. Aussi, capturer des atomes, figés dans leurs positions a toujours été un rêve: ce dernier se réalise enfin.

Pour la dimension spatiale, les évolutions du synchrotron de ces dernières décenniesa doté la science d'un outil indispensable à l'observation de la matière de manière à contempler les atomes individuellement. Malheureusement, en ce qui concerne la résolution  temporelle, c'est un échec! Les images de notre film atomique seraient alors floues.

Similairement, les lasers ont doté la science, plus rapidement que n'importe quel autre outil, d'une résolution temporelle sans précèdent; les femtosecondes (et plus récemment les attosecondes) sont une réalité dans les laboratoires. Cependant, la résolution spatiale est trop grossière pour visualiser l'échelle atomique.

Au bout du compte, les tentatives de réduction des impulsions synchrotroniques de résolution spatiale à l'échelle des femtosecondes (par exemple, par femtodivision) n'ont pas donné de meilleurs résultats que celles visant à réduire les impulsions laser de résolution temporelle à l'échelle atomique (par exemple, harmoniques à haute génération); dans les deux cas, la technologie existe mais le nombre de photons est insuffisant pour procéder à une réelleétudeexpérimentale. Ce n'est que depuis peu (2009) qu'il est possible de créer des images àrésolution spatio-temporelle satisfaisantes à cette échelle. Les sources de lumière de quatrièmegénération (lasers à rayon X) ont enrichi la science avec des résultats merveilleux ces dernièresannées et nous sommes tout près de voir danser les atomes dans leurs échelles spatio-temporelles.

La liste des progrès réalisés ces dernièresannées est longue: dynamique de l'aimantation, chimie des solutions et de la catalyse de surface, diffraction cohérente, ultra biochimie et matériaux à électrons corrélés n'en sont que quelques exemples.

Le magnétisme a boulversé la civilisation humaine; la boussole est l'une des plus anciennes inventions et la mémoire magnétique soutient la révolution de l'information d'aujourd'hui. Cependant, «une des leçons à tirer de l'histoire du magnétisme est que la compréhension fondamentale de la science n'est pas forcement une condition préalable au progrès technologique. Bien que, la compréhension fondamentale aide [1]». Le magnétisme est un phénomène spatio-temporel: il suffit de s'attarder aussi bien sur sa période temporelle que spatiale (la figure 2) pour être fasciné par les différents spectacles auquel le magnétisme nous convie à chacun de leurs égards. À chaque détour on découvre une merveille de la science et des applications pratiques potentielles. L'excitation d'une interaction magnétique nécessite une stimulation avec de l'énergie E. Après que le stimulus soit parti, le système magnétique récupère son état d'équilibre dans un temps τ, τ ≈ donnée par h / E, où h est la constante de Planck. Différents phénomènes sur l'échelle de temps peuvent alors être excités sélectivement en fonction des énergies impliquées [2]. Par conséquent, cela nécessite le stimulus (champ magnétique) pour avoir tous les accessoires spectraux (ν = 1 / τ) et d'amplitude. Avec l'augmentation de ν, l'amplitude requise augmente aussi. Ce champ magnétique peut simplement provenir d'une impulsion de courant dans un fil. Ces sources de champ magnétique à faible fréquence (jusqu'au gigahertz) sont bien connues. Alors que nous approchons de la placetérahertz, la plupart des sources commencent à perdre de leur efficacité. De plus, les phénomènes magnétiques correspondants exigent des champs toujours plus élevés au fur et à mesure que la fréquence augmente.

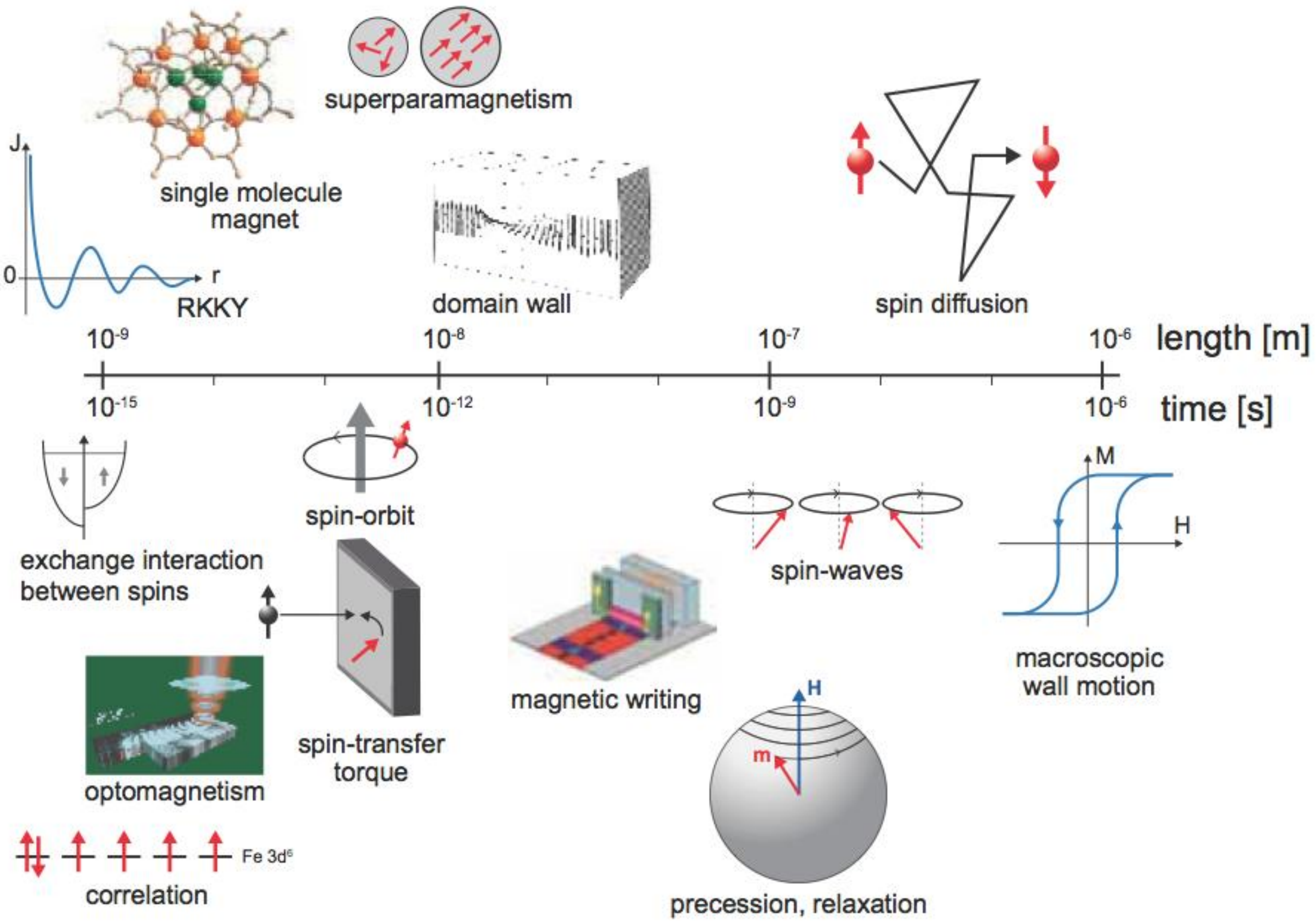

**Figure I.2 Échelles de temps et longueur du magnétisme [2].**

En plus d'un courant dans un fil, l'inversion du champ magnétique peut provenir d'ondes électromagnétiques. Dans le premier cas, à des fréquences térahertz, le courant doit être énorme. Pour une expérience typique, on doit avoir une pulsation caractérisée par 1nCet 2-10 ps. Ceci correspond à $10^{10}$ électrons livrés par un accélérateur [3]. En effet, cette variation magnétique ultrarapide à l'échelle des picosecondes atteint un nouveau record; c'est la variation magnétique la plus rapide à avoir été réalisée jusqu'à ce jour. Par contre, le rayon détruit la partie centrale de l'échantillon et cette source unique n'est pas accessible à la communauté scientifique et son échelle ne peut pas être réduite.

L'autre façon d'obtenir une transition de champ magnétique ultrarapide est d'utiliser des ondes électromagnétiques à l'échelle des picosecondes, par exemple des ondes térahertz. La science térahertz est un domaine de recherche en émergence depuis une décennie, grâce au développement de sources intenses et de détecteurs sensibles.

La bande térahertz est située entre deux bandes bien connues: la bande microondes et la bande optique. Le mot térahertz a toujours été associé à des défis et à des promesses. La génération de radiation térahertz n'est pas difficile! Nos corps, comme toute entité qui dégage une température de plus de 10 Kelvin, émettent de la radiationtérahertz. Mais cette source est trop faible pour avoir une utilité quelconque. Trouver une bonne source n'est pas une tâche facile. L'extension des sources microondes ou infrarouges aux fréquences térahertz n'eut pas de succès, puisque qu'elles tendent à perdre de leur efficacité quand elles approchent de la fréquence térahertz. Les détecteurs térahertz furent longtemps inefficaces. En bref, les térahertz furent contraints d'évoluer comme science distincte relevant d'une technologie indépendante. Toutefois, le développement de sources laser femtosecondes a apporté la plus grande contribution à la science térahertz. La plupart des techniques de génération et de détection de térahertz bien établies sont basées sur ce type de sorces de lumière pulsée.

Le térahertz trouve une grande variété d'applications qui s'étendent du déclenchement de phénomènes non-linéaires [4-8] à des applications linéaires telles que l'imagerie [9,10], les communications [11] et la spectroscopie de substances chimiques et d'explosifs [12,13].

La notion de magnéto-térahertz est utilisée depuis peu en science, avec très peu de connaissances sur le comportement des matériaux magnétiques exposés à des fréquences térahertz. Il y a peu de temps, l'existence même d'une telle chose était débattue.

La Figure I.3 montre un exemple d'impulsion térahertz dans le domaine temporel et fréquentiel.

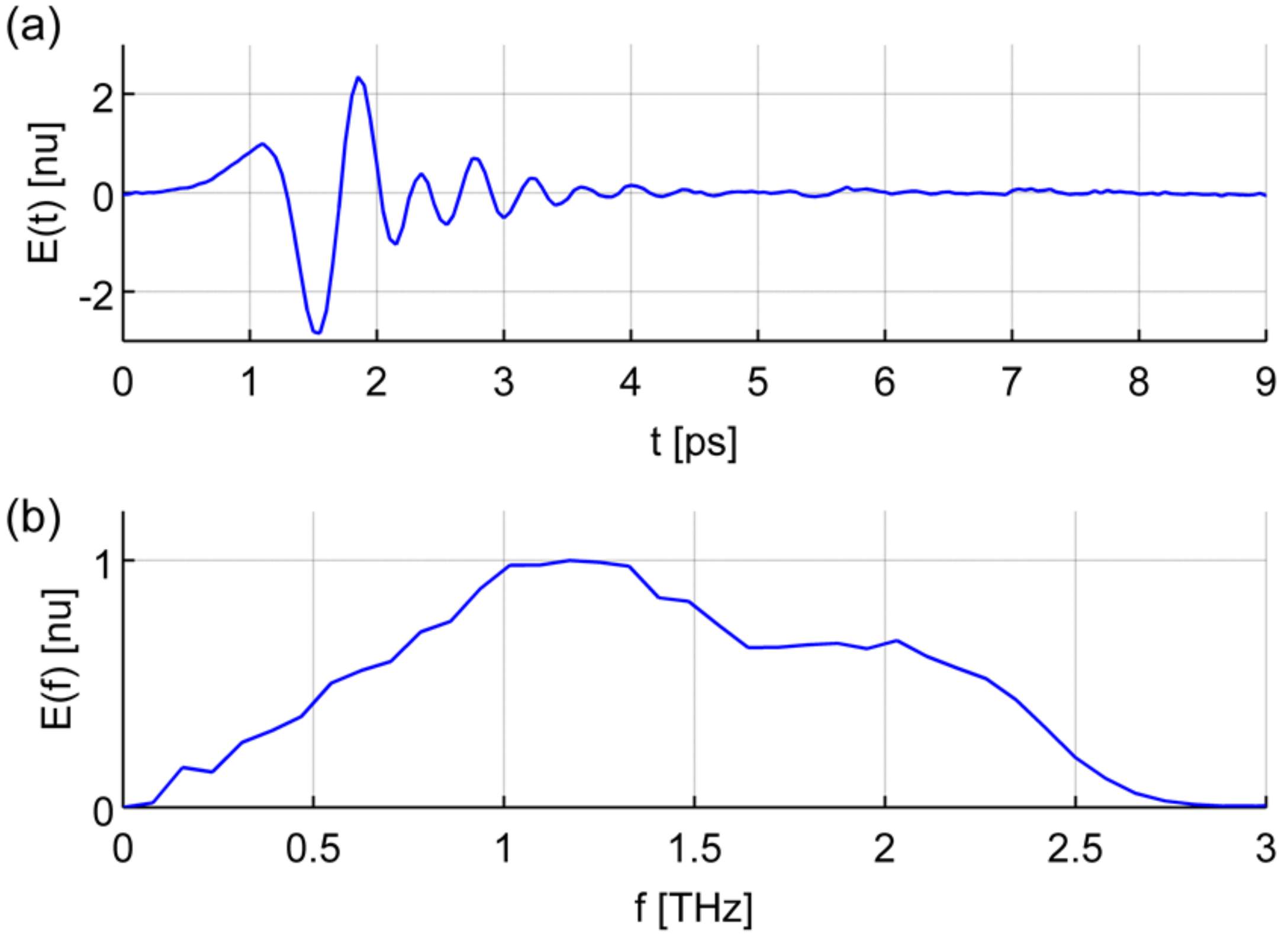

**Figure I.3** **Exemple d'impulsion térahertz, (a) Trace temporelle et (b) spectre de l'impulsion térahertz générée par le redressement optique d'un train d'impulsion laser centré à 800 nm, d'une durée de 35 fs et à un taux de répétition de 2.5 kHz dans un cristal de ZnTe d'une épaisseur de 2 mm. Un autre cristal de ZnTe, d'une épaisseur de 0.5 mm, a été utilisé pour réaliser la détection d'ondes Térahertz par échantillonnage EO.**

## Magnétisations aux fréquences térahertz

Les équations de Maxwell permettent de décrire la propagation des ondes magnétiques aux fréquences térahertz et l'équation de Landau-Lifshitz-Gilbert (LLG) décrit l'aspect dynamique des magnétisations[2]. Ce terme dynamique est obtenu par la résolution du système de ces équations. La simulation suivante décrit la propagation d'une impulsion térahertz (figure 4 (a)) à travers un film de matériaumagnétique (figure 4(b)).

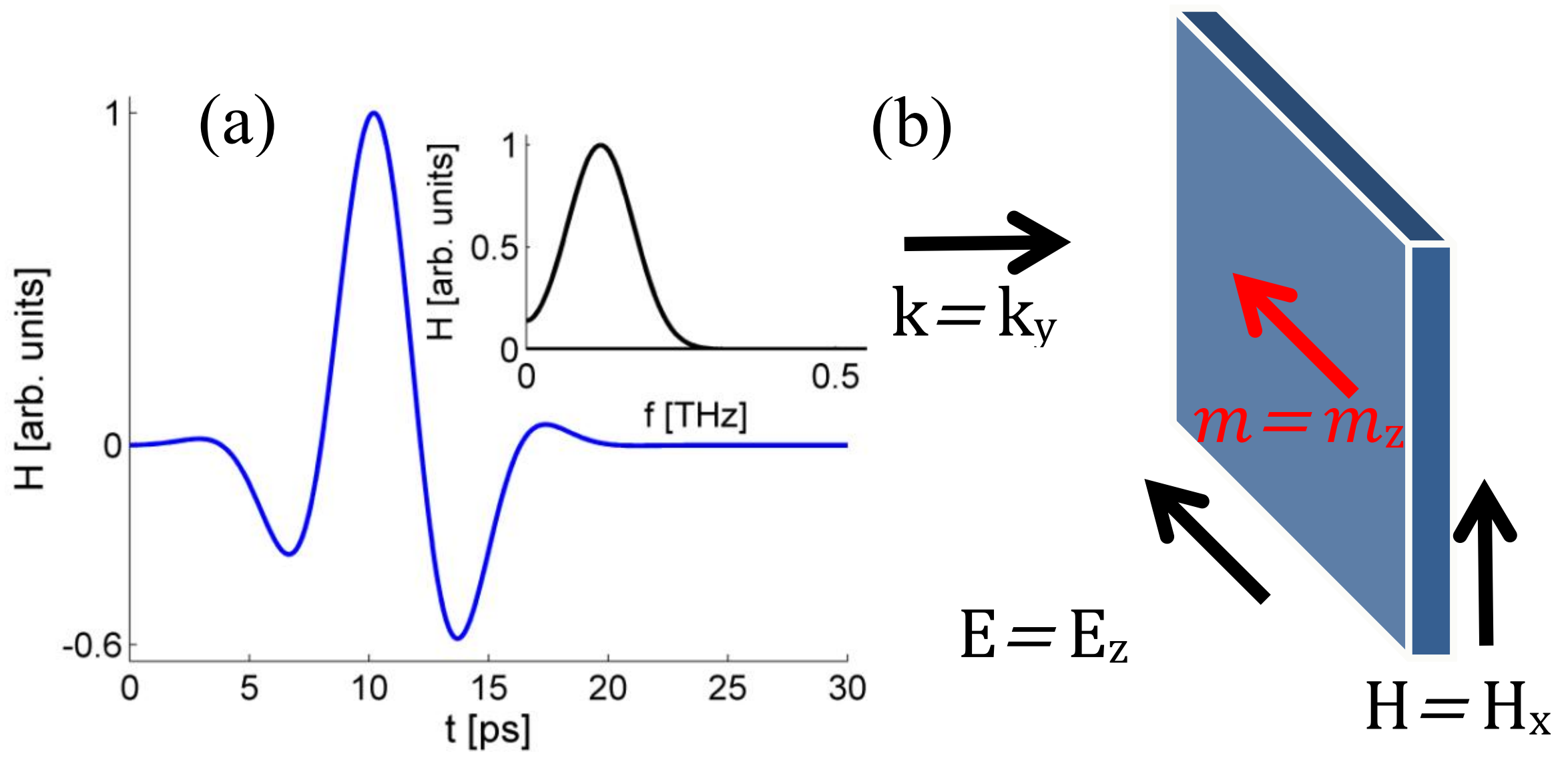

**Figure 4** Géométrie de l'intéraction magnétique. (a) Forme de l'impulsion térahertz dans le domaine temporel et spectral). Schéma de la couche magnétique.

Le matériau est décrit par le modèle de Drude

$$\varepsilon(\omega) = \varepsilon_\infty \left(\frac{\omega_p^2}{\omega\,(i\nu_c + \omega)}\right) \tag{1}$$

Où $\varepsilon_\infty = 1$ et $\varepsilon_\infty = 1$ est la fréquence de collision[14]. La fréquence de plasma $\omega_p$ est calculée en utilisant $\omega_p = (N\, e^2/\varepsilon_o m_o)^{1/2}$ où $N$, $e$, et $m_o$ sont respectivement la densité d'électrons libres, la charge de l'électron et la masse de l'électron.

Le modèle d'interaction est décrit par[14]

$$\mathbf{E}\,(z) = \left(\frac{\mathrm{D}\,(z)}{\varepsilon_o}\right)\left(1 + \left(\frac{\omega_p^2 \Delta t}{\upsilon_c}\right)\left(\frac{1}{1 - z^{-1}} - \frac{1}{1 - e^{-\upsilon_c \Delta t} z^{-1}}\right)\right)^{-1}. \tag{2}$$

$$\frac{\partial \mathrm{B}}{\partial t} = - \nabla \times \mathbf{E} \tag{3}$$

$$\mathbf{H} = \frac{\mathbf{B}}{\mu_o} - \mathbf{M} \tag{4}$$

$$\frac{\partial \mathrm{D}}{\partial t} = \nabla \times \mathbf{H} \tag{5}$$

$$\frac{\partial \mathrm{M}}{\partial t} = \gamma(\mathbf{M} \times \mathbf{H}_{eff}) - \frac{\alpha}{\mathrm{M_s}}\left[\mathbf{M} \times \frac{\partial \mathrm{M}}{\partial t}\right]. \tag{6}$$

Où E, B, D, H, M, $\mathbf{H}_{eff}$ et Δt sont respectivement le champ électrique, l'induction magnétique, le champ de déplacement, la magnétisation, le champ magnétique effectif et le délai temporel. L'équation (6) représente l'Équation de LLG. Ce modèle est illustré dans la figure 5.

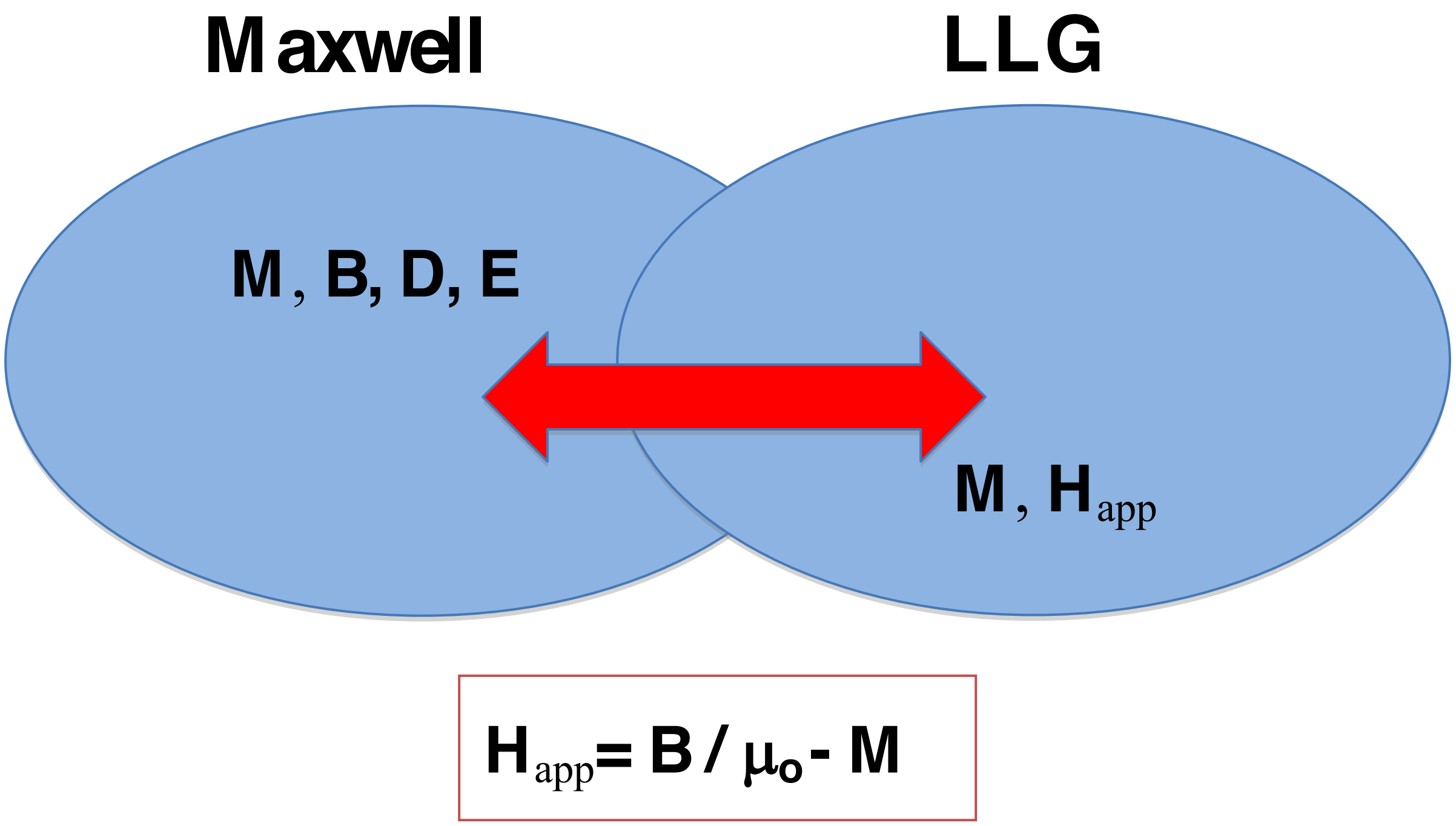

**Figure 5** **Le problème couplé de LLG-Maxwell. Le modèle de la magnétisation térahertz dépend des solutions simultanées des équations de LLG et Maxwell où M est la solution des paramètres mis à jour dans le problème couplé.**

IL est pratique d'utiliser la représentation : $\mathbf{M} = \mathrm{M_s}\mathbf{m} = \mathrm{M_s}(\mathrm{m_x}\tilde{x}+\mathrm{m_y}\tilde{y}+\mathrm{m_z}\tilde{z})$ avec $\mathrm{m_x}$, $\mathrm{m_y}$ et $\mathrm{m_z}$ qui sont les composantes cartésiennes de vecteur unitaire m. $\mathrm{M_s}$ est la magnétisation saturée. Le champ magnétique effectif contient les composantes internes et externes (**H**) du champ magnétique. Il est décrit par

$$\mathbf{H}_{eff}(\mathbf{r},t) = -\frac{1}{\mu_o}\frac{\delta U}{\delta \mathbf{M}(\mathbf{r},t)} \tag{7}$$

Où $U$ est l'énergie totale qui est donnée par :

$$U = U_{\mathrm{app}} + U_{\mathrm{ex}} + U_{\mathrm{d}} + U_{\mathrm{an}} \tag{8}$$

Et où, dans l'ordre respectif, $U_{\mathrm{app}}$, $U_{\mathrm{ex}}$, $U_{\mathrm{d}}$, et $U_{\mathrm{an}}$ sont les composantes appliqué associées au champ au terme d'échange de démagnétisation et d'anisotropie. Les composantes de l'énergie et le facteur champ correspondant peuvent être écrits par [16].

$$U_{\text{app}} = \text{-} \int \mathbf{M.H}\text{d}\mathbf{r}\text{and}\mathbf{H}_{\text{app}} = \text{-}\delta U_{app} / \delta\mathbf{M} \quad (9)$$

$$U_{\text{ex}} = A \int ((\nabla m_{\text{x}})^2 + (\nabla m_{\text{y}})^2 + (\nabla m_{\text{z}})^2)\text{d}\mathbf{r}\ \text{and}\mathbf{H}_{\text{ex}} = (2A)\nabla^2\mathbf{m} \quad (10)$$

$$U_{\text{d}} = \text{-}\frac{1}{2} \int \mathbf{M.H}_{\text{d}}\text{d}\mathbf{r}\text{and}\mathbf{H}_{\text{d}} = \text{-}\delta U_{\text{d}} / \delta\mathbf{M} \quad (11)$$

$$U_{\text{a}} = \text{-}(K/2) \int (m_x^4 + m_y^4 m_z^4)\text{d}\mathbf{r}\text{and}\mathbf{H}_{\text{a}} = (2K/\text{M}_s)(m_x^3\tilde{x} + m_y^3\tilde{y} + m_z^3\tilde{z}) \quad (12)$$

Où A et K sont respectivement les constantes d'échange et d'anisotropie. Les résultats des simulations sont montrés dans la figure 6.

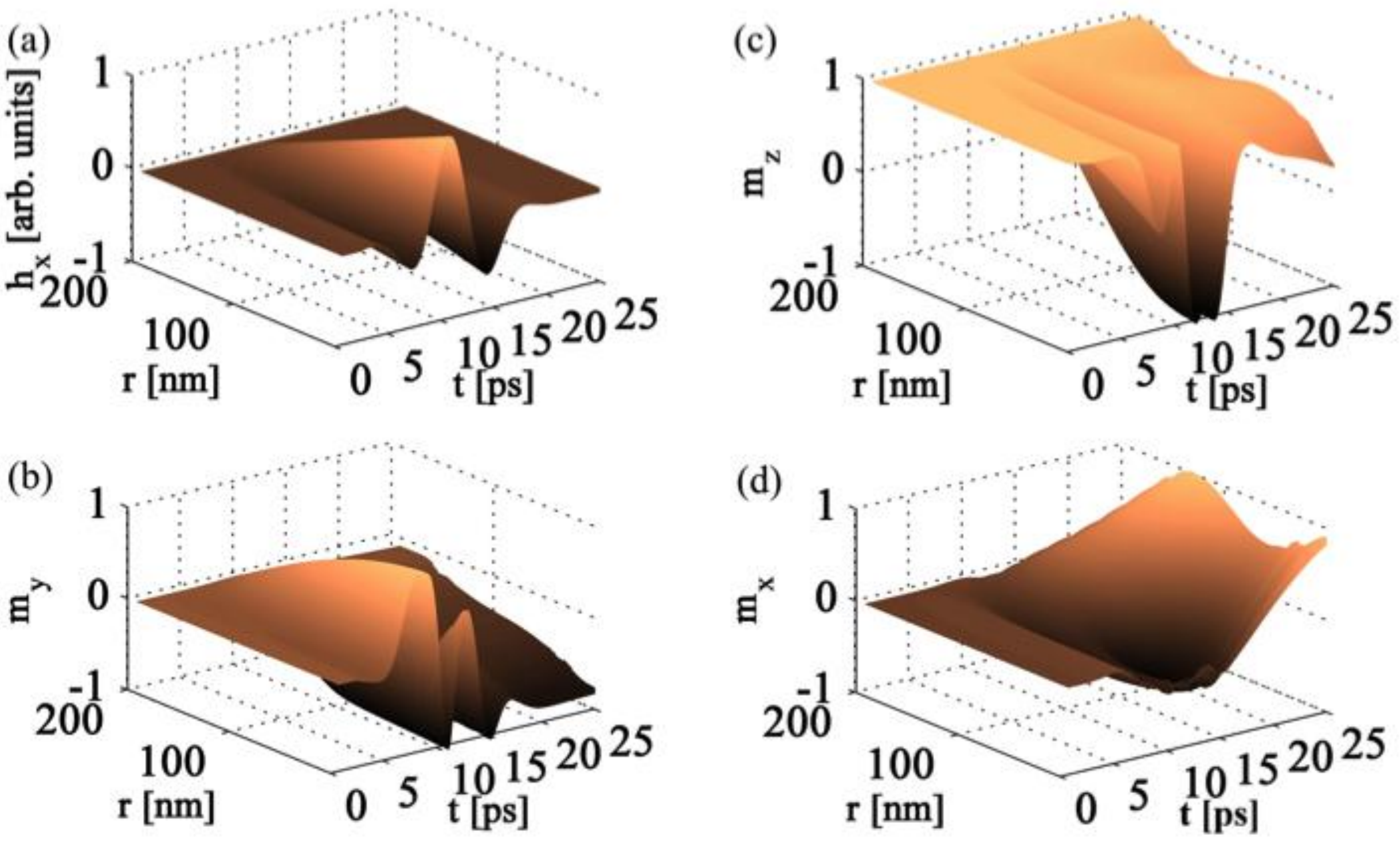

**Figure 6** **Dynamique spatio-temporelle de la magnétisation térahertz. (a) Le champ magnétique de l'impulsion térahertz. (b) La composante perpendiculaire de la magnétisation. (c) et (d) sont les composantes dans le plan de la magnétisation.**

L'application d'une impulsion térahertz donne naissance à des perturbations transitoires ultra rapides de la magnétisation au sein du matériau. Ce résultat est fondamental dans la compréhension des systèmesmagnétiques et le développement de mémoiresmagnétiques rapides.

## Protocole expérimental pour les études magnétiques térahertz

Les simulations ont montré la nécessité de contrôler la forme des impulsions térahertz tout en pouvant atteindre de très grandes amplitudes. Les expériences ont donc été conduites en prenant en compte ces contraintes.

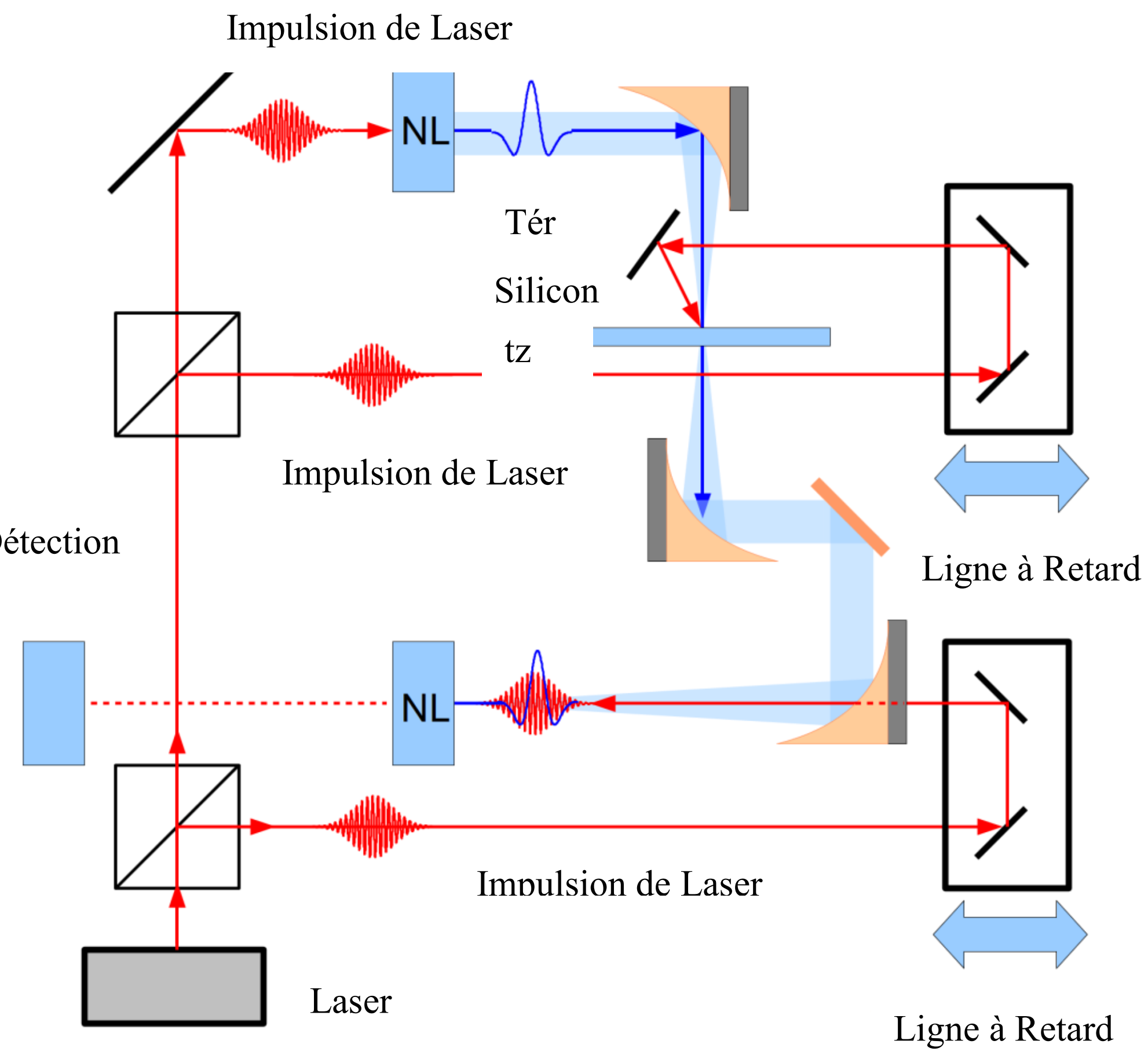

**Figure 6** **Modulation de la forme de l'impulsion térahertz**

Le dispositif expérimental permettant de contrôler la forme des impulsions térahertz est montré à la figure I.6. Le principe est le suivant: une impulsion laser sur un bloc de silicon permet de contrôler sa capacité conductrice et par conséquent de moduler la forme de l'impulsion térahertz la traversant. Le champ électrique térahertz transmis ($E_t$) à travers une couche d'épaisseur $d$ est [17]

$$E_t = \left(2Y_0 E_i \text{-} J d\right) / \left(Y_0 \text{-} Y_s\right) \quad (13)$$

Où $E_i$ est le champ électrique incident. $Y_0$ et $Y_s$ sont respectivement les admittances du vide et del'échantillon. J= *nev* est la densité de courant avec *n*, *v*, et *e* respectivement la densitéélectronique, la vitesse de déplacement des charges et la charge élémentaire.

Sous une excitation optique, la conductivité de l'échantillon augmente, ce qui fait décroitre sa capacité à transmettre l'impulsion térahertz. Lorsque l'on fait varier la durée entre les impulsions optiques et les impulsions térahertz, on module la forme de l'impulsion térahertz comme dans la figure 7 et 8.

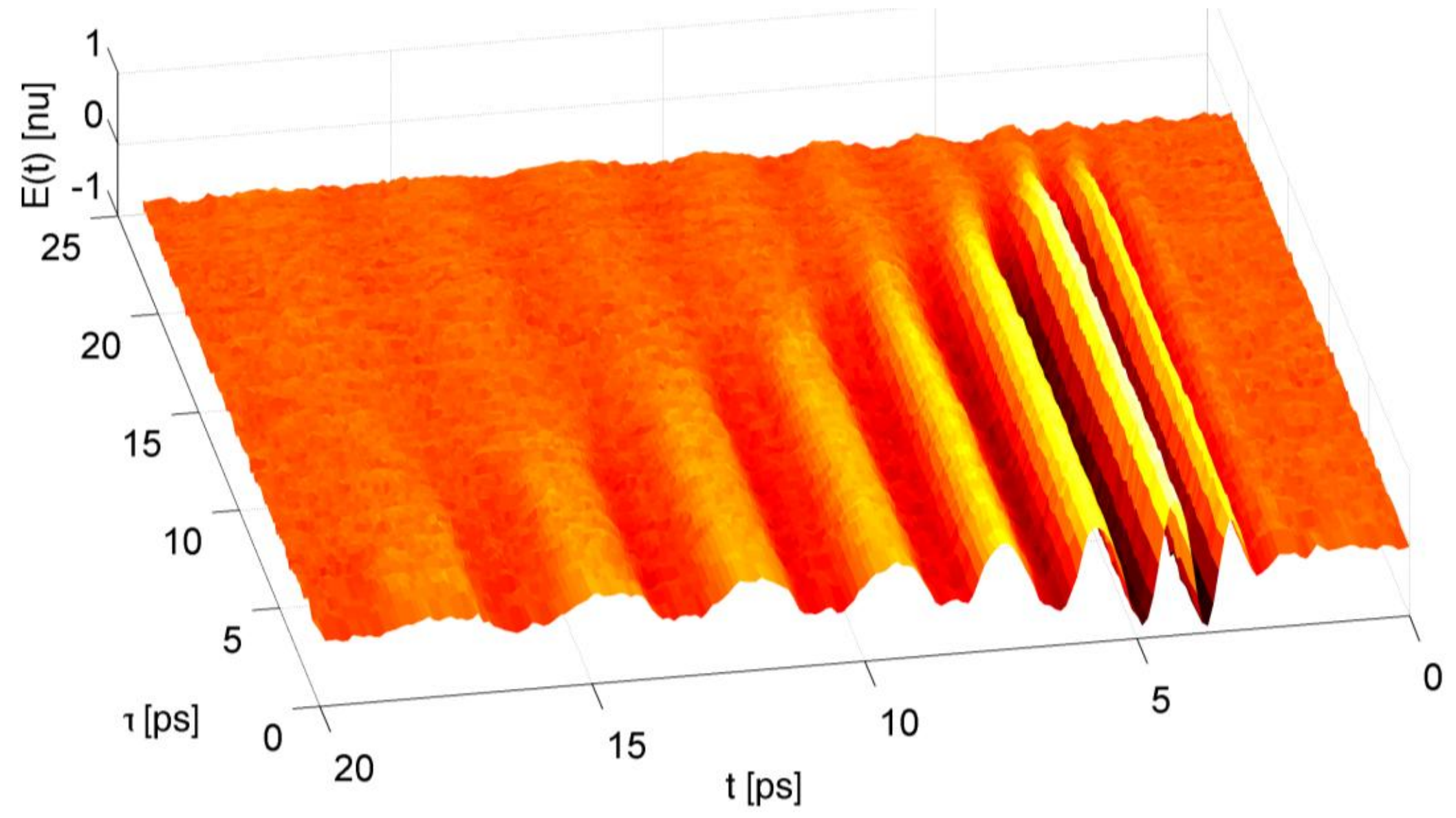

**Figure 7 Traces temporelles de modulation de la forme de l'impulsion térahertz**

Dans un second temps, en ce qui concerne l'intensité de la radiation térahertz, il s'agit de concentrer la propagation en utilisant l'effet plasmonique à travers un réseau de fentes rectangulaires à l'échelle du nanomètre[18,19]. La structure suivante a étéconçue (figure 9). C'est un réseau de fentes tracé dans une pellicule d'or dont les caractéristiques sont les suivantes (longueur l= 2 mm, espacement d= 10μm, largeur des fentes a= 40 nm, épaisseur de la pellicule d'or h= 60 nm). Lorsque la radiation térahertz percute la pellicule d'or, il en résulte un courant surfacique concentrant les charges au niveau des fentes, ce qui augmente l'intensité de la radiation térahertz traversant le réseau.

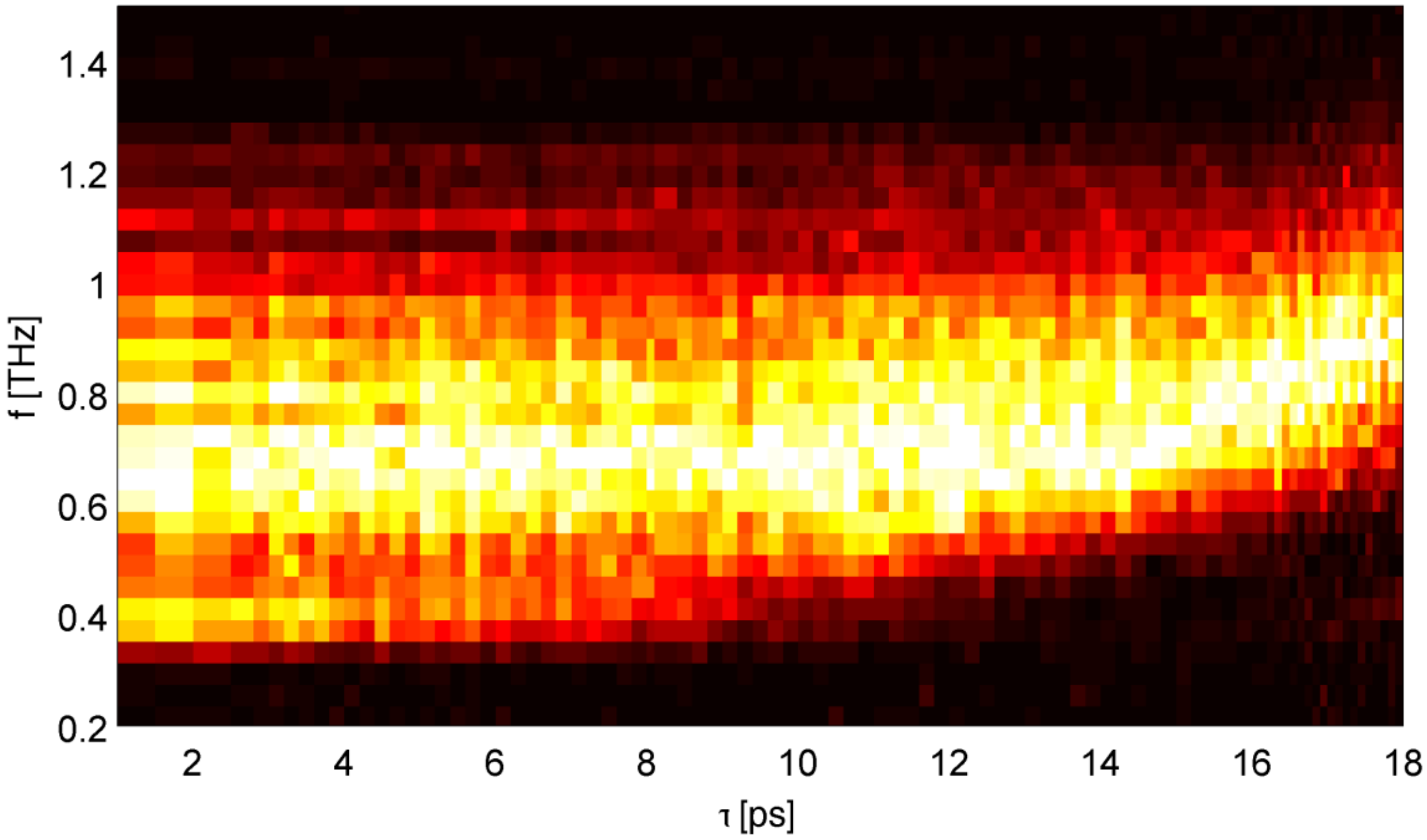

**Figure 8** **Traces spectres de modulation de la forme de l'impulsion térahertz**

Nous avons effectué la caractérisation en plaçant l'échantillon entre les deux cristaux. Le premier cristal génère le térahertz et le second le détecte. Nous avons mesuré l'impulstiontérahertz pendant qu'elle se propage vers le cristal de détection à différents moments. La figure 10(a) montre la mesure de référence. La figure 10(b) et (c) montrent la caractérisation des deux échantillons (espacement d = 1 μm et 100μm) où l'effet des fentes est évident. Dans la figure 10(b), le début de l'onde du faisceau térahertz est plat. Au contraire, dans la figure10(c), il y a interférence entre les ondes de térahertz. La figure 10(d) montre les simulations numériques du cas (c). En plus, nous avons fait la mesure de champ lointain pour obtenir les valeurs quantitatives du champ térahertz. Une estimation de 26 MV/cm a été trouvée.

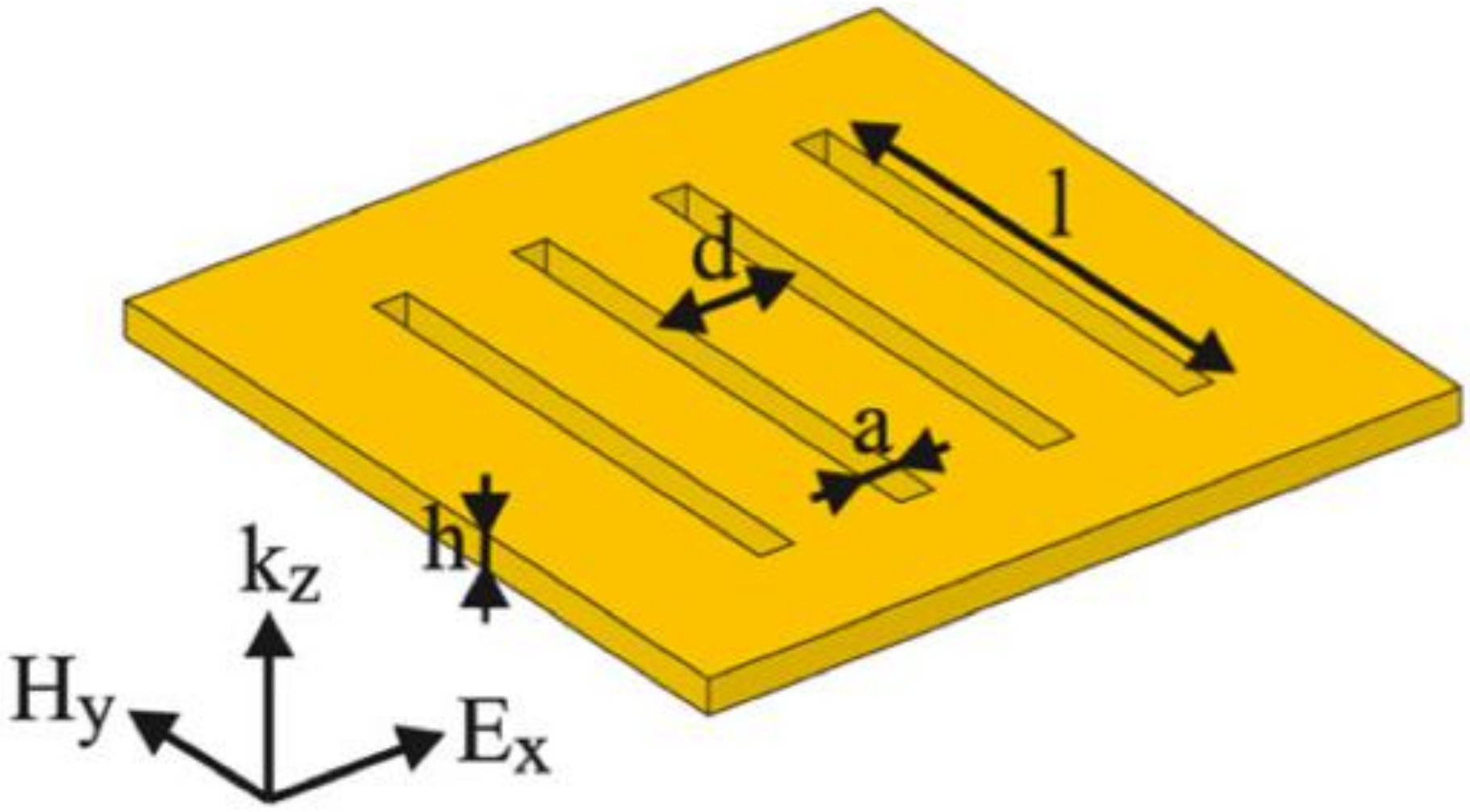

Figure 9 Diagramme schématique de la structure utilisée pour concentrer laradioationtérahertz.

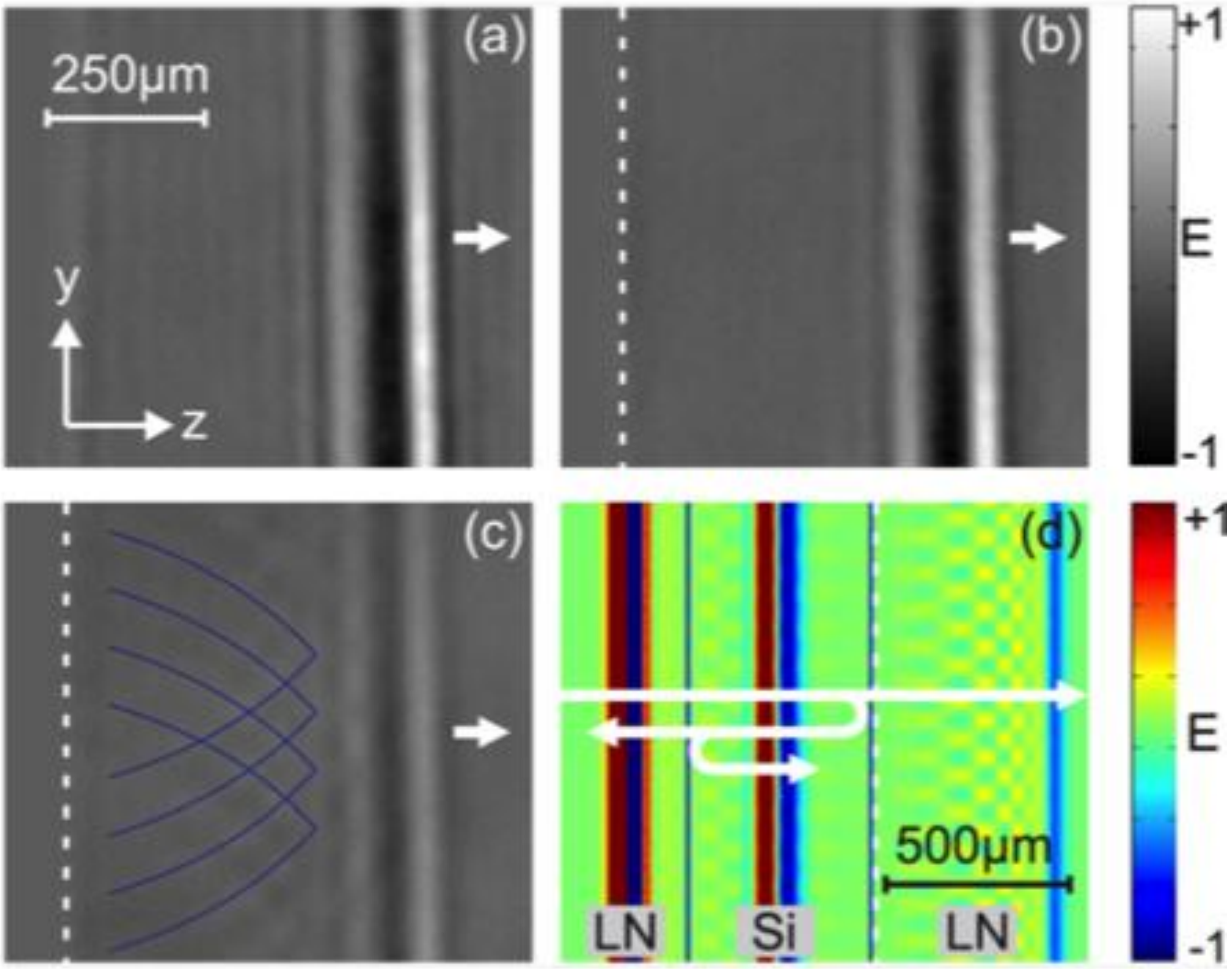

Figure 10 Caractérisation de l' échantillon. (a) référence. (b) matrice d'espacement 100 μm. (c) matrice d'espacement 1 μm. (d) simulations de (c).

## Appareils magnétiques térahertz

Dans cette dernière partie, nous présentons nos résultats sur le développement d'appareils magnétiques. Nous avons d'abord étudié le liquide magnétique et construit un rotateur de Faraday et un modulateur magnétique. Ensuite, nous avons utilisé leferrite (un aimant permanant) pour construire un isolateur pour la radioationtérahertz.

Les ferrofluides sont des particules magnétiques à l'échelle nanométrique. Normalement, elles sont orientées au hasard[20]. Quand un champ magnétique externe est appliqué, elles s'alignent vers la direction du champ, et elles forment des amas. Ceci est illustré dans la figure 11.

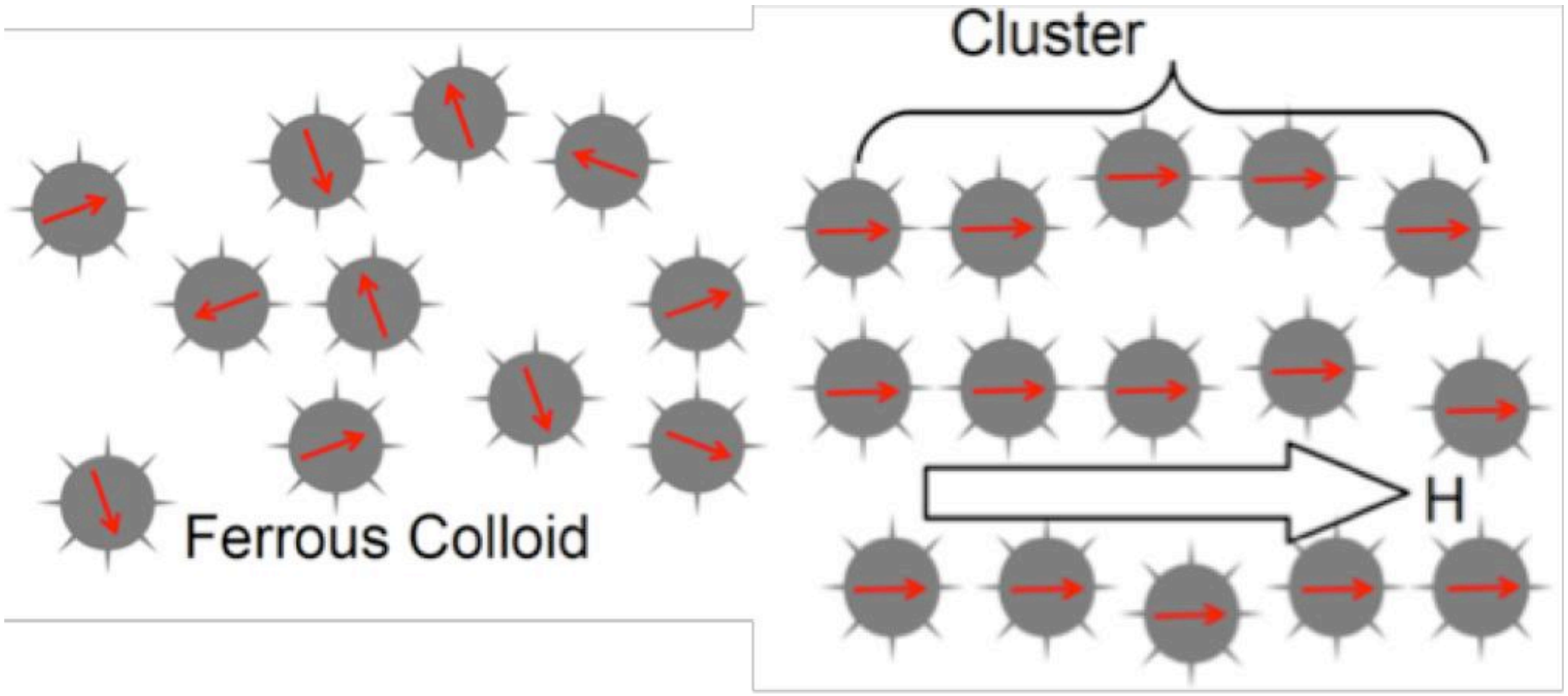

**Figure 11** **L'alignement des particules magnétiques selon la direction du champ magnétique externe.**

Un champ magnétique externe est alors responsable de leurs propriétés magnétiques, comme indiqué dans la figure 12. Il est important de noter que la moitié du niveau de saturation est obtenue en utilisant un petit champ magnétique de 30 mT.

Nous avons tout d'abord pratiqué la spectroscopie linéaire d'un ferrofluide commercial (EFH1) à la fréquence térahertz. L'indice de réfraction et le coefficient d'absorption sont illustrés à la figure 13. L'indice de réfraction du matériel est presque constant et le coefficient d'absorption est petit.

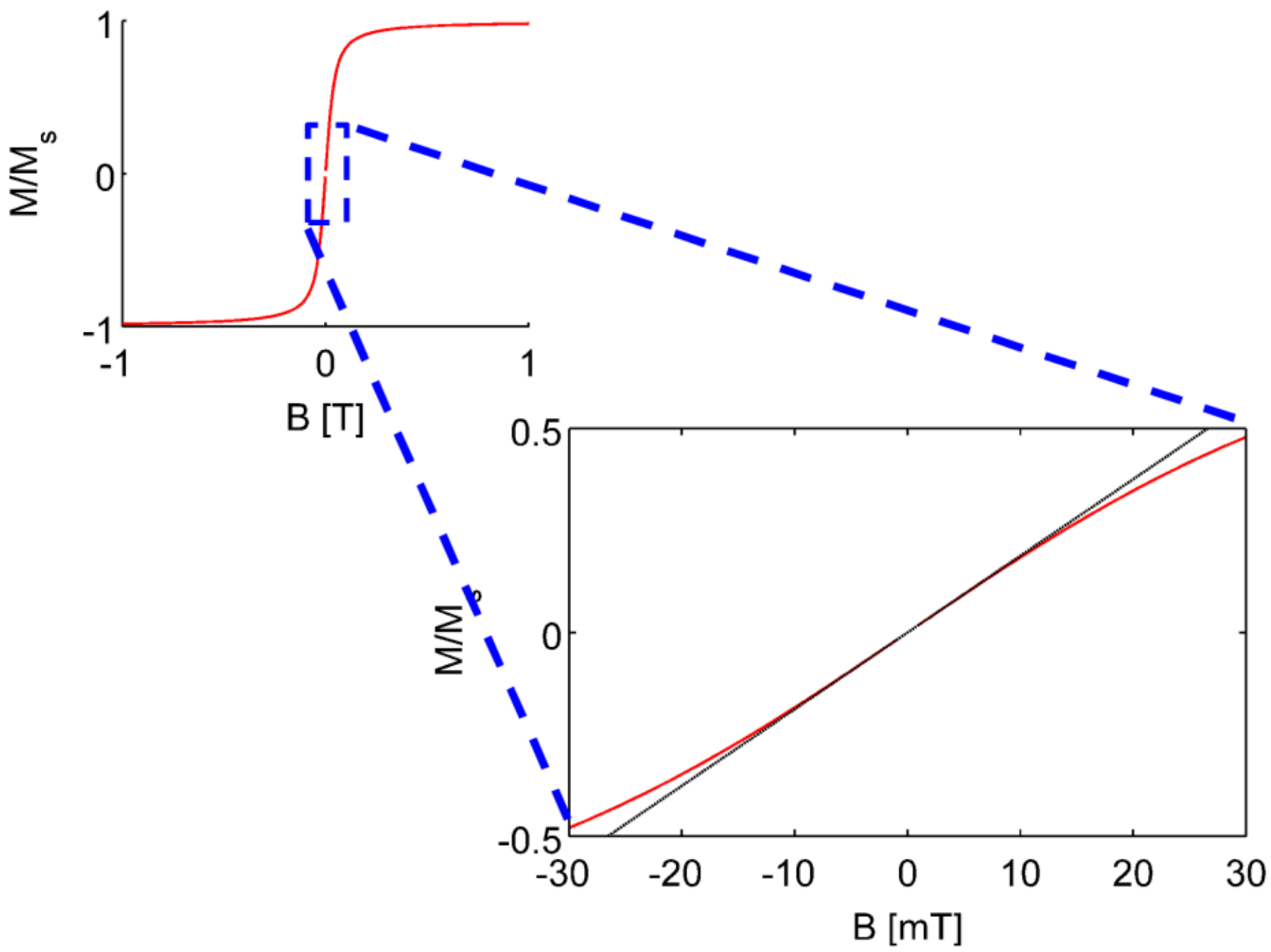

**Figure 12 Comportement des particules magnétiques sous l'application d'un champ magnétique externe.**

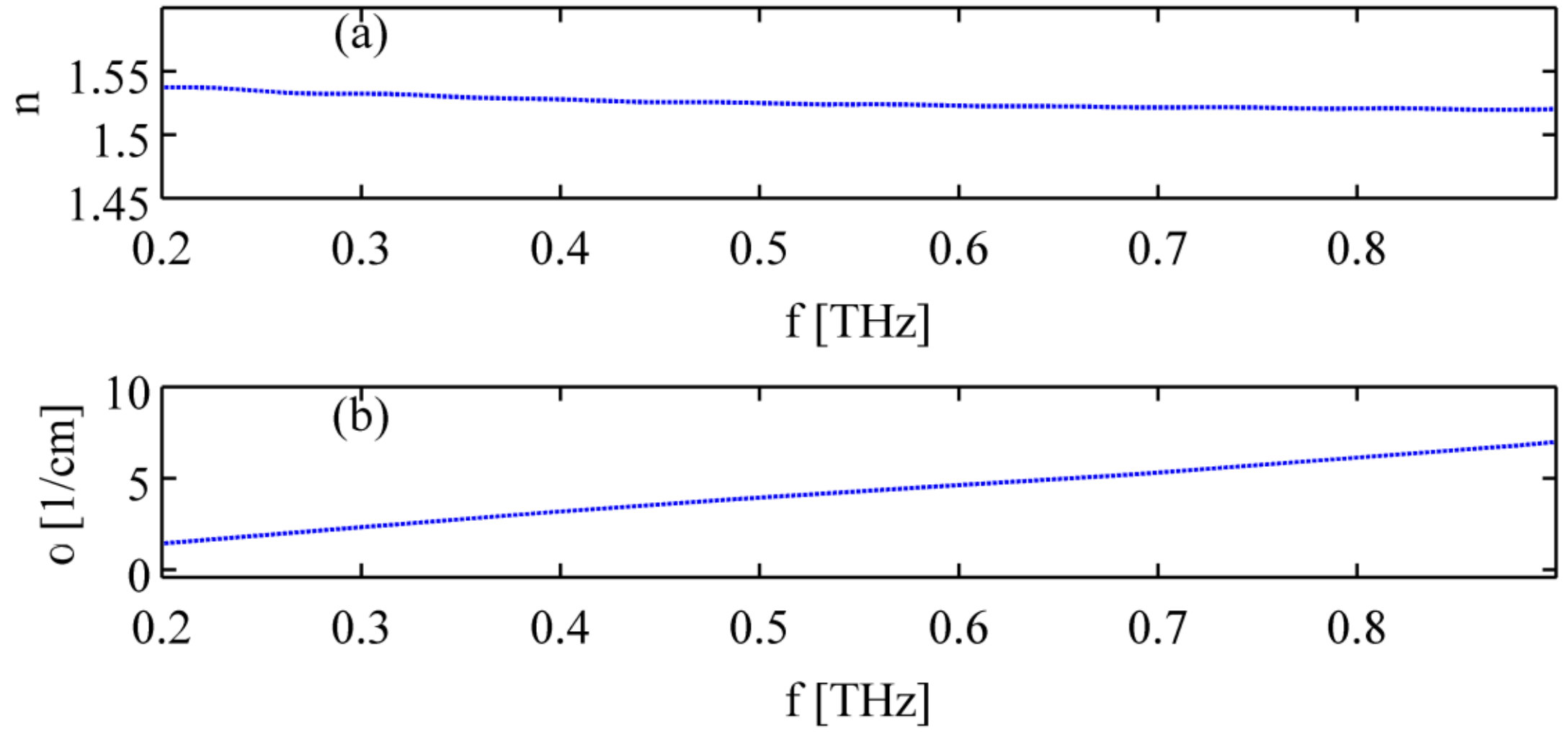

**Figure 11 Réponse diélectrique d'un ferrofluide commercial (EFH1) à température ambiante. (a) Indice de réfraction. (b) Coefficient d'absorption.**

Afin de mesurer la rotation Faraday, on a utilisé un montage ellipsométrique, tel qu'illustré à la figure 14. On a utilisé trois polarisateurs (WGP1 = 0°, WGP3= 0°). On a prits les deux mesures E+45o et E-45o qui correspondent à WGP1 = 45° et - 45°.

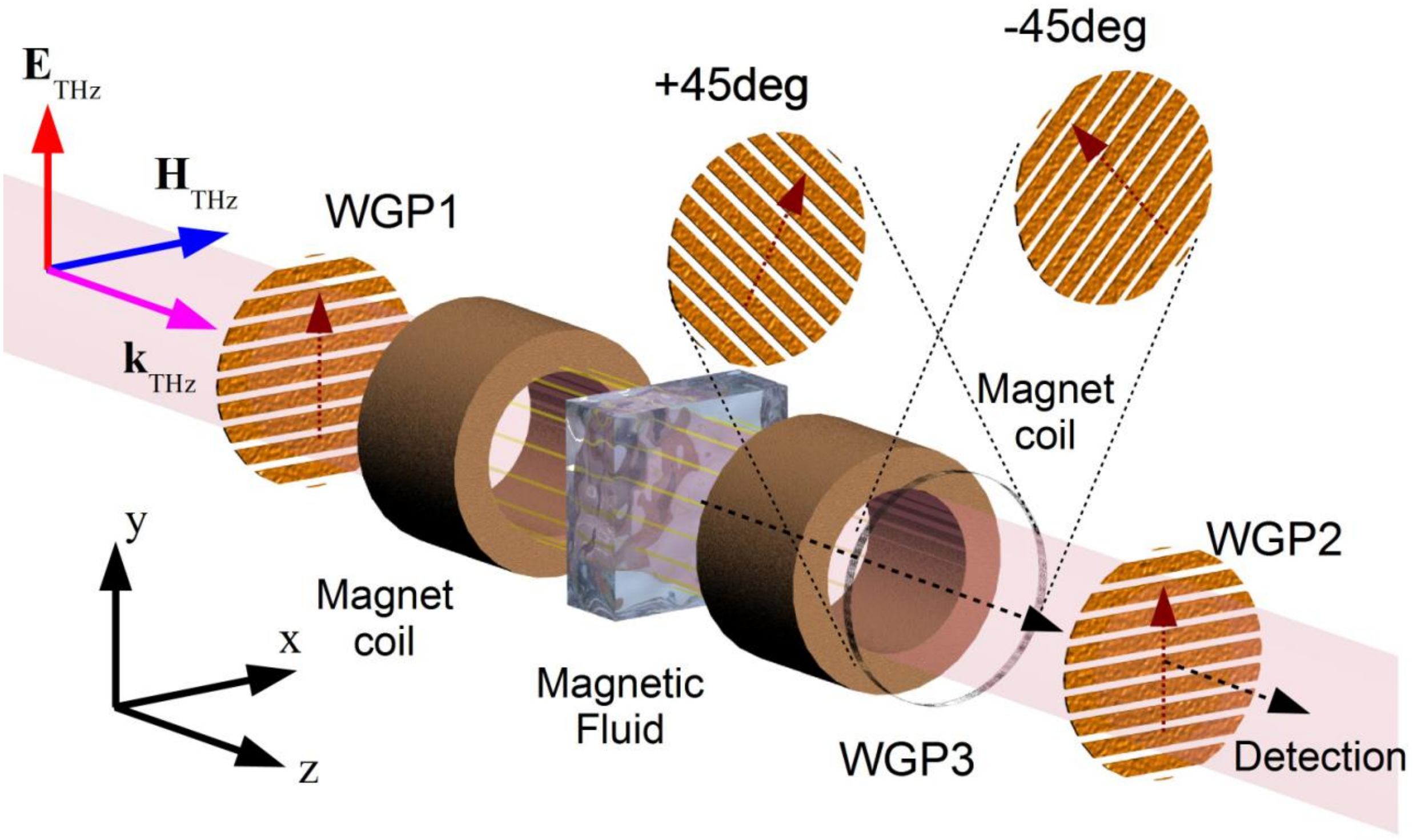

**Figure 14** **Ellipsométrietérahertz.**

Ensuite, en utilisant les équations suivantes, on a obtenu la rotation [21]

$$\begin{pmatrix} E_l(f) \\ E_r(f) \end{pmatrix} = \frac{1}{2}\begin{pmatrix} -1+i & 1+i \\ 1+i & -1+i \end{pmatrix}\begin{pmatrix} E_{+45^o}(f) \\ E_{-45^o}(f) \end{pmatrix} \quad (14)$$

$$\emptyset(f) = \frac{\arg[E_r(f)] - \arg[E_l(f)]}{2} \quad (15)$$

Les résultats sont illustrés à la figure 15, où la rotation était de 11 mrad/mm à 30 mT. Il s'agit d'une rotation très élevée pour un petit champ magnétique. De plus, elle ne dépend pas de la fréquence.

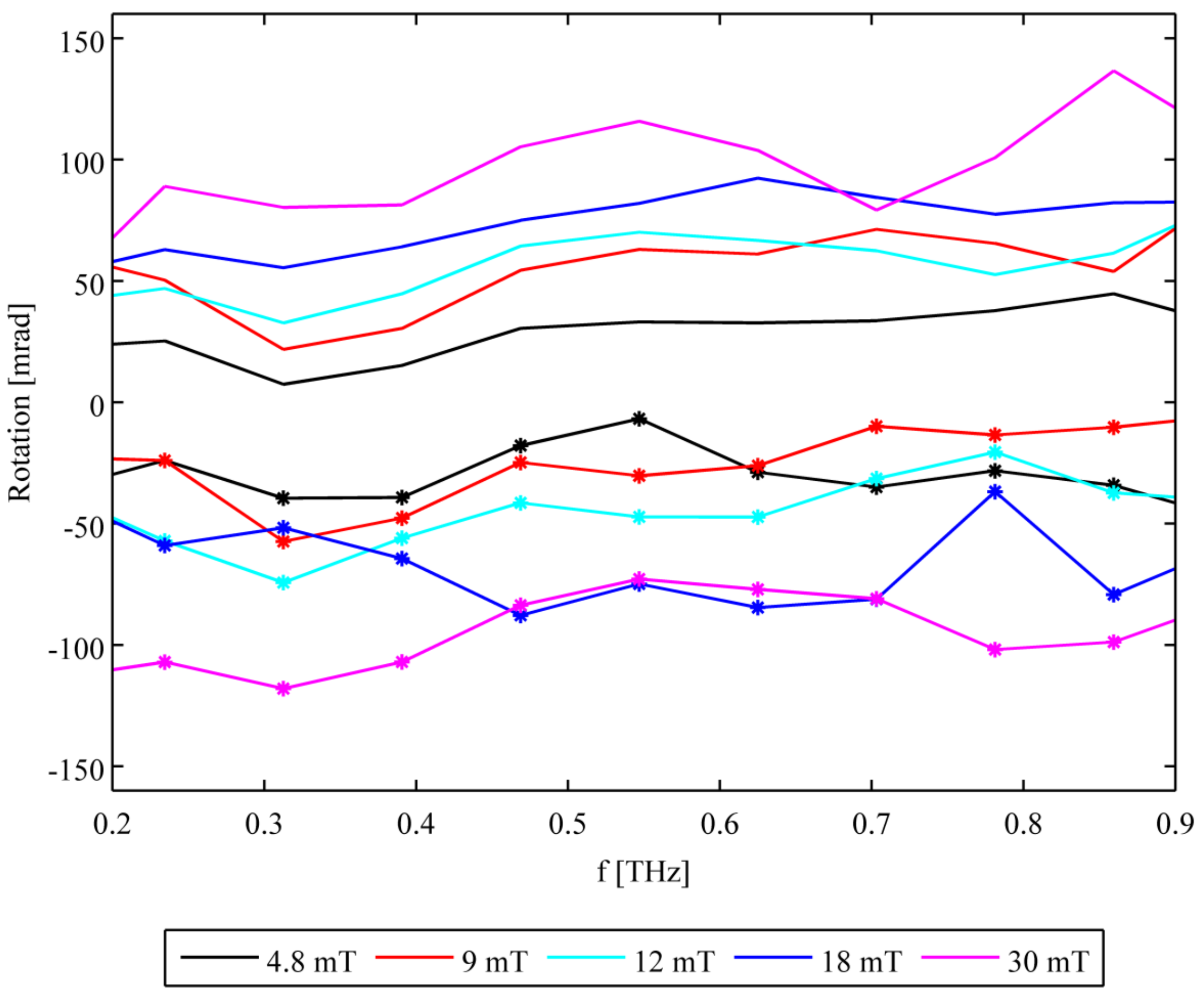

**Figure 15** **Mesure de la rotation Faraday.**

Nous passons maintenant à la deuxième expérience, la modulation magnétique. Dans un ferrofluide, si la direction de la propagation de la radioationtérahertz est perpendiculaire au canal du ferrofluide, on obtient une forte absorption des ondes térahertz. Le niveau d'absorption dépend de l'angle du canal par rapport au champ magnétique. S'ils sont parallèles, la transmission décroit. S'ils sont perpendiculaires, elle augmente. Ce comportement est illustré à la figure I6. En l'absence d'un champ magnétique externe, le coefficient d'absorption est calculé par[22,23]

$$\propto (0) = (36\pi\phi_M/c)\sqrt{\varepsilon_1}\sigma/(\varepsilon+2)^2 \qquad (16)$$

Où $\phi_M$, $\varepsilon_1$, et $\sigma$ sont respectivement la concentration, le constant diélectrique relatif et la conductivité des nanoparticules magnétiques. ε est la constante diélectrique relative du liquide porteur.

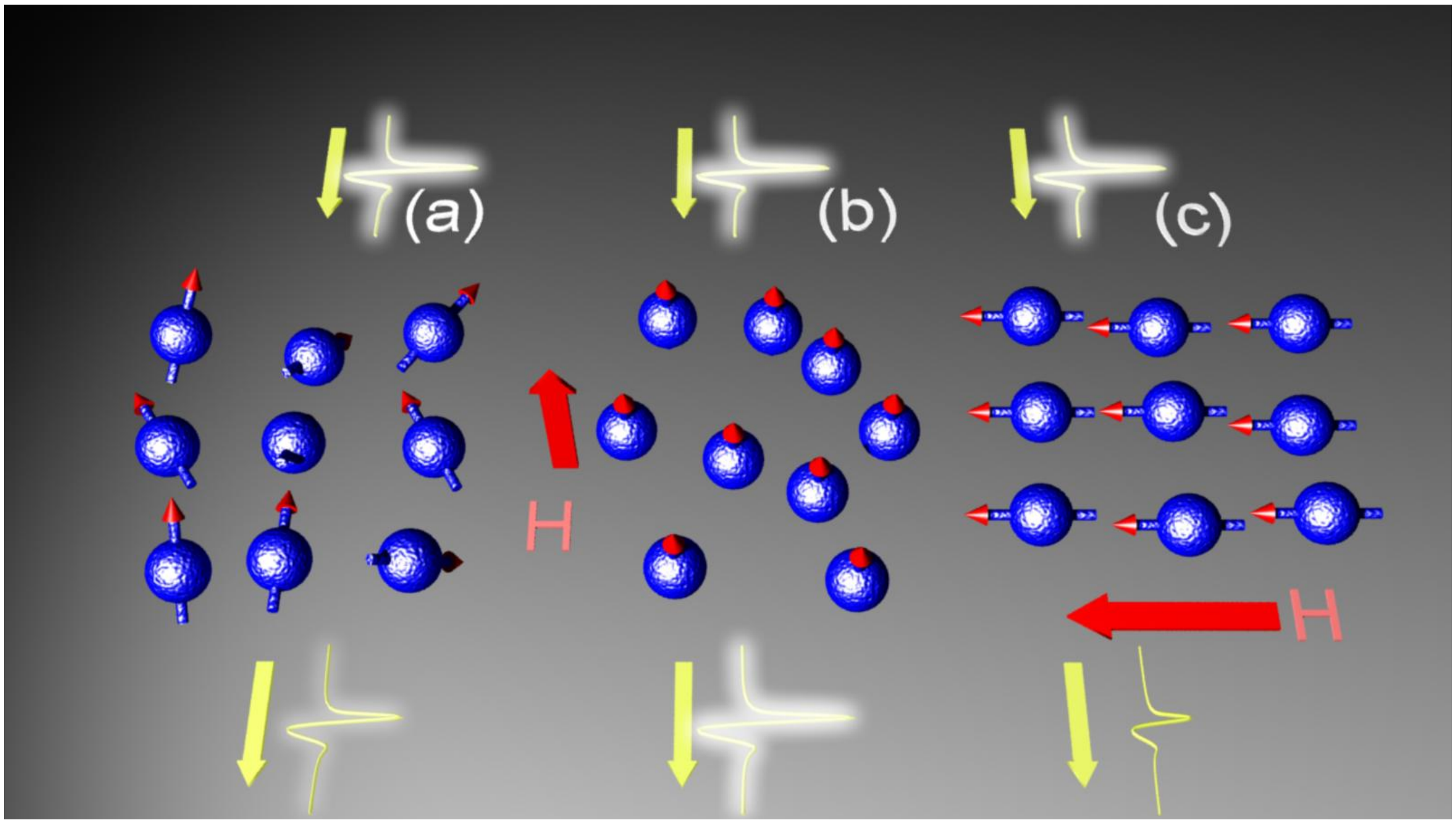

**Figure 16** **Transmission du térahertz par un ferrofluide.**

Lorsqu'on applique un champ magnétique externe, les coefficients d'absorption sont

$$\propto_{//}= 4\pi\sigma\sqrt{\varepsilon_1}\phi_M/c[1+\langle N\rangle(\varepsilon\text{-}1)]^2 \quad (17)$$

$$\propto_{\perp}= 16\pi\sigma\sqrt{\varepsilon_1}\phi_M/c[(\varepsilon+1)\text{-}\langle N\rangle(\varepsilon\text{-}1)]^2 \quad (18)$$

Où $\langle N\rangle=1/3\text{-}\Delta N$ et $\Delta N$ sont le facteur de déviation. Le changement de $\Delta\alpha_{//} = \alpha_{//}\text{-}\alpha(0)$ et $\Delta\alpha_{\perp} = \alpha_{\perp}\text{-}\alpha(0)$ peut ensuite être donné par

$$\Delta \propto_{//}= 6[(\varepsilon\text{-}1)/(\varepsilon+1)] \propto (0)\Delta N \quad (19)$$

$$\Delta \propto_{\perp}= \text{-}3[(\varepsilon\text{-}1)\text{-}(\varepsilon+2)] \propto (0)\Delta N \quad (20)$$

et les deux coefficients d'absorption peuvent être donnés par

$$\Delta \propto_{//}= \text{-}2\Delta \propto_{\perp} \quad (21)$$

La caractérisation du modulateur est montrée dans la figure 17.

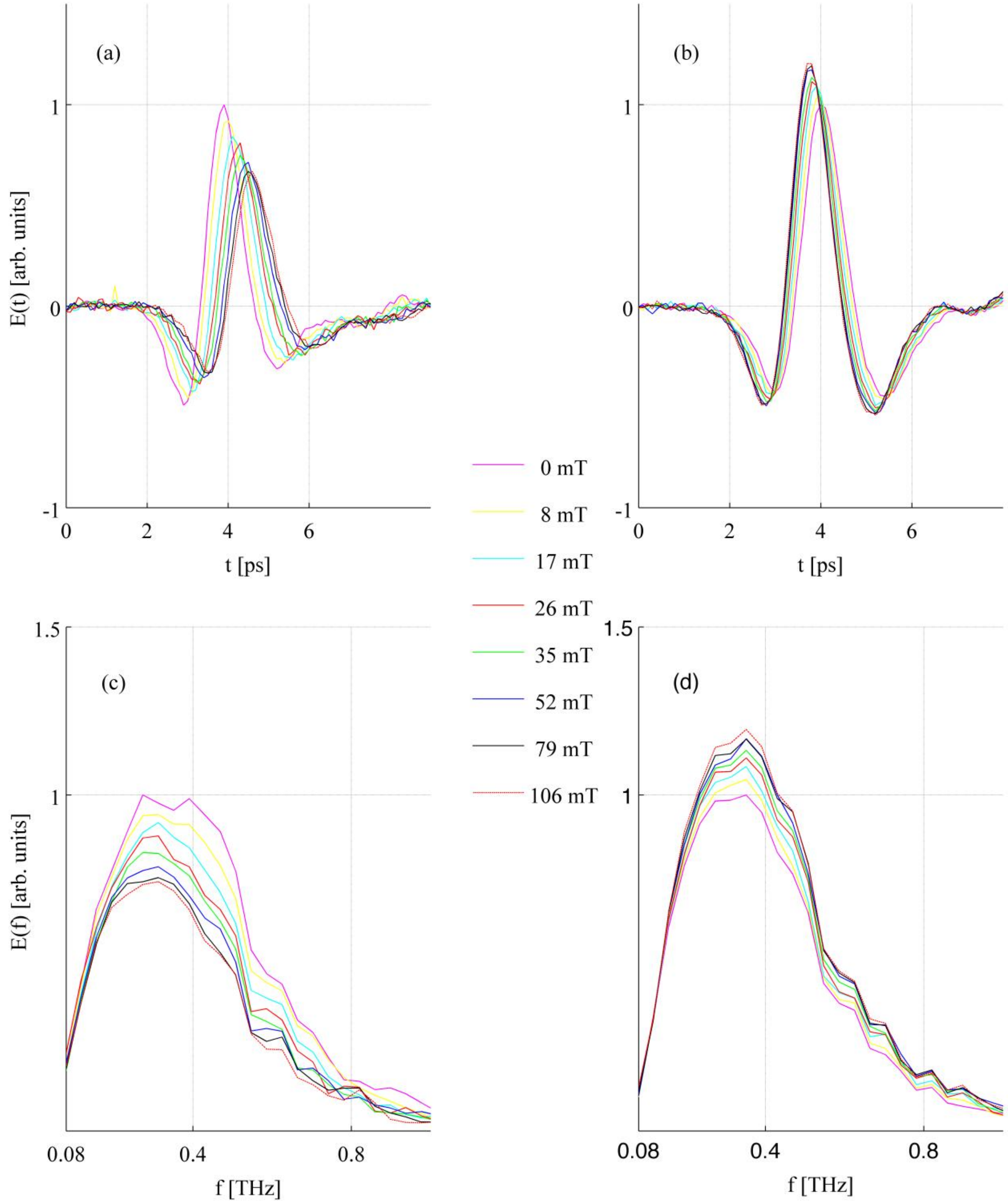

**Figure 17** **La caractérisation du modulateur**

Nous présentons finalement l'isolateur térahertz. L'isolateur est un dispositif qui permet aux ondes électromagnétiques de voyager dans seulement une direction [24]. Le principe est basé sur l'effet Faraday. Dans cette expérience, nous avons utilisé un ferrite (aimant permanent). La mesure de rotation Faraday est montrée dans la figure 18 où des rotations atteignant 105° ont été obtenues.

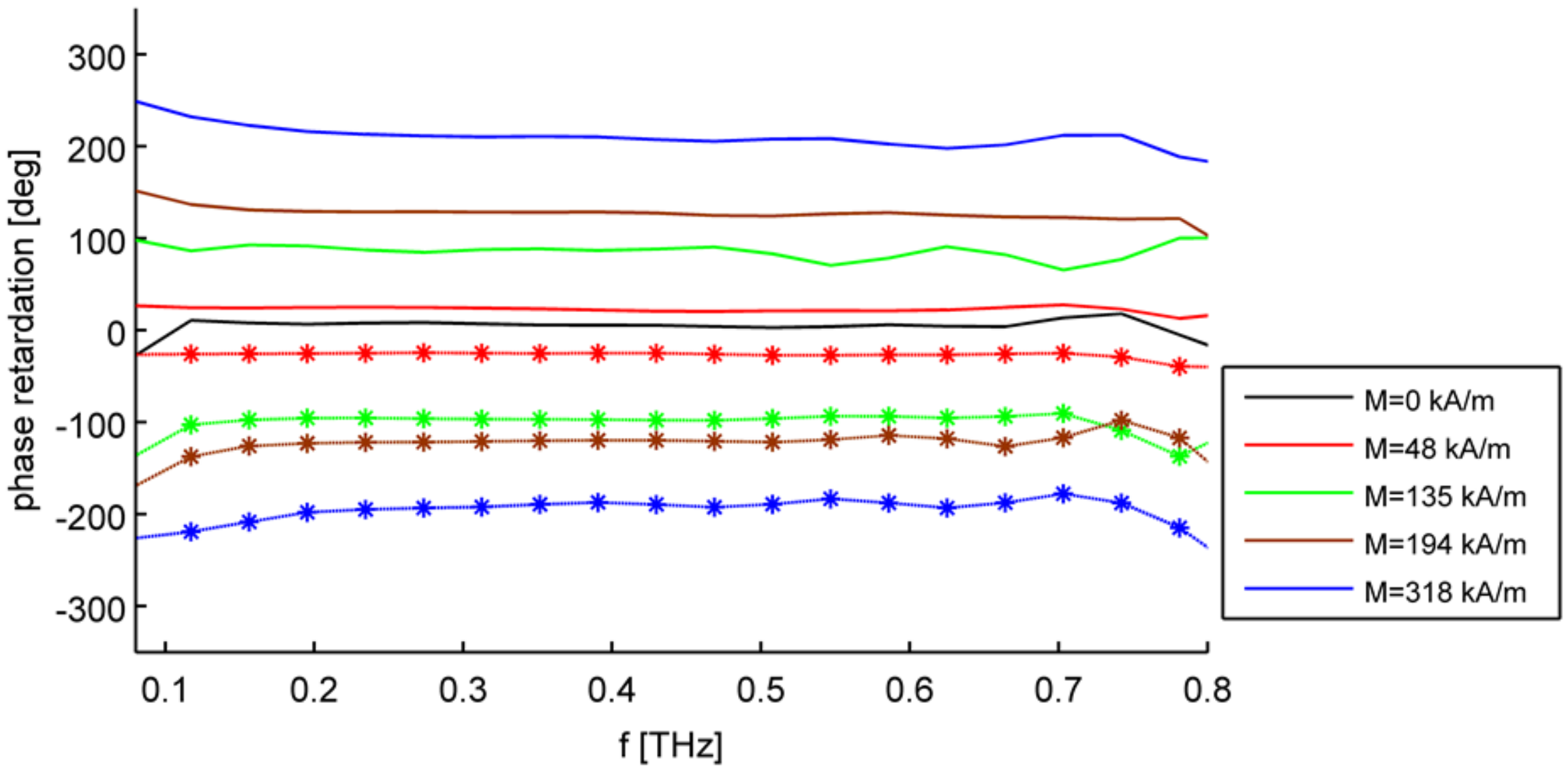

**Figure 18** **Retard (rotation de Faraday x2) dans un ferrite. Les astérisques montrent les mesures quand le champ magnétique est inversé.**

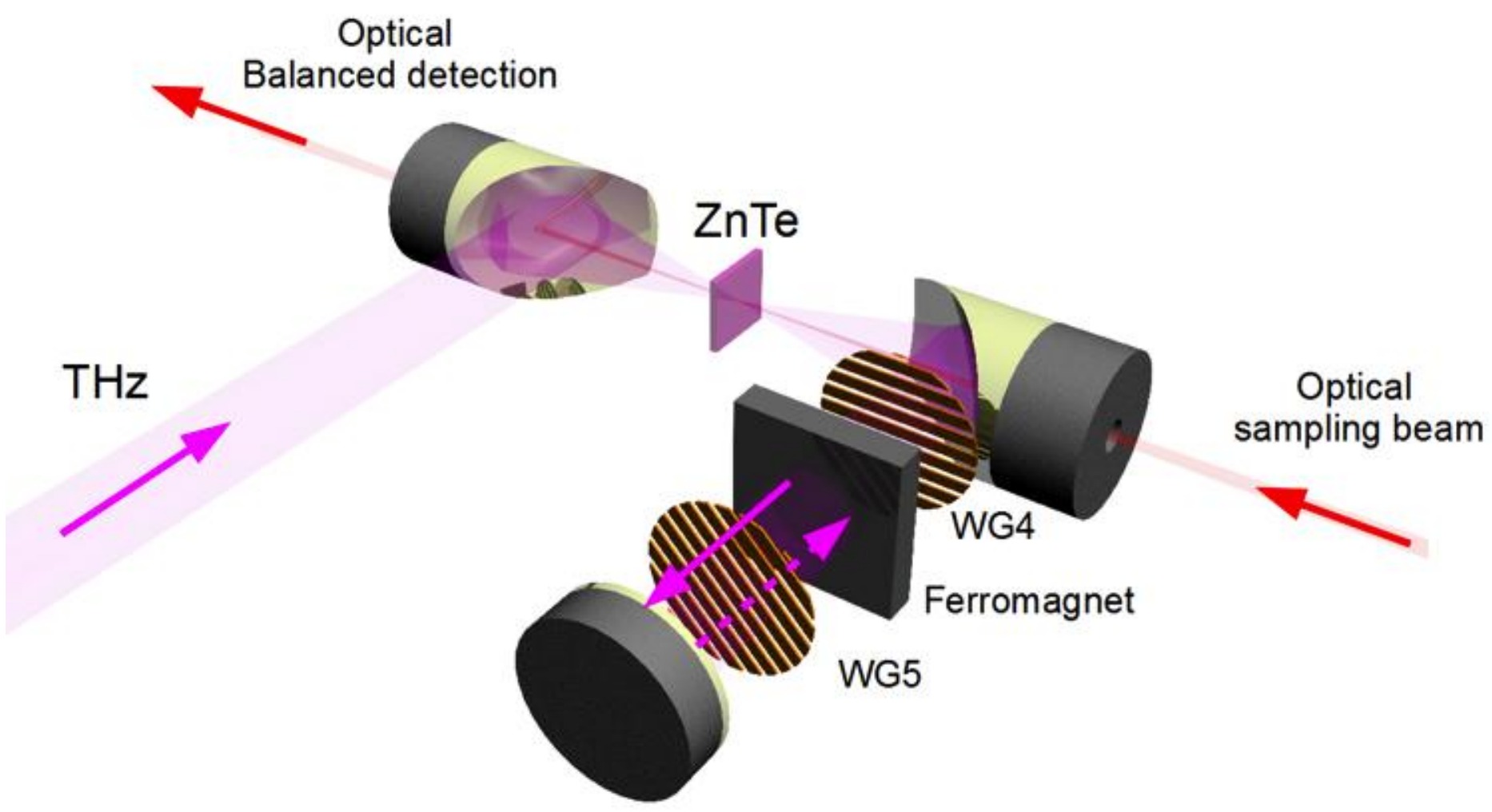

**Figure 19** **Organisation de l'isolateur.**

La figure 19 montre la construction qui a été utilisée pour mesurer les ondestérahertz rétro-réflechives. Les résultats sont montrés dans la figure 20 pour différents niveaux de champ magnétique.

Pour construire l'isolateur, nous avons aimanté l'échantillon de façon à obtenir une rotation de Faraday de 45°. Les résultats sont montrés dans la figure 21. Quand l'isolateur fonctionne, la radiation térahertz est complètement bloquée et ne revient pas vers la source.

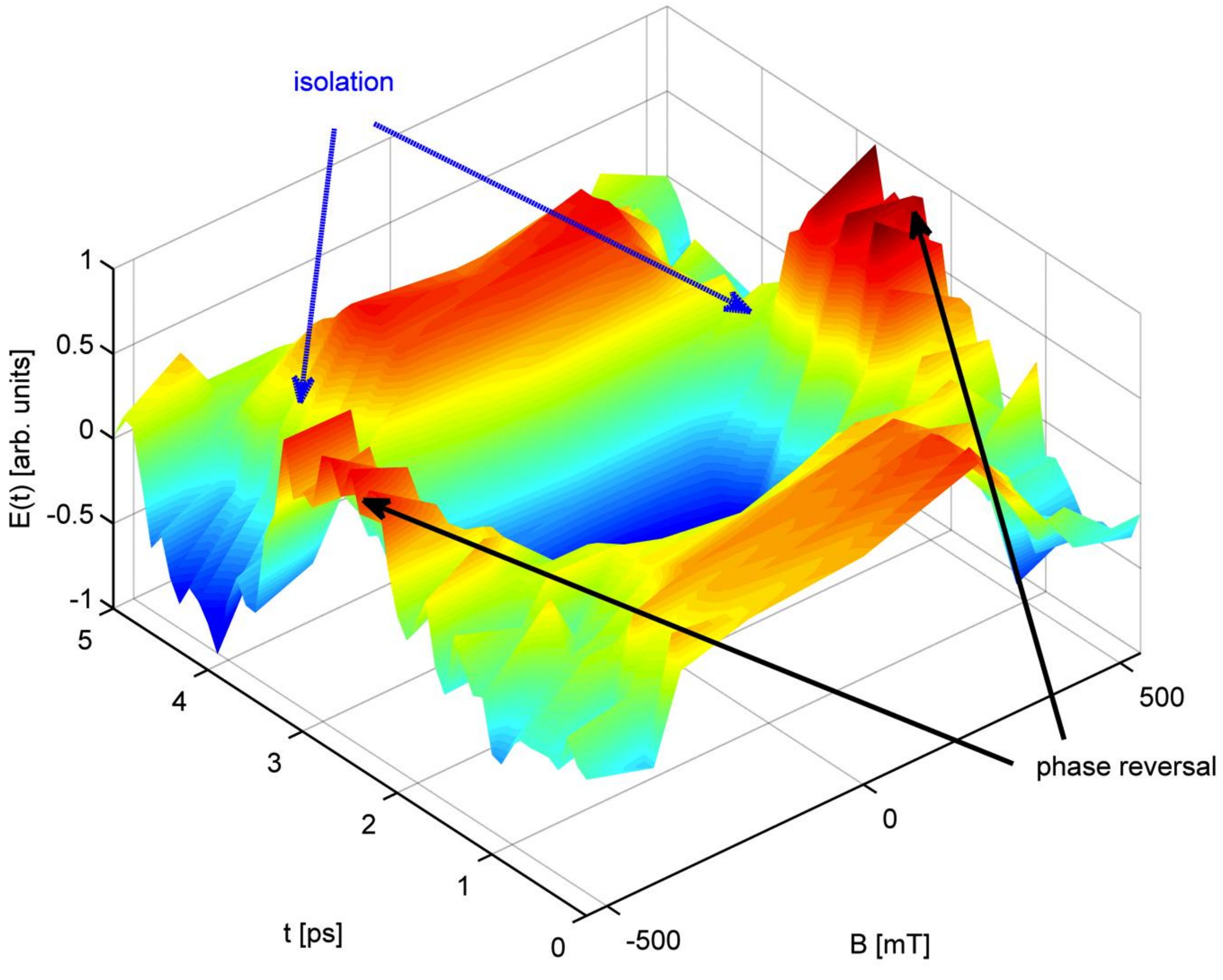

**Figure 20 La mesure des arrières signes reflétés pour de différents niveaux de champ magnétique.**

En conclusion, la thèse s'est concentrée sur les effets magnétiques sur la radiation térahertz. Nous avons exécuté des simulations numériques des actions réciproques non linéaires. Alors, nous avons exécuté des expériences pour surmonter les restrictions technologiques. Finalement, nous avons construit trois dispositifs magnétiques-térahertz : le rotateur, le modulateur, et l'isolateur.

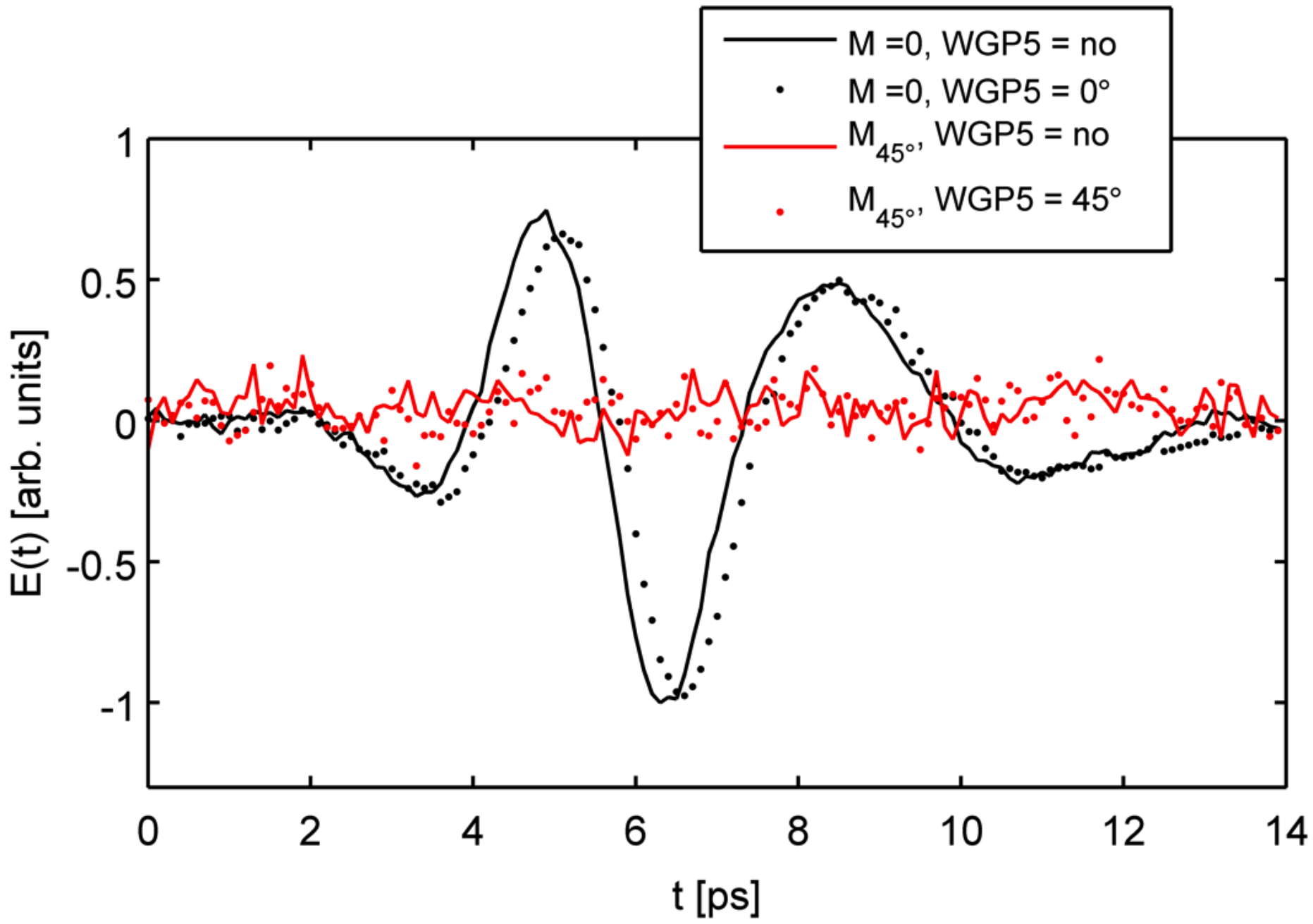

**Figure 21** **Caractérisation de l'isolateur.**

[1] B. D. Patterson, *et al.*, *UltrafastPhenomenaat the Nanoscale: Science opportunitiesat the SwissFEL X-ray Laser* (PSI, Switzerland, 2009)

[2] J. M. D. Coey, *Magnetism and Magnetic Materials* (Cambridge University Press, 2010)

[3] C. H. Back, *et al.*, *Science* **6**, 864 (1999)

[4] K. Tanaka, H. Hirori, and M. Nagai, *IEEE Trans. TerahertzSci. Technol.* **1**, 301 (2011)

[5] T. Kampfrath, *et al.*, *Nat. Photonics* **5**, 31 (2011)

[6] L. Razzari, *et al.*, *Phys. Rev. B* **79**, 193204 (2009)

[7] M. C. Hoffmann, *et al.*, *Appl. Phys. Lett.* **95**, 231105 (2009)

[8] M. Liu, *et al.*, *Nature* **487**, 345 (2012)

[9] T. Kleine-Ostmann and T. Nagatsuma, *J Infrared Milli Terahz Waves* **32**, 143 (2011)

[10] B. B. Hu and M. C. Nuss, *Opt. Lett.* **20**, 1716 (1995)

[11] W. L. J. Deibel, and D. M. Mittleman, *Rep. Prog. Phys.* **70**, 1325 (2007)

[12] X. C. Zhang, *Phys. Med. Biol.* **47**, 3667 (2002)

[13] M. R. Leahy-Hoppa, M. J., Fitch, X. Zheng, and L. M. Hayden, *Chem. Phys. Lett.* **434**, 227 (2007)

[14] M. Fox, *Optical Properties of Solids* (Oxford University Press, 2001)

[15] D. M. Sullivan, *Electromagnetic Simulation Using the FDTD Method* (IEEE Press, 2000)

[16] T. L. Gilbert, *IEEE Trans. Magn.* **40**, 3443 (2004)

[17] M. C. Nuss and J. Orenstein, *Terahertz time-domain spectroscopy in Millimter and Submillimeter Wave Spectroscopy of Solids* (Springer-Verlag, 1998)

[18] T. Ebbesen, H. Lezec, H. Ghaemi, T. Thio, and P. Wolff, *Nature* **391**, 667 (1998) [19] M. Seo, *et al.*, *Nat. Photonics* **3**, 152 (2009)

[20] T. A. Franklin, *Ferrofluid flow phenomena*, Master's thesis (Massachusetts Institute of Technology, 2003)

[21] E. D. Palik and J. K. Furdyna, *Rep. Prog. Phys.* **33**, 1193 (1970)

[22] N. Inaba, *et al., IEEE Trans. Magn.* **25**, 3866 (1989)

[23] S. Taketomi, *et al., IEEE Translation J. on Magn. in Japan* **4**, 384 (1989)

[24] L. Rayleigh, *Nature* **64**, 577 (1901).